\documentclass[11pt]{article}

\RequirePackage[OT1]{fontenc}
\RequirePackage{amsthm,amsmath}
\RequirePackage[numbers]{natbib}
\RequirePackage[colorlinks,citecolor=blue,urlcolor=blue]{hyperref}

\usepackage{hyperref}
\usepackage{url}
\usepackage{xcolor,ulem}
\usepackage{tikz}
\usepackage{amsfonts,amsmath,amssymb,tabularx,ifthen,twoopt,enumerate,color}
\usepackage{graphicx}
\usepackage[latin1]{inputenc}
\usepackage[figuresright]{rotating}
\usepackage{dsfont}
\usepackage{scalefnt}
\usepackage{placeins}
\usepackage{natbib}
\usepackage{multicol}
\usepackage{multirow}
\usepackage[ruled,vlined]{algorithm2e}
\definecolor{mblue}{RGB}{94,135,173}
\colorlet{mred}{red!80!black}

\allowdisplaybreaks
	
\newcommand{\bzr}{\mathbf{0}}
\newcommand{\set}[1]{\left\{#1\right\}}
\newcommand{\supp}{\mathcal{A}}
\newcommand{\Y}{\mathbf{Y}}
\newcommand{\U}{\mathbf{U}}
\newcommand{\E}{\mathbf{E}}
\newcommand{\X}{\mathbf{X}}
\newcommand{\TT}{\mathbf{T}}
\newcommand{\B}{\mathbf{B}}
\newcommand{\A}{\mathbf{A}}
\newcommand{\C}{\mathbf{C}}
\newcommand{\Ks}{K^\star}
\newcommand{\ts}{t^\star}
\newcommand{\that}{\widehat{t}}
\newcommand{\bth}{\mathbf{\widehat{t}}}

\newcommand{\mus}{\mu^\star}
\newcommand{\bts}{\mathbf{t}^\star}
\newcommand{\calY}{\mathcal{Y}}
\newcommand{\calX}{\mathcal{X}}
\newcommand{\calB}{\mathcal{B}}
\newcommand{\calE}{\mathcal{E}}
\newcommand{\calU}{\mathcal{U}}
\newcommand{\Ve}{\textrm{Vec}}
\newcommand{\Argmin}{\mathop{\mathrm{Argmin}}}
\newcommand{\rset}{\mathbb{R}}
\newcommand{\PP}{\mathbb{P}}
\newcommand{\Imindeux}{I_{\textrm{min},2}^\star}
\newcommand{\Jmin}{J_{\textrm{min}}^\star}
\newcommand{\Imin}{I_{\textrm{min}}^\star}
\newcommand{\PE}{\mathbb{E}}

\newcommand{\n}{n}
\newcommand{\ba}{\mathbf{a}}
\newcommand{\cor}{\mathbf{c}}

\newcommand{\ii}{i}

	\renewcommand{\t}[1]{{#1}^\top}

	\newcounter{hypA}
	\newenvironment{hypA}{\refstepcounter{hypA}\begin{itemize}
	  \item[({\bf A\arabic{hypA}})]}{\end{itemize}}

\newtheorem{theo}{Theorem}
\newtheorem{proposition}[theo]{Proposition}
\newtheorem{lemma}[theo]{Lemma}

\theoremstyle{remark}
\newtheorem*{remark}{Remark}

\begin{document}

\title{Fast Detection of Block Boundaries in Block Wise Constant
  Matrices: An Application to HiC data}

\author{Vincent Brault\footnote{The authors would like to thank the French National Research Agency ANR, which partly supported this research
through the ABS4NGS project (ANR-11-BINF-0001-06).}, Julien Chiquet and C\'eline L\'evy-Leduc}


\maketitle

\begin{center}UMR MIA-Paris, AgroParisTech, INRA, Universit\'{e} Paris-Saclay\end{center}



\begin{abstract}
We propose a novel approach for estimating the location of block boundaries (change-points) in a random matrix consisting of a block wise constant matrix observed in white noise. Our method consists in rephrasing this task as a variable selection issue. We use a penalized least-squares criterion with an $\ell_1$-type penalty for dealing with this 
issue. We first provide some theoretical results ensuring the consistency of our change-point estimators. Then, we explain how to implement our method in a very efficient way. Finally, we provide some empirical evidence to support our claims and apply our approach to HiC data which are used in molecular biology for better understanding 
the influence of the chromosomal conformation on the cells functioning.
\end{abstract}


\section{Introduction}

Detecting automatically the block boundaries in large block wise constant matrices
corrupted with noise is a very important issue which may have several applications.
One of the main situations in which this problem occurs is in the study of HiC data.
It corresponds to one of the most recent chromosome conformation capture technologies
that have been developed to better understand the influence of the chromosomal
conformation on the cells functioning.
This technology is based on a deep sequencing approach
and provides read pairs corresponding to pairs of genomic loci that
physically interacts in the nucleus, see \cite{lieberman2009comprehensive} for more details.
The raw measurements provided by HiC data are often summarized as a square matrix where each entry
at row $i$ and column $j$ stands for the total number of read pairs matching in position
$i$ and position $j$, respectively, see \cite{dixon2012topological} for further details. Positions refer here to a sequence
of non-overlapping windows of equal sizes covering the genome.

Blocks of different intensities arise among this matrix, revealing
interacting genomic regions among which some have already been confirmed to host co-regulated
genes. The purpose of the statistical analysis is then to provide a fully automated and efficient
strategy to determine a decomposition of the matrix in non-overlapping blocks, which gives, as a by-product,
a list of non-overlapping interacting chromosomic regions. In the following, our goal will thus be to design
an efficient and fully automated method to find the block boundaries, also called change-points, of non-overlapping blocks in
very large matrices which can be modeled as block wise constant matrices corrupted with white noise.


An abundant literature is dedicated to the change-point detection issue for one-dimensional data
both from a theoretical and practical point of view. 
From a practical point of
view, the standard approach for estimating the change-point locations is based on least-
square fitting, performed via a dynamic programming algorithm (DP).
Indeed, for a given number of change-points $K$, the
dynamic programming algorithm, proposed by \cite{Bellman:1961} and \cite{fisher:1958}, takes advantage of the intrinsic additive 
nature of the least-square objective to recursively compute
the optimal change-points locations with a complexity of $O(Kn^2)$ in time, see \cite{kay:1993}.
This complexity has recently been improved by \cite{rigaill2016} in some specific cases.

However, in general one-dimensional situations, the computational
burden of these methods is prohibitive to handle very large data sets.
In this situation, \cite{harchaoui:levyleduc:2012}  proposed   to  rephrase   the
change-point estimation issue as a variable  selection problem. 
This approach has also been extended  by \cite{vert2010fast} to find shared
change-points  between  several signals.  In the two-dimensional case, namely
when matrices have to be processed, 
no method has been proposed, to the best of our knowledge, for providing  the block boundaries of non overlapping blocks of very large $n\times n$ matrices.
Typically, we aim at being able to handle  $5000\times 5000$ matrices, which corresponds to matrices having $2.5\times 10^7$ entries.
The only statistical approach proposed
for retrieving such non-overlapping block boundaries in this two-dimensional framework is the one devised by \cite{levy2014}
but it is limited to the case where the block wise matrix is assumed to be block wise constant on the diagonal
and constant outside the diagonal blocks.



\begin{figure}[!h]
\centering
\includegraphics[width=.3\linewidth]{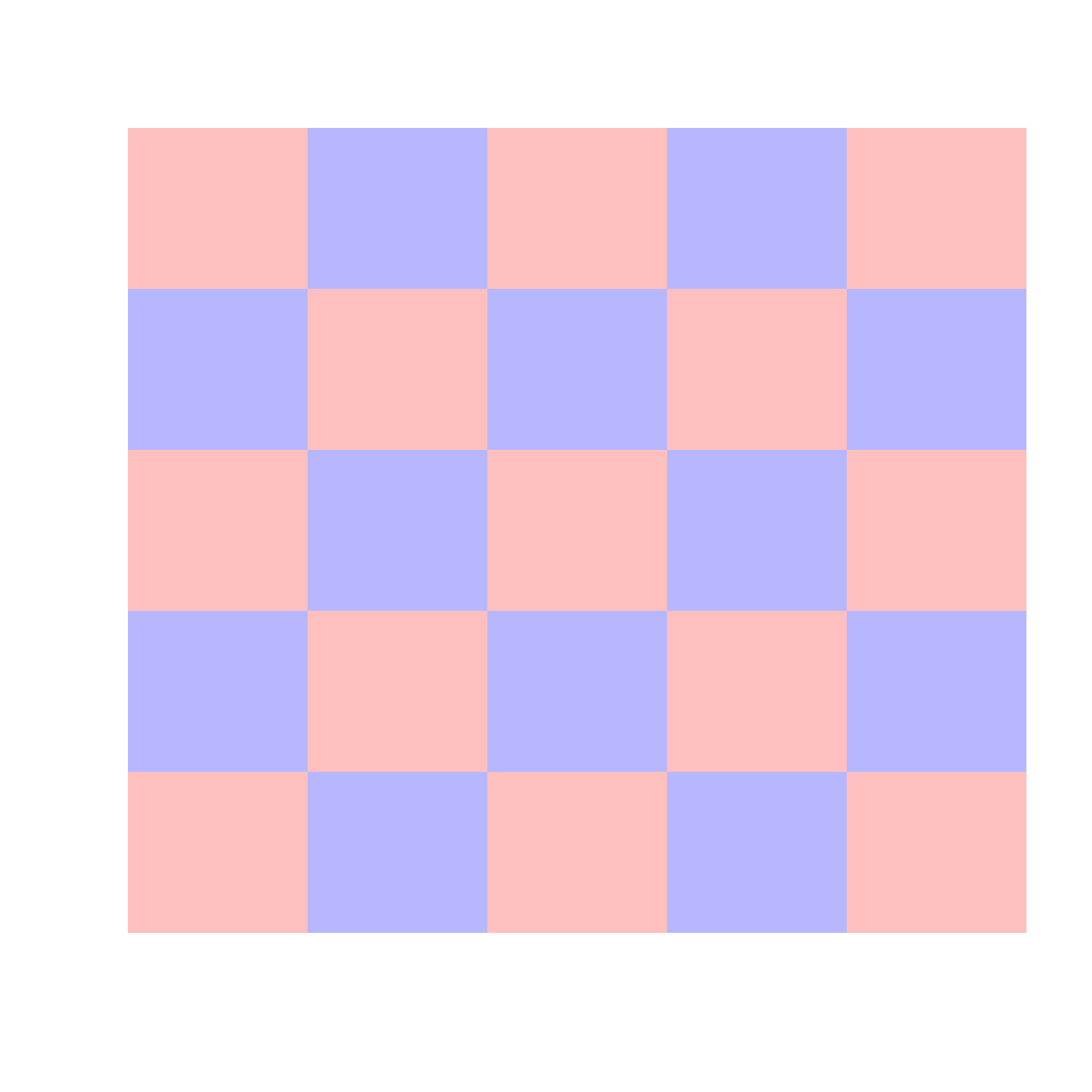}
\caption{\label{fig:data:intro} Block wise constant matrix without noise.}
\end{figure}

\begin{figure}[!h]
  \centering
  \begin{tabular}{cccc}
    \rotatebox{90}{\hspace{3.5em}\small $\sigma=1$}
    & \includegraphics[width=.3\linewidth]{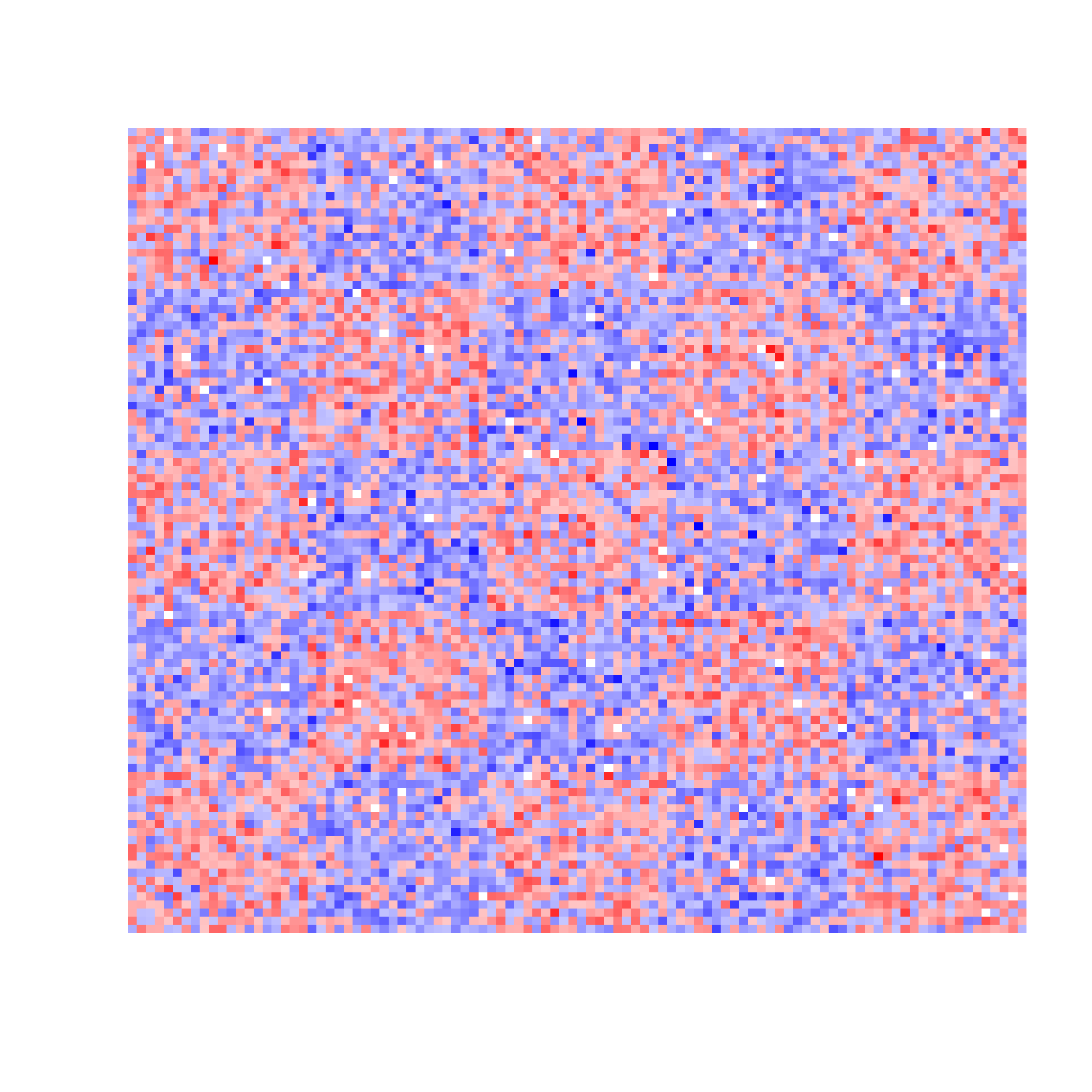}
    & \includegraphics[width=.3\linewidth]{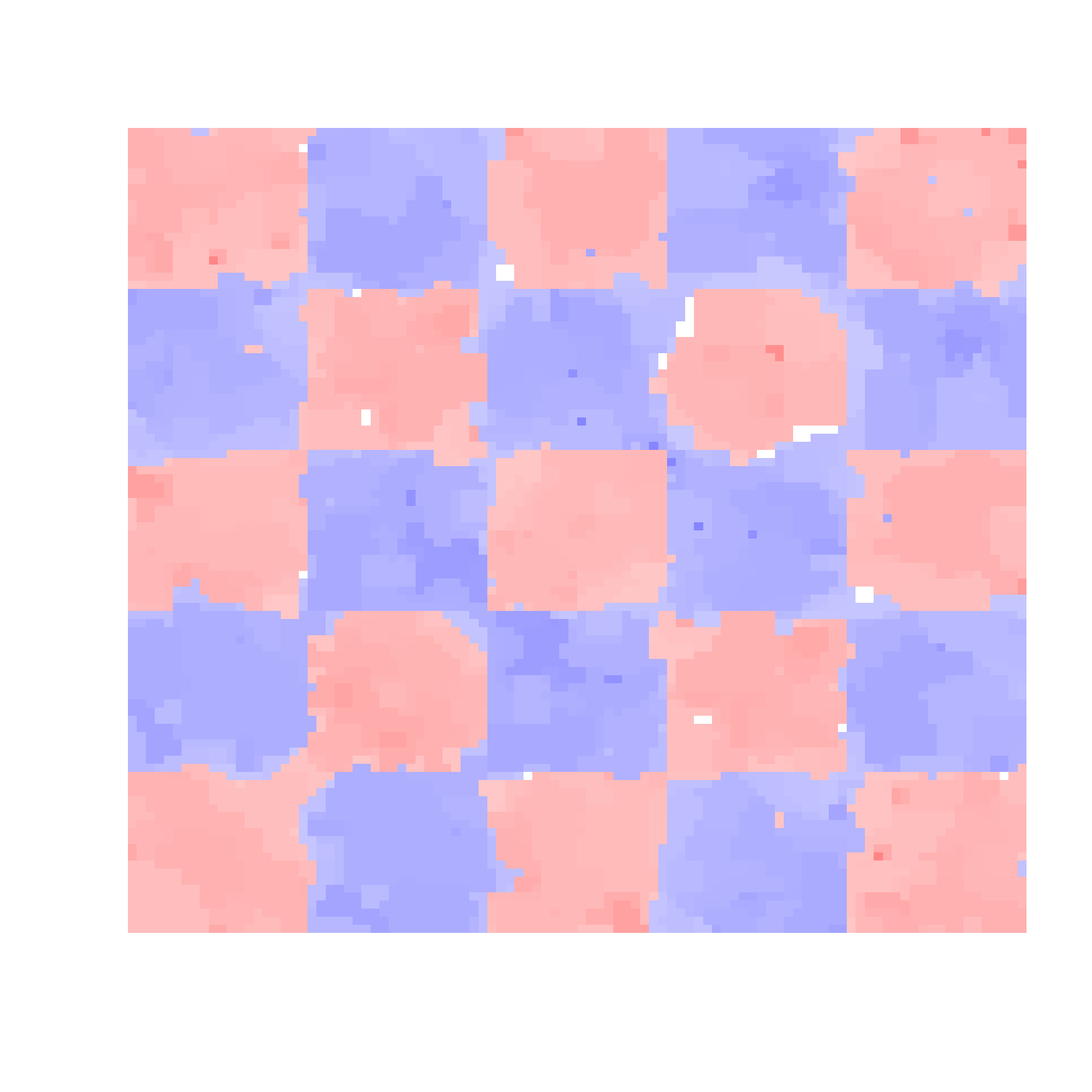}
    & \includegraphics[width=.3\linewidth]{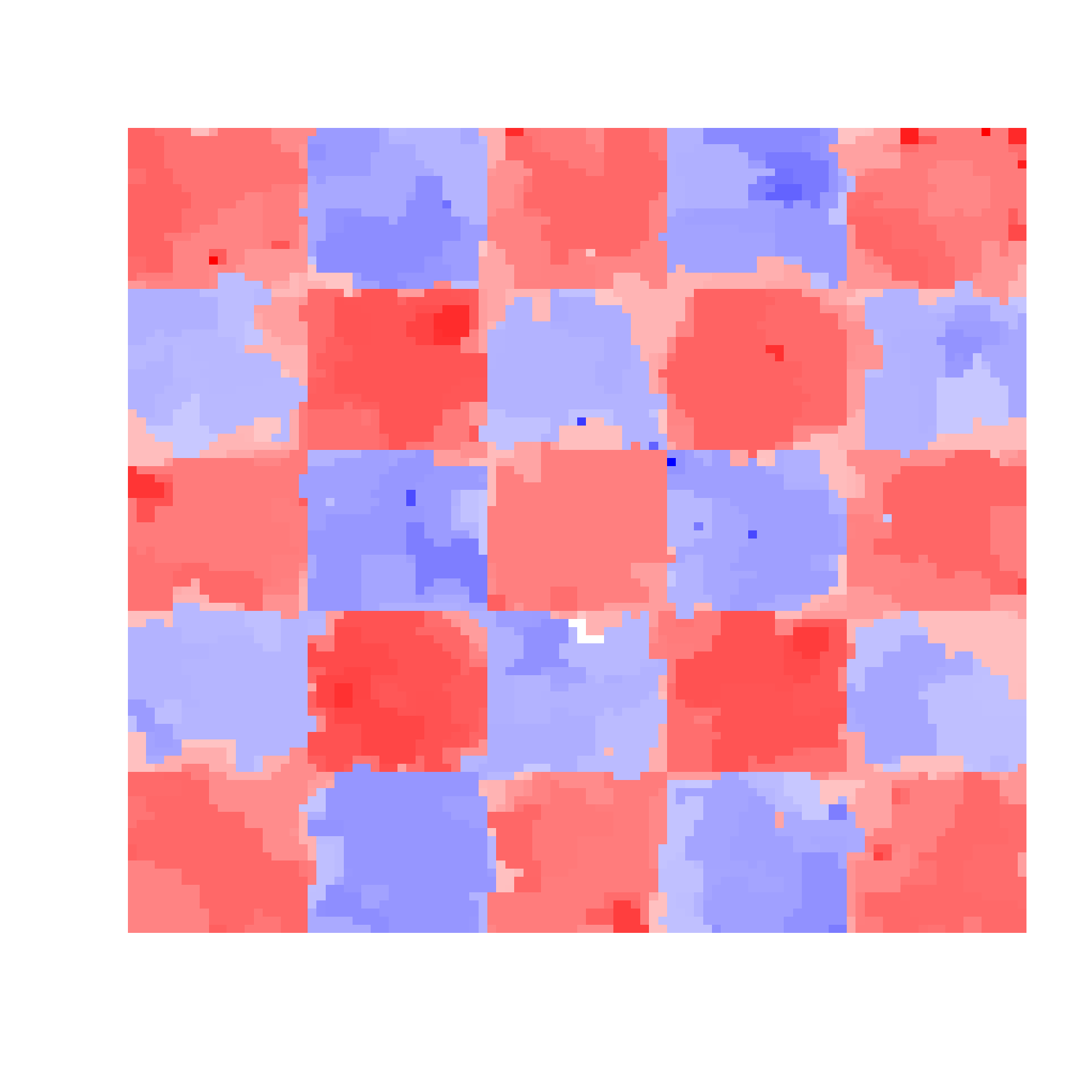} \\
    \rotatebox{90}{\hspace{3.5em}\small $\sigma=2$}
    & \includegraphics[width=.3\linewidth]{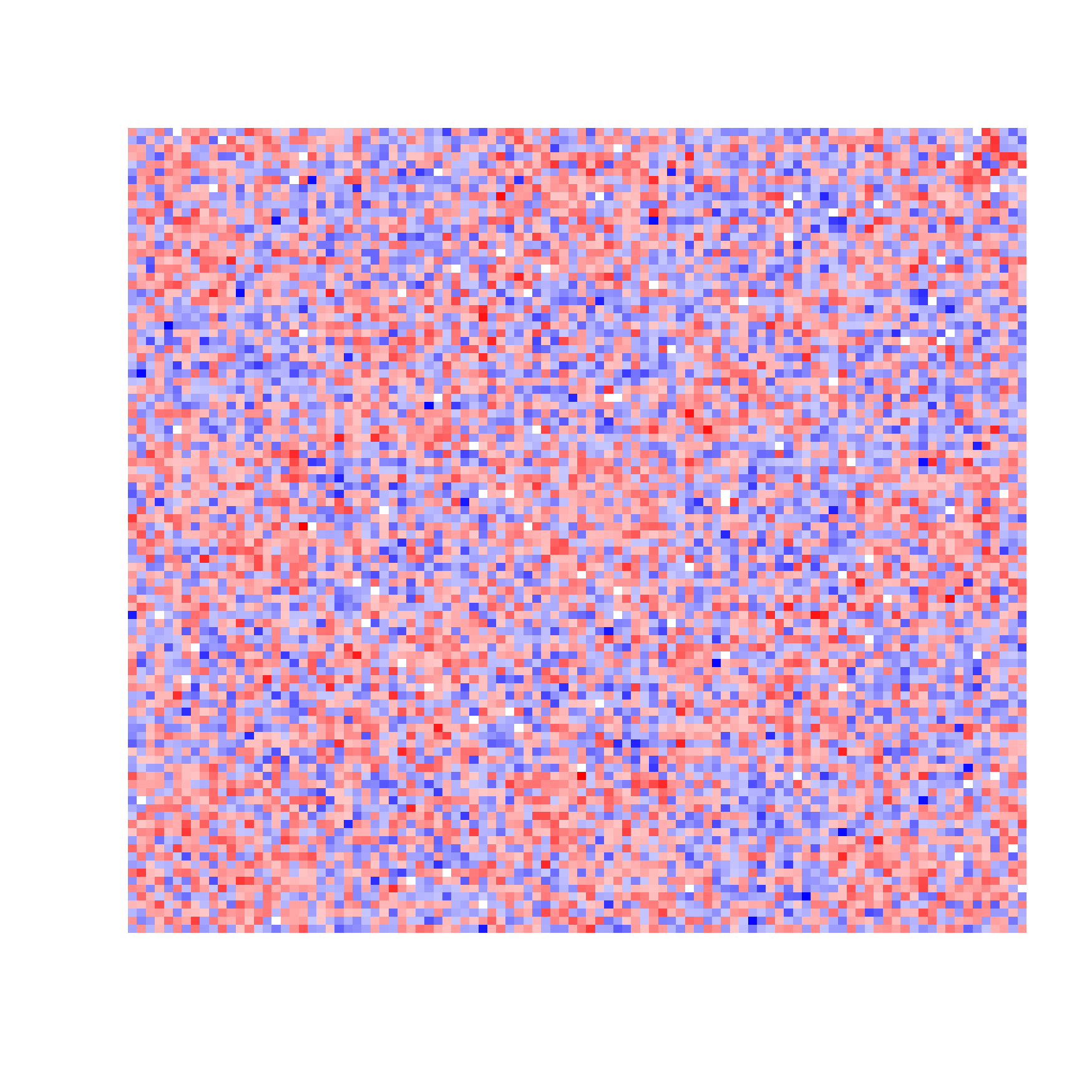}
    & \includegraphics[width=.3\linewidth]{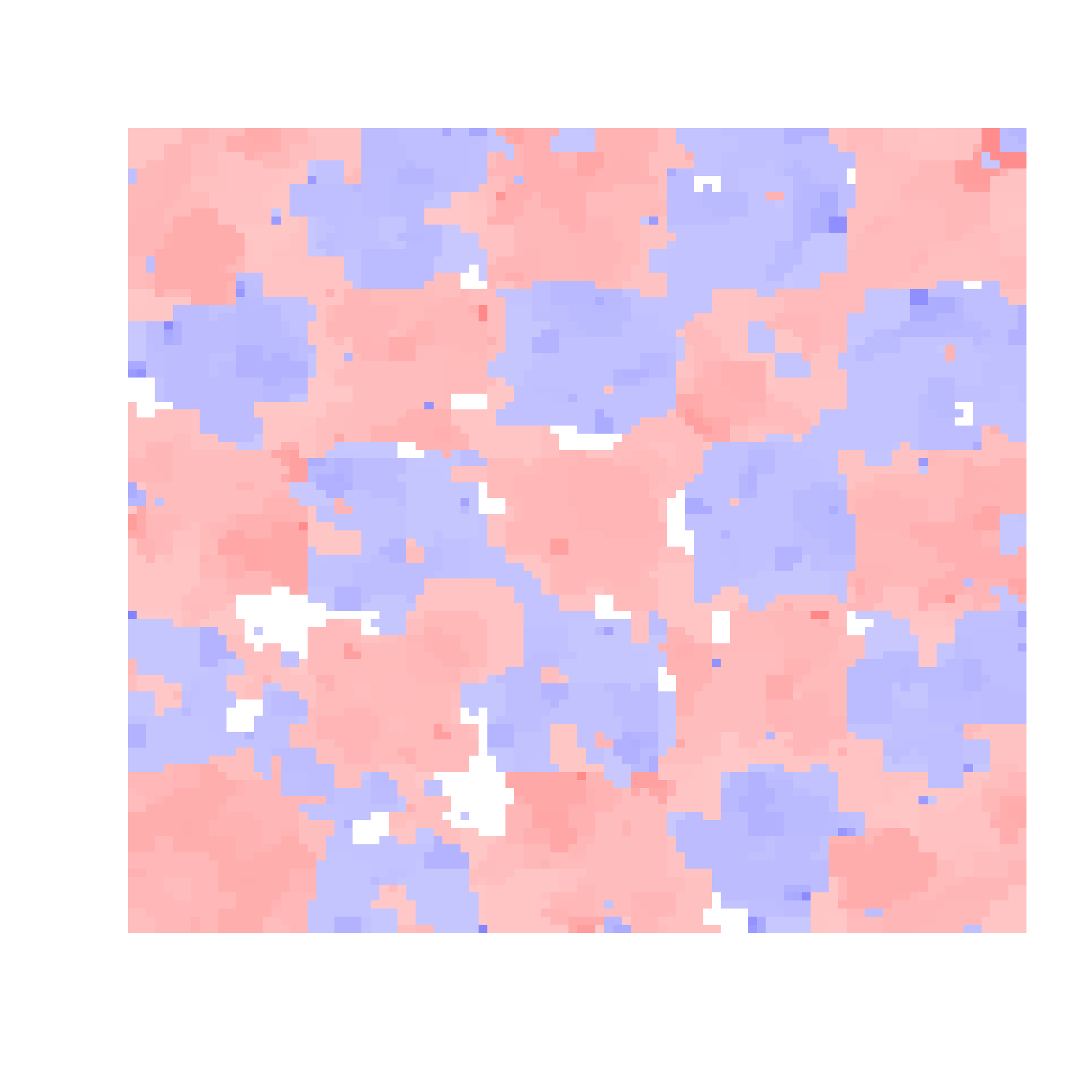}
    & \includegraphics[width=.3\linewidth]{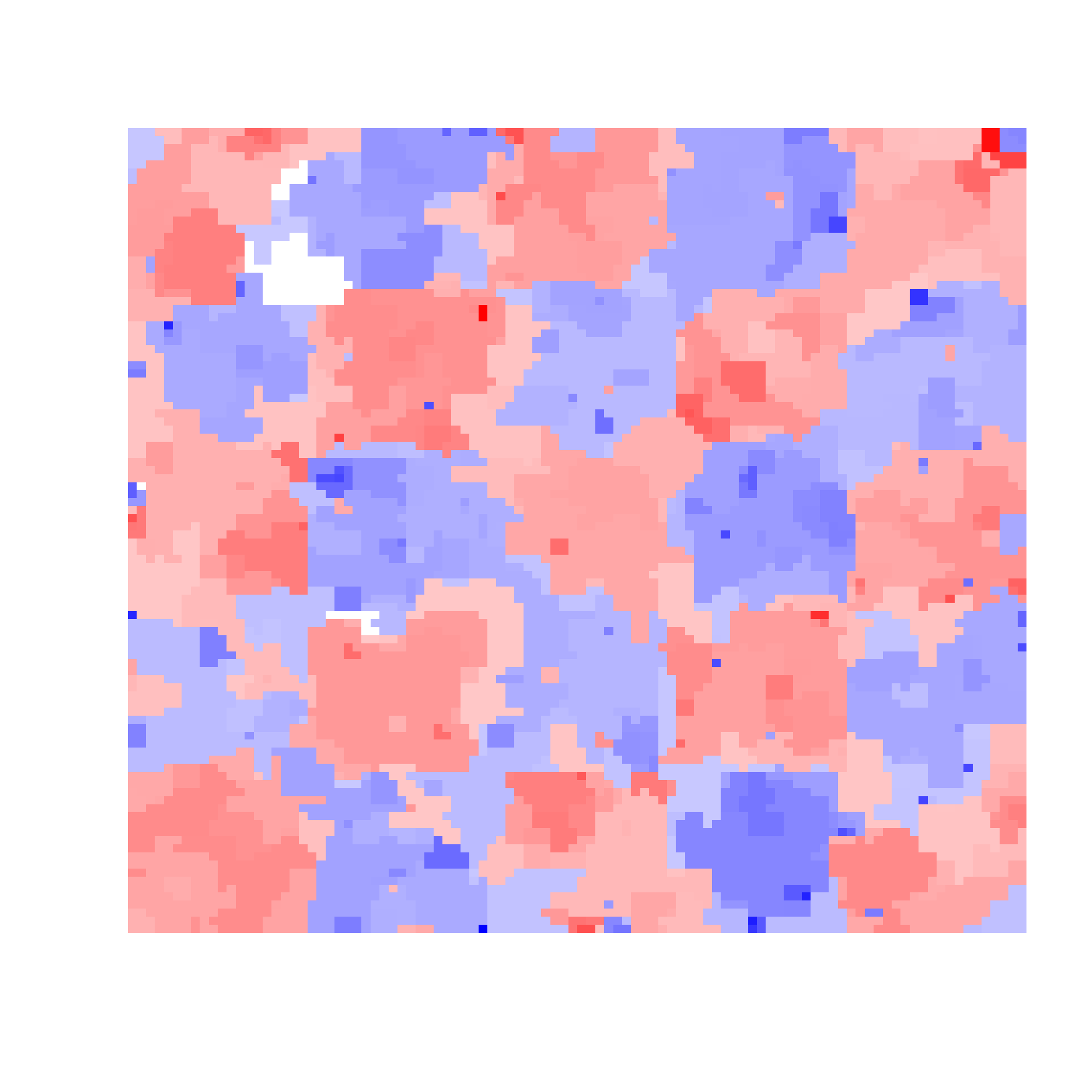} \\
    & Original data & \texttt{genlasso} & \texttt{flsa} \\
  \end{tabular}
  \caption{\label{fig:comp:intro} Left: Matrix of
      Figure \ref{fig:data:intro} corrupted with Gaussian white noise
      of variance $\sigma$. Middle: Denoising obtained with 
      \texttt{genlasso}. Right:  Denoising obtained with 
      \texttt{flsa}.}
\end{figure}

The difficulties that we have to face with in the
  two-dimensional framework are the following.
Firstly, it has to  be noticed  that the
classical dynamic  programming algorithm cannot  be applied in  such a
framework since  the Markov  property does not  hold anymore.  Secondly,
the group-lars approach of \cite{vert2010fast}
cannot be used in this framework since it would only provide
change-points in  columns and  not in rows. Thirdly,  although very efficient for image denoising, neither the generalized Lasso approach
devised  by  \cite{tibshirani2011} nor the  
fused Lasso  signal approximator of  \cite{hoefling2010path}, which are implemented in the R packages \texttt{genlasso} and \texttt{flsa},  respectively,
give access to the boundaries of non-overlapping blocks of a noisy block wise constant matrix. 
This fact is illustrated in Figure \ref{fig:comp:intro}. The first
column of this figure contains the block wise constant matrix of
Figure \ref{fig:data:intro}  corrupted with additional noise in high
signal to noise ratio contexts.
The denoising of these noisy matrices obtained by the packages
\texttt{genlasso} and \texttt{flsa} is displayed in the second and
third columns of Figure \ref{fig:data:intro}, respectively. Note that,
for obtaining these results, we used the default parameters of these
packages and for the parameter
$\lambda$ we used the one giving the denoised matrix being the closest to the original one
in terms of recovered blocks.


In this paper, our goal is  thus to design a  statistical method
for estimating the location of the boundaries of non-overlapping
blocks from a block wise constant matrix corrupted with white noise.
To the best of our knowledge,  there is  indeed no statistical
procedure for answering  this specific question  in the
  literature  that  is  both  computationally   and   statistically
  efficient.

The    paper     is    organized     as    follows.      In    Section
\ref{sec:stat_framework},  we  first  describe  how  to  rephrase  the
problem  of   two-dimensional  change-point   estimation  as   a  high
dimensional  sparse linear  model  and give  some theoretical  results
which   prove  the   consistency  of   our  change-point   estimators.
In Section \ref{sec:implementation},  we describe how
  to efficiently  implement our  method.  Then,  we provide  in Section
  \ref{sec:sim_study} experimental  evidence of  the relevance  of our
  approach   on    synthetic   data.     We   conclude    in   Section
  \ref{sec:app_hic} by a thorough analysis of a HiC dataset.

\section{Statistical framework}\label{sec:stat_framework}

\subsection{Statistical modeling}

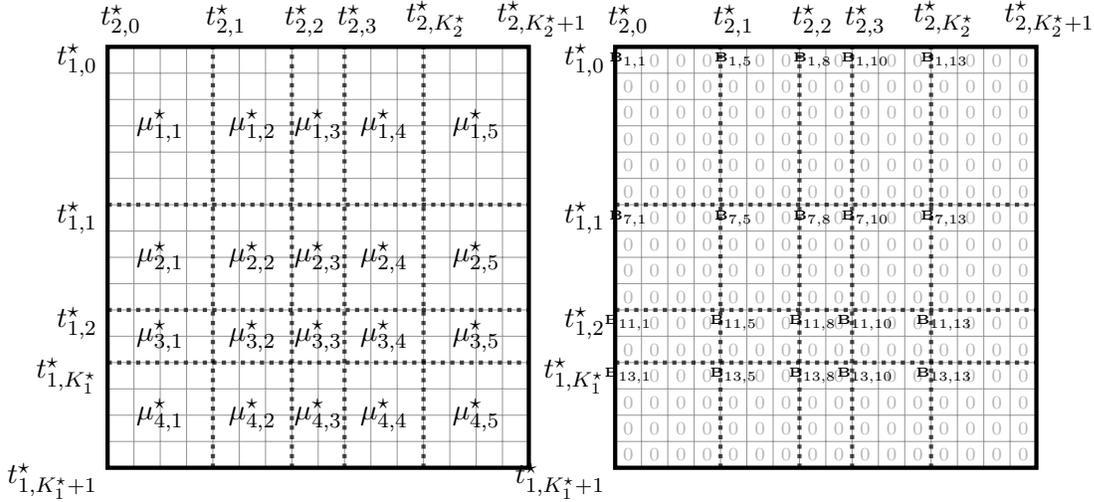
\begin{figure}[!h]
\hspace{-2cm}
\begin{tabular}{cc}
\begin{minipage}[c]{0.5\textwidth}
    \begin{center}
\begin{tikzpicture}[scale=0.7]
\draw[opacity=0.4] (0,0) grid[step=0.5] (8,8);

\draw (2.25,8) node[above]{$\ts_{2,1}$};
\draw (0,4.75) node[left]{$\ts_{1,1}$};
\draw[opacity=0.7,ultra thick,dotted] (2,0) -- (2,8);
\draw[opacity=0.7,ultra thick,dotted] (0,5) -- (8,5);
\draw (3.75,8) node[above]{$\ts_{2,2}$};
\draw (0,2.75) node[left]{$\ts_{1,2}$};
\draw[opacity=0.7,ultra thick,dotted] (3.5,0) -- (3.5,8);
\draw[opacity=0.7,ultra thick,dotted] (0,3) -- (8,3);
\draw (4.75,8) node[above]{$\ts_{2,3}$};
\draw (0,1.75) node[left]{$\ts_{1,\Ks_{1}}$};
\draw[opacity=0.7,ultra thick,dotted] (4.5,0) -- (4.5,8);
\draw[opacity=0.7,ultra thick,dotted] (0,2) -- (8,2);
\draw (6.25,8) node[above]{$\ts_{2,\Ks_{2}}$};
\draw[opacity=0.7,ultra thick,dotted] (6,0) -- (6,8);
\draw (8.25,8) node[above]{$\ts_{2,\Ks_{2}+1}$};
\draw (0,-0.25) node[left]{$\ts_{1,\Ks_{1}+1}$};

\draw (0.25,8) node[above]{$\ts_{2,0}$};
\draw (0,7.75) node[left]{$\ts_{1,0}$};
\draw[ultra thick] (0,0) rectangle (8,8);

\draw (1,6.5) node{$\mus_{1,1}$};
\draw (2.75,6.5) node{$\mus_{1,2}$};
\draw (4,6.5) node{$\mus_{1,3}$};
\draw (5.25,6.5) node{$\mus_{1,4}$};
\draw (7,6.5) node{$\mus_{1,5}$};
\draw (1,4.0) node{$\mus_{2,1}$};
\draw (2.75,4.0) node{$\mus_{2,2}$};
\draw (4,4.0) node{$\mus_{2,3}$};
\draw (5.25,4.0) node{$\mus_{2,4}$};
\draw (7,4.0) node{$\mus_{2,5}$};
\draw (1,2.5) node{$\mus_{3,1}$};
\draw (2.75,2.5) node{$\mus_{3,2}$};
\draw (4,2.5) node{$\mus_{3,3}$};
\draw (5.25,2.5) node{$\mus_{3,4}$};
\draw (7,2.5) node{$\mus_{3,5}$};
\draw (1,1.0) node{$\mus_{4,1}$};
\draw (2.75,1.0) node{$\mus_{4,2}$};
\draw (4,1.0) node{$\mus_{4,3}$};
\draw (5.25,1.0) node{$\mus_{4,4}$};
\draw (7,1.0) node{$\mus_{4,5}$};

\end{tikzpicture}
\end{center}
\end{minipage}&
\begin{minipage}[c]{0.5\textwidth}
    \begin{center}
\begin{tikzpicture}[scale=0.7]
\draw[opacity=0.4] (0,0) grid[step=0.5] (8,8);

\draw (2.25,8) node[above]{$\ts_{2,1}$};
\draw (0,4.75) node[left]{$\ts_{1,1}$};
\draw[opacity=0.7,ultra thick,dotted] (2,0) -- (2,8);
\draw[opacity=0.7,ultra thick,dotted] (0,5) -- (8,5);
\draw (3.75,8) node[above]{$\ts_{2,2}$};
\draw (0,2.75) node[left]{$\ts_{1,2}$};
\draw[opacity=0.7,ultra thick,dotted] (3.5,0) -- (3.5,8);
\draw[opacity=0.7,ultra thick,dotted] (0,3) -- (8,3);
\draw (4.75,8) node[above]{$\ts_{2,3}$};
\draw (0,1.75) node[left]{$\ts_{1,\Ks_{1}}$};
\draw[opacity=0.7,ultra thick,dotted] (4.5,0) -- (4.5,8);
\draw[opacity=0.7,ultra thick,dotted] (0,2) -- (8,2);
\draw (6.25,8) node[above]{$\ts_{2,\Ks_{2}}$};
\draw[opacity=0.7,ultra thick,dotted] (6,0) -- (6,8);
\draw (8.25,8) node[above]{$\ts_{2,\Ks_{2}+1}$};
\draw (0,-0.25) node[left]{$\ts_{1,\Ks_{1}+1}$};

\draw (0.25,8) node[above]{$\ts_{2,0}$};
\draw (0,7.75) node[left]{$\ts_{1,0}$};
\draw[ultra thick] (0,0) rectangle (8,8);

\draw (0.25,7.75) node{{\scalefont{0.5}\selectfont $\B_{\text{{\scalefont{0.8}\selectfont 1,1}}}$}};
\draw (2.25,7.75) node{{\scalefont{0.5}\selectfont $\B_{\text{{\scalefont{0.8}\selectfont 1,5}}}$}};
\draw (3.75,7.75) node{{\scalefont{0.5}\selectfont $\B_{\text{{\scalefont{0.8}\selectfont 1,8}}}$}};
\draw (4.75,7.75) node{{\scalefont{0.5}\selectfont $\B_{\text{{\scalefont{0.8}\selectfont 1,10}}}$}};
\draw (6.25,7.75) node{{\scalefont{0.5}\selectfont $\B_{\text{{\scalefont{0.8}\selectfont 1,13}}}$}};
\draw (0.25,4.75) node{{\scalefont{0.5}\selectfont $\B_{\text{{\scalefont{0.8}\selectfont 7,1}}}$}};
\draw (2.25,4.75) node{{\scalefont{0.5}\selectfont $\B_{\text{{\scalefont{0.8}\selectfont 7,5}}}$}};
\draw (3.75,4.75) node{{\scalefont{0.5}\selectfont $\B_{\text{{\scalefont{0.8}\selectfont 7,8}}}$}};
\draw (4.75,4.75) node{{\scalefont{0.5}\selectfont $\B_{\text{{\scalefont{0.8}\selectfont 7,10}}}$}};
\draw (6.25,4.75) node{{\scalefont{0.5}\selectfont $\B_{\text{{\scalefont{0.8}\selectfont 7,13}}}$}};
\draw (0.25,2.75) node{{\scalefont{0.5}\selectfont $\B_{\text{{\scalefont{0.8}\selectfont 11,1}}}$}};
\draw (2.25,2.75) node{{\scalefont{0.5}\selectfont $\B_{\text{{\scalefont{0.8}\selectfont 11,5}}}$}};
\draw (3.75,2.75) node{{\scalefont{0.5}\selectfont $\B_{\text{{\scalefont{0.8}\selectfont 11,8}}}$}};
\draw (4.75,2.75) node{{\scalefont{0.5}\selectfont $\B_{\text{{\scalefont{0.8}\selectfont 11,10}}}$}};
\draw (6.25,2.75) node{{\scalefont{0.5}\selectfont $\B_{\text{{\scalefont{0.8}\selectfont 11,13}}}$}};
\draw (0.25,1.75) node{{\scalefont{0.5}\selectfont $\B_{\text{{\scalefont{0.8}\selectfont 13,1}}}$}};
\draw (2.25,1.75) node{{\scalefont{0.5}\selectfont $\B_{\text{{\scalefont{0.8}\selectfont 13,5}}}$}};
\draw (3.75,1.75) node{{\scalefont{0.5}\selectfont $\B_{\text{{\scalefont{0.8}\selectfont 13,8}}}$}};
\draw (4.75,1.75) node{{\scalefont{0.5}\selectfont $\B_{\text{{\scalefont{0.8}\selectfont 13,10}}}$}};
\draw (6.25,1.75) node{{\scalefont{0.5}\selectfont $\B_{\text{{\scalefont{0.8}\selectfont 13,13}}}$}};

\foreach  \x in {0.25,0.75,...,7.75}
\foreach  \y in {0.25,0.75,1.25,2.25,3.25,3.75,4.25,5.25,5.75,6.25,6.75,7.25}
\draw[opacity=0.3] (\x,\y) node{{\scalefont{0.6}\selectfont 0}};
\foreach  \x in {0.75,1.25,1.75,2.75,3.25,4.25,5.25,5.75,6.75,7.25,7.75}
\foreach  \y in {1.75,2.75,4.75,7.75}
\draw[opacity=0.3] (\x,\y) node{{\scalefont{0.6}\selectfont 0}};

\end{tikzpicture}
\end{center}
\end{minipage}\\
\end{tabular}
    \caption{\label{fig:matrice_moy} Left: An example of a matrix $\U$ with $n=16$, $\Ks_{1}=3$ and $\Ks_{2}=4$. Right: The matrix $\B$ associated
to this matrix $\U$.}
\end{figure}

In this section, we explain how the two-dimensional retrospective change-point estimation issue can be seen as a variable selection problem.
Our goal 
is to estimate $\bts_1=(\ts_{1,1},\dots,\ts_{1,\Ks_1})$
and $\bts_2=(\ts_{2,1},\dots,\ts_{2,\Ks_2})$ from the random matrix $\Y=(Y_{i,j})_{1\leq i,j\leq n}$ defined by
\begin{equation}\label{eq:model1}
\Y=\U+\E,
\end{equation}
where $\U=(U_{i,j})$ is a blockwise constant matrix such that
$$
U_{i,j}=\mus_{k,\ell}\quad \textrm{ if } \ts_{1,k-1}\leq i\leq\ts_{1,k}-1 \textrm{ and } \ts_{2,\ell-1}\leq j\leq\ts_{2,\ell}-1,
$$
with       the       convention      $\ts_{1,0}=\ts_{2,0}=1$       and
$\ts_{1,\Ks_1+1}=\ts_{2,\Ks_2+1}=n+1$.  An  example of such a matrix  $\U$ is
displayed  in Figure~\ref{fig:matrice_moy}.   The
entries $E_{i,j}$  of the matrix $\E=(E_{i,j})_{1\leq  i,j\leq n}$ are
iid zero-mean random variables.  With  such a definition the $Y_{i,j}$
are  assumed  to be  independent  random  variables with  a  blockwise
constant mean.

Let $\TT$ be a $n\times n$ lower triangular matrix with nonzero elements equal to one and $\B$
a sparse matrix containing null entries except for the $\B_{i,j}$ such
that $(i,j)\in \{\ts_{1,0},\dots,\ts_{1,\Ks_1}\}\times\{\ts_{2,0},\dots,\ts_{2,\Ks_2}\}$. Then, \eqref{eq:model1} can be rewritten as follows:
\begin{equation}\label{eq:model2}
\Y=\TT\B\t{\TT}+\E,
\end{equation}
where $\t{\TT}$ denotes the transpose of the matrix $\TT$.
For an example of a matrix $\B$, see Figure~\ref{fig:matrice_moy}.
Let $\Ve(\X)$ denotes the vectorization of the matrix $\X$ formed by stacking the columns of $\X$ into a single column vector then
$\Ve(\Y)=\Ve(\TT\B\t{\TT})+\Ve(\E)$.   Hence,   by   using   that
$\Ve(\A\X\C)=(\t{\C}\otimes \A) \Ve(\X)$, where $\otimes$ denotes the Kronecker product,
\eqref{eq:model2} can be rewritten as:
\begin{equation}\label{eq:model3}
\calY=\calX\calB+\calE,
\end{equation}
where  $\calY=\Ve(\Y)$,  $\calX=\TT\otimes   \TT$,  $\calB=\Ve(\B)$  and
$\calE=\Ve(\E)$. Thanks to these transformations, Model~\eqref{eq:model1}
has thus been rephrased as a sparse high dimensional linear model where $\calY$ and $\calE$ are $n^2  \times 1$ column vectors,
$\calX$ is a $n^2\times n^2$ matrix and $\calB$ is $n^2  \times 1$ sparse column vectors.
%
Multiple change-point estimation Problem~\eqref{eq:model1} can thus be addressed as a variable selection problem:
\begin{equation}\label{eq:crit_lasso}
\widehat{\calB}(\lambda_n)=\Argmin_{\calB\in\rset^{n^2}}\left\{\|\calY-\calX\calB\|_2^2 +\lambda_n\|\calB\|_1\right\},
\end{equation}
where $\|u\|_2^2$ and $\|u\|_1$ are defined for a vector $u$ in $\rset^{N}$ by $\|u\|_2^2=\sum_{i=1}^{N} u_i^2$
and $\|u\|_1=\sum_{i=1}^{N} |u_i|$. Criterion (\ref{eq:crit_lasso}) is related to the popular Least Absolute Shrinkage and Selection Operator (LASSO)
in least-square regression. Thanks to the sparsity enforcing property of the $\ell_1$-norm, the estimator
$\widehat{\calB}$ of $\calB$ is expected to be sparse and to have non-zero elements matching with those of
$\calB$. Hence, retrieving the positions of the non zero
elements of $\widehat{\calB}$ thus provides estimators of $(\ts_{1,k})_{1\leq k\leq\Ks_1}$ and of $(\ts_{2,k})_{1\leq k\leq\Ks_2}$.
More precisely, let us define by $\widehat{\mathcal{A}}(\lambda_n)$ the set of active variables:
\begin{displaymath}
\widehat{\mathcal{A}}(\lambda_n)=\left\{j\in\{1,\dots,n^2\}:\widehat{\calB}_j(\lambda_n)\neq 0\right\}.
\end{displaymath}
For  each  $j$  in  $\widehat{\mathcal{A}}(\lambda_n)$,  consider  the
Euclidean division of $(j-1)$ by $n$, namely $(j-1)=n q_j+r_j$ then
\begin{multline}\label{eq:t_hat}
\bth_1=({\that}_{1,k})_{1\leq k\leq |\widehat{\mathcal{A}}_1(\lambda_n)|} \in\{r_j+1 : j\in\widehat{\mathcal{A}}(\lambda_n)\},\\
\bth_2=({\that}_{2,\ell})_{1\leq \ell\leq |\widehat{\mathcal{A}}_2(\lambda_n)|}\in\{q_j+1 : j\in\widehat{\mathcal{A}}(\lambda_n)\}\\
\textrm{ where }
{\that}_{1,1}<{\that}_{1,2}<\dots<{\that}_{1,|\widehat{\mathcal{A}}_1(\lambda_n)|},\quad
{\that}_{2,1}<{\that}_{2,2}<\dots<{\that}_{2,|\widehat{\mathcal{A}}_2(\lambda_n)|}.
\end{multline}
In \eqref{eq:t_hat}, $|\widehat{\mathcal{A}}_1(\lambda_n)|$ and $|\widehat{\mathcal{A}}_2(\lambda_n)|$ correspond to the number of
distinct elements in $\{r_j : j\in\widehat{\mathcal{A}}(\lambda_n)\}$ and $\{q_j : j\in\widehat{\mathcal{A}}(\lambda_n)\}$, respectively.

As far as we know, neither thorough practical implementation nor theoretical
grounding have been given so far to support such an approach
for change-point estimation in the two-dimensional case. In the following section, we give theoretical results supporting the use
of such an approach. 

\subsection{Theoretical results}

In order to establish the consistency of the estimators $\bth_{1}$ and
$\bth_{2}$ defined in \eqref{eq:t_hat}, we shall use assumptions ({\bf
  A1--A4}). These assumptions involve the two following quantities
\begin{align*}
  \Imin & = \min_{0\leq
    k\leq\Ks_1}|\ts_{1,k+1}-\ts_{1,k}|\wedge\min_{0\leq
    k\leq\Ks_2}|\ts_{2,k+1}-\ts_{2,k}|,\\
  \Jmin & = \min_{1\leq k\leq\Ks_1,1\leq\ell\leq\Ks_2+1}|\mus_{k+1,\ell}-\mus_{k,\ell}|\wedge\min_{1\leq k\leq\Ks_1+1,1\leq\ell\leq\Ks_2}|\mus_{k,\ell+1}-\mus_{k,\ell}|,
\end{align*}
which corresponds  to the smallest  length between two consecutive change-points and to the smallest jump size between  two consecutive
blocks, respectively.

\begin{hypA}\label{hyp:noise}
The random variables $(E_{i,j})_{1\leq i,j\leq n}$ are iid zero mean random variables such that there exists a positive constant $\beta$
such that for all $\nu$ in $\rset$, $\PE[\exp(\nu E_{1,1})]\leq\exp(\beta\nu^2)$.
\end{hypA}
\begin{hypA}\label{hyp:Jmin}
The sequence $(\lambda_n)$ appearing in (\ref{eq:crit_lasso}) is such that $(n\delta_n\Jmin)^{-1}\lambda_n\to 0$, as $n$ tends to infinity.
\end{hypA}
\begin{hypA}\label{hyp:delta_n}
The sequence $(\delta_n)$ is a non increasing and positive sequence tending to zero such that $n\delta_n{\Jmin}^2/\log(n)\to\infty$, as $n$
tends to infinity.
\end{hypA}
\begin{hypA}\label{hyp:Imin}
$\Imin\geq n\delta_n$.
\end{hypA}


\begin{proposition}\label{prop:consist}
Let $(Y_{i,j})_{1\leq i,j\leq n}$ be defined by (\ref{eq:model1}) and $\that_{1,k}$, $\that_{2,k}$ be defined by (\ref{eq:t_hat}).
Assume that $|\widehat{\mathcal{A}}_1(\lambda_n)|=\Ks_1$
and that $|\widehat{\mathcal{A}}_2(\lambda_n)|=\Ks_2$, with probabilty tending to one, then,
\begin{multline}\label{eq:consist}
\PP\left(\left\{\max_{1\leq k\leq\Ks_1}\left|\that_{1,k}-\ts_{1,k}\right|\leq n\delta_n\right\}\cap
\left\{\max_{1\leq k\leq\Ks_2}\left|\that_{2,k}-\ts_{2,k}\right|\leq n\delta_n\right\}\right)\to 1,\\
\textrm{ as } n\to\infty.
\end{multline}
\end{proposition}

The proof of Proposition \ref{prop:consist} is based on the two following lemmas.
The first one comes from the Karush-Kuhn-Tucker conditions of the optimization problem stated in (\ref{eq:crit_lasso}).
The second one allows us to control the supremum of the empirical mean of the noise.

\begin{lemma}\label{lem:KKT}
Let $(Y_{i,j})_{1\leq i,j\leq n}$ be defined by (\ref{eq:model1}). Then, $\widehat{\calU}=\calX\widehat{\calB}$, where $\calX$ and $\widehat{\calB}$
are defined in (\ref{eq:model3}) and (\ref{eq:crit_lasso}) respectively, is such that
\begin{align}\label{eq:kkt_1}
&\sum_{k=r_j+1}^n\,\sum_{\ell=q_j+1}^n Y_{k,\ell}-\sum_{k=r_j+1}^n\,\sum_{\ell=q_j+1}^n \widehat{\calU}_{k,\ell}=\frac{\lambda_n}{2}\textrm{sign}(\widehat{\calB}_j),
\textrm{ if } \widehat{\calB}_j\neq 0,\\\label{eq:kkt_2}
&\left|\sum_{k=r_j+1}^n\,\sum_{\ell=q_j+1}^n Y_{k,\ell}-\sum_{k=r_j+1}^n\,\sum_{\ell=q_j+1}^n \widehat{\calU}_{k,\ell}\right|
\leq\frac{\lambda_n}{2}, \textrm{ if } \widehat{\calB}_j =0,
\end{align}
where $q_j$ and $r_j$ are the quotient and the remainder of the Euclidean division of $(j-1)$ by $n$, respectively, that is
$(j-1)=n q_j+r_j$. In (\ref{eq:kkt_1}), $\textrm{sign}$ denotes the function which is defined by
$\textrm{sign}(x)=1$, if $x>0$, $-1$, if $x<0$ and 0 if $x=0$.
Moreover, the matrix $\widehat{\U}$, which is such that $\widehat{\calU}=\Ve(\widehat{\U})$, is blockwise constant and satisfies
$\widehat{U}_{i,j}=\widehat{\mu}_{k,\ell}$, if $\that_{1,k-1}\leq i\leq\that_{1,k}-1$ and $\that_{2,\ell-1}\leq j\leq\that_{2,\ell}-1$,
$k\in\{1,\dots,|\widehat{\mathcal{A}}_1(\lambda_n)|\}$, $\ell\in\{1,\dots,|\widehat{\mathcal{A}}_2(\lambda_n)|\}$, where
the $\that_{1,k}$, $\that_{2,k}$, $\widehat{\mathcal{A}}_1(\lambda_n)$ and $\widehat{\mathcal{A}}_2(\lambda_n)$ are defined in
(\ref{eq:t_hat}).
\end{lemma}

\begin{lemma}\label{lem:noise}
Let $(E_{i,j})_{1\leq i,j\leq n}$ be random variables satisfying (A\ref{hyp:noise}). Let also $(v_n)$ and $(x_n)$ be two positive
sequences such that $v_n x_n^2/\log(n)\to\infty$, then
$$
\PP\left(\max_{\stackrel{1\leq r_n<s_n\leq n}{|r_n-s_n|\geq v_n}}\left|(s_n-r_n)^{-1}\sum_{j=r_n}^{s_n-1}E_{n,j}\right|\geq x_n\right)\to 0,\textrm{ as } n\to\infty,
$$
the result remaining valid if $E_{n,j}$ is replaced by $E_{j,n}$.
\end{lemma}

The proofs of Proposition \ref{prop:consist}, Lemmas \ref{lem:KKT} and \ref{lem:noise} are given in Section \ref{sec:proofs}.

\begin{remark}
If $\Y$ is a non square matrix having $n_1$ rows and $n_2$ columns, with $n_1\neq n_2$, the result of
Proposition \ref{prop:consist}
remains valid if in Assumption (A\ref{hyp:delta_n}) $\delta_n$ is replaced by $\delta_{n_1,n_2}$ satisfying
$\n_1\delta_{\n_1,\n_2}{\Jmin}^2/\log(\n_2)\to\infty$ and $\n_2\delta_{\n_1,\n_2}{\Jmin}^2/\log(n_1)\to\infty$, as $\n_1$ and $\n_2$ tend to infinity.

\end{remark}

\section{Implementation}\label{sec:implementation}

In order to identify a series  of change-points
we  look  for   the  whole  path  of  solutions  in
\eqref{eq:crit_lasso},     \textit{i.e.},     $\{\hat{\calB}(\lambda),
\lambda_{\min}<\lambda      <     \lambda_{\max}\}$      such     that
$|\hat{\supp}(\lambda_{\max})|           =            0$           and
$|\hat{\supp}(\lambda_{\min})|=s$ with $s$ a predefined maximal number
of activated variables.  To this end it is natural to adopt the famous
homotopy/LARS strategy of \cite{osborne2000new,efron2004least}.  Such an
algorithm identifies  in Problem~\eqref{eq:crit_lasso}  the successive
values  of  $\lambda$ that  correspond  to  the  activation of  a  new
variable, or the deletion of one that became irrelevant.  
However, the existing implementations do not apply here since the size
of  the  design matrix  $\calX$  --  even  for  reasonable $n$  --  is
challenging  both  in terms  of  memory  requirement and  computational
burden.  To overcome these limitations,  we need to take advantage of
the  particular structure  of the  problem.  In  the following  lemmas
(which are  proved in Section \ref{sec:proofs}),  we show  that the
most  involving  computations  in  the  LARS  can  be  made  extremely
efficiently thanks to the particular structure of $\calX$.

\begin{lemma}\label{lem:Xtx}   For   any    vector   $\mathbf{v}   \in
  \rset^{n^2}$,   computing   $\calX   \mathbf{v}$   and   $\calX^\top
  \mathbf{v}$ requires at worse $2n^2$ operations.
\end{lemma}

\begin{lemma}\label{lem:cholXtX} 
Let $\supp=\{a_1,\dots,a_K\}$ and for each $j$ in $\supp$ let us consider the Euclidean division of $j-1$ by $n$ given by
$j-1=n q_j+r_j$, then
  \begin{equation}
    \label{eq:XtXij}
    \left(\left(\calX^\top  \calX\right)_{\supp,\supp}\right)_{1\leq k,\ell\leq K}
=\left(\left(n -  (q_{a_k}\vee q_{a_\ell})\right) \times \left(n - (r_{a_k}\vee r_{a_\ell})\right)\right)_{1\leq k,\ell\leq K}.
  \end{equation}
  Moreover,  for   any  non  empty   subset  $\supp$  of distinct indices  in
  $\set{1,\dots,  n^2}$,  the  matrix  $\calX^\top_\supp\calX_\supp$  is
  invertible.
\end{lemma}

\begin{lemma}  \label{lem:cholupdate}  Assume  that  we  have  at  our
  disposal   the    Cholesky   factorization    of   $\calX^\top_\supp
  \calX_\supp$.   The  updated  factorization   on  the  extended  set
  $\supp  \cup   \set{j}$   only   requires  solving   a
  $|\supp|$-size      triangular      system,     with      complexity
  $\mathcal{O}(|\supp|^2)$.  Moreover, the  downdated factorization on
  the restricted set  $\supp \backslash  \set{j}$ requires a
  rotation with negligible cost to preserve the triangular form of the
  Cholesky factorization after a column deletion.
\end{lemma}

\begin{remark} We were able to  obtain a closed-form expression of the
  inverse $(\calX^\top_\supp \calX_\supp)^{-1}$ for some special cases
  of the subset $\supp$,  namely, when the quotients/ratios associated
  with the Euclidean divisions of the elements  of $\supp$  are endowed  with a  particular ordering.
  Moreover, for addressing  any general problem, we rather  solve systems
  involving  $\calX^\top_\supp \calX_\supp$  by  means  of a  Cholesky
  factorization which is updated  along the homotopy algorithm.  These
  updates correspond  to adding or  removing an  element at a  time in
  $\supp$    and   are    performed   efficiently    as   stated    in
  Lemma~\ref{lem:cholupdate}.
\end{remark}

These lemmas are the building blocks for our LARS implementation given
in Algorithm~\ref{algo:lars2D}, where we detail the leading complexity
associated with each part.  The global complexity is in $\mathcal{O}(m
n^2 +  m s^2)$ where  $m$ is  the final number  of steps in  the while
loop.  These steps include all  the successive additions and deletions
needed to reach $s$, the final targeted number of active variables. At
the  end  of  day,  we   have  $m$  block wise  prediction  $\hat{\Y}$
associated    with    the    series     of    $m$    estimations    of
$\hat{\calB}(\lambda)$.  The above complexity  should be compared with
the  usual  complexity  of  the LARS  algorithm,  when  no  particular
structure is at play in Problem (\ref{eq:crit_lasso}): in such a case,
a implementation  of the LARS as  in \cite{bach2012optimization} would
be at least in $\mathcal{O}(m n^4 + m s^2)$.


Concerning the memory requirements, we only need to store the $n\times
n$ data  matrix $\mathbf{Y}$  once.  Indeed, since  we have at our disposal
the analytic form of  any sub matrix extracted  from $\calX^\top\calX$, we
never  need to  compute  neither  store this  large  $n^2 \times  n^2$
matrix.  This paves the way for quickly processing data with thousands
of rows and columns.

\begin{algorithm}[htbp!]
{
  \SetSideCommentLeft
  \DontPrintSemicolon
  \KwIn{data matrix $\Y$, maximal number of active variables $s$.}
  \BlankLine
  \tcp{\textcolor{mred}{Initialization}}
  Start with no change-point $\supp \leftarrow \emptyset$, $\hat{\calB} = \bzr$\;
  Compute current  correlations $\hat{\cor} = \calX^\top \calY$ with
  Lemma \ref{lem:Xtx} \tcp*[r]{\textcolor{mblue}{\small $\mathcal{O}(n^2)$}}

  \While{$\lambda >0$ or $|\supp|<s$} {
    \BlankLine

    \tcp{\textcolor{mred}{Update the set of active variables}}
    Determine next change-point(s) by setting $\lambda \leftarrow \| \hat{\cor} \|_\infty $ and
    $\supp \leftarrow \set{j: \hat{\cor}_j =  \lambda}$\;
    Update the Cholesky  factorization of $\calX^\top_{\supp} \calX_{\supp}$
    with Lemma~\ref{lem:cholXtX}\tcp*[r]{\textcolor{mblue}{$\mathcal{O}(|\supp|^2)$}}
    \BlankLine
    \BlankLine

    \tcp{\textcolor{mred}{Compute the direction of descent}}
    Get the unormalized direction $\tilde{w}_\supp \leftarrow \left(\calX^\top_{\cdot\supp} \calX_{\cdot\supp}\right)^{-1} \mathrm{sign}(\hat{c}_{\supp})$  \tcp*[r]{\textcolor{mblue}{\small $\mathcal{O}(|\supp|^2)$}}
    Normalize $w_\supp \leftarrow \alpha \tilde{w}_\supp$ with $\alpha
    \leftarrow 1/\sqrt{\tilde{w}_\supp^\top \mathrm{sign}(\hat{c}_\supp)}$\;
    Compute the  equiangular vector $u_{\supp}  = \calX_{\supp}  w_{\supp} $ and  $\ba = \calX^\top u_{\supp}$ with Lemma~\ref{lem:Xtx}\tcp*[r]{\textcolor{mblue}{$\mathcal{O}(n^2)$}}
    \BlankLine
    \BlankLine

    \tcp{\textcolor{mred}{Compute the direction step}}
    Find the maximal step preserving equicorrelation $\gamma_{\textrm{in}}\leftarrow \min_{j\in\supp^c}^+\left\{\frac{\lambda-\cor_j}{\alpha-a_j},\frac{\lambda+\cor_j}{\alpha+a_j}\right\}$\;
    Find the maximal step preserving the signs $\gamma_{\textrm{out}} \leftarrow \min_{j\in\supp}^+ \left\{-\hat{\calB}_\supp/w_\supp\right\}$ \;
    The direction step that preserves both is $\hat{\gamma} \leftarrow \min(\gamma_{\textrm{in}},\gamma_{\textrm{out}})$\;
    Update the correlations $\hat{\cor} \leftarrow \hat{\cor} -
    \hat{\gamma} \ba$ and $\hat{\calB}_\supp \leftarrow
    \hat{\calB}_\supp + \hat{\gamma} w_\supp$ accordingly \tcp*[r]{\textcolor{mblue}{\small $\mathcal{O}(n)$}}

    \BlankLine
    \BlankLine
    \tcp{\textcolor{mred}{Drop variable crossing the zero line}}
    \If{$\gamma_{\textrm{out}} < \gamma_{\textrm{in}}$}{
      Remove   existing change-point(s)  $\supp  \leftarrow
      \supp \backslash \set{j\in\supp :\hat{\calB}_j = 0}$\;
      Downdate  the  Cholesky   factorization  of  $\calX^\top_{\supp}
      \calX_{\supp}$\tcp*[r]{\textcolor{mblue}{\small
          $\mathcal{O}(|\supp|)$}}
    }
  }
  \KwOut{Sequence of triplet $(\supp,\lambda,\hat{\calB})$ recorded at each iteration.}
}
  \caption{Fast LARS for two-dimensional change-point estimation}
  \label{algo:lars2D}
\end{algorithm}


\section{Simulation study}
\label{sec:sim_study}

  In this  Section, we conduct a  set of simulation studies  to assess
  the   performances  of   our   proposal.   First,   we  report   the
  computational performances of Algorithm~\ref{algo:lars2D} and of its
  practical implementation in terms of timings.  Second, we report the
  statistical  performances  of  our  estimators~\eqref{eq:t_hat}  for
  recovering  the true  change-points by  means of  Receiver Operating
  Characteristic (ROC) curves.

\subsection{Data generation}
\label{sec:sim_settings}

All  synthetic data  are generated  from Model~\eqref{eq:model1}.   We
control the  computational difficulty  of the  problem by  varying the
sample size $n$.  The statistical  difficulty is controlled by varying
$\sigma$, the standard deviation of the Gaussian noise $\E$.  We chose
different patterns  for the true  matrix $\U^\star$ designed  to mimic
the  variety of  block  matrix  structures met  in  Hi-C data.   These
patterns     are    obtained     by     changing    the     parameters
$\mu^\star_{k,\ell}$s, each of whom controlling the intensity in block
$(k,\ell)$ of  $\U^\star$.  We  consider four different  scenarii, all
with $K_1^\star = 4$ change-points along  the rows and $K_2^\star = 4$
change-points along the columns.
\begin{equation}\label{eq:model:mustar}
  \begin{array}{@{}cc@{}}
    \left({\mu}_{k,\ell}^{\star,(1)}\right) =\begin{pmatrix} 1& 0 & 1 & 0 & 1 \\ 0 & 1 & 0 & 1 & 0 \\
      1& 0 & 1 & 0 & 1 \\ 0 & 1 & 0 & 1 & 0 \\ 1& 0 & 1 & 0 & 1 \\
    \end{pmatrix}, &
    \left({\mu}_{k,\ell}^{\star,(2)}\right)=\begin{pmatrix} 1& 0 & 0 & 0 & 0 \\ 0 & 1 & 0 & 0 & 0 \\
      0& 0 & 1 & 0 & 0 \\ 0 & 0 & 0 & 1 & 0 \\ 0& 0 & 0 & 0 & 1 \\
    \end{pmatrix},\\[2em]
    
    \left({\mu}_{k,\ell}^{\star,(3)}\right)=\begin{pmatrix} 1& 0 & 0 & 0 & 0 \\ 0 & 1 & 1 & 1 & 1 \\
      0& 1 & 1 & 0 & 0 \\ 0 & 1 & 0 & 1 & 0 \\ 0& 1 & 0 & 0 & 1 \\
    \end{pmatrix}, &
    \left({\mu}_{k,\ell}^{\star,(4)}\right)=\begin{pmatrix} 0& -1 & -1 & -1 & -1 \\ -1 & -1 & 0 & -1 & 0 \\
      -1& 0 & 1 & 0 & 1 \\ -1 & -1 & 0 & -1 & 0 \\ -1& 0 & 1 & 0 & 1 \\
    \end{pmatrix}.
  \end{array}
\end{equation}

The    first    ($\mu_{k,\ell}^{\star,(1)}$)    corresponds    to    a
``checkerboard-shaped''  matrix, that  is, a  natural two  dimensional
extension  of  a one  dimensional  piece-wise  constant problem.   The
second  ($\mu_{k,\ell}^{\star,(2)}$) defines  a  block diagonal  model
that   mimics  the   \textit{cis-interactions}  in   the  human   Hi-C
experiments: these are the most  usual interactions found in the cell,
which  occur between  nearby  elements along  the  genome.  The  third
($\mu_{k,\ell}^{\star,(3)}$)  and fourth  ($\mu_{k,\ell}^{\star,(4)}$)
configurations describe  more complex  patterns that  can be  found in
Hi-C   experiments,   which   also  correspond   to   more   difficult
change-points problems.

\begin{figure}[htbp!]
  \centering
  \begin{tabular}{cccc}
    $\mu^{\star,(1)}$ & $\mu^{\star,(2)}$ & $\mu^{\star,(3)}$ & $\mu^{\star,(4)}$ \\
    \includegraphics[width=.225\linewidth]{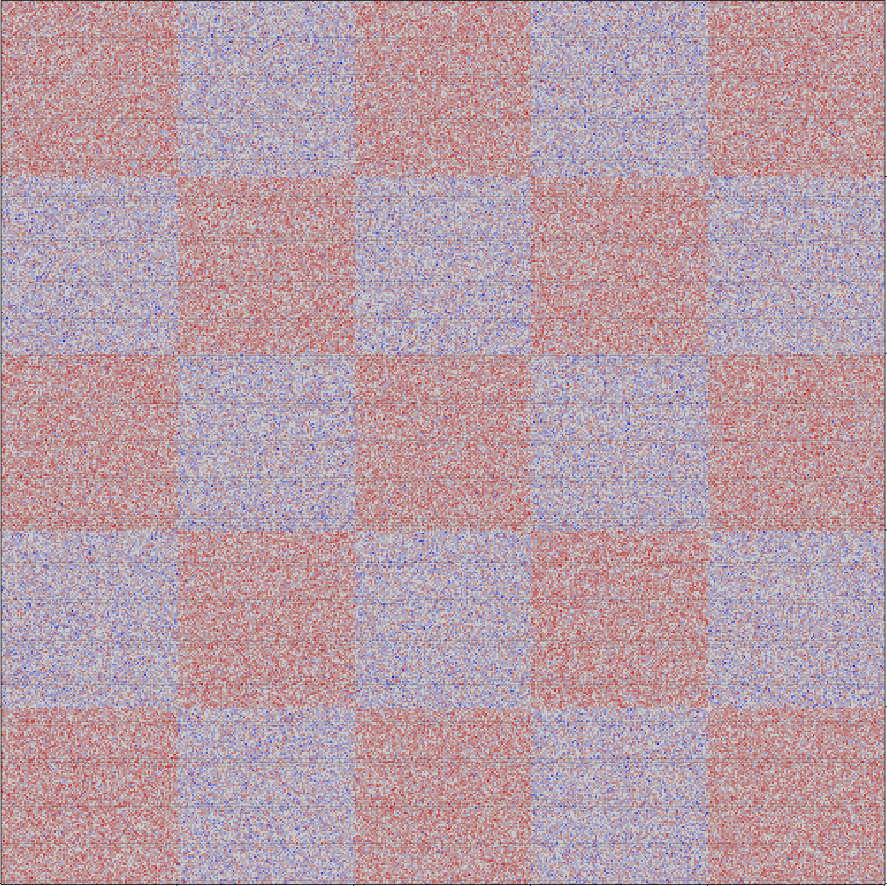}
    & \includegraphics[width=.225\linewidth]{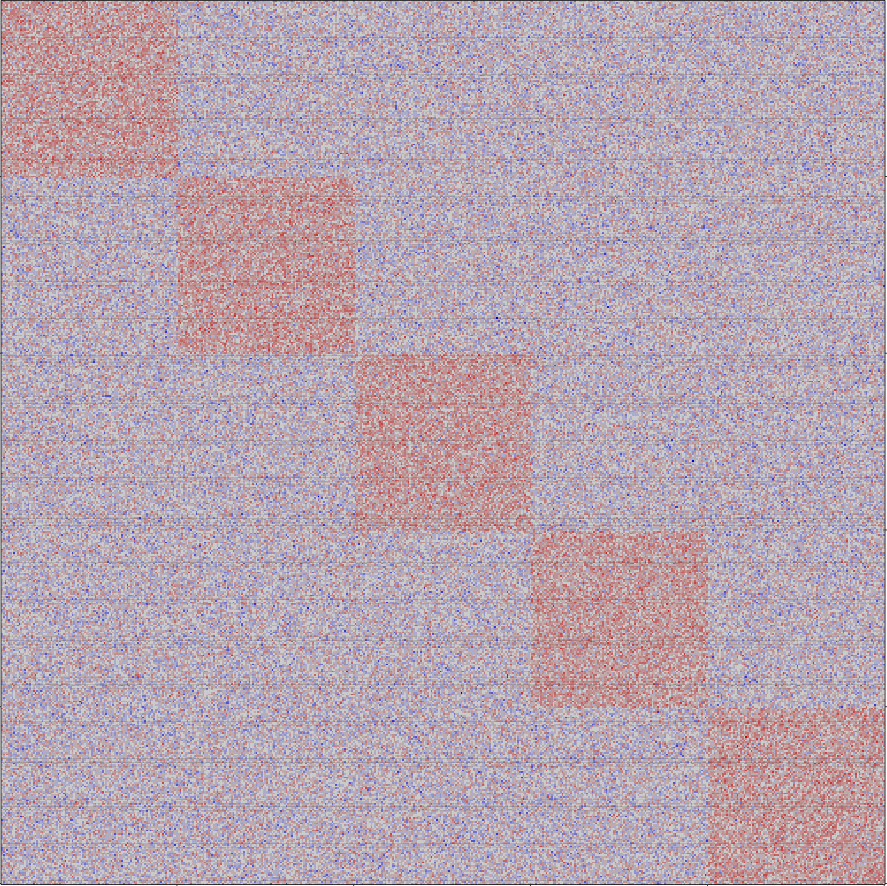}
    & \includegraphics[width=.225\linewidth]{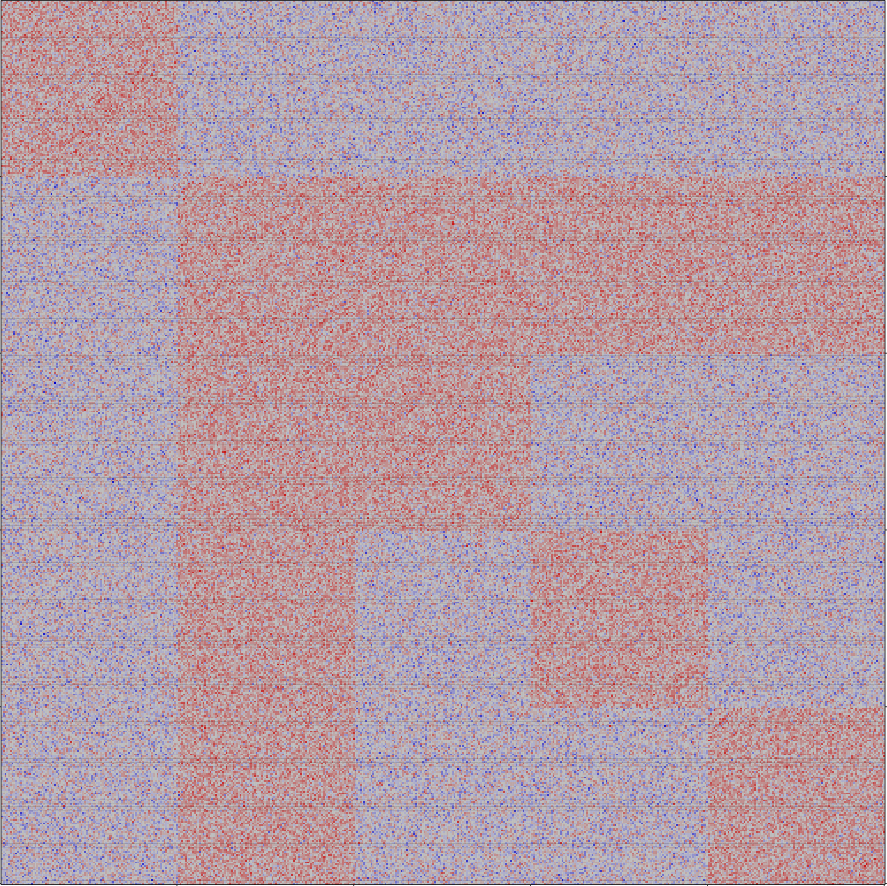}
    & \includegraphics[width=.225\linewidth]{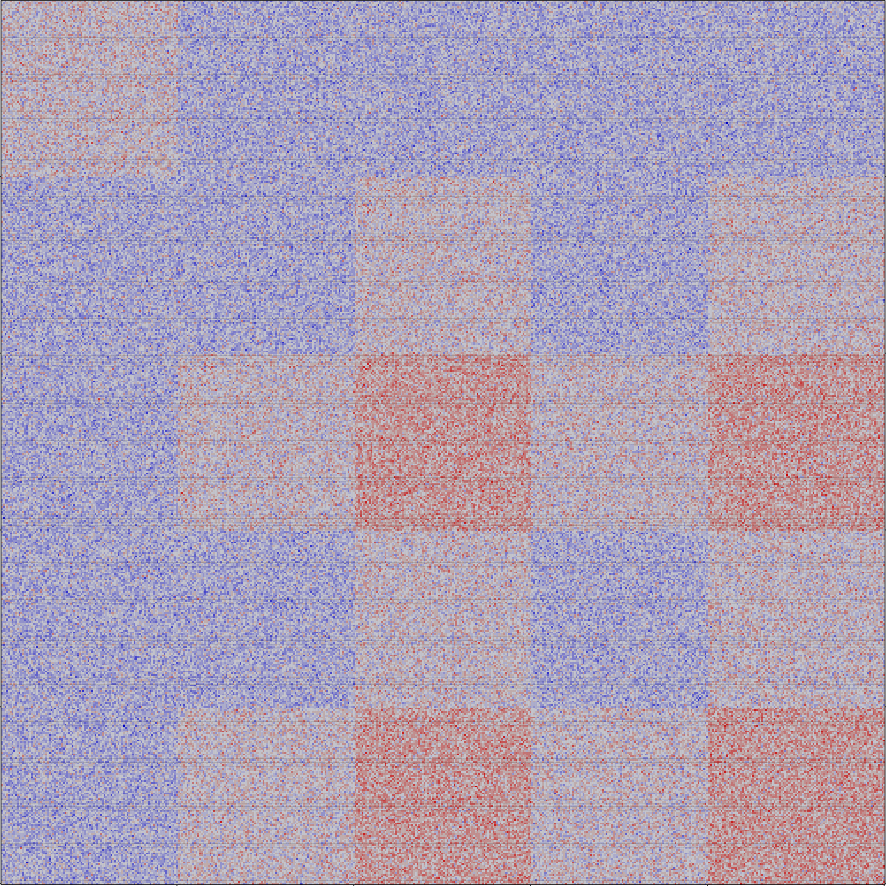} \\
  \end{tabular}
  \caption{Data matrices $\Y$ drawn  from Model \ref{eq:model1} for
    $\sigma=1,   n=100$   and   various  block wise   pattern   for
    $\U^\star$.}
  \label{fig:examplefigure}
\end{figure}

Examples    of    matrices    $\Y$    are    displayed    in    Figure
\ref{fig:examplefigure}  for these  four  scenarii,  with $n=100$  and
$\sigma=1$ which corresponds  to a relatively small level  of noise in
this problem.

\subsection{Competitors and implementation details}

In our  experiments, we compare  our methodology with  popular methods
for  segmentation  and  variable  selection that  we  adapted  to  the
specific problem of two-dimensional change-points detection:
\begin{enumerate}
\item First,  we adapt \citeauthor{breiman:1984}'s  classification and
  regression  trees \cite{breiman:1984}  (hereafter called \texttt{CART})  by
  using the  successive boundaries  provided by CART  as change-points
  for the two-dimensional data.  We use the implementation provided by
  the publicly available R package \texttt{rpart}.
\item Second, we  adapt \citeauthor{harchaoui:levyleduc:2012}'s method
  \cite{harchaoui:levyleduc:2012}  (hereafter  \texttt{HL}), which  is
  the exact  one-dimensional counterpart of our  approach.  To analyse
  two-dimensional data, we apply this procedure to each row of $\Y$ in
  order to recover  the change-points of each  row.  The change-points
  appearing in the different rows  are claimed to be change-points for
  the two-dimensional data  either if they appear at least  in one row
  (variant  \texttt{HL1})  or  if  they  appear  in  $([n/2]+1)$  rows
  (variant \texttt{HL2}). This approach is fitted by solving $n$ Lasso
  problems  (one  per  row  of  $\Y$)   by  means  of  the  R  package
  \texttt{glmnet}.
\item Third, we  consider an adaptation of  the fused-Lasso (hereafter
  \texttt{FL2D}).   Indeed, as  illustrated in  the introduction,  the
  basic  2-dimensional  fused-Lasso  for signal  approximator  is  not
  tailored  for  recovering  change  points.   We  thus  consider  the
  following  variant,  which  applied  a fused-Lasso  penalty  on  the
  following linear model:
  \begin{equation*}
    \calY=\underbrace{\left(\begin{tabular}{cccccc}
          $\mathds{1}_{\n}$&$0_{\n}$         &$\cdots$&$\cdots$                &$0_{\n}$                  &$\mathbb{I}_{\n}$\\
          $0_{\n}$         &$\mathds{1}_{\n}$&$\ddots$&                        &$\vdots$                         &$\vdots$\\
          $\vdots$                &$\ddots$                &$\ddots$&$\ddots$                &$\vdots$                         &$\vdots$\\
          $\vdots$                &                        &$\ddots$&$\mathds{1}_{\n}$&$0_{\n}$                  &$\vdots$\\
          $0_{\n}$         &$\cdots$                &$\cdots$&$0_{\n}$         &$\mathds{1}_{\n}$         &$\mathbb{I}_{\n}$\\
        \end{tabular}\right)}_{\calX^{(FL)}}
    \underbrace{\begin{pmatrix}
        \beta_1^{(FL)}\\\vdots\\\beta_{\n}^{(FL)}\\\beta_{\n+1}^{(FL)}\\\vdots\\\beta_{2\n}^{(FL)}\\
      \end{pmatrix}}_{\calB^{(FL)}}+ \ \calE
  \end{equation*}

  where  $\mathds{1}_{\n}$  (resp.   $0_{\n}$) is  a  size-$n$  column
  vector    of    ones    (resp.     zeros),    $\mathbb{I}_{\n}$    a
  $\n\times\n$-diagonal matrix of ones  and $\calY, \calE$ are defined
  as in
  Equation~(\ref{eq:model3}).  
  The \texttt{FL2D}  method detects  a change-point in  columns (resp.
  in   row)  if   two  successive   values  $\beta_{\ii}^{(FL)}$   and
  $\beta_{\ii+1}^{(FL)}$      with      $1\leq\ii\leq\n-1$      (resp.
  $\n+1\leq\ii\leq2\n-1$) are  different.  To  solve this  problem, we
  must   fit  a   general  fused-Lasso   problem.  We   rely  on   the
  R package \texttt{genlasso} for this task.

\item Finally, our  own procedure, that we  call \texttt{blockseg}, is
  implemented  in the  R package  \texttt{blockseg} which  is
  available   from  the   Comprehensive  R   Archive  Network   (CRAN,
  \cite{rbase}). Most of the computation are performed in \texttt{C++}
  using   the   library    \texttt{armadillo}   for   linear   algebra
  \citep{armadillo}.
\end{enumerate}

In what follows, all experiments were conducted on a Linux workstation
with Intel Xeon 2.4 GHz processor and 8 GB of memory.

\subsection{Numerical performances}

\begin{figure}[htbp!]
  \begin{center}
    \centering
    \begin{tabular}{@{}l@{\hspace{.1em}}c@{}c@{}}
      \rotatebox{90}{\hspace{3.5em}\small Procedures}
      & \includegraphics[width=.45\linewidth]{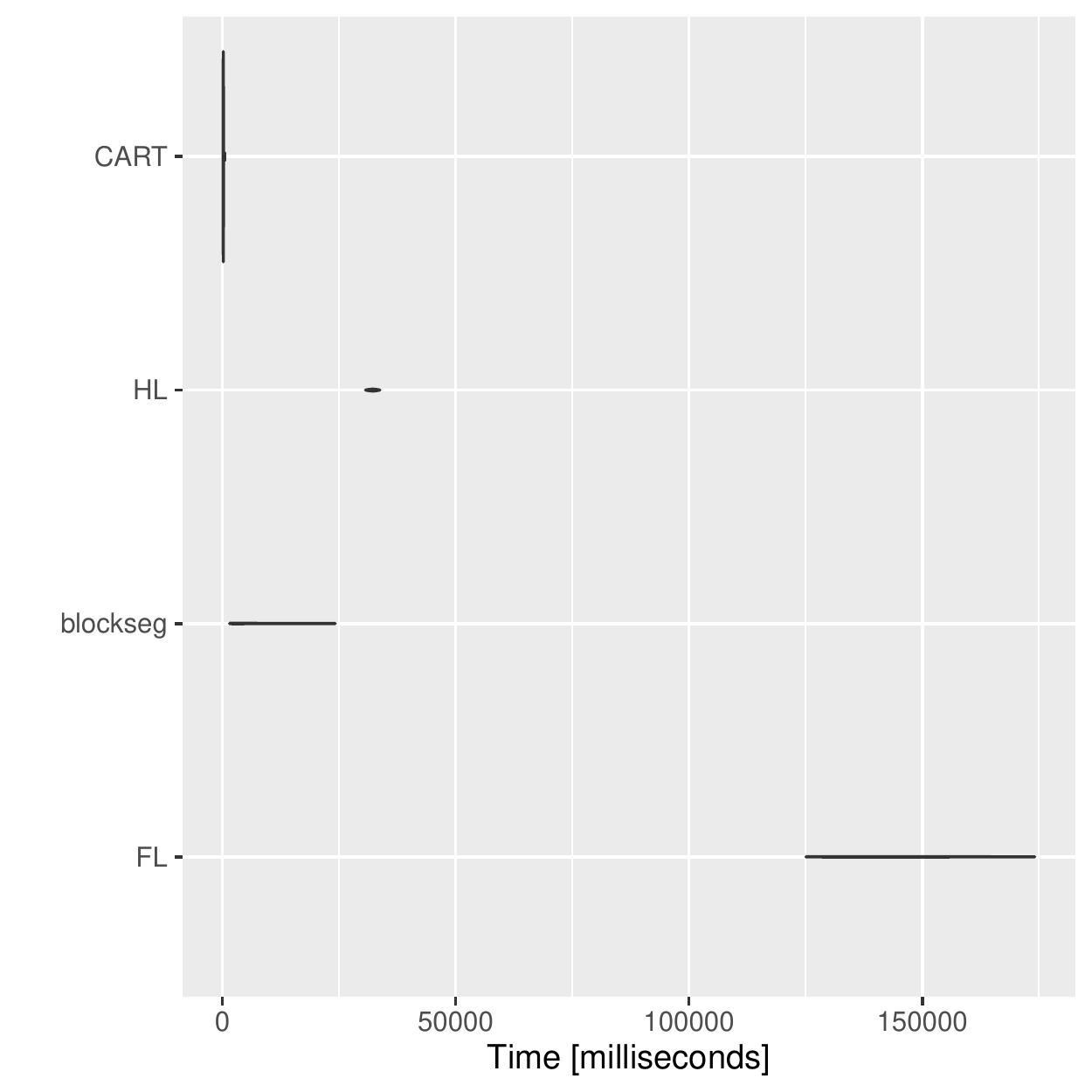}
      &   \includegraphics[width=.45\linewidth]{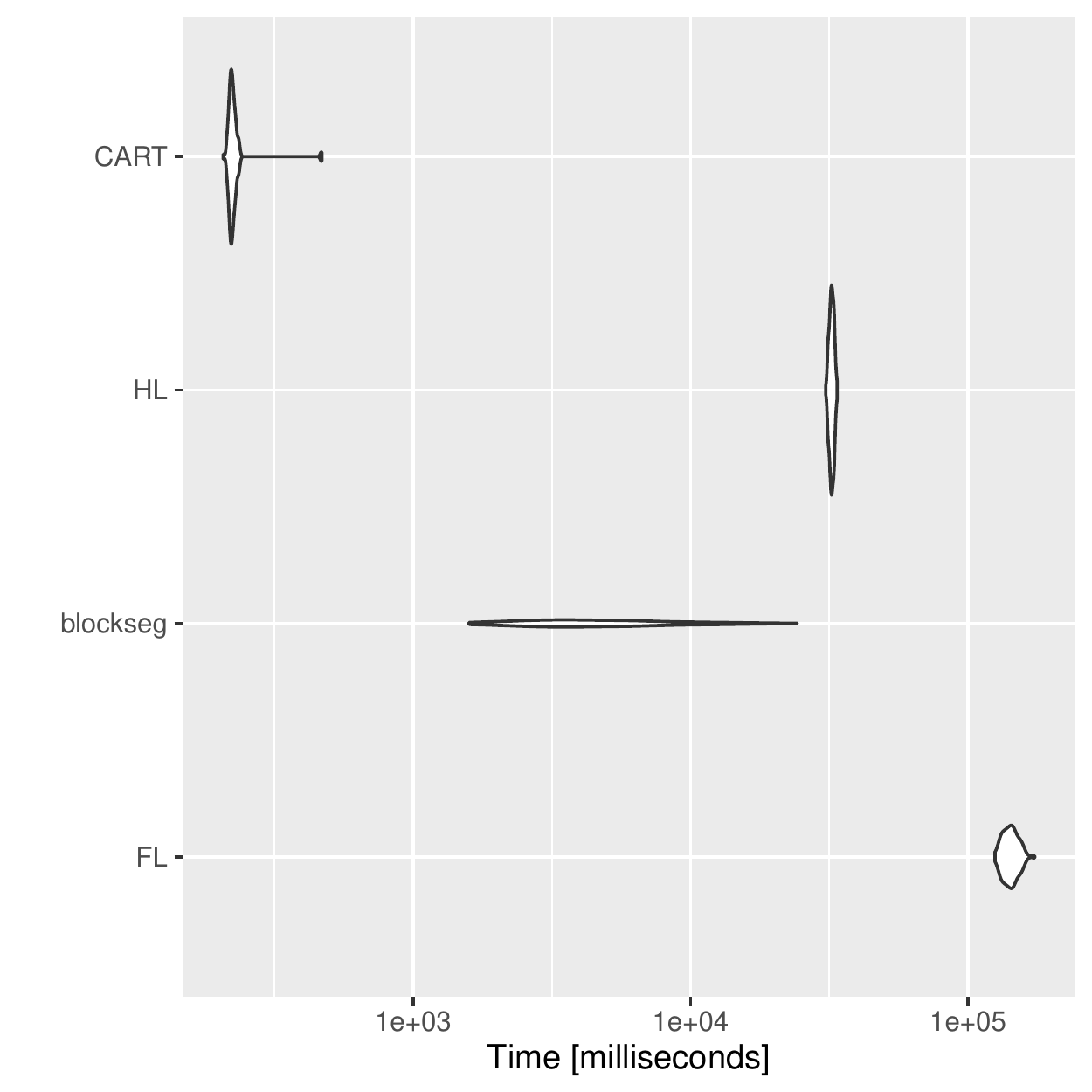}
      \\
      & linear scale & logarithm scale\\
    \end{tabular}
    \caption{\label{fig:timecomparison}  Violin
        plots of the computational times  for each procedure with a linear
        scale  (left) and  logarithm scale  (right): CART  methodology
        (\texttt{CART}), adaptation of \cite{harchaoui:levyleduc:2012}
        (\texttt{HL}),  our method
        (\texttt{blockseg}) and  fused LASSO
        (\texttt{FL}).}
  \end{center}
\end{figure}

We   start  by   presenting  in   Figure~\ref{fig:timecomparison}  the
computational  time for  $100$ runs  of each  method for  finding $\n$
change-points in  a matrix  drawn from the  ``checkerboard'' scenario,
with $n=100$ and $\sigma=5$.
Independent of its  statistical performance, we can see  on this small
problem  that the  adaptation of  the fused-Lasso  cannot be  used for
analyzing real  Hi-C problems.  On  the other hand, our  modified CART
procedure  is  extremely   fast.   However,  we  will   see  that  its
statistical performances  are quite poor. Finally,  our implementation
\texttt{blockseg}  is  quite  efficient   as  it  clearly  outperforms
\texttt{HL}.    This  should   be  emphasized   since
  \texttt{blockseg} is a two-dimensional method dealing with data with
  size $n^2$,  while \texttt{HL}  is a  1-dimensional approach that addresses two
  univariate problems of size $n$.

\begin{figure}[htbp!]
  \begin{center}
    \centering
    \begin{tabular}{@{}l@{\hspace{.1em}}c@{}c@{}}
      \rotatebox{90}{\hspace{1.5em}\small timings (seconds, log-scale)}
      & \includegraphics[width=.45\linewidth]{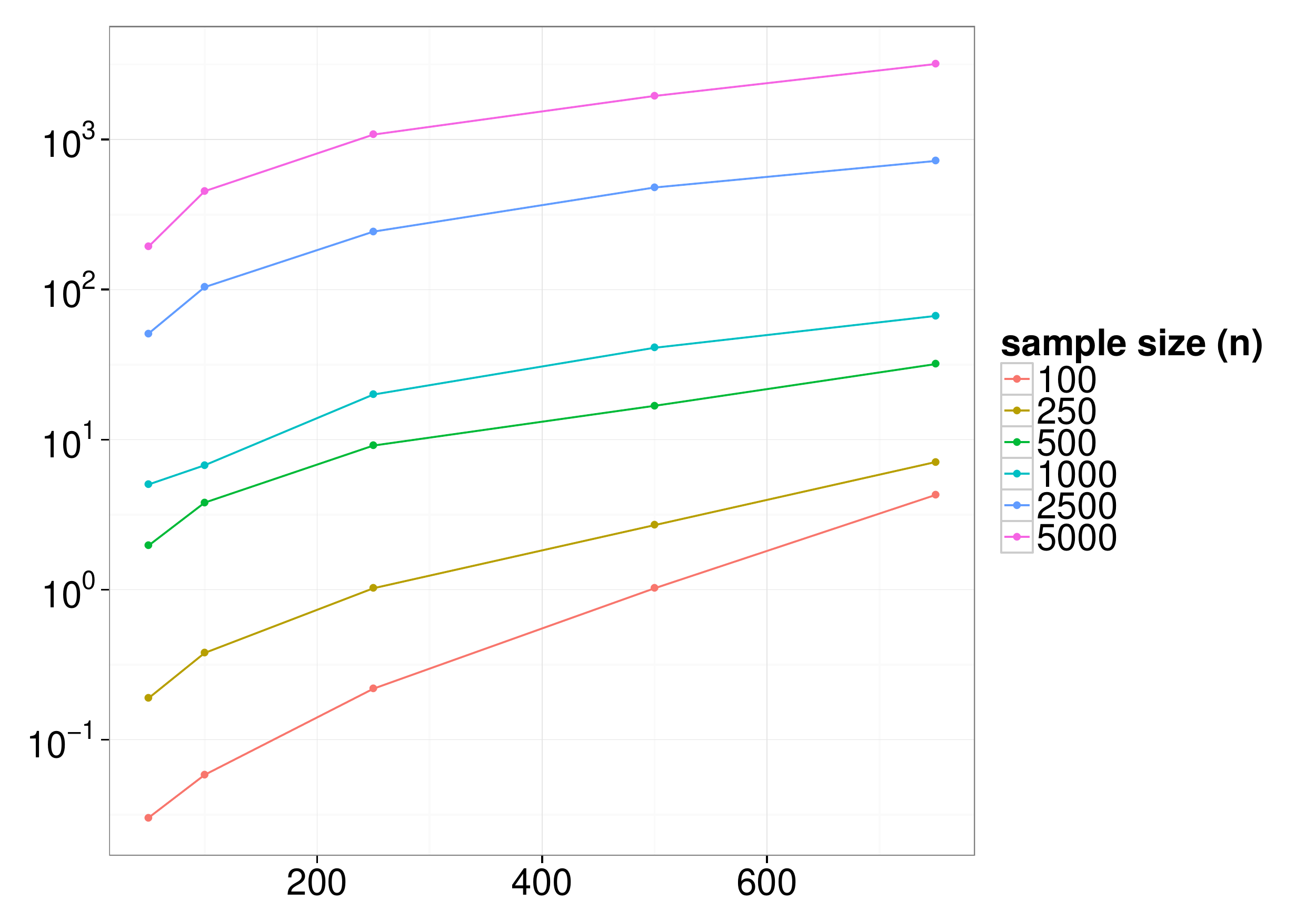}
      &   \includegraphics[width=.45\linewidth]{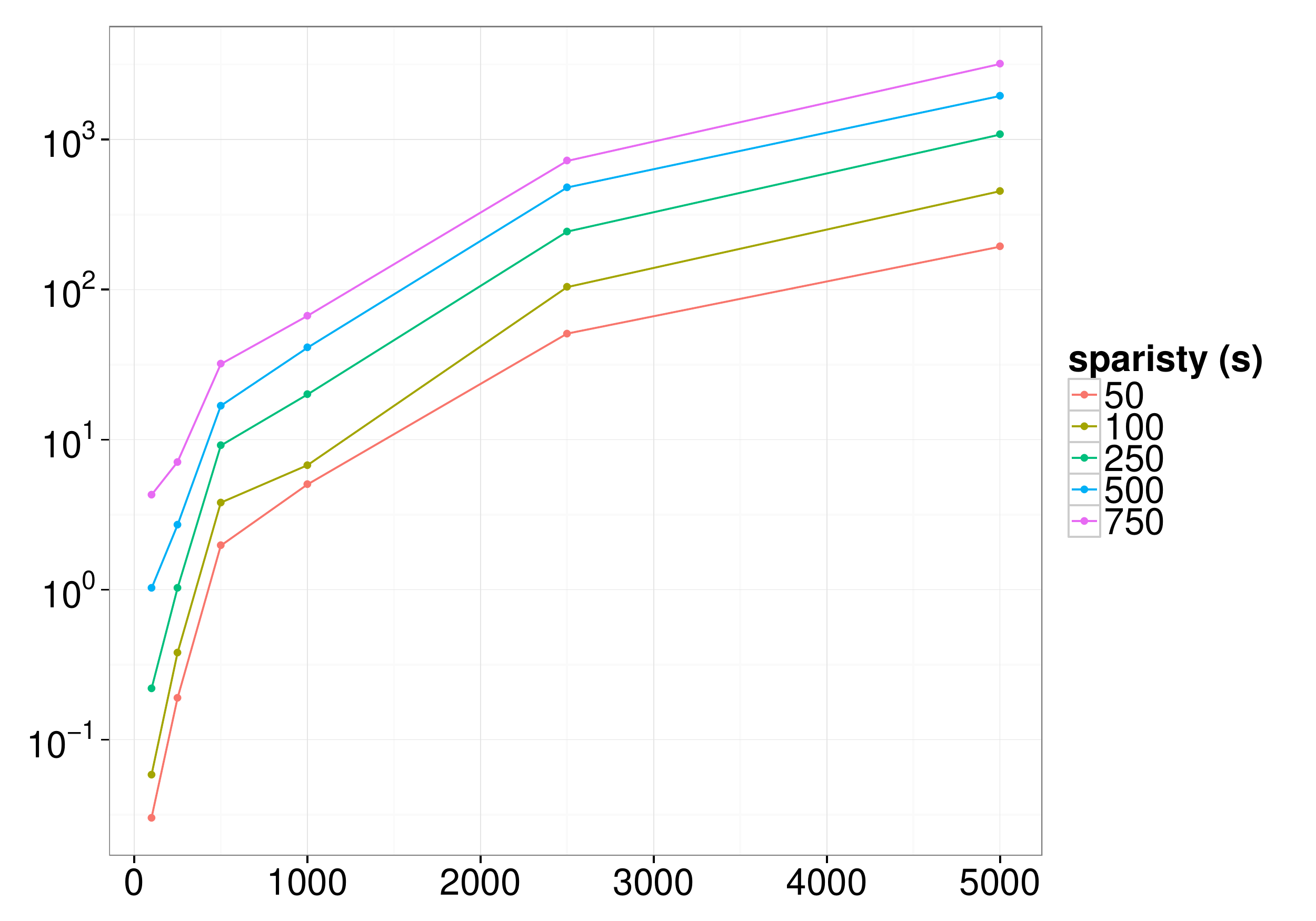}
      \\
      & sparsity level ($s$) & sample size $(n)$ \\
    \end{tabular}
    \caption{\label{fig:timeestimation}  Left:  Computational  time  (in
      seconds) for various values of $n$ as a function of the sparsity
      level $s=|\supp|$ reached  at the end of  the algorithm.  Right:
      Computation time (in seconds) as a function of sample size $n$.}
  \end{center}
\end{figure}

We now  consider \texttt{blockseg} on  its own  in order to  study the
scalability of our  approach regarding the problem  dimension. To this
end,        we         generated        ``checkerboard''        matrix
$\left({\mu}_{k,\ell}^{\star,(1)}\right)$           given           in
(\ref{eq:model:mustar}) with various sizes $n$  (from 100 to 5000) and
various values of the maximal  number of activated variables $s$ (from
50 to 750).  The median runtimes obtained from 4 replications (+ 2 for
warm-up) are  reported in Figures~\ref{fig:timeestimation}.   The left
(resp.  the right)  panel gives the runtimes in seconds  as a function
of $s$ (resp.  of $n$).   These results give experimental evidence for
the  theoretical  complexity $\mathcal{O}(m  n^2  +  m s^2)$  that  we
established  in  Section  \ref{sec:implementation} and  thus  for  the
computational efficiency  of our approach:  applying \texttt{blockseg}
to matrices  containing $10^7$ entries  takes less than 2  minutes for
$s=750$.

\subsection{Statistical performances}  We evaluate the  performance of
the different competitors for recovering the true change-points in the
4 scenarii defined in Section \ref{sec:sim_settings} for an increasing
level of  difficulty.  We draw 1000  datasets for each scenario  for a
varying level  of noise $\sigma\in  \set{1,2,5,10}$ and for  a problem
size of $n=100$.  Note that we  use this relatively small problem size
to  allow the  comparison with  methods \texttt{HL}  and \texttt{FL2D}
that would not work for greater values of $n$.

\begin{figure}[htbp!]
  \centering
  \begin{tabular}{@{}l@{}cccc@{}r}
    & $\sigma=1$ & $\sigma=2$
    & $\sigma=5$ & $\sigma=10$\\
    \rotatebox{90}{\hspace{1.25em}\scriptsize True positive rate}
    & \includegraphics[width=.225\linewidth]{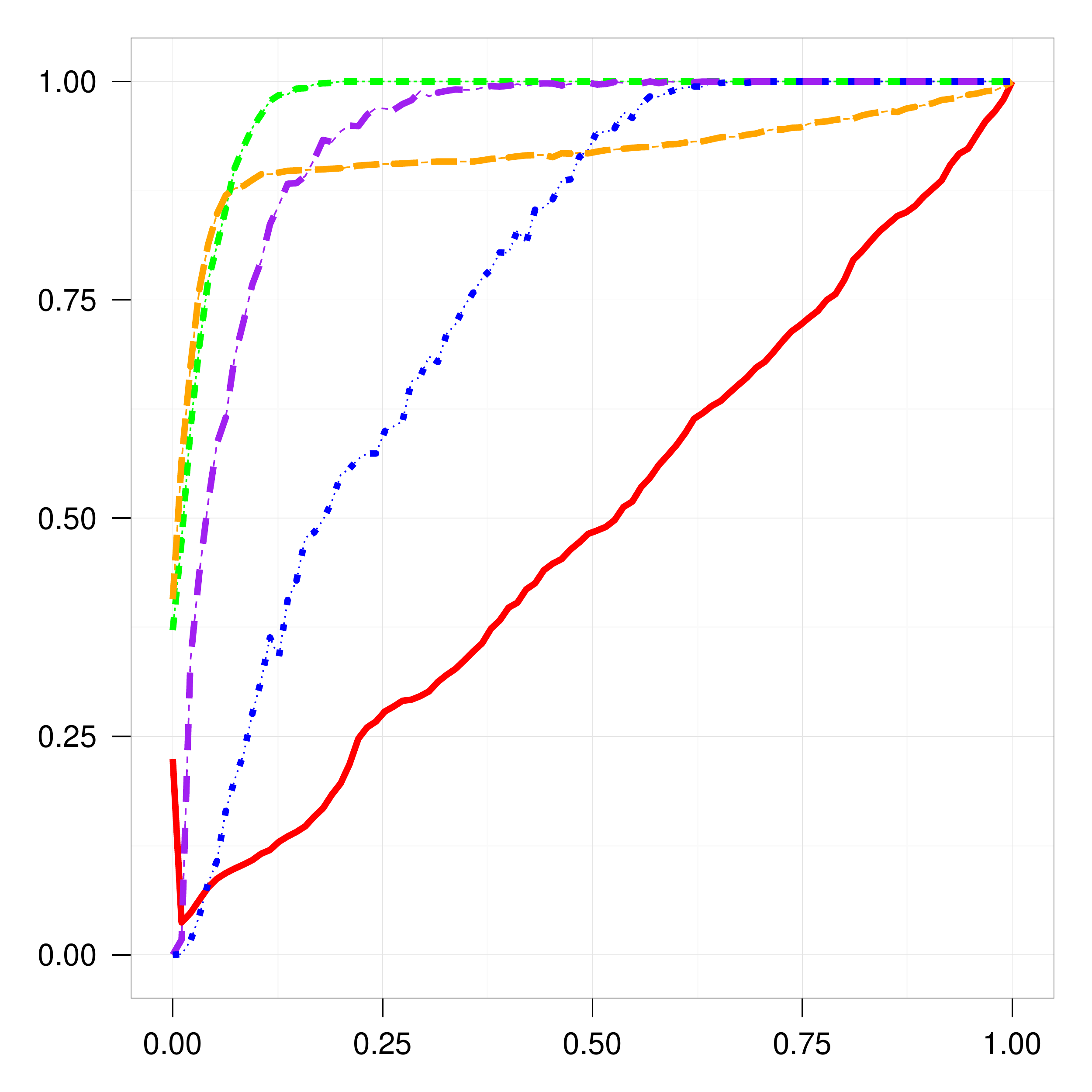}
    & \includegraphics[width=.225\linewidth]{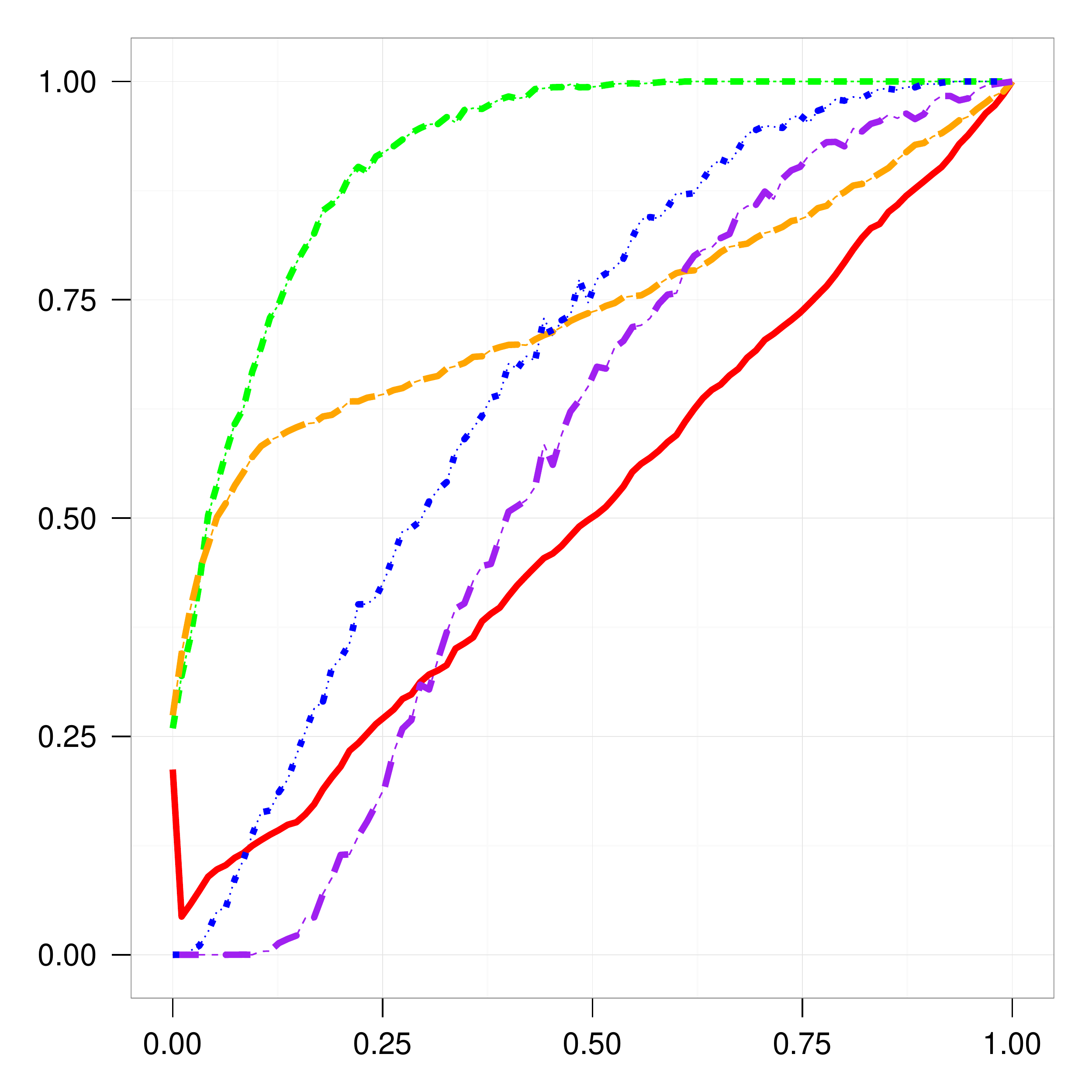}
    & \includegraphics[width=.225\linewidth]{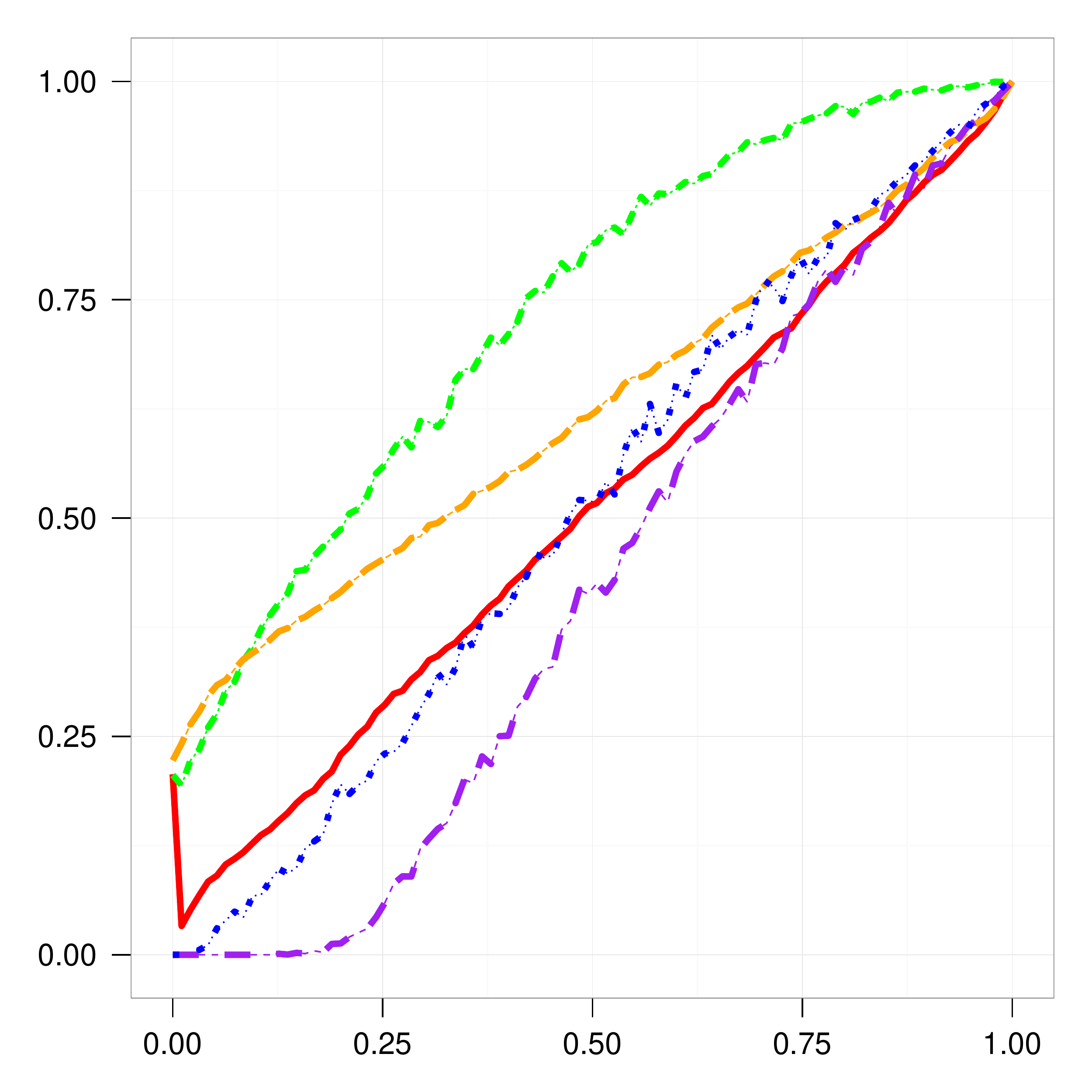}
    & \includegraphics[width=.225\linewidth]{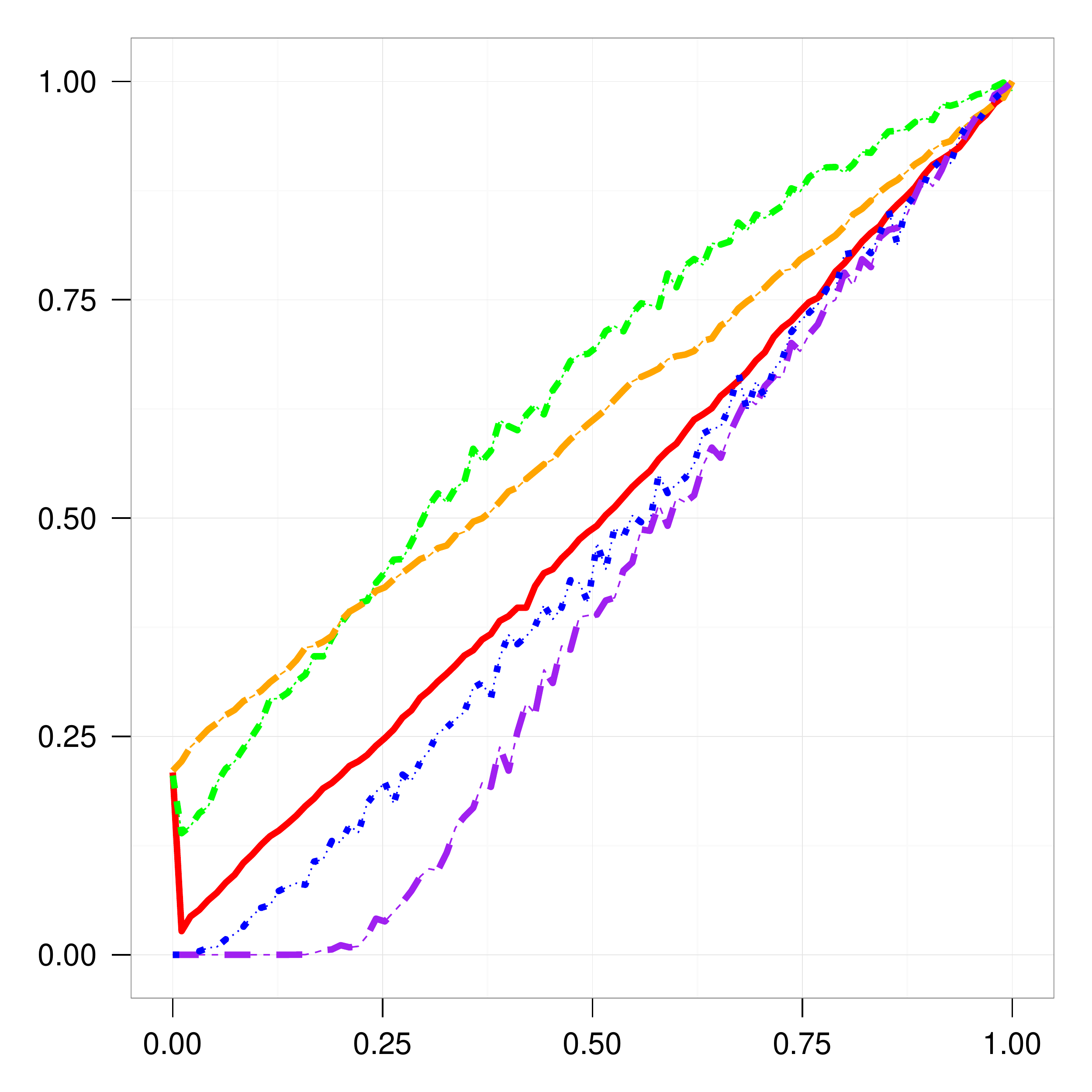}
    & \rotatebox{90}{\hspace{2em}\scriptsize scenario 1}
    \\
    & \multicolumn{4}{c}{\scriptsize False positive rate} \\
    \rotatebox{90}{\hspace{1.25em}\scriptsize True positive rate}
    & \includegraphics[width=.225\linewidth]{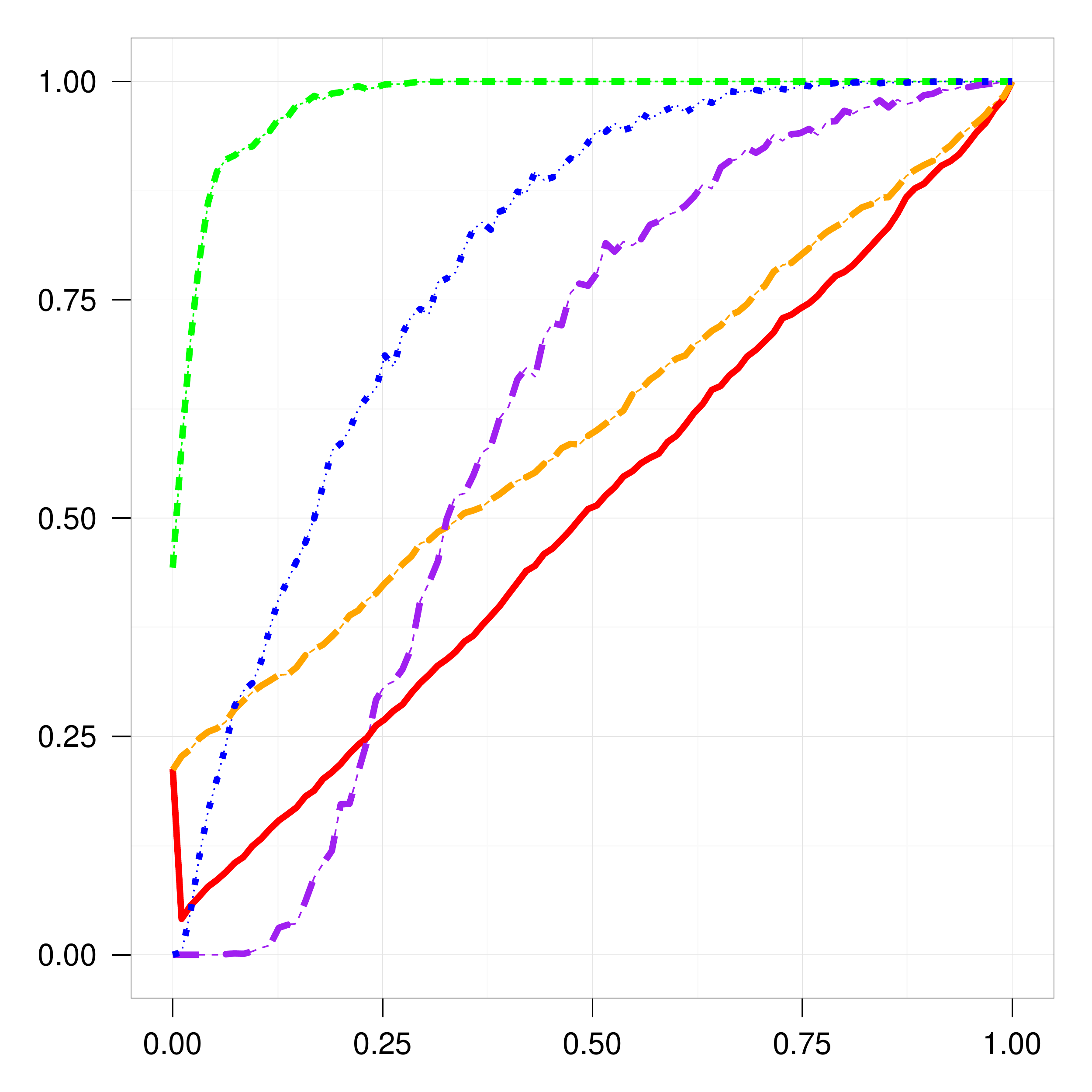}
    & \includegraphics[width=.225\linewidth]{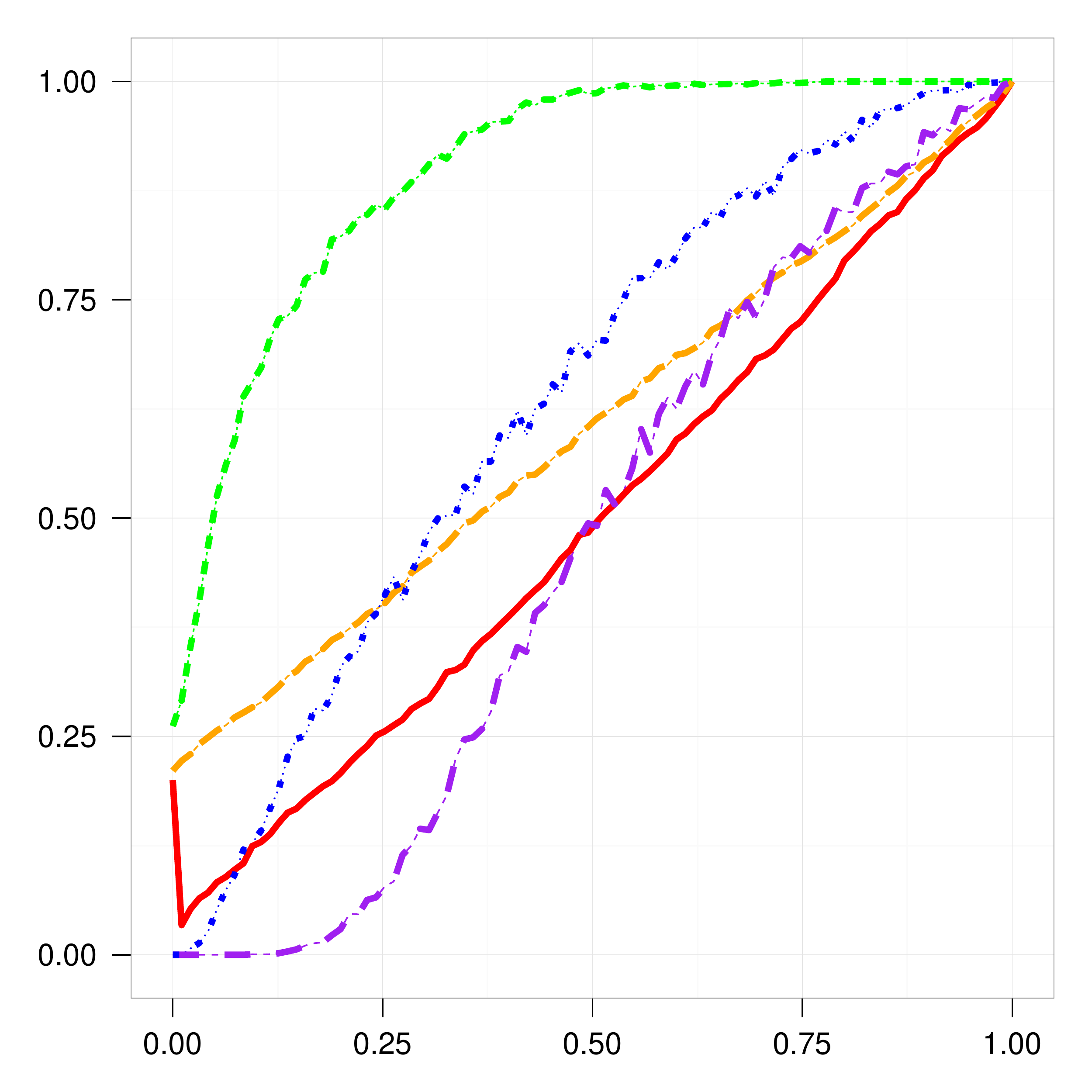}
    & \includegraphics[width=.225\linewidth]{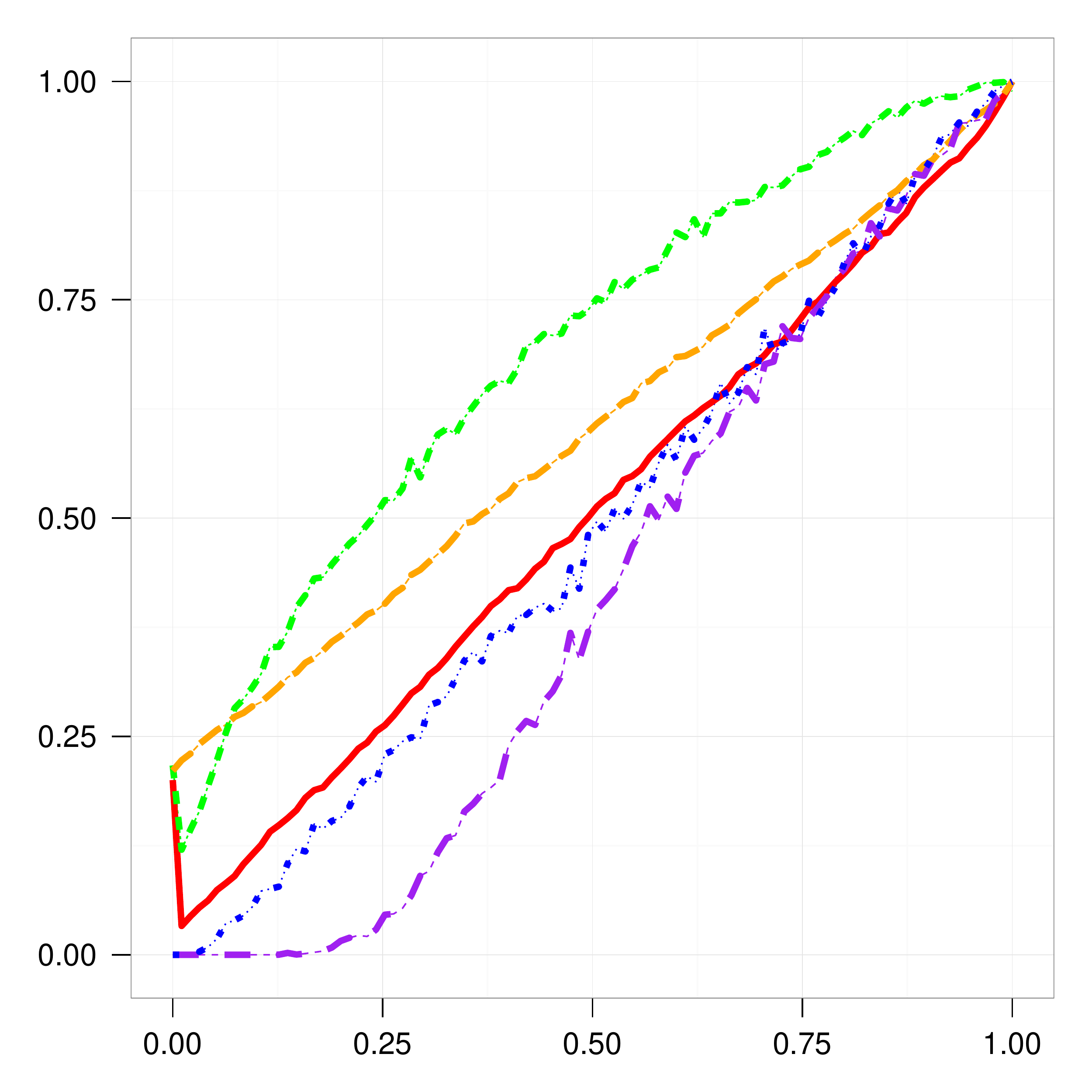}
    & \includegraphics[width=.225\linewidth]{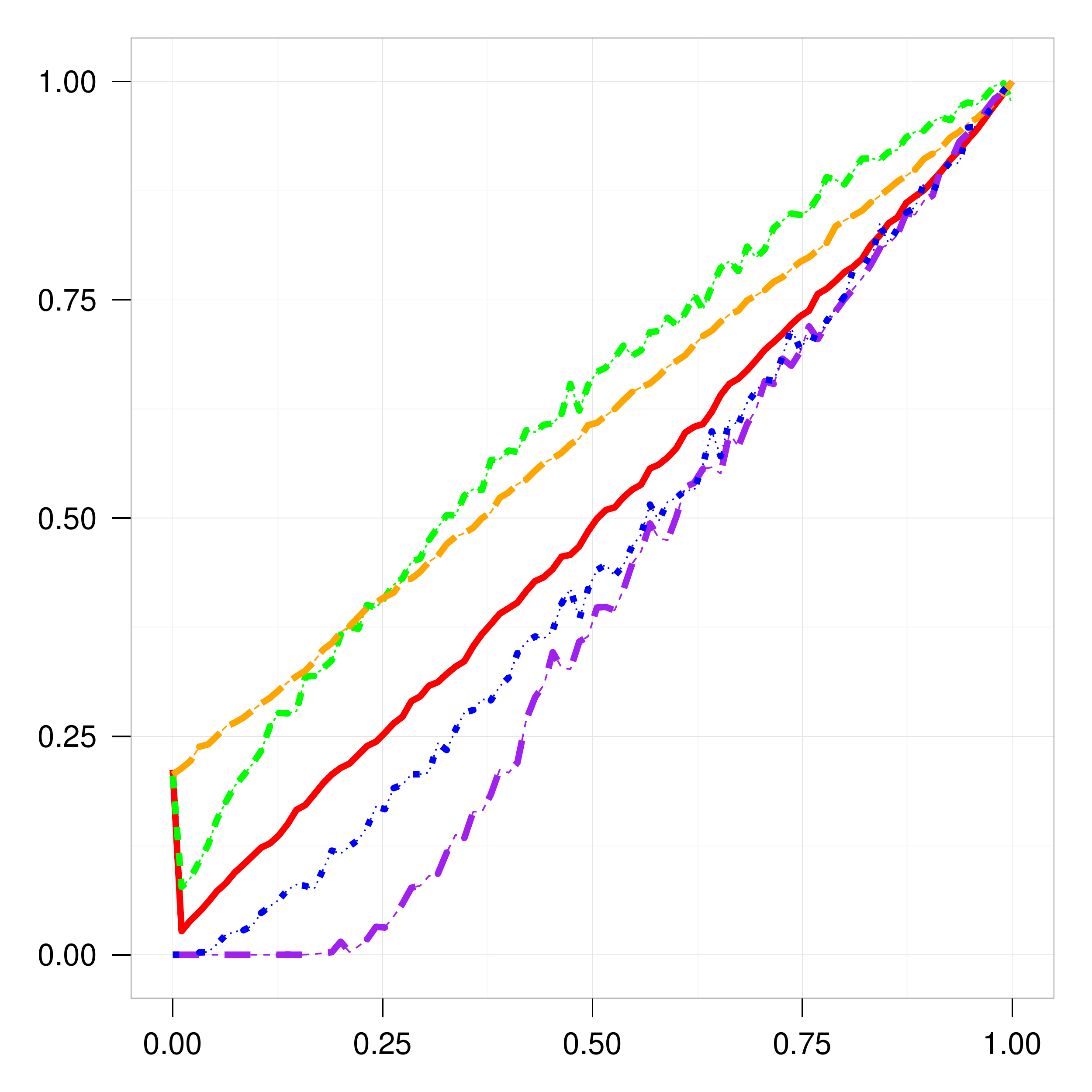}
    & \rotatebox{90}{\hspace{2em}\scriptsize scenario 2}
    \\
    & \multicolumn{4}{c}{\scriptsize False positive rate} \\
    \rotatebox{90}{\hspace{1.25em}\scriptsize True positive rate}
    & \includegraphics[width=.225\linewidth]{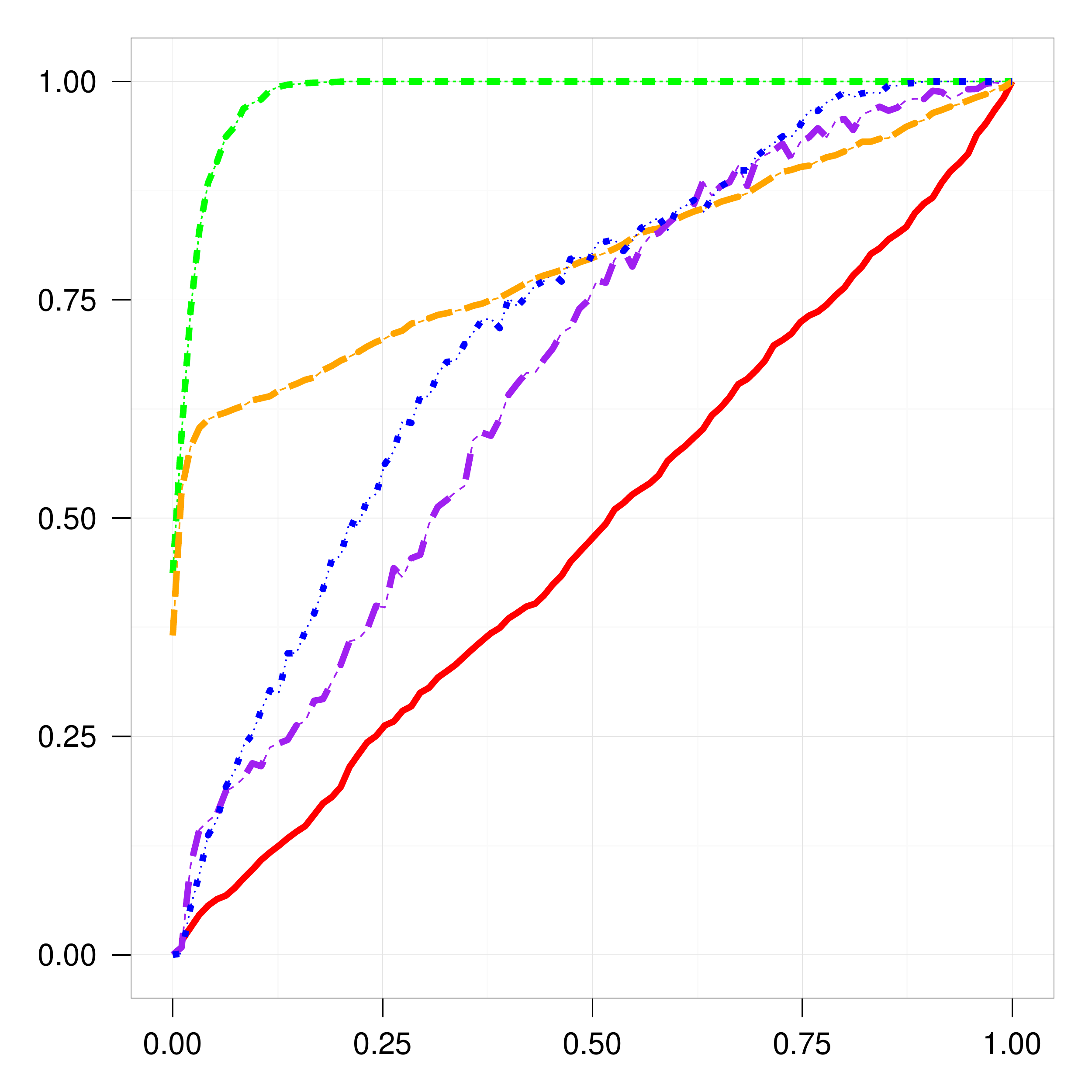}
    & \includegraphics[width=.225\linewidth]{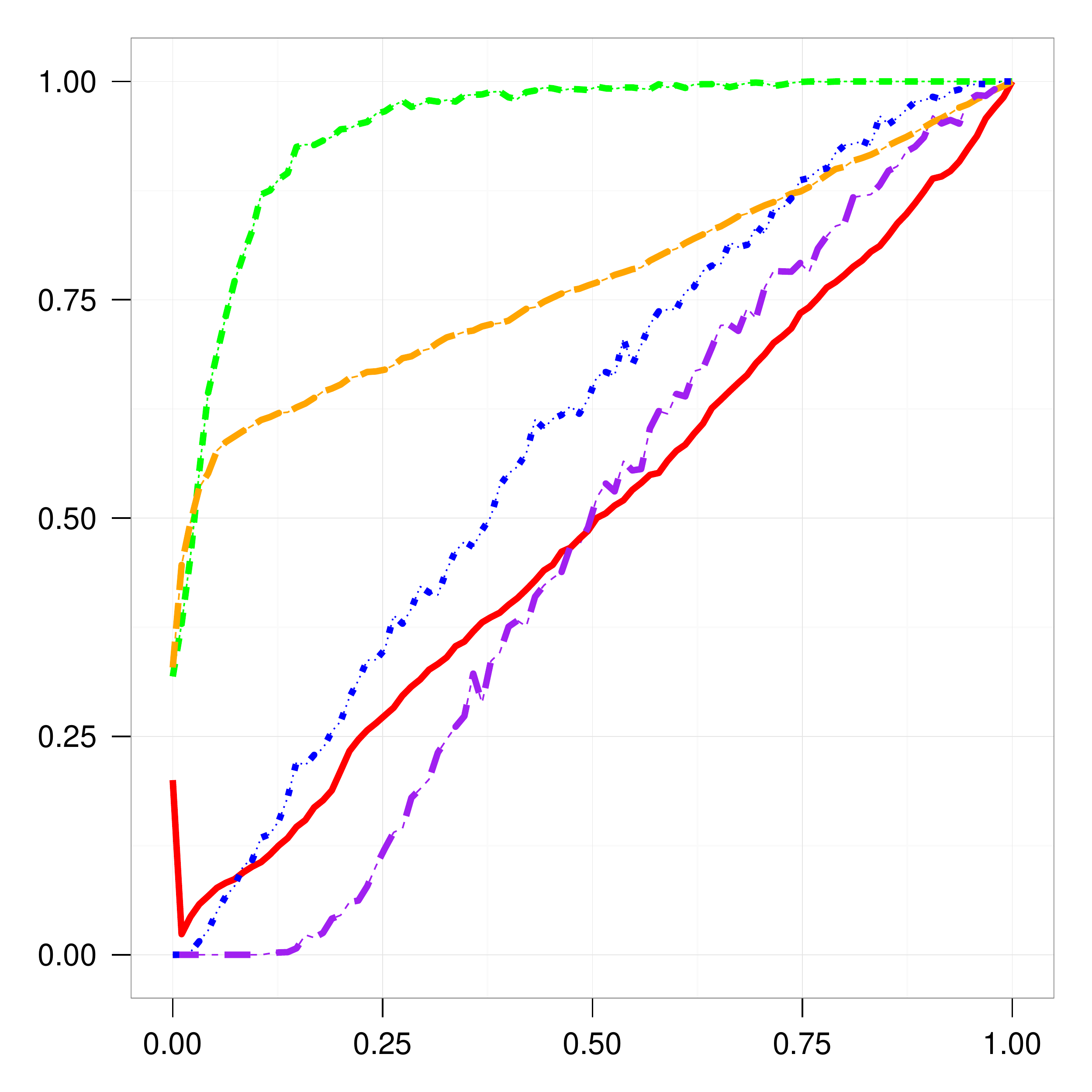}
    & \includegraphics[width=.225\linewidth]{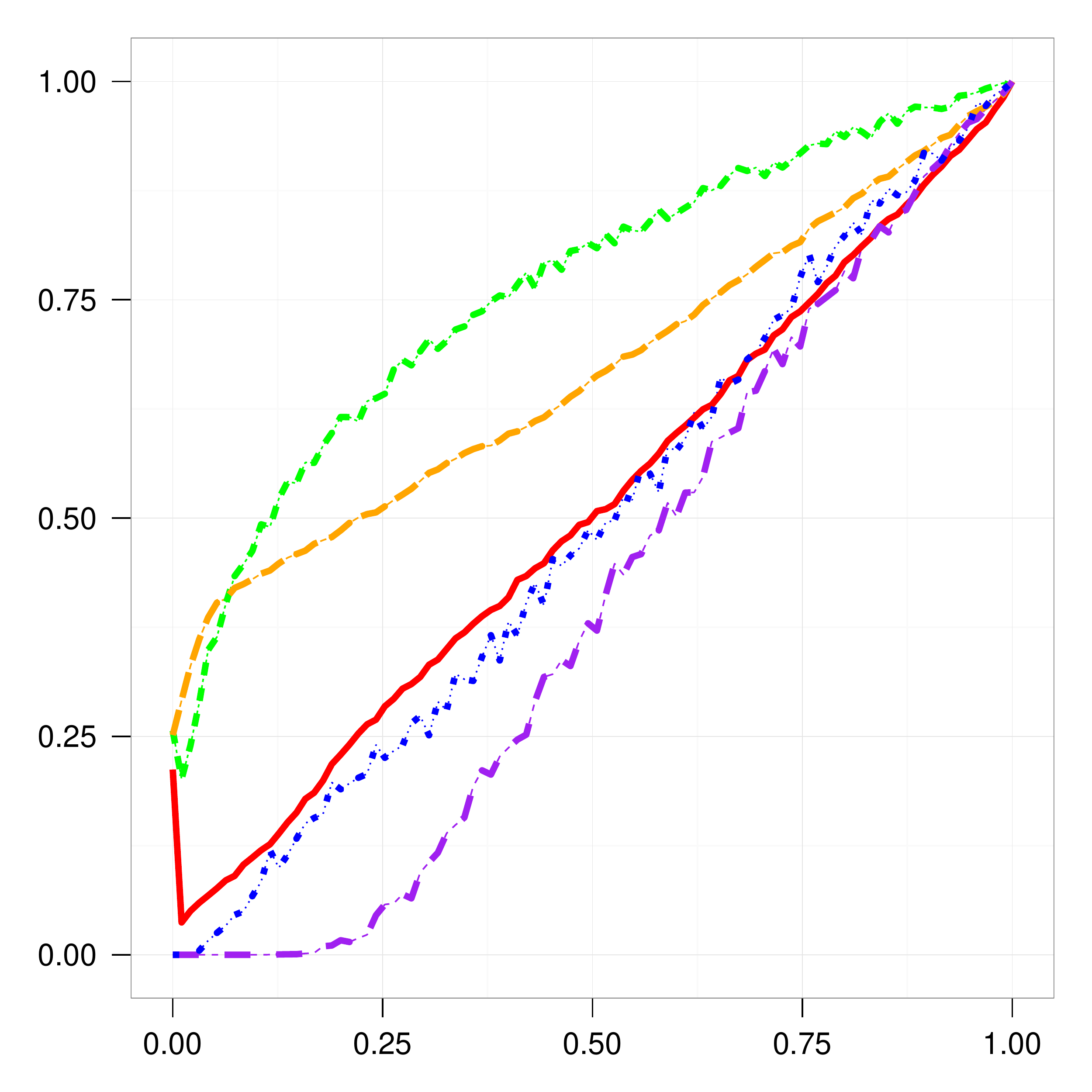}
    & \includegraphics[width=.225\linewidth]{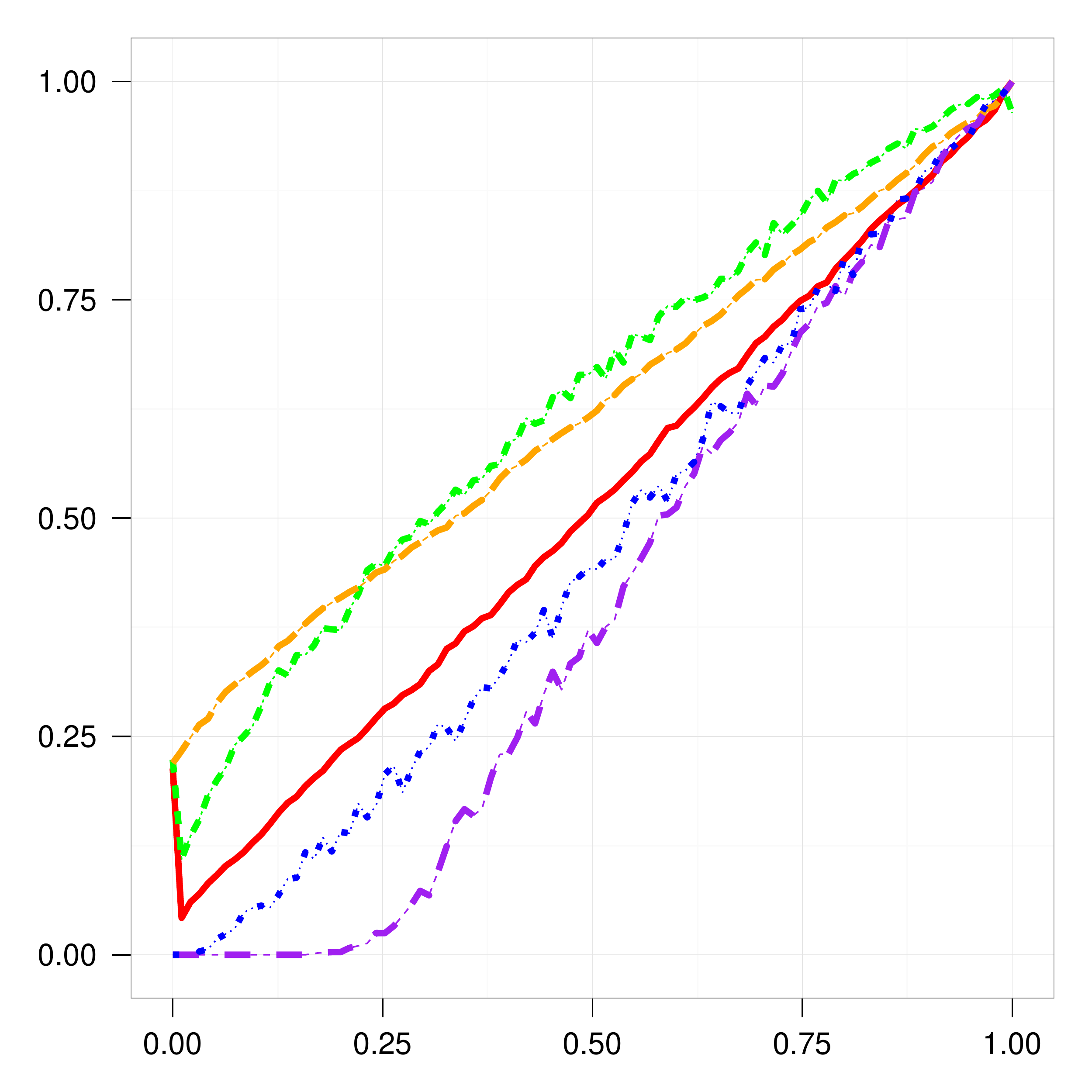}
    & \rotatebox{90}{\hspace{2em}\scriptsize scenario 3}
    \\
    & \multicolumn{4}{c}{\scriptsize False positive rate} \\
    \rotatebox{90}{\hspace{1.25em}\scriptsize True positive rate}
    & \includegraphics[width=.225\linewidth]{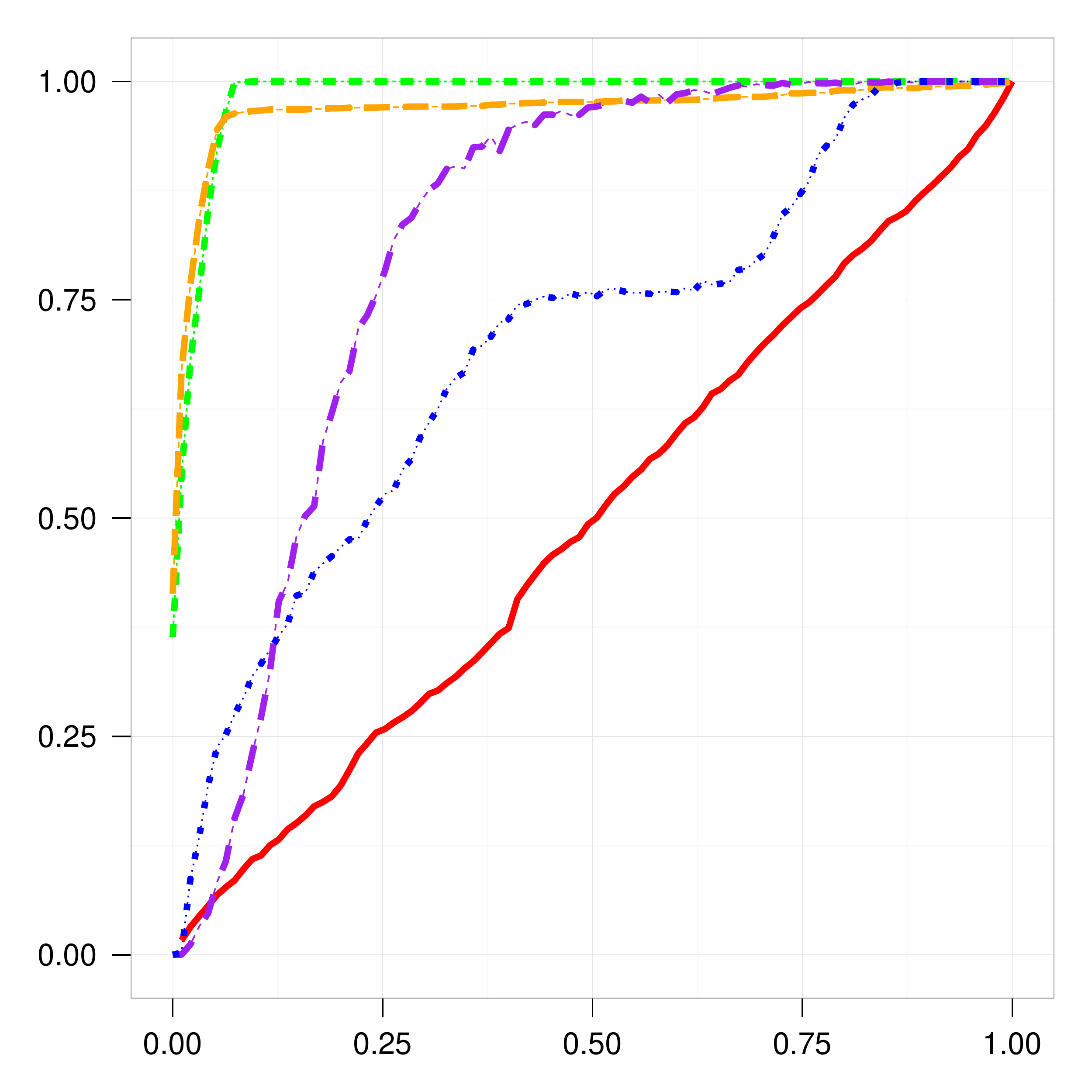}
    & \includegraphics[width=.225\linewidth]{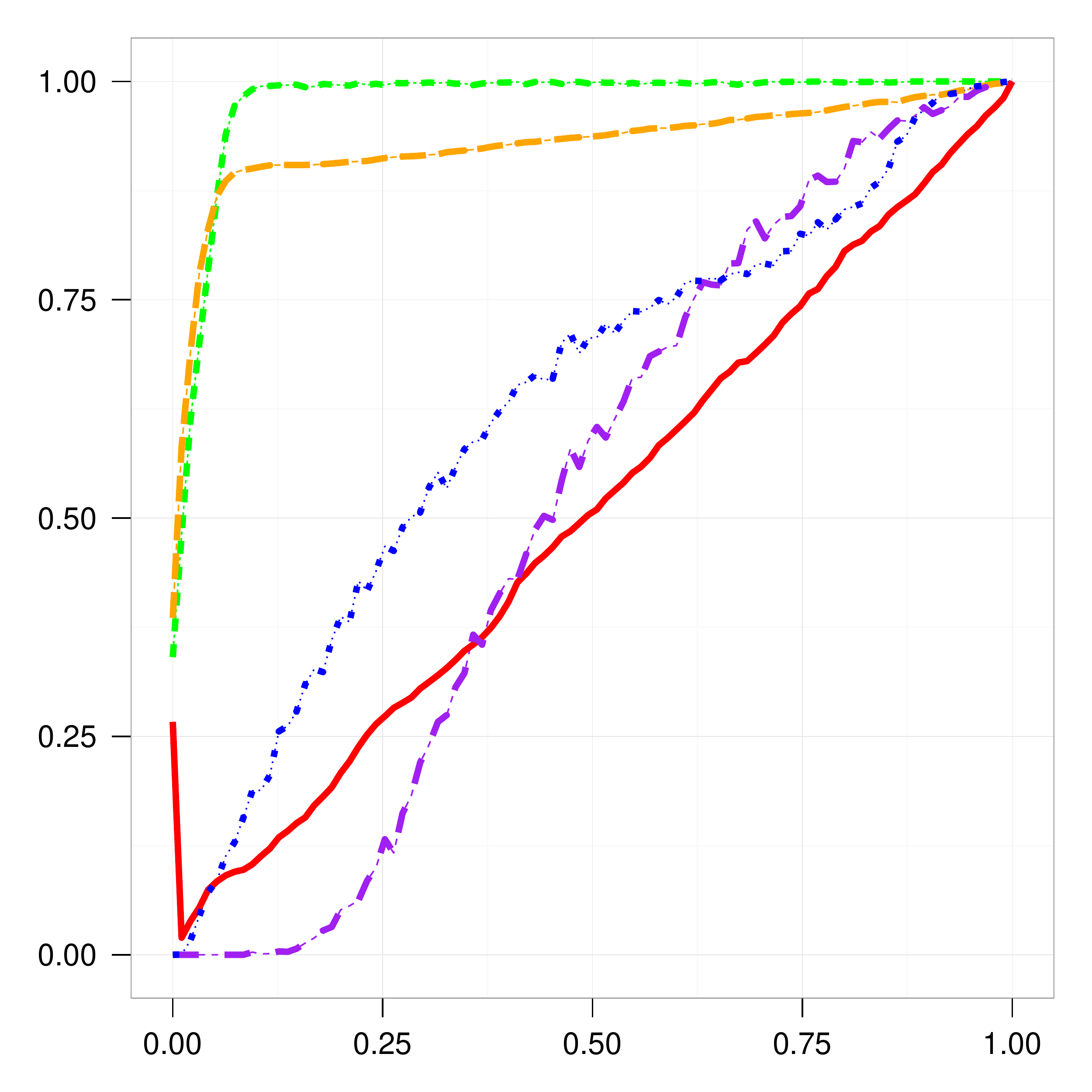}
    & \includegraphics[width=.225\linewidth]{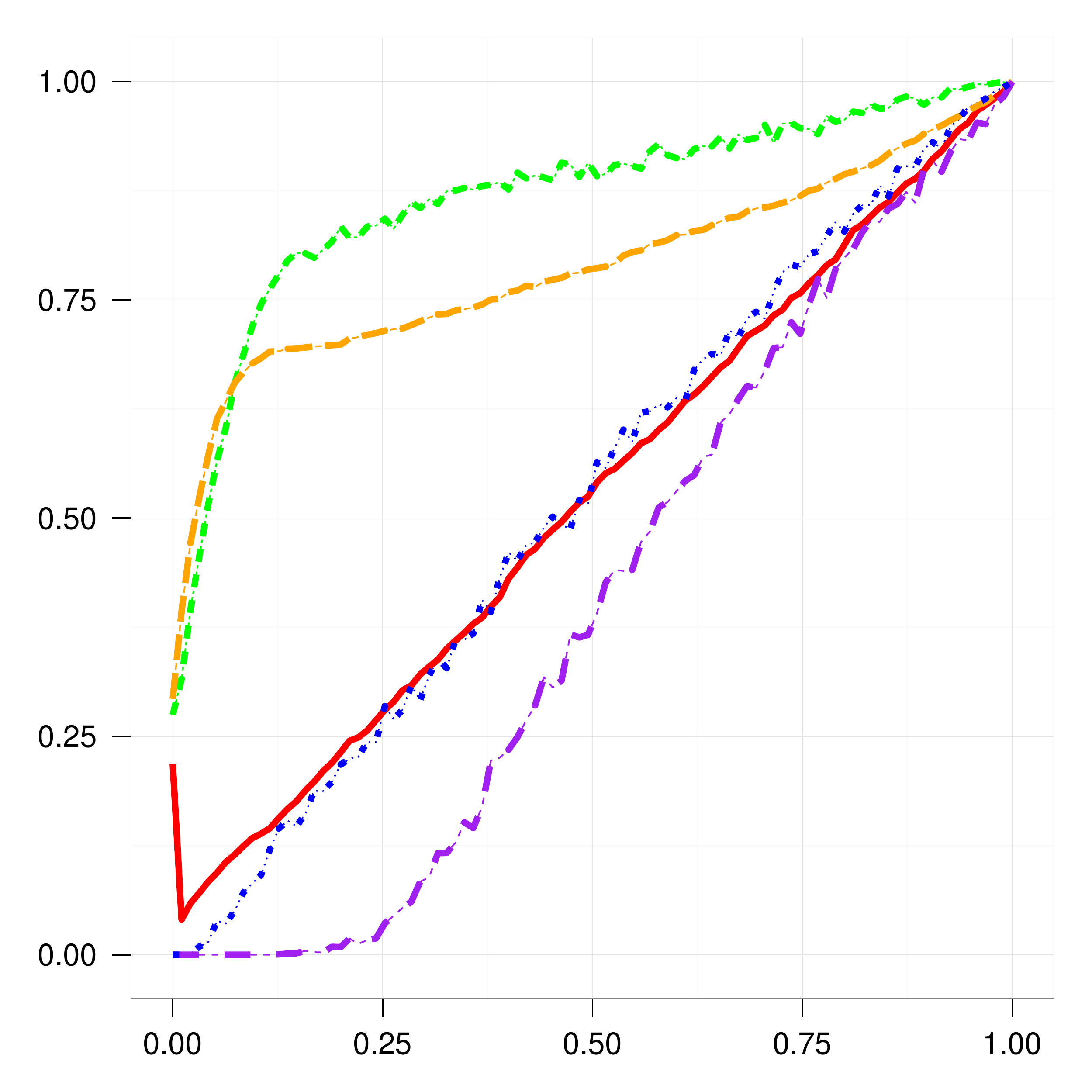}
    & \includegraphics[width=.225\linewidth]{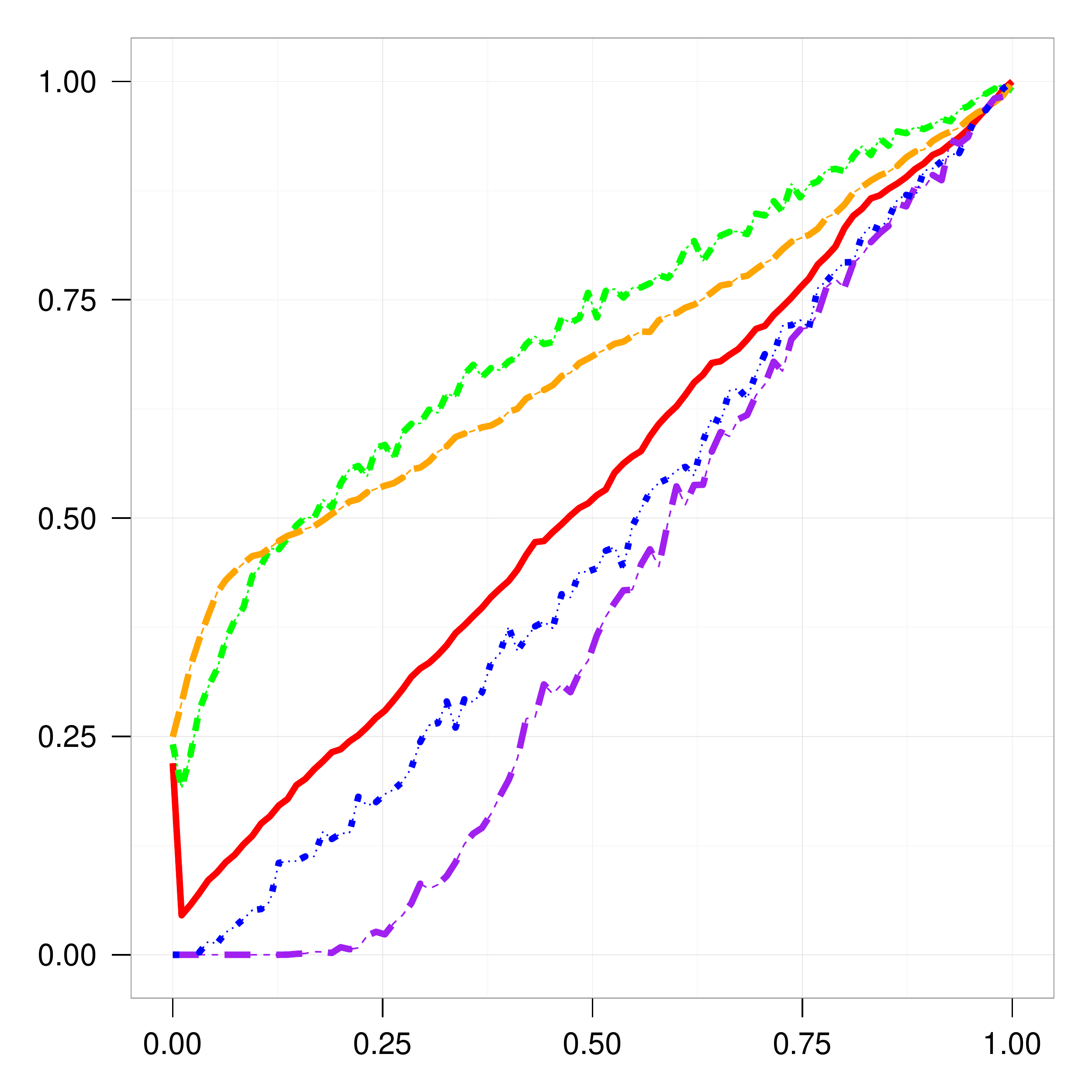} 
    & \rotatebox{90}{\hspace{2em}\scriptsize scenario 4}
    \\
    & \multicolumn{4}{c}{\scriptsize False positive rate} \\
  \end{tabular}
  \caption{ROC  curves for  the  estimated change-points  in rows  for
    \texttt{blockseg} (dotted green), \texttt{HL1} (double-dashed
    purple), \texttt{HL2} (in dotted blue), \texttt{CART}
    (solid red) and \texttt{FL2D} (long-dashed orange).  Each row
    is  associated   to  a scenario  depicted   in  Section
    \ref{sec:sim_settings}.}
  \label{fig:qualitecomparaison}
\end{figure}

\begin{figure}[h!]
  \centering
  \begin{tabular}{cc}
    Scenario 1 & Scenario 2 \\
    \includegraphics[width=.475\linewidth,trim= 1.1cm 1.1cm 1.1cm 1.1cm,clip=true]{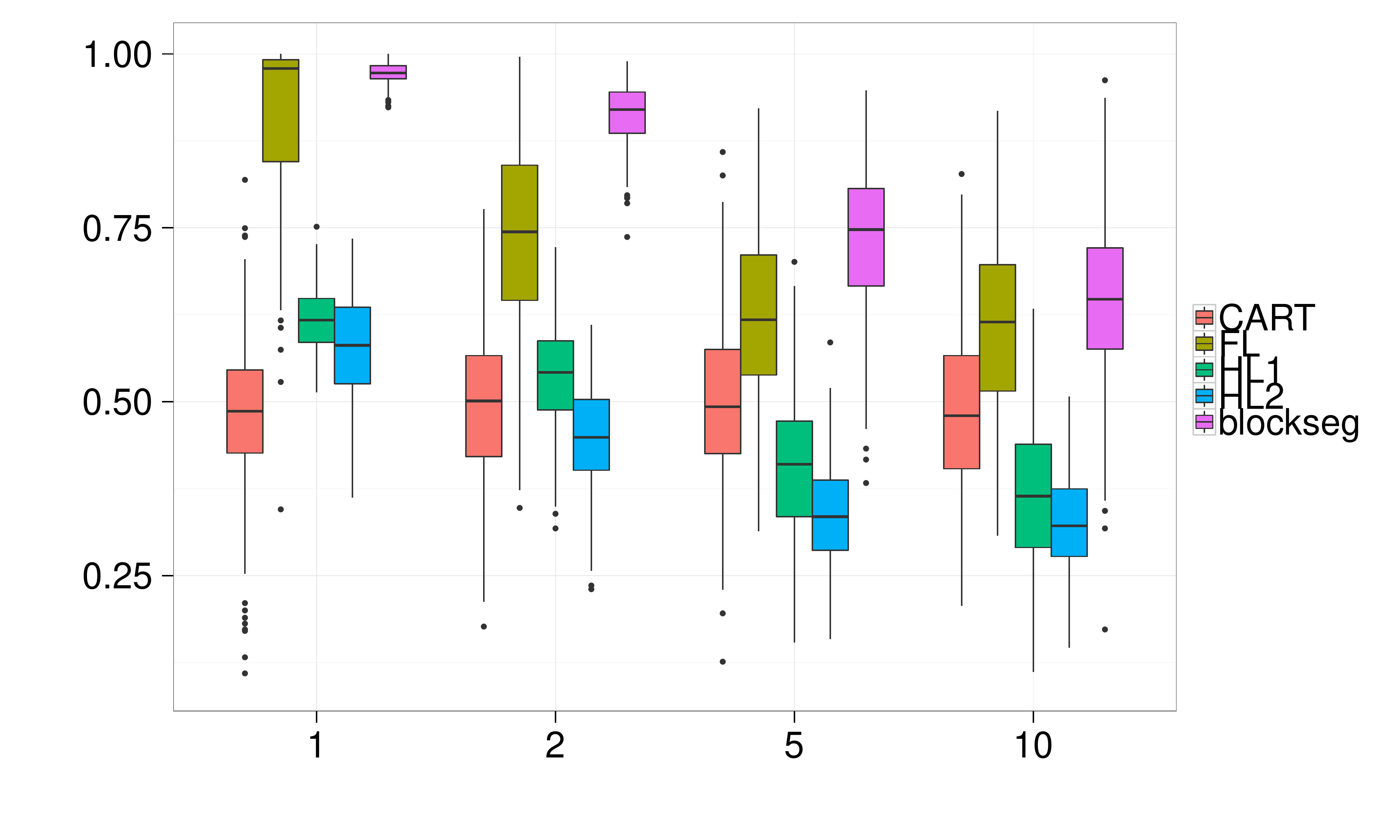}
    & \includegraphics[width=.475\linewidth,trim= 1.1cm 1.1cm 1.1cm 1.1cm,clip=true]{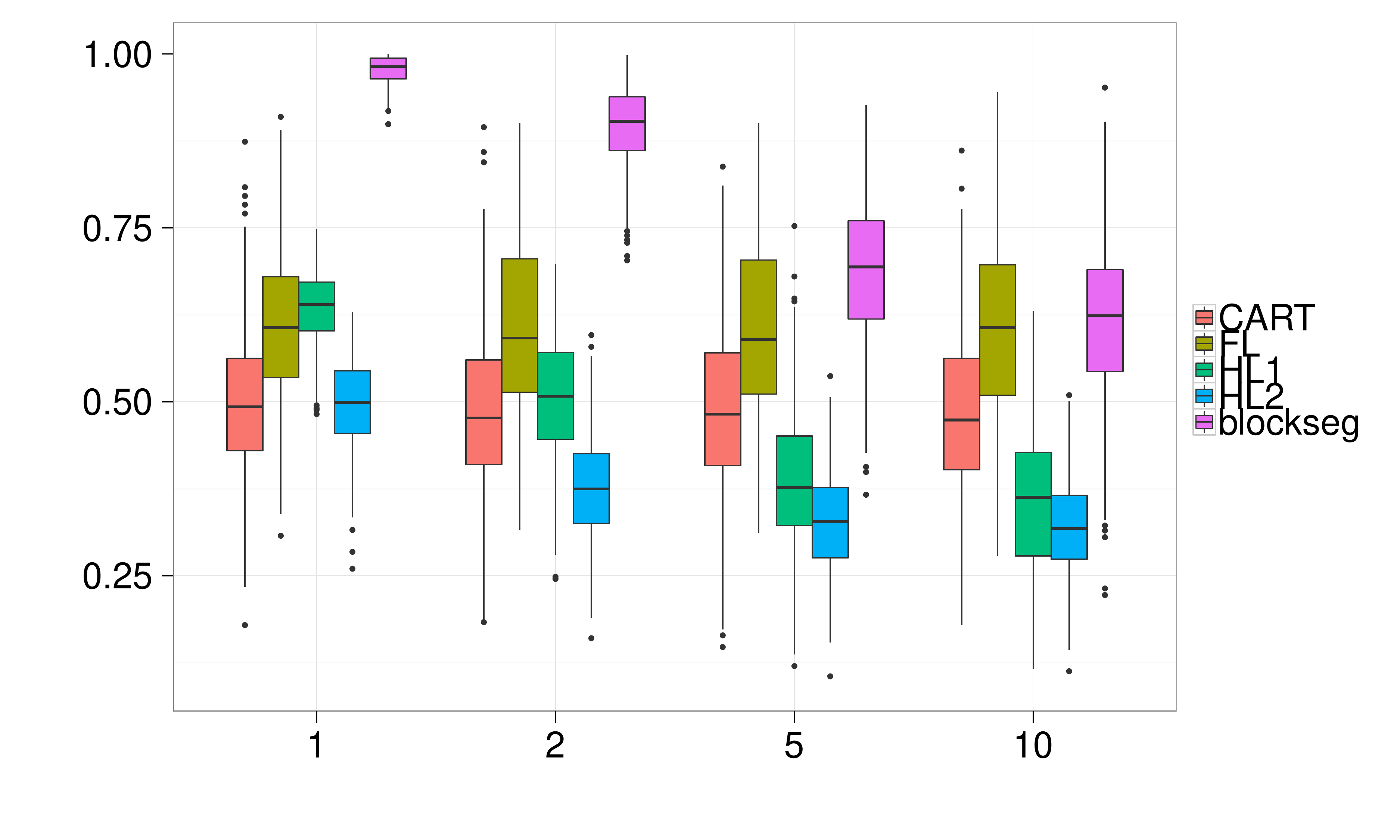}\\
    $\sigma$&$\sigma$\\
    &\\
    Scenario 3 & Scenario 4 \\
    \includegraphics[width=.475\linewidth,trim= 1.1cm 1.1cm 1.1cm 1.1cm,clip=true]{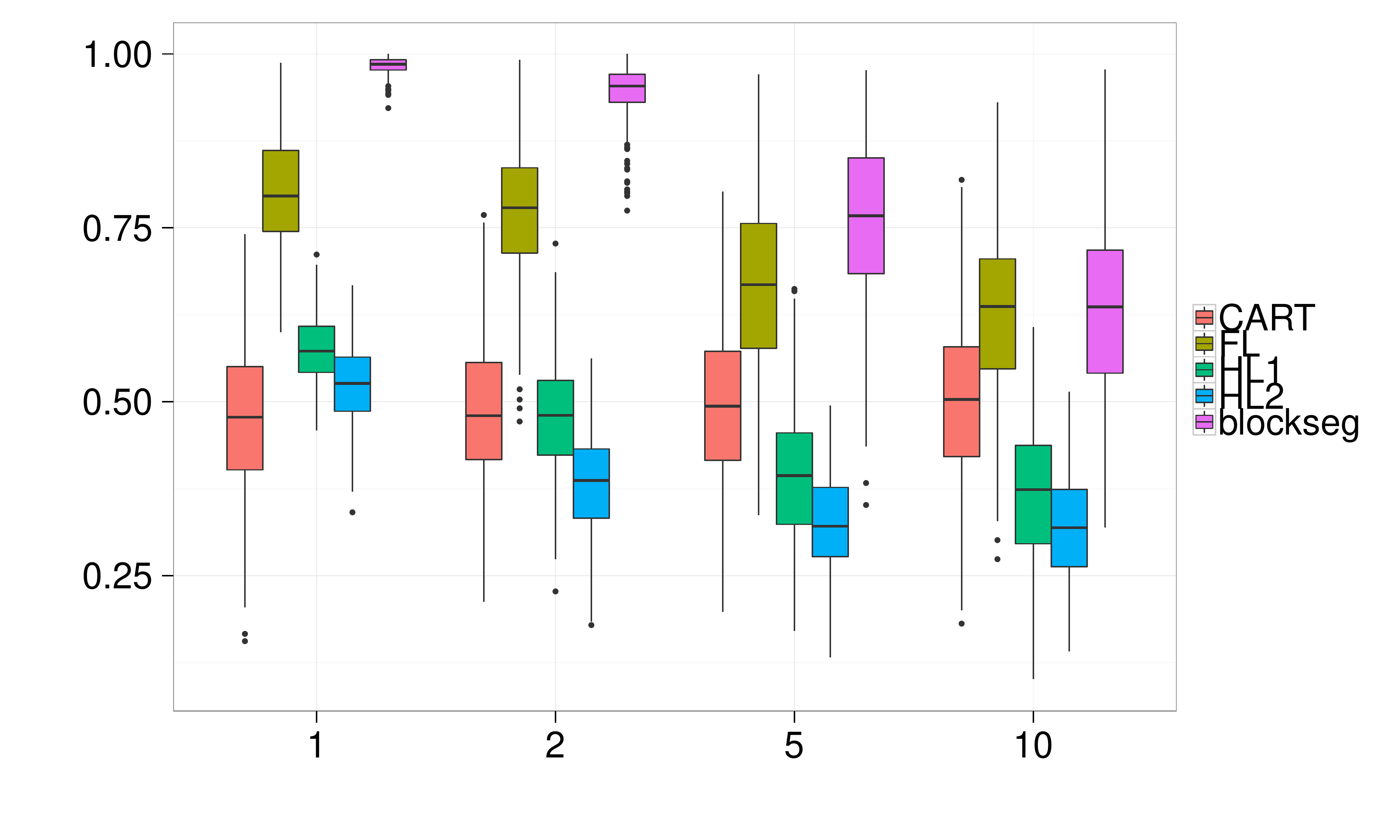}
    & \includegraphics[width=.475\linewidth,trim= 1.1cm 1.1cm 1.1cm 1.1cm,clip=true]{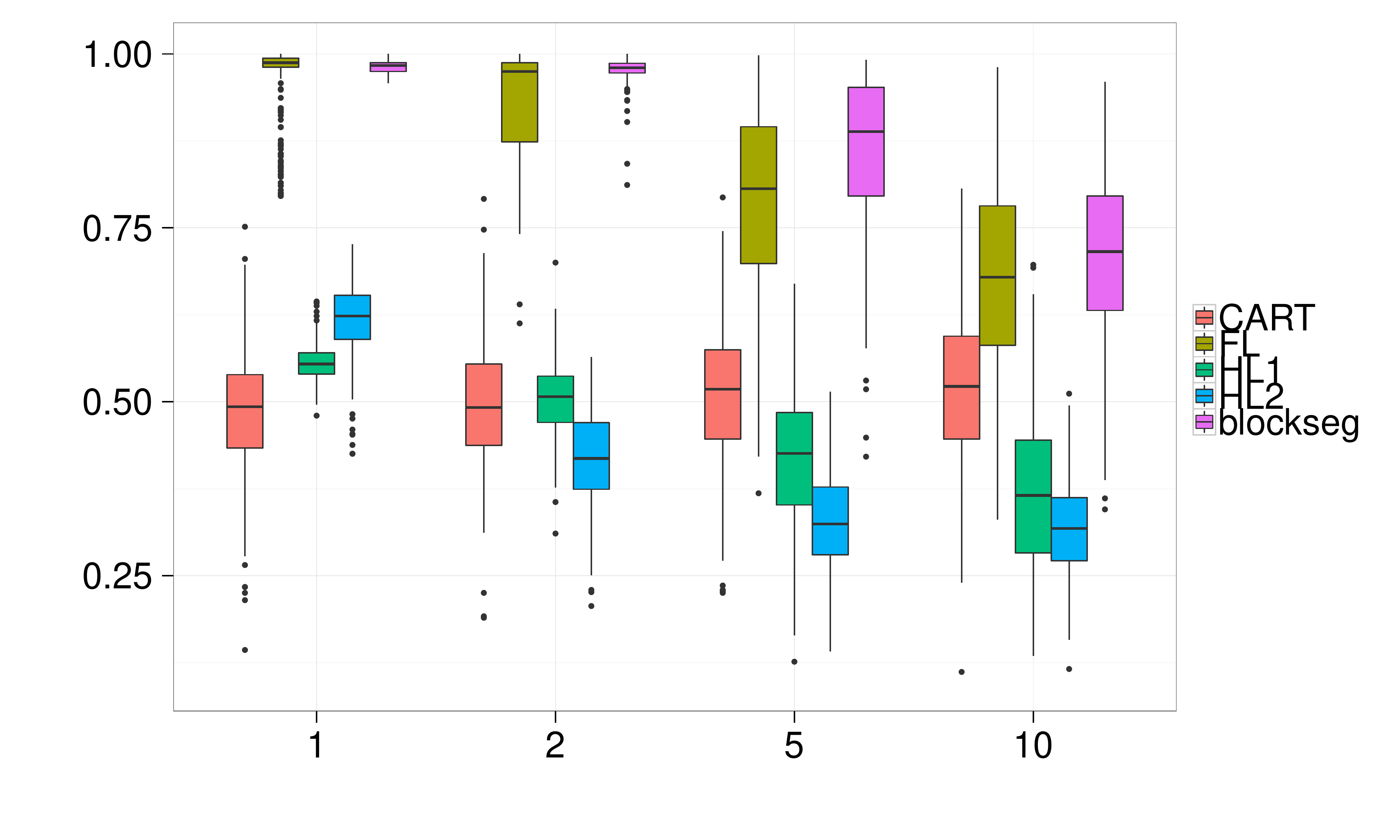}\\
    $\sigma$&$\sigma$\\
  \end{tabular}
  \caption{\label{fig:comp} Boxplots of the area under the ROC curve
    for the different scenarii and the
    different algorithms as a function of the noise variance.}
\end{figure}

Figure~\ref{fig:qualitecomparaison}  shows  the  results in  terms  of
receiver  operating characteristic  (ROC)  curves  for recovering  the
change-points in rows,  averaged over the 1000  runs.  Similar results
hold for the  change-points in columns. This Figure  exhibits the very
good performance  of our method,  which outperforms its  competitors by
retrieving the change-points with a very small error rate even in high
noise level frameworks. Moreover, our method seems to be less sensitive to the
block pattern shape in matrix $\U$ than the other ones.
In order to further assess our approach we give in Figure
\ref{fig:comp} the boxplots of the Area Under Curve (AUC) for the
different ROC curves. We also give in Table \ref{tab:comp} the mean of
the AUC and the associated standard deviation. 

\begin{table}[!h]

\begin{tabular}{c|c|c|c|c||c|c|c|c}
&\multicolumn{4}{c||}{Scenario 1}&\multicolumn{4}{c}{Scenario 2}\\
\cline{2-9}
&$\sigma=1$&$\sigma=2$&$\sigma=5$&$\sigma=10$&$\sigma=1$&$\sigma=2$&$\sigma=5$&$\sigma=10$\\
\hline
\multirow{2}{*}{\tiny blockseg}&\textcolor{red!97!blue}{0.972}&\textcolor{red!91!blue}{0.913}&\textcolor{red!73!blue}{0.733}&\textcolor{red!64!blue}{0.644}&\textcolor{red!97!blue}{0.977}&\textcolor{red!89!blue}{0.896}&\textcolor{red!68!blue}{0.689}&\textcolor{red!61!blue}{0.617}\\
&{\tiny(0.0145)}&{\tiny(0.0421)}&{\tiny(0.0988)}&{\tiny(0.118)}&{\tiny(0.0206)}&{\tiny(0.0555)}&{\tiny(0.107)}&{\tiny(0.123)}\\
\hline
\multirow{2}{*}{\tiny FL2D}&\textcolor{red!91!blue}{0.918}&\textcolor{red!73!blue}{0.738}&\textcolor{red!62!blue}{0.623}&\textcolor{red!60!blue}{0.608}&\textcolor{red!60!blue}{0.608}&\textcolor{red!60!blue}{0.603}&\textcolor{red!60!blue}{0.601}&\textcolor{red!60!blue}{0.603}\\
&{\tiny(0.102)}&{\tiny(0.139)}&{\tiny(0.127)}&{\tiny(0.13)}&{\tiny(0.116)}&{\tiny(0.125)}&{\tiny(0.127)}&{\tiny(0.127)}\\
\hline
\multirow{2}{*}{\tiny HL1}&\textcolor{red!61!blue}{0.618}&\textcolor{red!53!blue}{0.535}&\textcolor{red!40!blue}{0.407}&\textcolor{red!36!blue}{0.363}&\textcolor{red!63!blue}{0.635}&\textcolor{red!50!blue}{0.505}&\textcolor{red!38!blue}{0.382}&\textcolor{red!35!blue}{0.351}\\
&{\tiny(0.0427)}&{\tiny(0.0708)}&{\tiny(0.102)}&{\tiny(0.108)}&{\tiny(0.0535)}&{\tiny(0.0874)}&{\tiny(0.105)}&{\tiny(0.107)}\\
\hline
\multirow{2}{*}{\tiny HL2}&\textcolor{red!57!blue}{0.576}&\textcolor{red!44!blue}{0.448}&\textcolor{red!33!blue}{0.337}&\textcolor{red!32!blue}{0.323}&\textcolor{red!49!blue}{0.498}&\textcolor{red!37!blue}{0.374}&\textcolor{red!32!blue}{0.326}&\textcolor{red!31!blue}{0.317}\\
&{\tiny(0.0744)}&{\tiny(0.0713)}&{\tiny(0.0734)}&{\tiny(0.072)}&{\tiny(0.0653)}&{\tiny(0.0777)}&{\tiny(0.0727)}&{\tiny(0.0745)}\\
\hline
\multirow{2}{*}{\tiny CART}&\textcolor{red!48!blue}{0.482}&\textcolor{red!49!blue}{0.497}&\textcolor{red!49!blue}{0.498}&\textcolor{red!48!blue}{0.486}&\textcolor{red!49!blue}{0.496}&\textcolor{red!48!blue}{0.487}&\textcolor{red!49!blue}{0.491}&\textcolor{red!48!blue}{0.484}\\
&{\tiny(0.107)}&{\tiny(0.107)}&{\tiny(0.117)}&{\tiny(0.119)}&{\tiny(0.112)}&{\tiny(0.124)}&{\tiny(0.126)}&{\tiny(0.118)}\\
\multicolumn{9}{c}{}\\
&\multicolumn{4}{c||}{Scenario 3}&\multicolumn{4}{c}{Scenario 4}\\
\cline{2-9}
&$\sigma=1$&$\sigma=2$&$\sigma=5$&$\sigma=10$&$\sigma=1$&$\sigma=2$&$\sigma=5$&$\sigma=10$\\
\hline
\multirow{2}{*}{\tiny blockseg}&\textcolor{red!98!blue}{0.983}&\textcolor{red!94!blue}{0.945}&\textcolor{red!75!blue}{0.758}&\textcolor{red!63!blue}{0.63}&\textcolor{red!98!blue}{0.983}&\textcolor{red!97!blue}{0.977}&\textcolor{red!86!blue}{0.866}&\textcolor{red!70!blue}{0.707}\\
&{\tiny(0.0114)}&{\tiny(0.0391)}&{\tiny(0.113)}&{\tiny(0.125)}&{\tiny(0.00927)}&{\tiny(0.0179)}&{\tiny(0.102)}&{\tiny(0.124)}\\
\hline
\multirow{2}{*}{\tiny FL2D}&\textcolor{red!79!blue}{0.799}&\textcolor{red!77!blue}{0.772}&\textcolor{red!66!blue}{0.667}&\textcolor{red!62!blue}{0.623}&\textcolor{red!96!blue}{0.969}&\textcolor{red!93!blue}{0.931}&\textcolor{red!78!blue}{0.789}&\textcolor{red!68!blue}{0.68}\\
&{\tiny(0.0855)}&{\tiny(0.0956)}&{\tiny(0.121)}&{\tiny(0.121)}&{\tiny(0.051)}&{\tiny(0.0722)}&{\tiny(0.135)}&{\tiny(0.134)}\\
\hline
\multirow{2}{*}{\tiny HL1}&\textcolor{red!57!blue}{0.575}&\textcolor{red!47!blue}{0.479}&\textcolor{red!39!blue}{0.391}&\textcolor{red!36!blue}{0.368}&\textcolor{red!55!blue}{0.556}&\textcolor{red!50!blue}{0.504}&\textcolor{red!41!blue}{0.418}&\textcolor{red!36!blue}{0.368}\\
&{\tiny(0.0458)}&{\tiny(0.0819)}&{\tiny(0.0981)}&{\tiny(0.105)}&{\tiny(0.0252)}&{\tiny(0.0514)}&{\tiny(0.0974)}&{\tiny(0.11)}\\
\hline
\multirow{2}{*}{\tiny HL2}&\textcolor{red!52!blue}{0.524}&\textcolor{red!38!blue}{0.384}&\textcolor{red!32!blue}{0.326}&\textcolor{red!31!blue}{0.319}&\textcolor{red!61!blue}{0.616}&\textcolor{red!41!blue}{0.416}&\textcolor{red!32!blue}{0.327}&\textcolor{red!31!blue}{0.316}\\
&{\tiny(0.0612)}&{\tiny(0.0711)}&{\tiny(0.0716)}&{\tiny(0.0738)}&{\tiny(0.0527)}&{\tiny(0.0696)}&{\tiny(0.0714)}&{\tiny(0.067)}\\
\hline
\multirow{2}{*}{\tiny CART}&\textcolor{red!47!blue}{0.474}&\textcolor{red!48!blue}{0.485}&\textcolor{red!49!blue}{0.495}&\textcolor{red!50!blue}{0.502}&\textcolor{red!48!blue}{0.484}&\textcolor{red!49!blue}{0.493}&\textcolor{red!51!blue}{0.512}&\textcolor{red!51!blue}{0.516}\\
&{\tiny(0.106)}&{\tiny(0.11)}&{\tiny(0.114)}&{\tiny(0.115)}&{\tiny(0.0905)}&{\tiny(0.0889)}&{\tiny(0.0985)}&{\tiny(0.111)}\\
\end{tabular}

\caption{\label{tab:comp} Mean and standard deviation of the area
  under the ROC curve for the different scenarii, different
  algorithms and different values of the noise variance.}
\end{table}

\begin{figure}[h!]
  \centering
  \begin{tabular}{cc}
    $\sigma=1$ & $\sigma=2$ \\
    \includegraphics[width=.475\linewidth,trim= 1.1cm 1.1cm 1.1cm 1.1cm,clip=true]{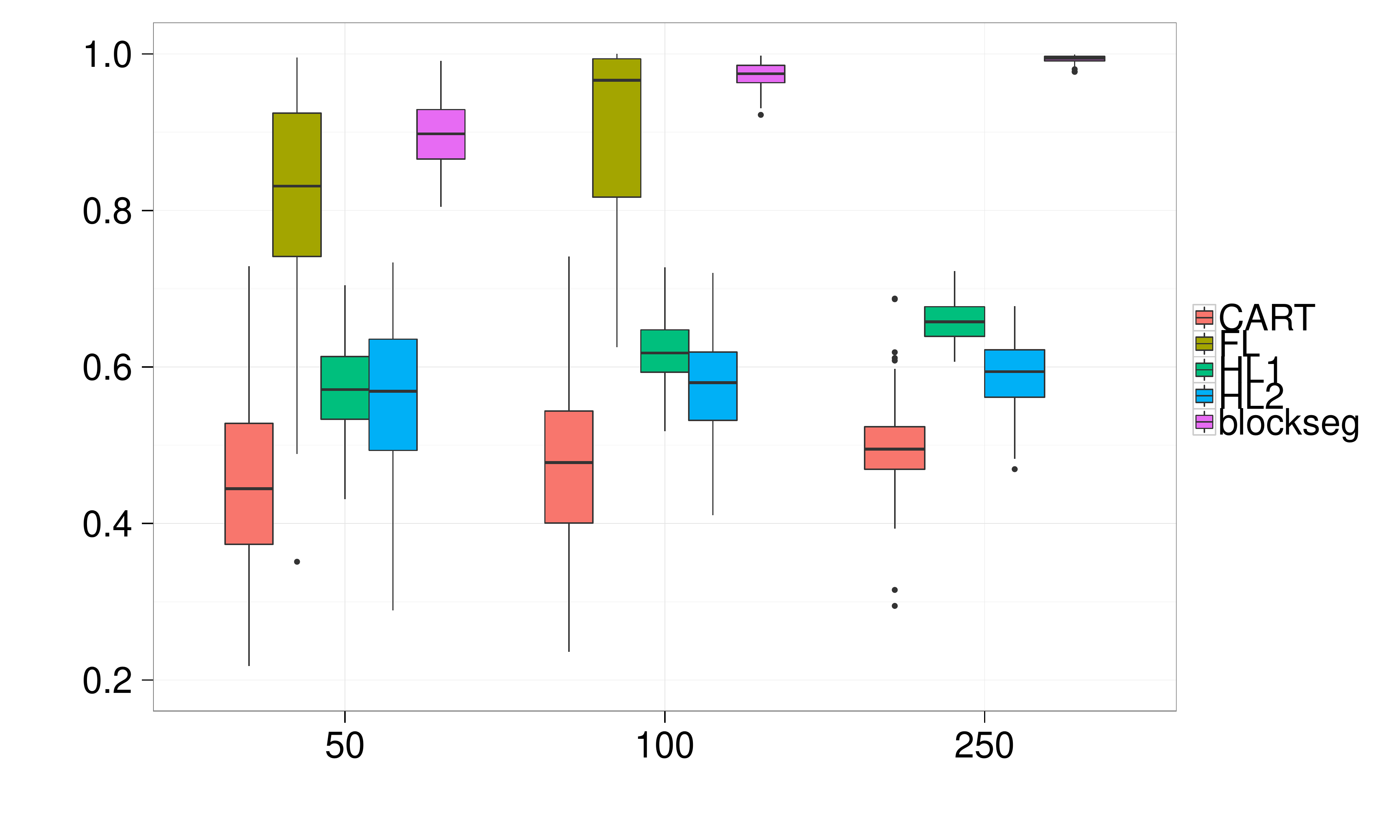}
    & \includegraphics[width=.475\linewidth,trim= 1.1cm 1.1cm 1.1cm 1.1cm,clip=true]{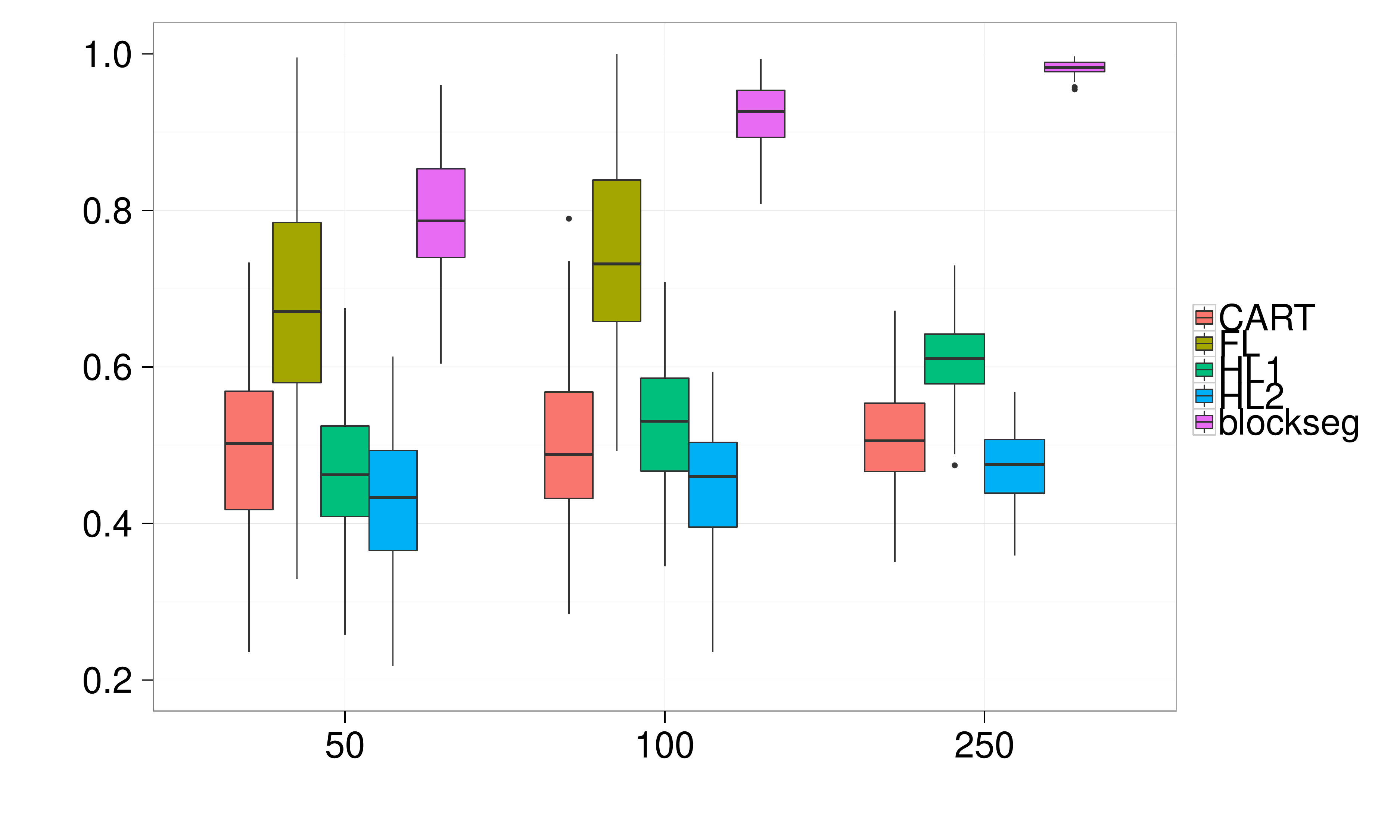}\\
    $n$&$n$\\
    &\\
    $\sigma=5$ & $\sigma=10$ \\
    \includegraphics[width=.475\linewidth,trim= 1.1cm 1.1cm 1.1cm 1.1cm,clip=true]{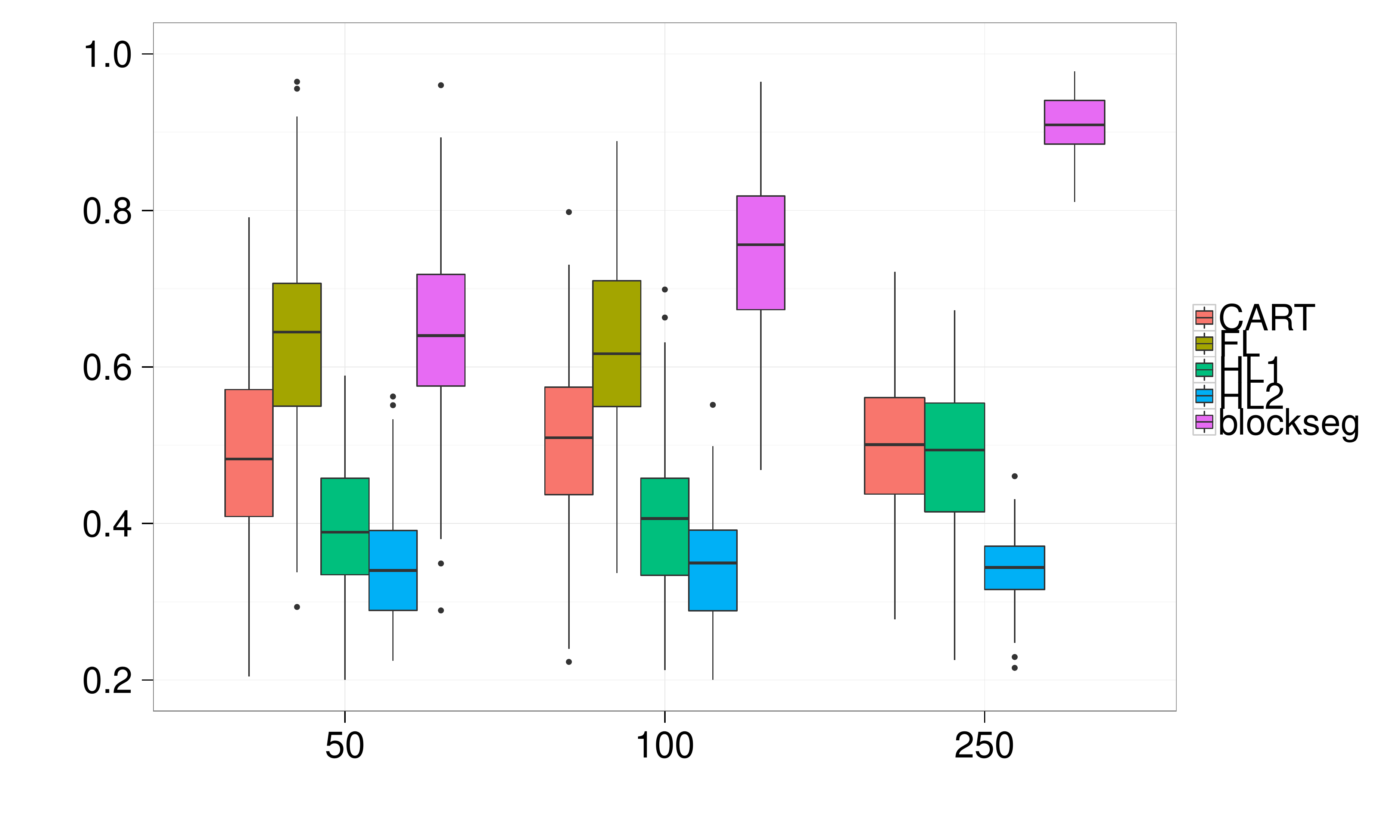}
    & \includegraphics[width=.475\linewidth,trim= 1.1cm 1.1cm 1.1cm 1.1cm,clip=true]{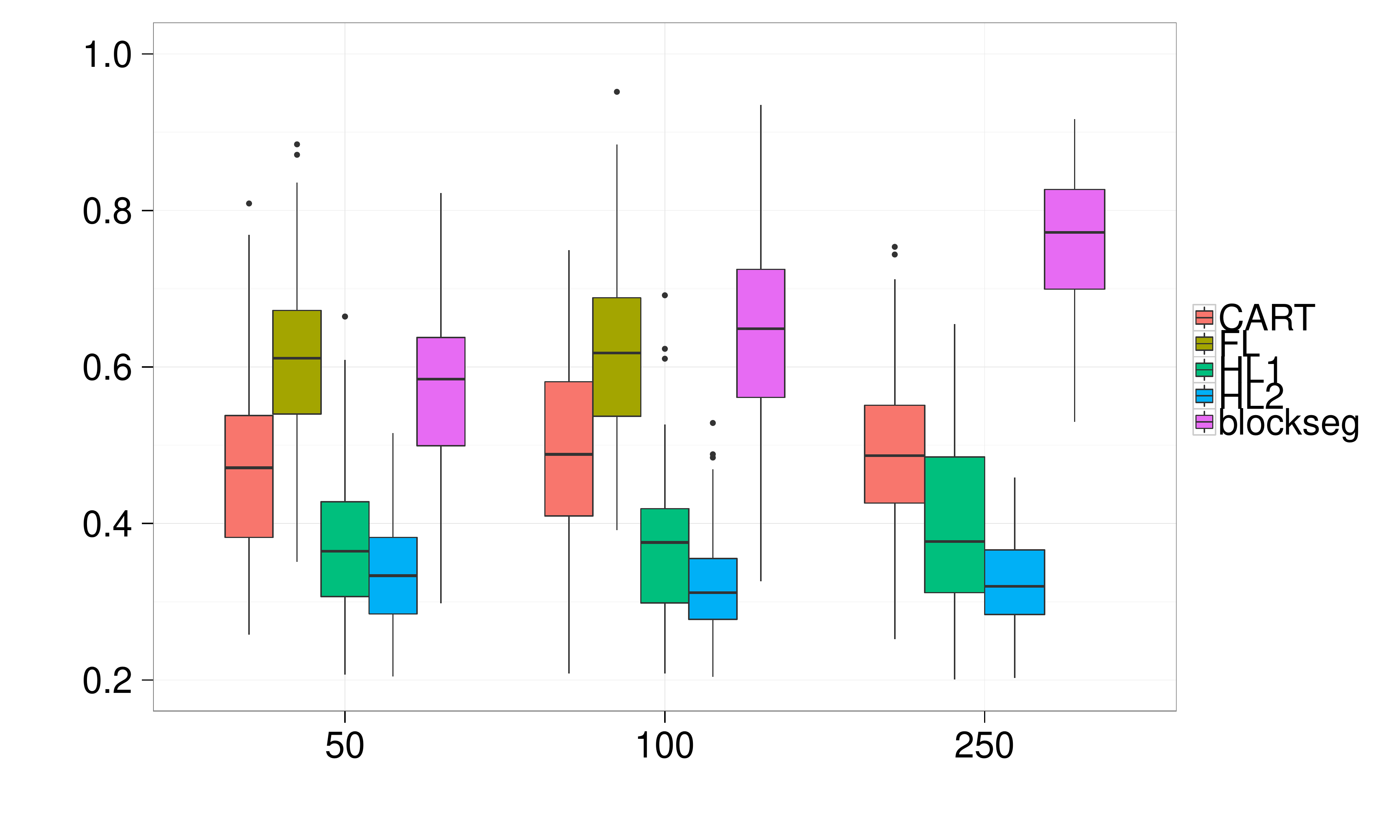}\\
    $n$&$n$\\
  \end{tabular}
  \caption{\label{fig:compN} Boxplots of the area under the ROC curve
    as a function of the standard deviation of the noise, the size of
    the matrices and the methods.}
\end{figure}

In order to further compare the different approaches we generated
matrices $\Y$ satisfying Model (\ref{eq:model1}) with  a ``checkerboard'' matrix
$\left({\mu}_{k,\ell}^{\star,(1)}\right)$  given in
(\ref{eq:model:mustar}) for $n\in\{50,100,250\}$. We observe from
Table \ref{tab:compN} that the performance of our method are on a par
with those of \texttt{FL2D} for $n=50$ and 100. However, for $n=250$
the computational burden of \texttt{FL2D} is so large that the results are not
available, see the blue crosses in Table \ref{tab:compN}. The AUC are
also displayed with boxplots in Figure \ref{fig:compN}.

\begin{table}[!h]

\begin{tabular}{c|c|c|c||c|c|c}
&\multicolumn{3}{c||}{$\sigma=1$}&\multicolumn{3}{c}{$\sigma=2$}\\
\cline{2-7}
&$\n=50$&$\n=100$&$\n=250$&$\n=50$&$\n=100$&$\n=250$\\
\hline
\multirow{2}{*}{\tiny blockseg}&\textcolor{red!89!blue}{0.896}&\textcolor{red!97!blue}{0.972}&\textcolor{red!99!blue}{0.993}&\textcolor{red!79!blue}{0.791}&\textcolor{red!92!blue}{0.923}&\textcolor{red!98!blue}{0.982}\\
&{\tiny(0.0425)}&{\tiny(0.0162)}&{\tiny(0.00463)}&{\tiny(0.0789)}&{\tiny(0.0398)}&{\tiny(0.00865)}\\
\hline
\multirow{2}{*}{\tiny FL2D}&\textcolor{red!81!blue}{0.814}&\textcolor{red!90!blue}{0.906}&\multirow{2}{*}{\textcolor{red!0!blue}{\Large X}}&\textcolor{red!67!blue}{0.679}&\textcolor{red!75!blue}{0.753}&\multirow{2}{*}{\textcolor{red!0!blue}{\Large X}}\\
&{\tiny(0.132)}&{\tiny(0.0997)}&&{\tiny(0.133)}&{\tiny(0.128)}&\\
\hline
\multirow{2}{*}{\tiny HL1}&\textcolor{red!57!blue}{0.574}&\textcolor{red!61!blue}{0.619}&\textcolor{red!66!blue}{0.66}&\textcolor{red!46!blue}{0.467}&\textcolor{red!52!blue}{0.527}&\textcolor{red!61!blue}{0.611}\\
&{\tiny(0.0598)}&{\tiny(0.0426)}&{\tiny(0.0255)}&{\tiny(0.0899)}&{\tiny(0.084)}&{\tiny(0.0513)}\\
\hline
\multirow{2}{*}{\tiny HL2}&\textcolor{red!56!blue}{0.56}&\textcolor{red!57!blue}{0.573}&\textcolor{red!59!blue}{0.59}&\textcolor{red!42!blue}{0.424}&\textcolor{red!45!blue}{0.451}&\textcolor{red!47!blue}{0.472}\\
&{\tiny(0.101)}&{\tiny(0.0642)}&{\tiny(0.0432)}&{\tiny(0.0972)}&{\tiny(0.0713)}&{\tiny(0.0467)}\\
\hline
\multirow{2}{*}{\tiny CART}&\textcolor{red!44!blue}{0.445}&\textcolor{red!47!blue}{0.479}&\textcolor{red!49!blue}{0.498}&\textcolor{red!48!blue}{0.487}&\textcolor{red!48!blue}{0.487}&\textcolor{red!51!blue}{0.512}\\
&{\tiny(0.123)}&{\tiny(0.108)}&{\tiny(0.0589)}&{\tiny(0.125)}&{\tiny(0.114)}&{\tiny(0.0708)}\\
\multicolumn{7}{c}{}\\
&\multicolumn{3}{c||}{$\sigma=5$}&\multicolumn{3}{c}{$\sigma=10$}\\
\cline{2-7}
&$\n=50$&$\n=100$&$\n=250$&$\n=50$&$\n=100$&$\n=250$\\
\hline
\multirow{2}{*}{\tiny blockseg}&\textcolor{red!64!blue}{0.646}&\textcolor{red!73!blue}{0.739}&\textcolor{red!91!blue}{0.91}&\textcolor{red!57!blue}{0.577}&\textcolor{red!64!blue}{0.642}&\textcolor{red!76!blue}{0.766}\\
&{\tiny(0.127)}&{\tiny(0.11)}&{\tiny(0.0394)}&{\tiny(0.112)}&{\tiny(0.124)}&{\tiny(0.0867)}\\
\hline
\multirow{2}{*}{\tiny FL2D}&\textcolor{red!63!blue}{0.631}&\textcolor{red!62!blue}{0.629}&\multirow{2}{*}{\textcolor{red!0!blue}{\Large X}}&\textcolor{red!60!blue}{0.602}&\textcolor{red!61!blue}{0.616}&\multirow{2}{*}{\textcolor{red!0!blue}{\Large X}}\\
&{\tiny(0.132)}&{\tiny(0.125)}&&{\tiny(0.118)}&{\tiny(0.115)}&\\
\hline
\multirow{2}{*}{\tiny HL1}&\textcolor{red!38!blue}{0.382}&\textcolor{red!39!blue}{0.397}&\textcolor{red!48!blue}{0.481}&\textcolor{red!36!blue}{0.364}&\textcolor{red!35!blue}{0.35}&\textcolor{red!38!blue}{0.386}\\
&{\tiny(0.106)}&{\tiny(0.107)}&{\tiny(0.0909)}&{\tiny(0.103)}&{\tiny(0.108)}&{\tiny(0.115)}\\
\hline
\multirow{2}{*}{\tiny HL2}&\textcolor{red!33!blue}{0.333}&\textcolor{red!34!blue}{0.342}&\textcolor{red!34!blue}{0.341}&\textcolor{red!32!blue}{0.325}&\textcolor{red!31!blue}{0.313}&\textcolor{red!31!blue}{0.317}\\
&{\tiny(0.0905)}&{\tiny(0.0775)}&{\tiny(0.0451)}&{\tiny(0.083)}&{\tiny(0.0729)}&{\tiny(0.0539)}\\
\hline
\multirow{2}{*}{\tiny CART}&\textcolor{red!48!blue}{0.488}&\textcolor{red!50!blue}{0.501}&\textcolor{red!49!blue}{0.497}&\textcolor{red!46!blue}{0.466}&\textcolor{red!48!blue}{0.483}&\textcolor{red!48!blue}{0.48}\\
&{\tiny(0.115)}&{\tiny(0.119)}&{\tiny(0.0917)}&{\tiny(0.129)}&{\tiny(0.131)}&{\tiny(0.117)}\\
\end{tabular}

\caption{\label{tab:compN} Mean and standard deviation of the area
  under the ROC curve as a function of the standard deviation of the
  noise, the algorithms and the size of the matrices.
The crosses correspond to cases where the results are not available.}
\end{table}

\section{Model selection}\label{sec:model_selection}



In the previous experiments we did not need to explain how to choose the number of estimated change-points since we used ROC curves for
comparing the methodologies. However, in real data applications, it is necessary to propose a methodology for estimating the number of
change-points. This is what we explain in the following.

In practice, we take $s=K_{\textrm{max}}^2$  where $K_{\textrm{max}}$ is an
upper bound for $\Ks_1$ and $\Ks_2$. For choosing the final change-points we
shall adapt the well-known \textit{stability selection} approach
devised by \cite{meinshausen2010stability}.
More precisely, we randomly choose $M$ times $n/2$ columns and $n/2$ rows of
the matrix $\Y$ and for each subsample
we select $s=K_{\textrm{max}}^2$ active variables. Finally, after the $M$ data resamplings, we keep the change-points
which appear a number of times larger than a given threshold. By the definition of the change-points given in (\ref{eq:t_hat}), a change-point $\that_{1,k}$ or
$\that_{2,\ell}$ may appear several times in a given set of resampled observations. Hence, the score associated with each change-point corresponds
to the sum of the number of times it appears in each of the $M$ subsamplings.

\begin{figure}[!h]
  \centering
  \begin{tabular}{@{}l@{}ccc}
    & $\sigma=1$ & $\sigma=2$
    & $\sigma=5$ \\
    & \includegraphics[width=.225\linewidth, height=2cm]{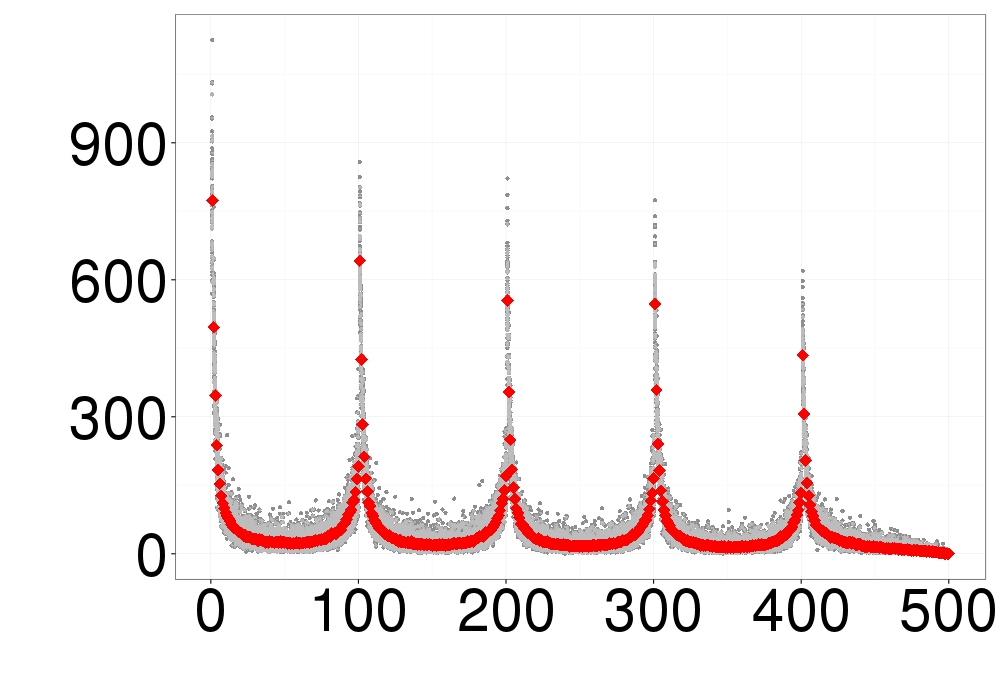}
    & \includegraphics[width=.225\linewidth, height=2cm]{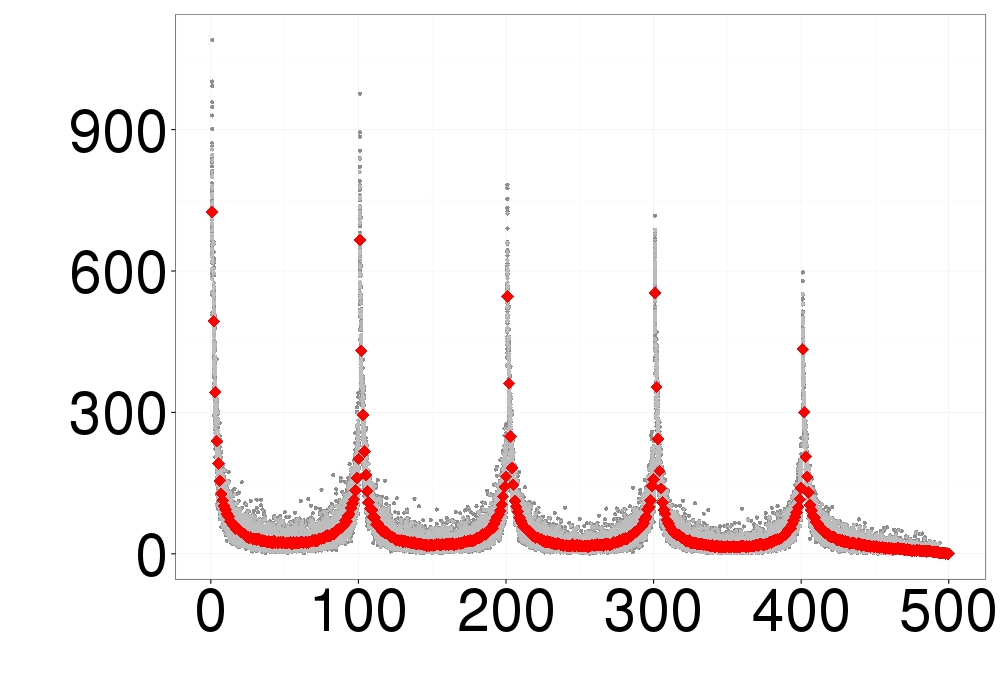}
    & \includegraphics[width=.225\linewidth, height=2cm]{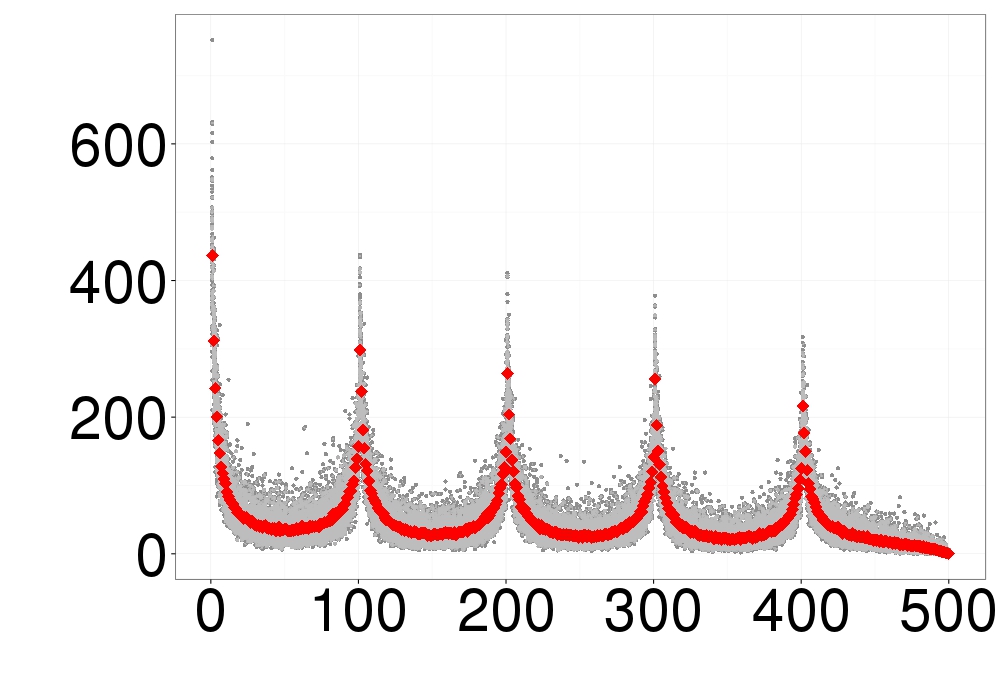} \\
    & \includegraphics[width=.225\linewidth, height=2cm]{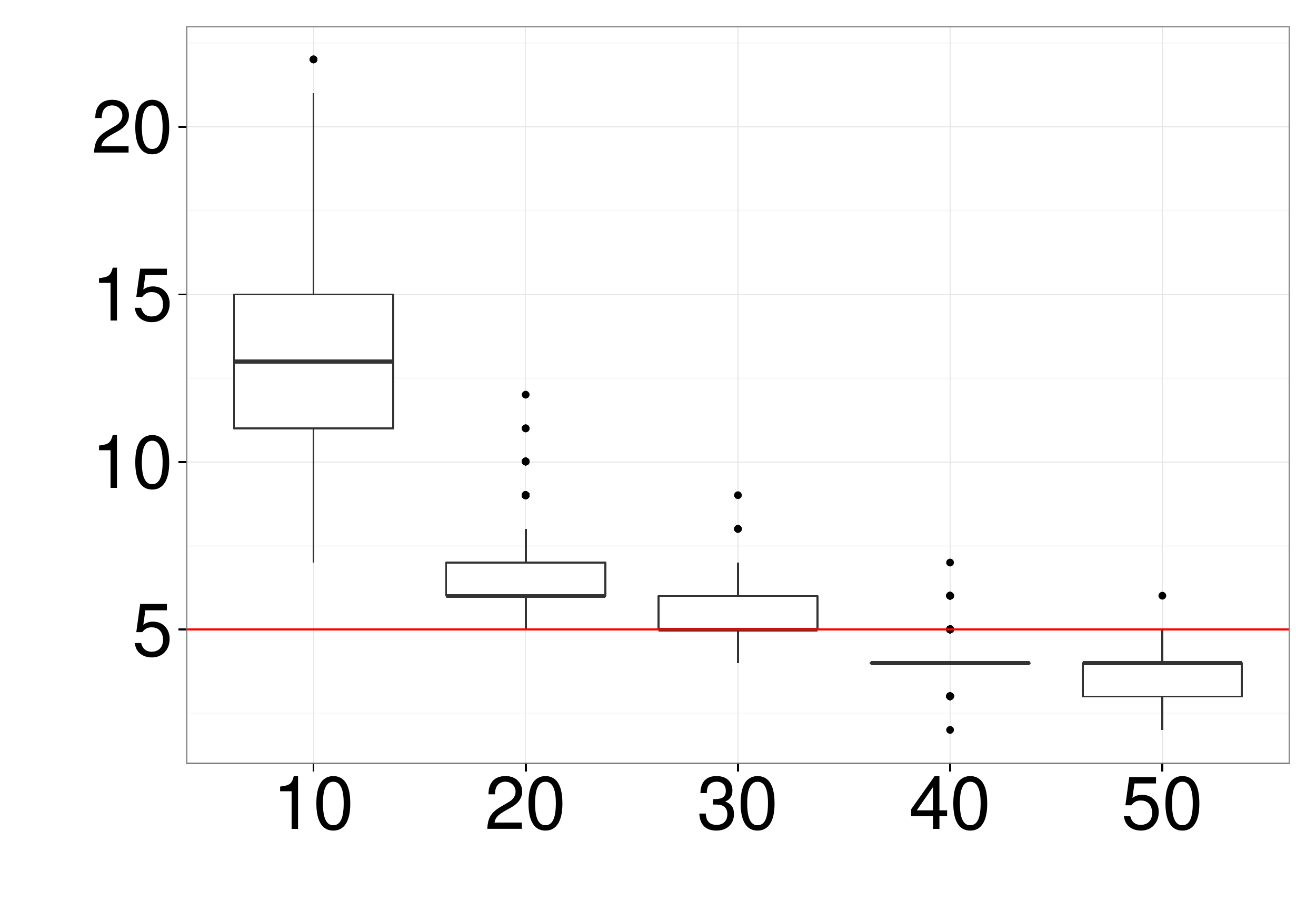}
    & \includegraphics[width=.225\linewidth, height=2cm]{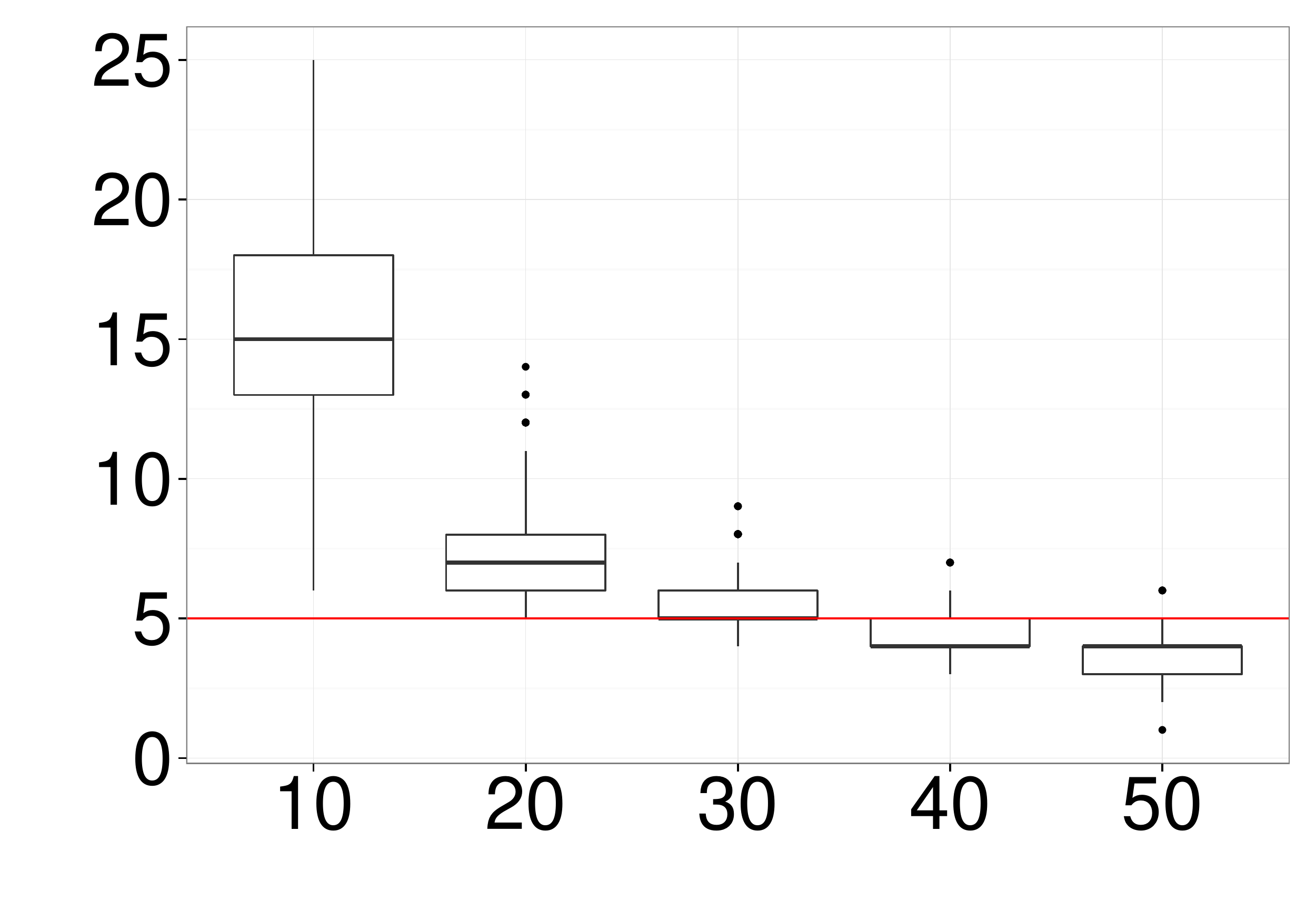}
    & \includegraphics[width=.225\linewidth, height=2cm]{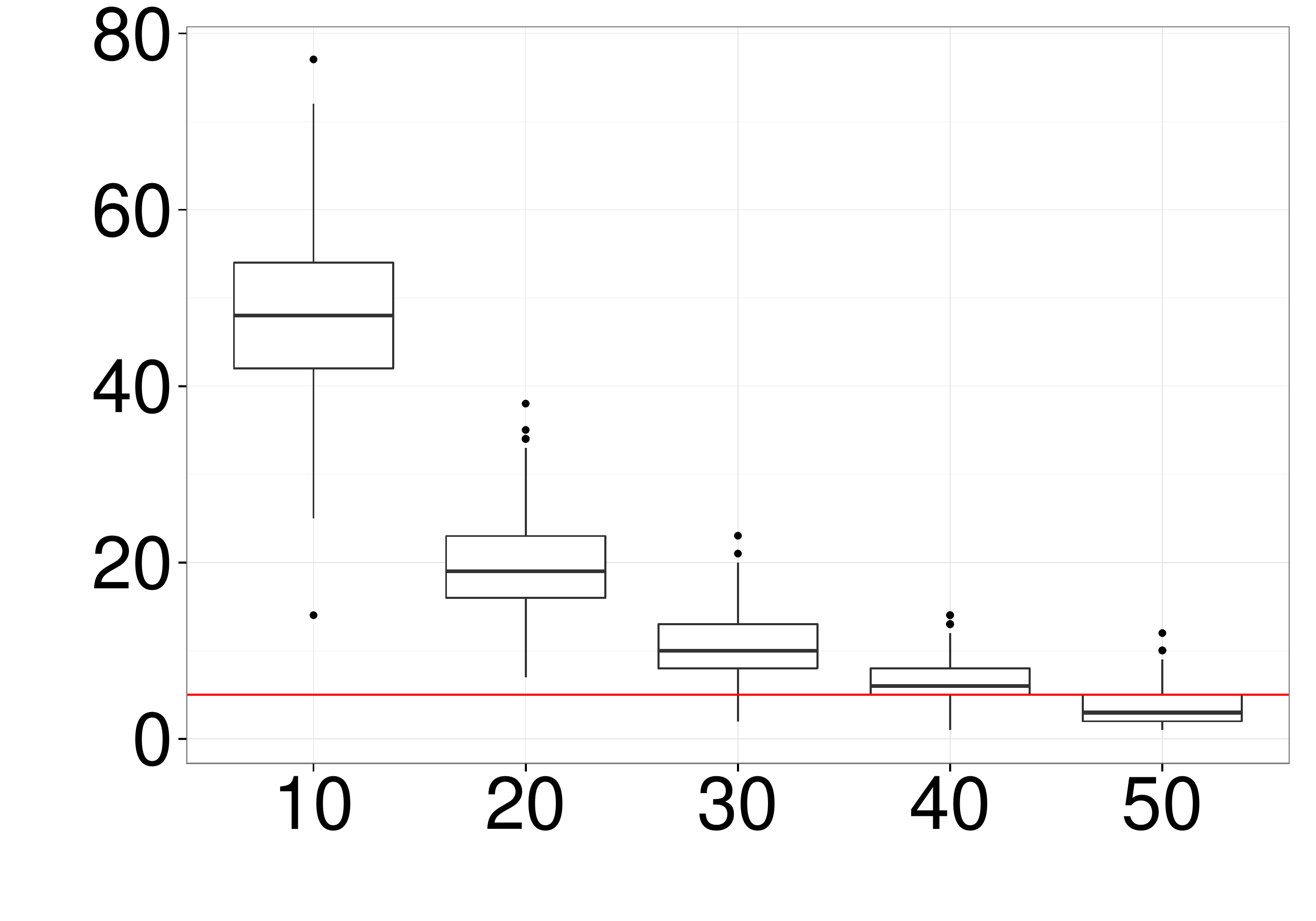} \\
  \end{tabular}
  \caption{\label{fig:SelectionModel}
Top: Scores associated to each estimated change-points for different
values of $\sigma$; the true change-point positions in rows and columns are located at 101, 201, 301 and 401.
Bottom:  Boxplots of the estimation of $\Ks_1$ for different values of $\sigma$ and thresh after the post-processing step.
}
\end{figure}

\begin{figure}[!h]
  \centering
  \begin{tabular}{@{}l@{}ccc}
    & $\sigma=1$ & $\sigma=2$
    & $\sigma=5$ \\
    \rotatebox{90}{\hspace{2.2em}\small 50$\%$}
    & \includegraphics[width=.225\linewidth, height=2cm]{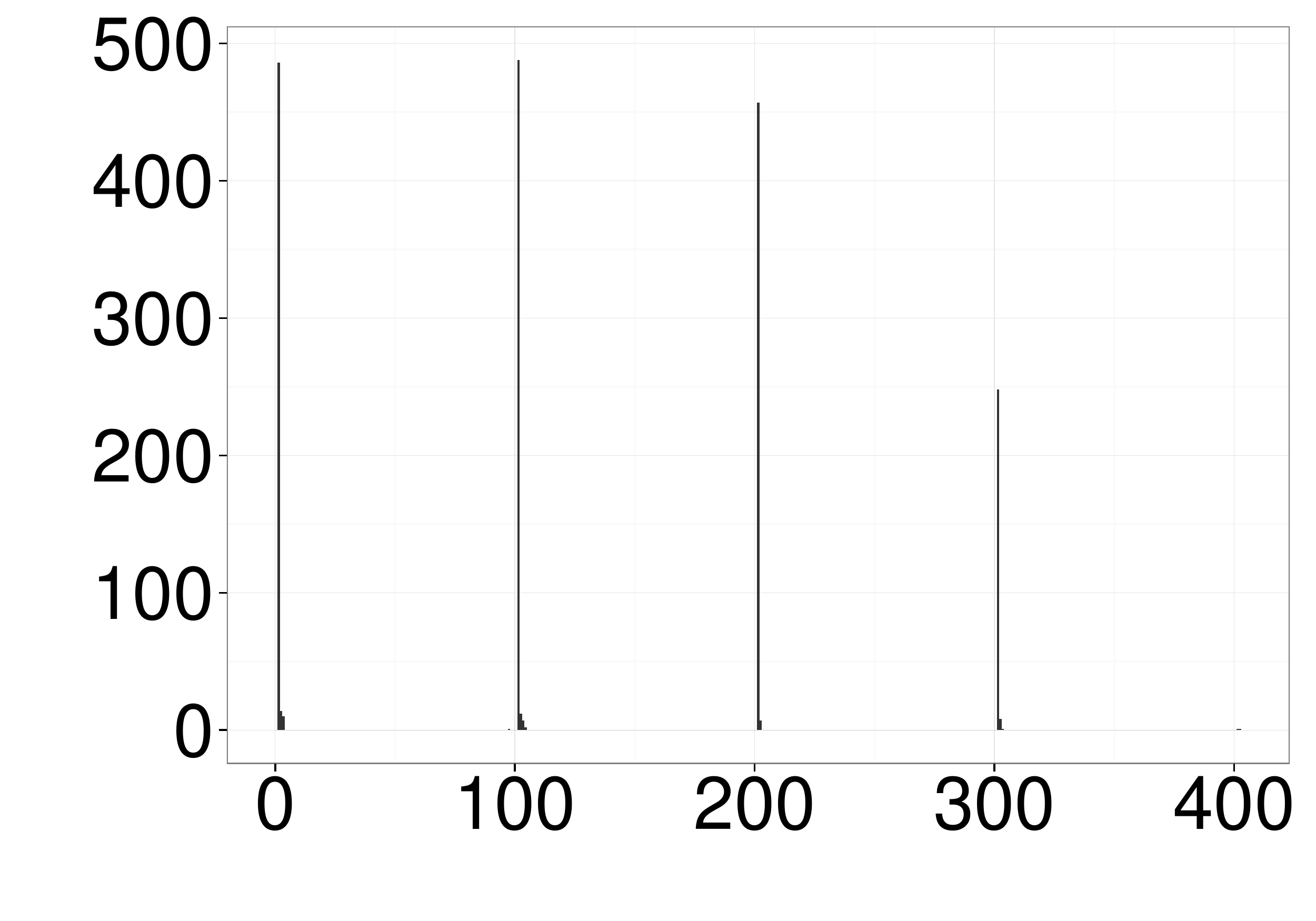}
    & \includegraphics[width=.225\linewidth, height=2cm]{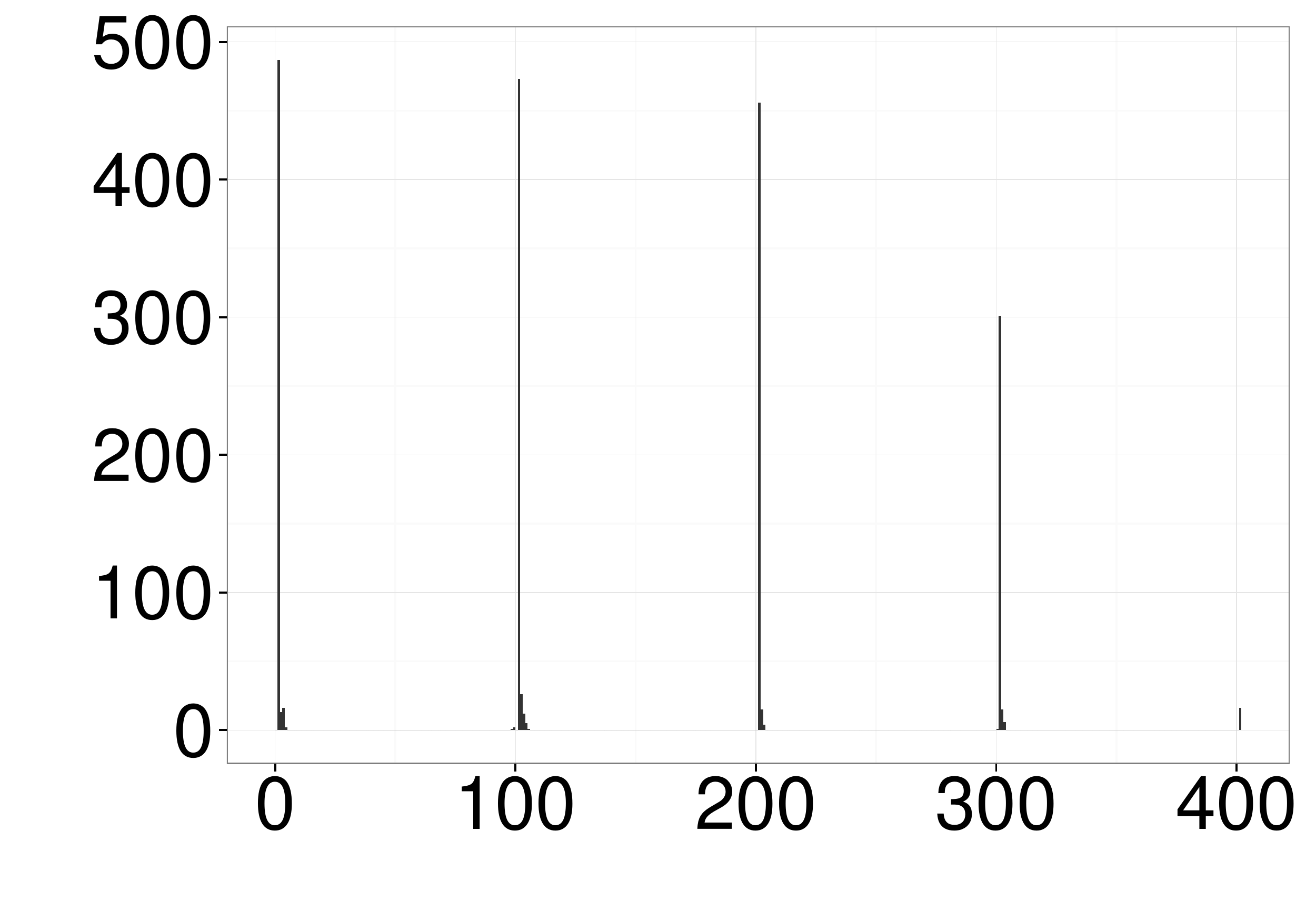}
    & \includegraphics[width=.225\linewidth, height=2cm]{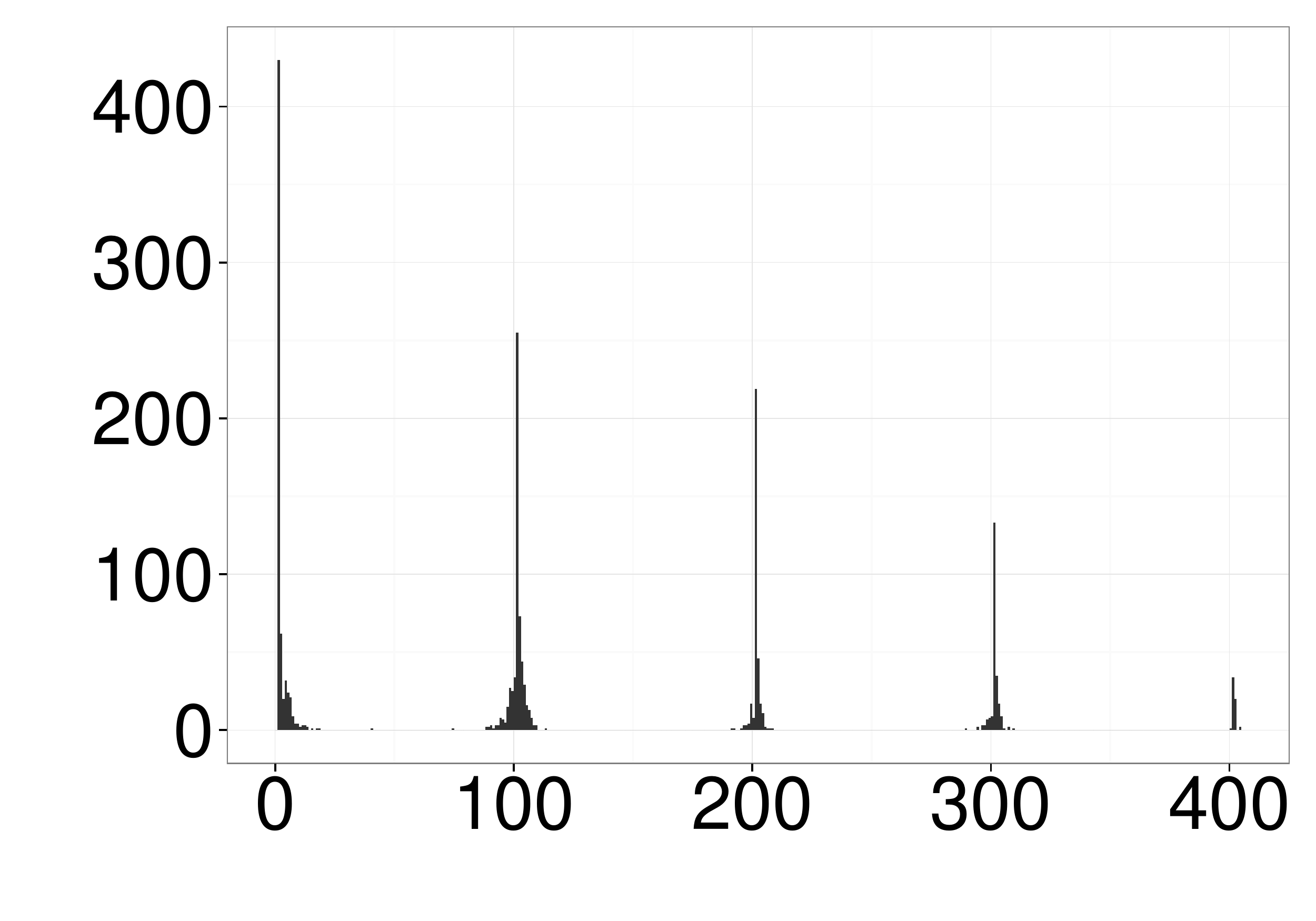} \\
    \rotatebox{90}{\hspace{2.2em}\small 40$\%$}
    & \includegraphics[width=.225\linewidth, height=2cm]{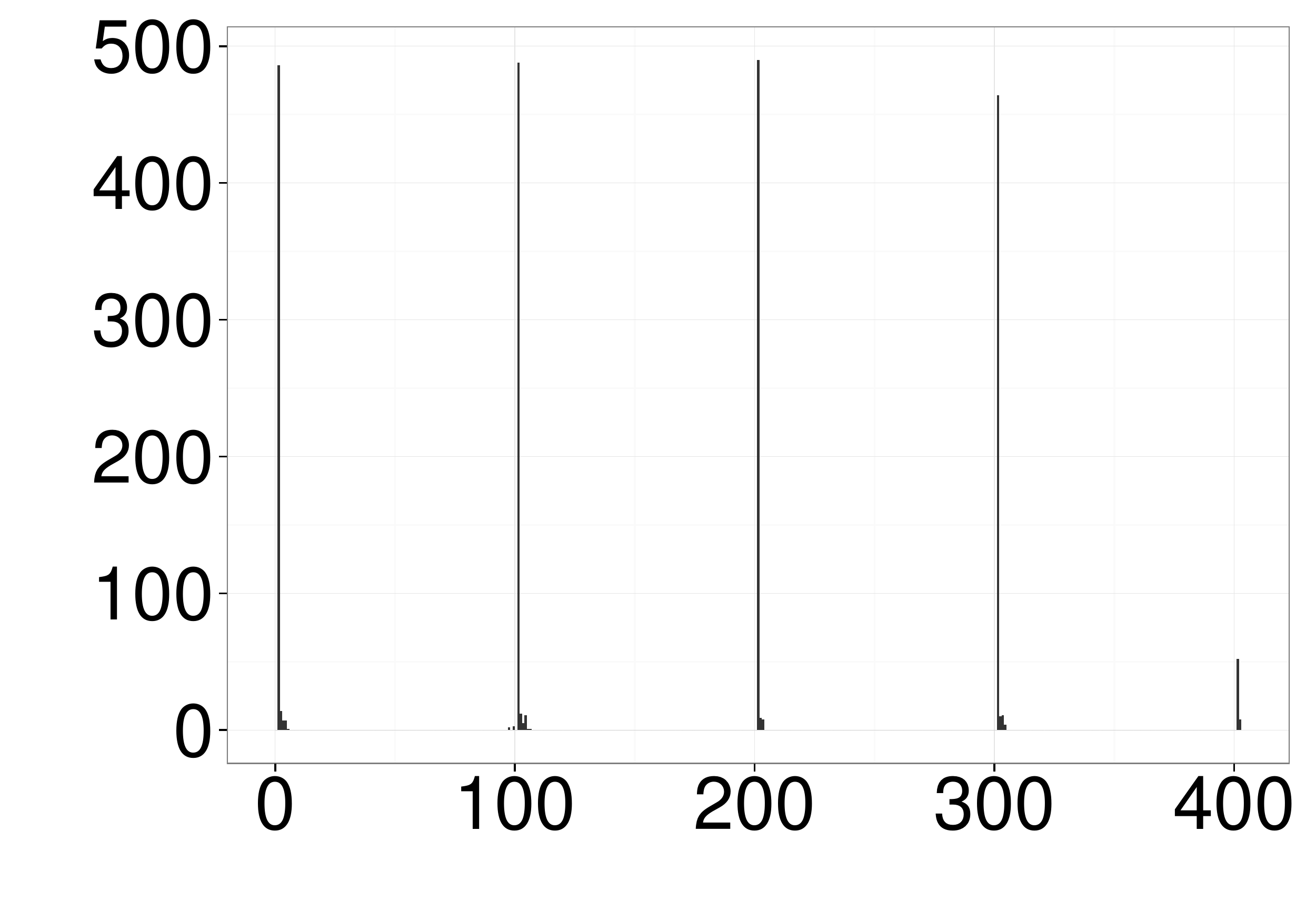}
    & \includegraphics[width=.225\linewidth, height=2cm]{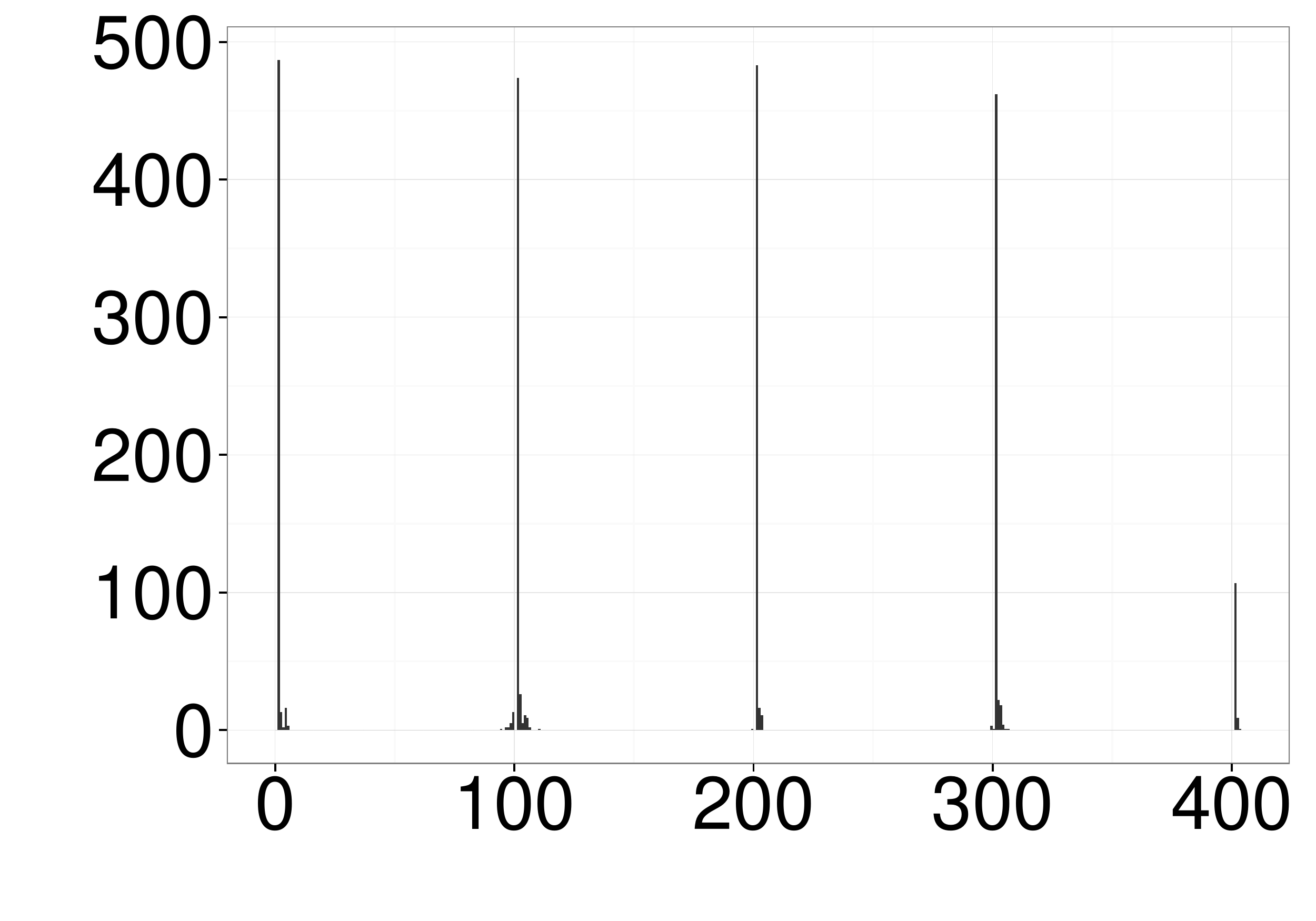}
    & \includegraphics[width=.225\linewidth, height=2cm]{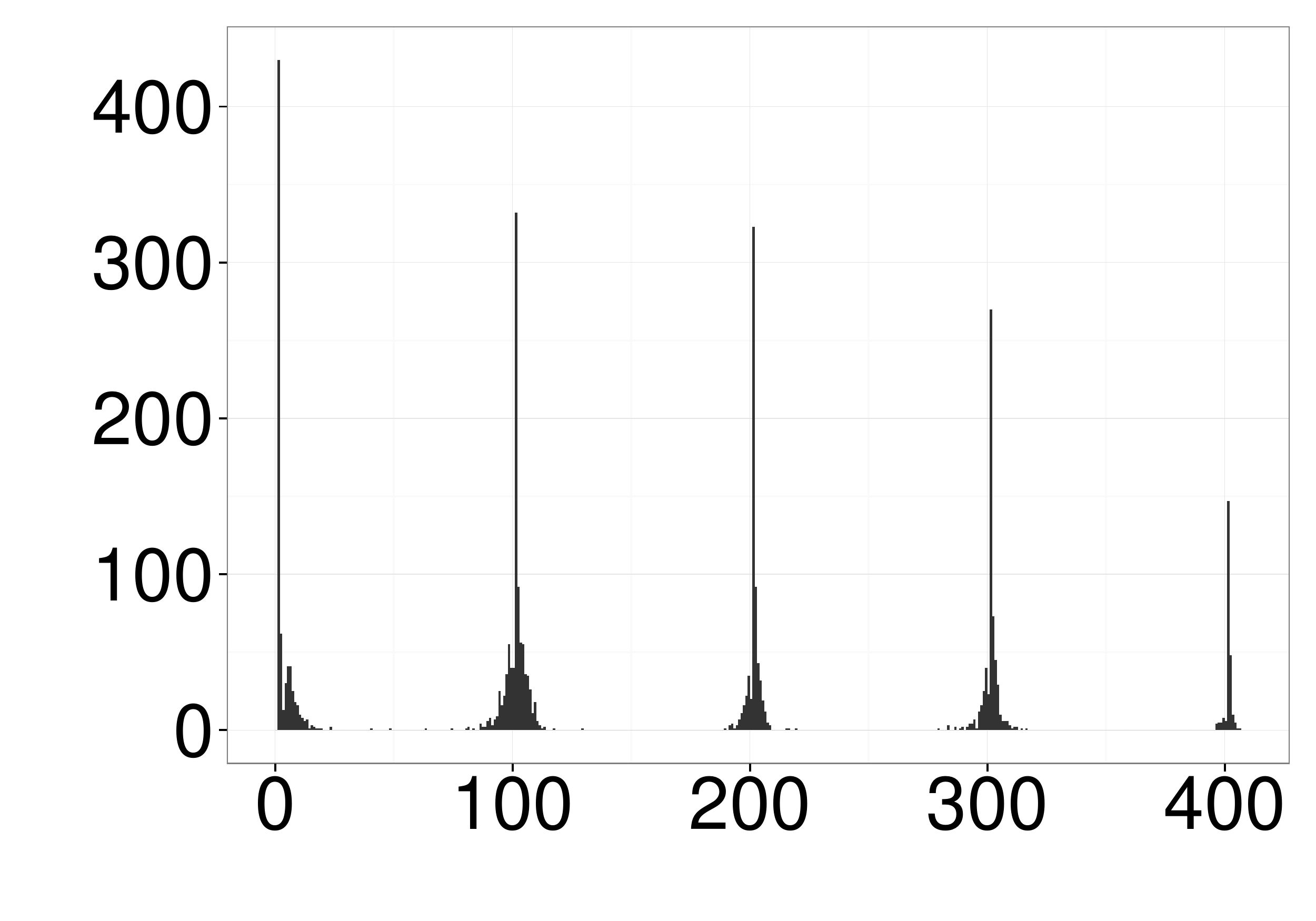} \\
    \rotatebox{90}{\hspace{2.2em}\small 30$\%$}
    & \includegraphics[width=.225\linewidth, height=2cm]{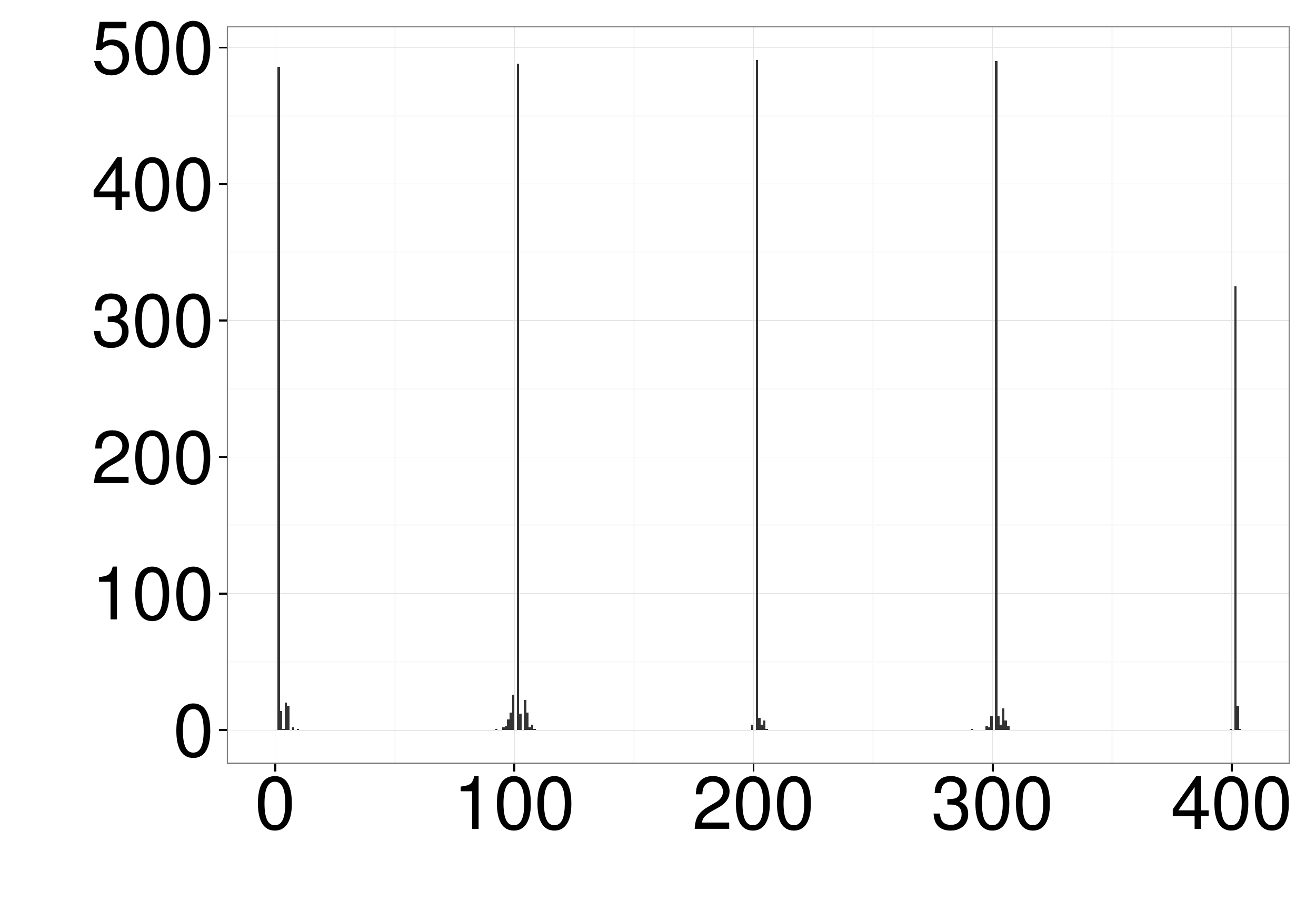}
    & \includegraphics[width=.225\linewidth, height=2cm]{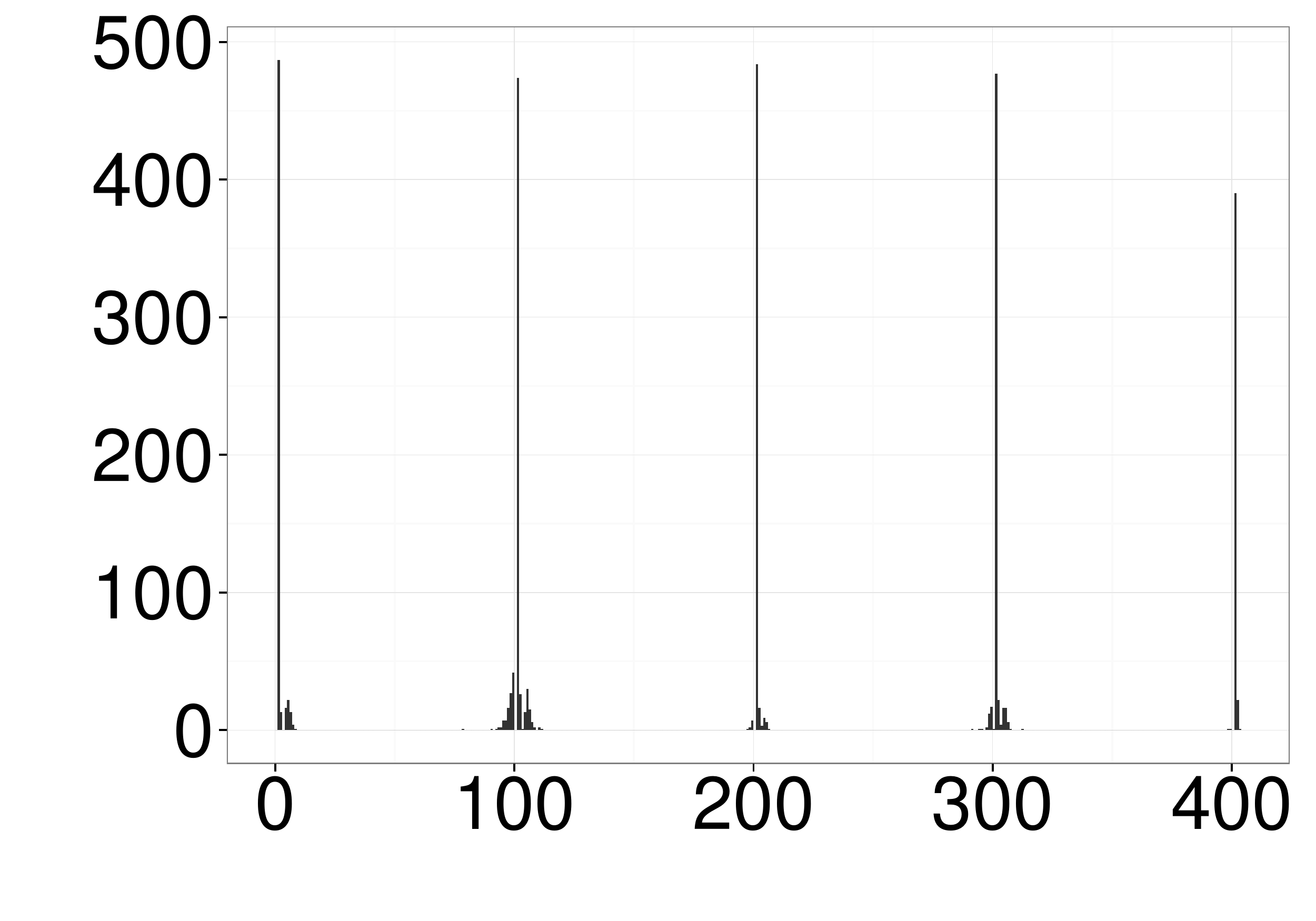}
    & \includegraphics[width=.225\linewidth, height=2cm]{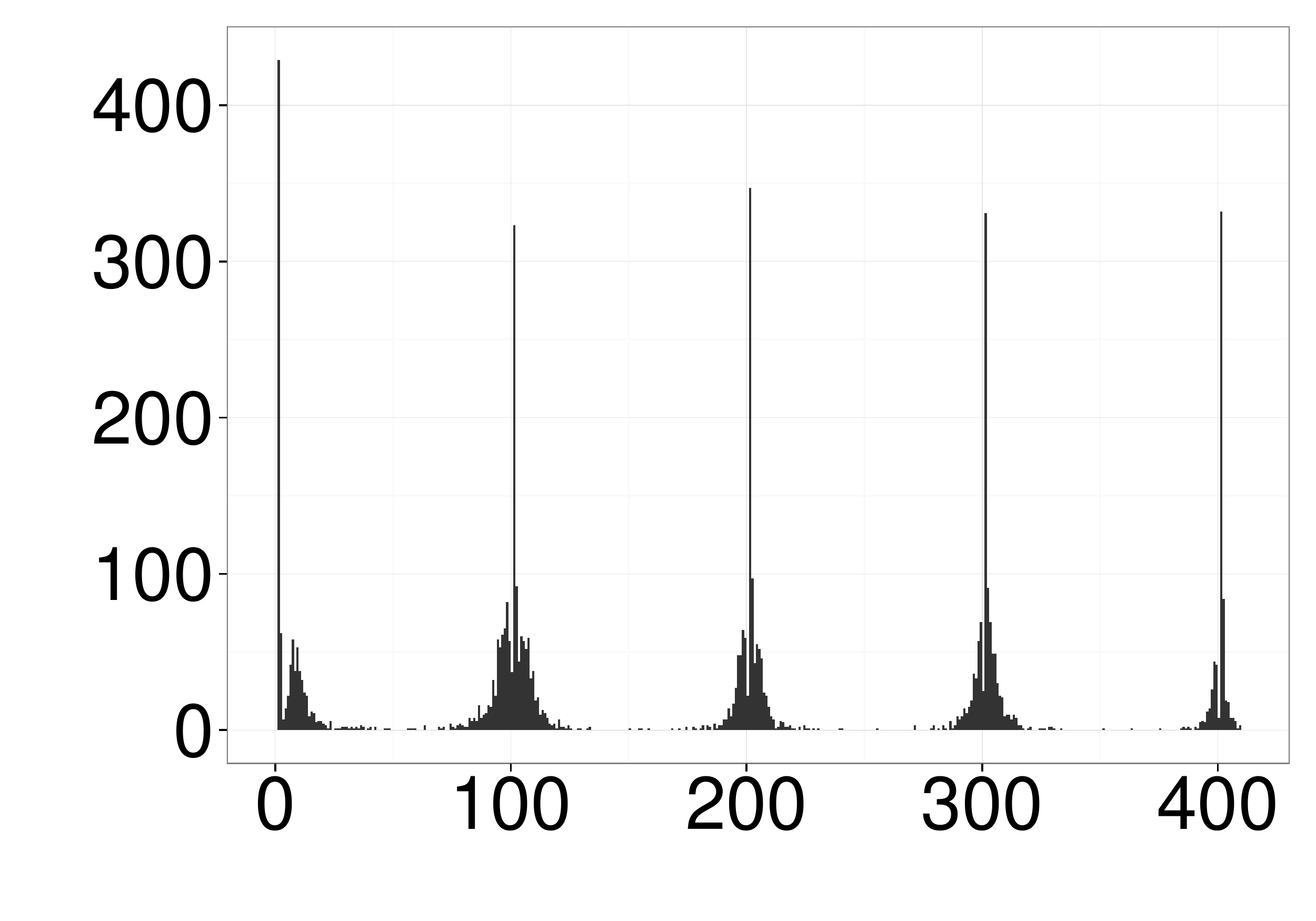} \\
    \rotatebox{90}{\hspace{2.2em}\small 20$\%$}
    & \includegraphics[width=.225\linewidth, height=2cm]{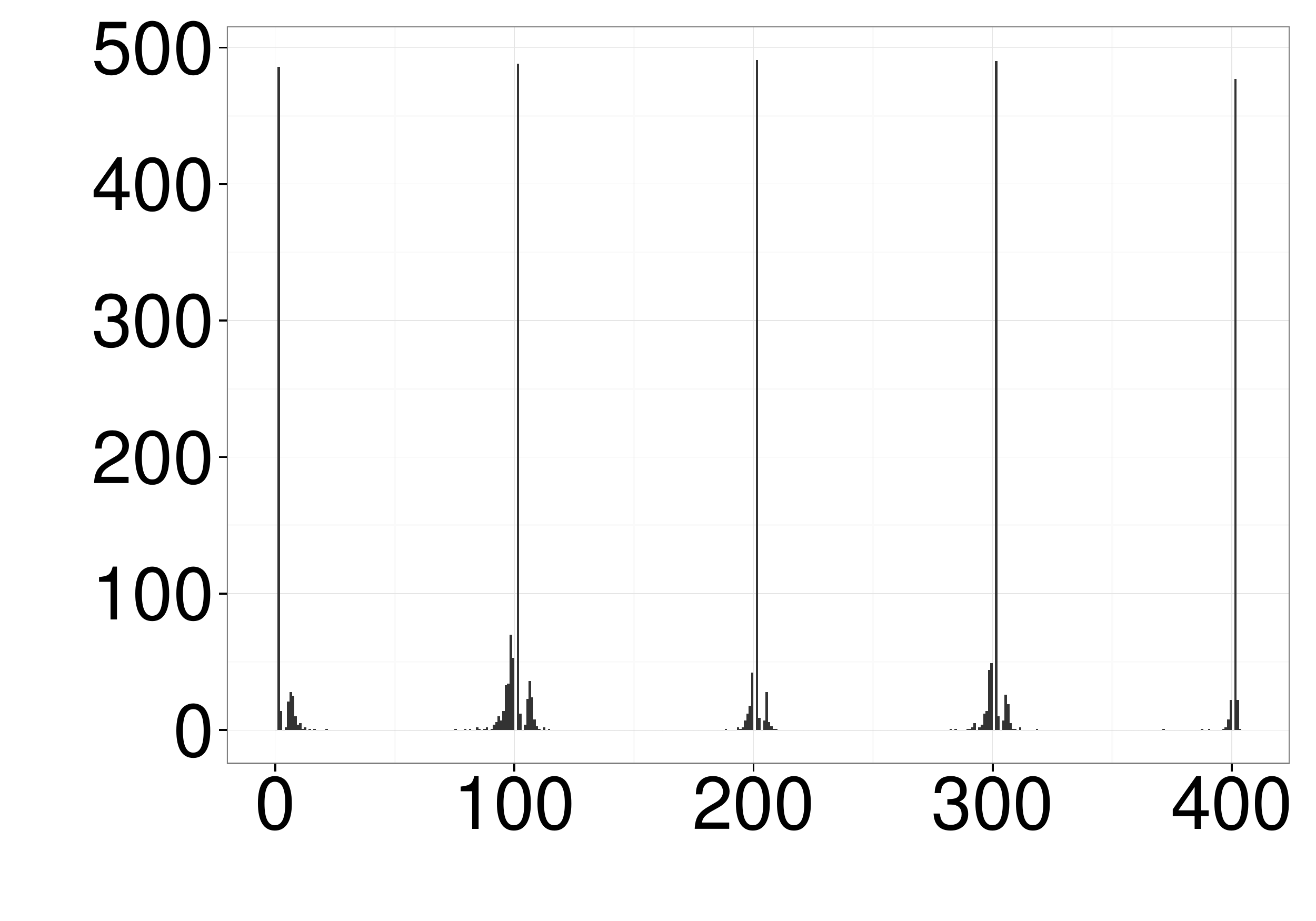}
    & \includegraphics[width=.225\linewidth, height=2cm]{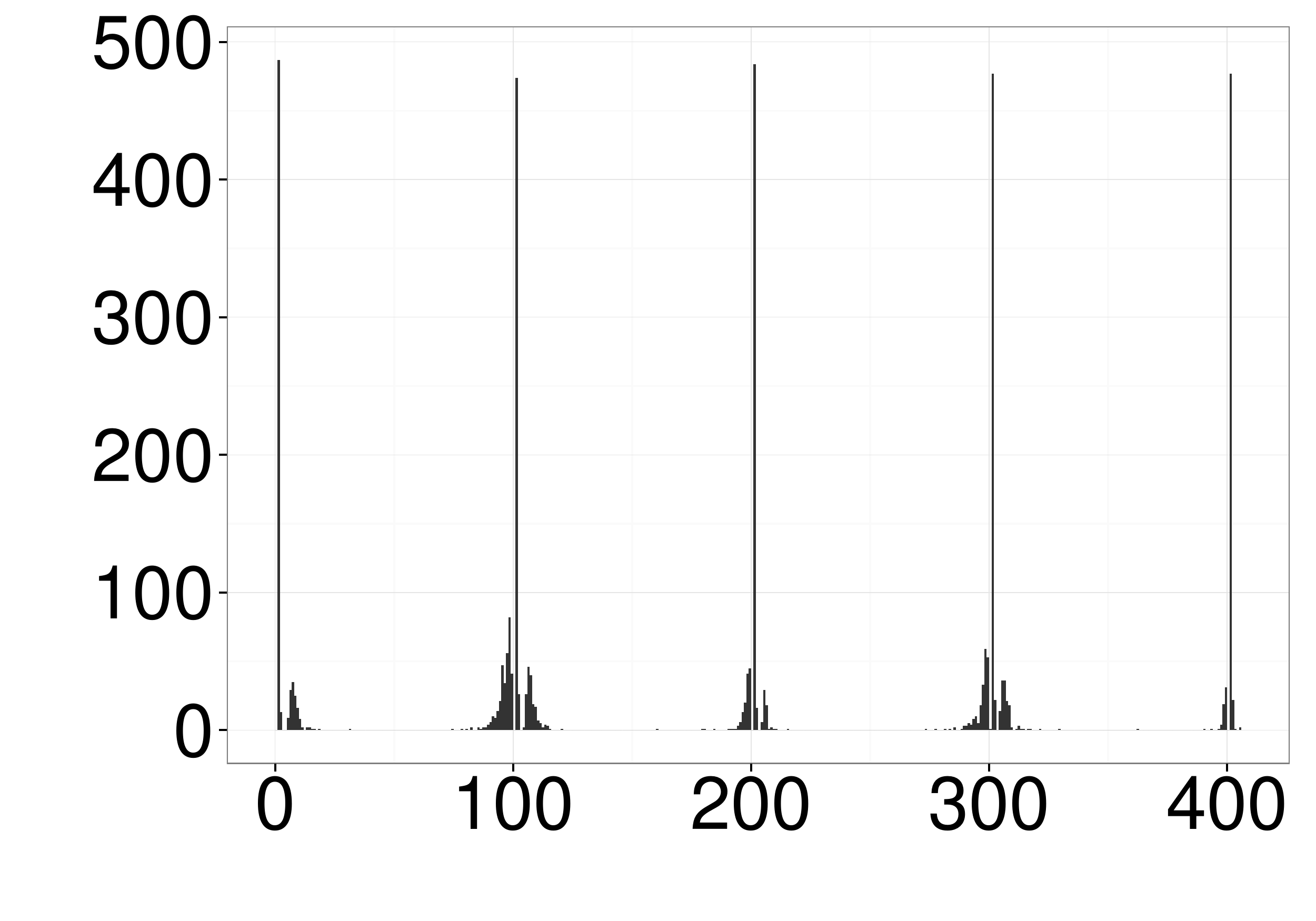}
    & \includegraphics[width=.225\linewidth, height=2cm]{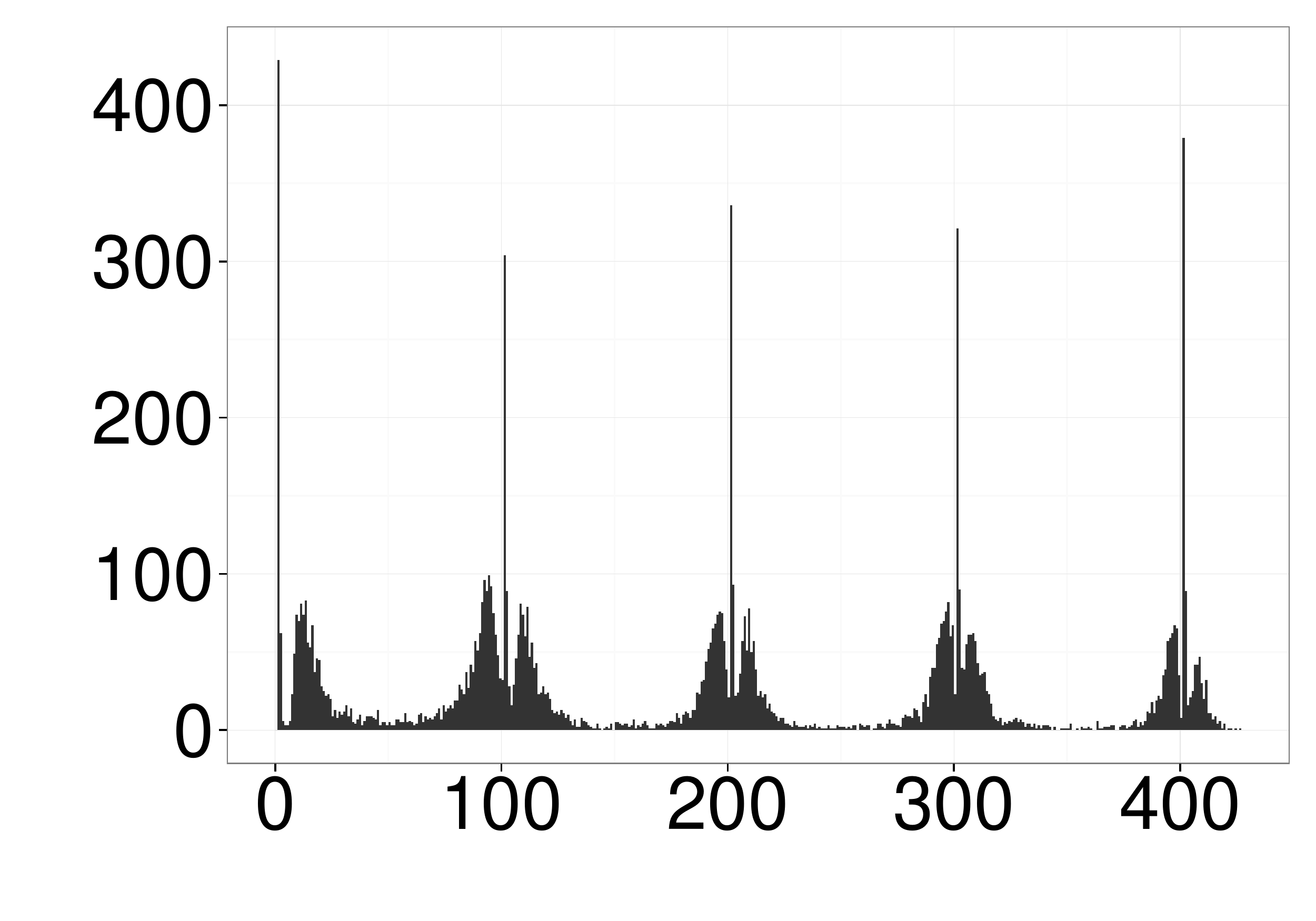} \\
    \rotatebox{90}{\hspace{2.2em}\small 10$\%$}
    & \includegraphics[width=.225\linewidth, height=2cm]{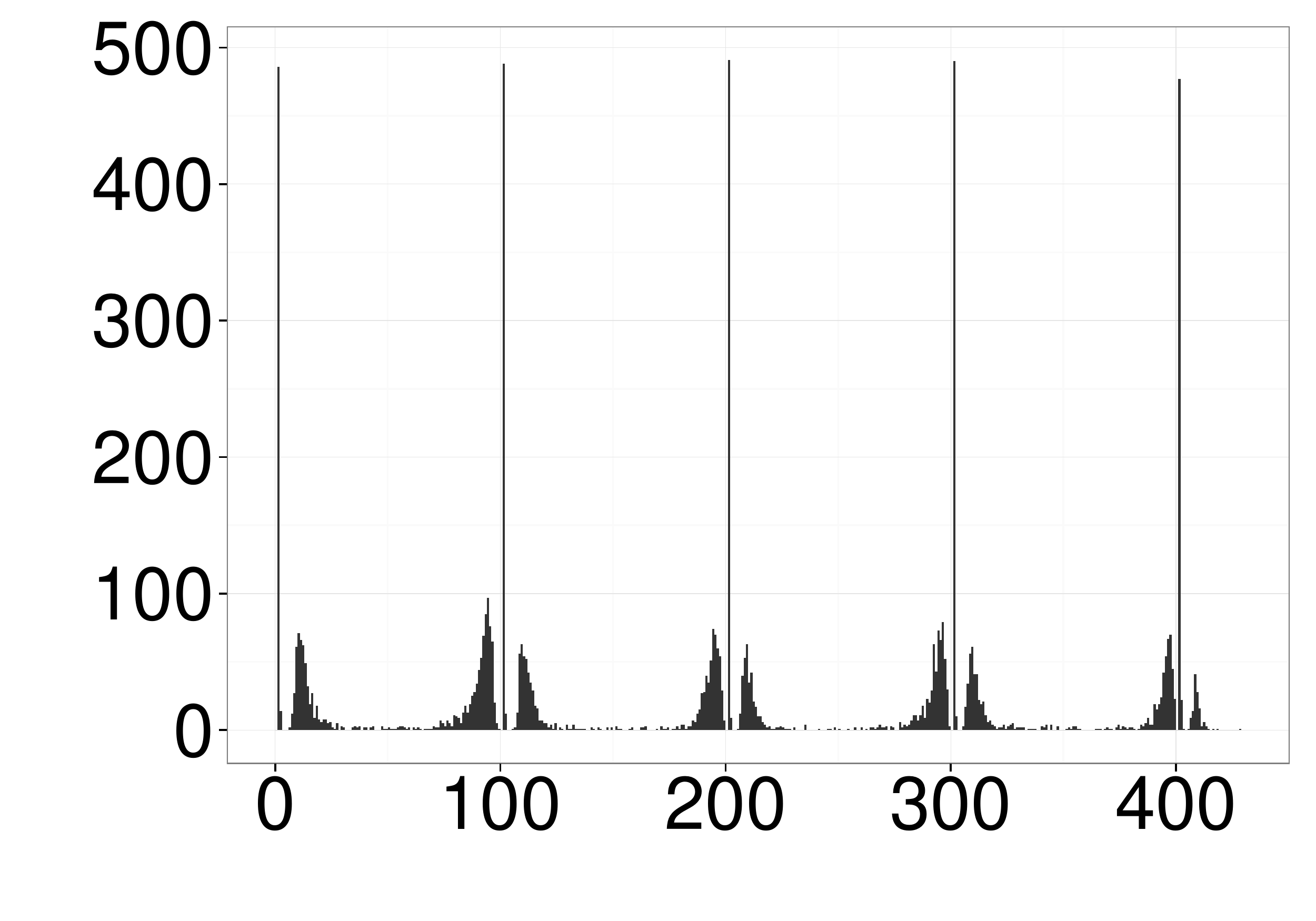}
    & \includegraphics[width=.225\linewidth, height=2cm]{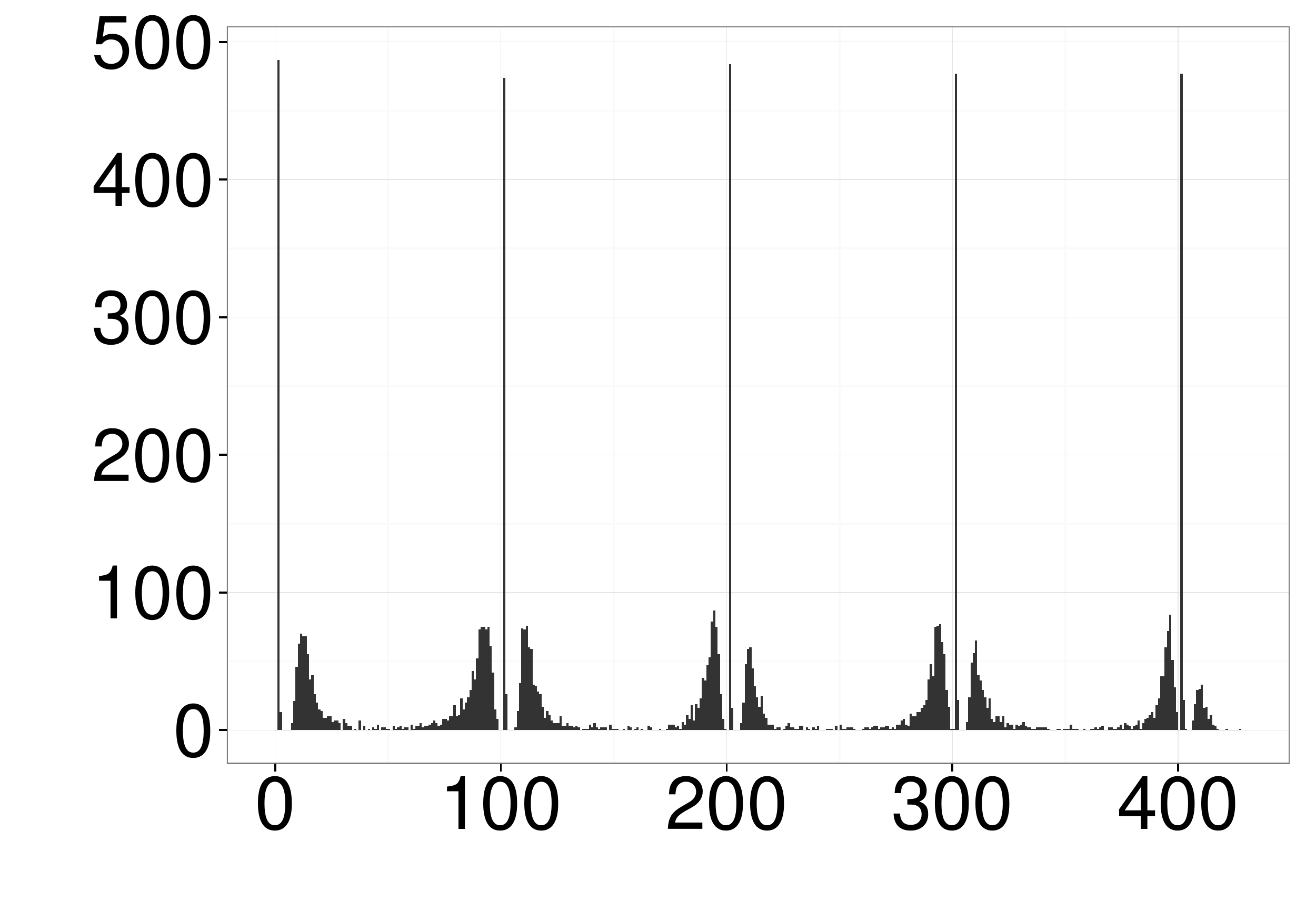}
    & \includegraphics[width=.225\linewidth, height=2cm]{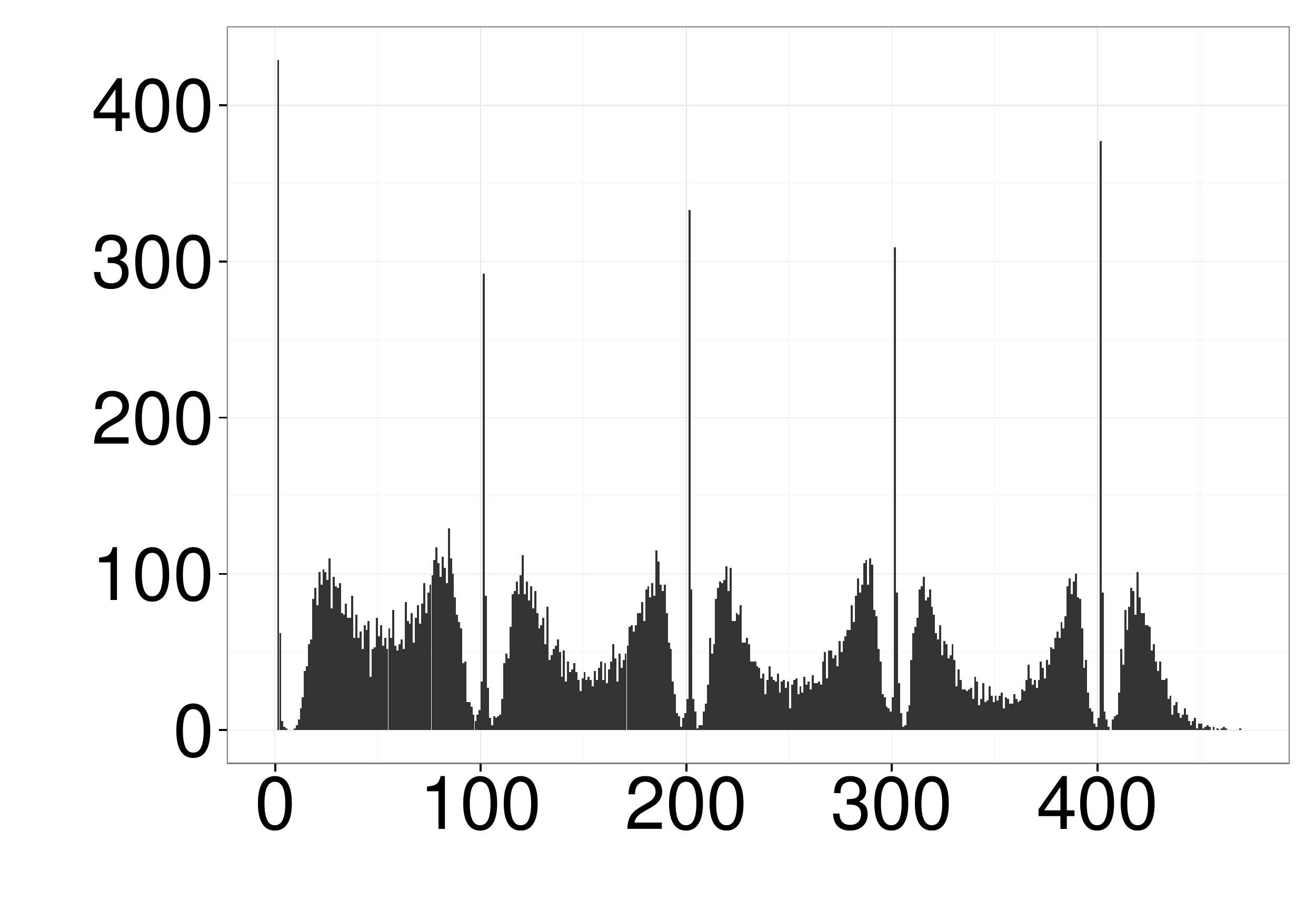} \\
  \end{tabular}
  \caption{\label{fig:SelectionModelHisto} Barplots of the estimated change-points for different variances (columns) and different
thresholds (rows) for the model $\mu^{\star,(1)}$.}
\end{figure}

To evaluate the performances of this methodology, we generated observations according to the ``checkerboard'' model defined in (\ref{eq:model1})
with $\left({\mu}_{k,\ell}^{\star,(1)}\right)$ defined in (\ref{eq:model:mustar}),
$s=225$ and $M=100$. The results are given in Figure \ref{fig:SelectionModel} which displays the score associated to each change-point for a given matrix $\Y$
(top). We can see from the top part of Figure \ref{fig:SelectionModel} some spurious change-points appearing near from the true
change-point positions.
In order to identify the most representative change-point in a given neighborhood, we keep the one with the largest score
among a set of contiguous candidates. The result of such a post-processing is displayed
in the bottom part of Figure \ref{fig:SelectionModel} and in Figure \ref{fig:SelectionModelHisto}.
More precisely the boxplots associated to the estimation of
$\Ks_1$ (resp. the histograms of the estimated change-points in rows) are displayed in the bottom part of Figure \ref{fig:SelectionModel}
(resp. in Figure \ref{fig:SelectionModelHisto}) for different values of $\sigma$ and different thresholds (thresh) expressed
as a percentage of the largest score. We can see from these figures that when thresh is in the interval $[20,40]$
the number and the location of the change-points are very well estimated even in the high noise level case.


\begin{figure}[!h]
  \centering
  \begin{tabular}{@{}l@{}ccc}
    & $\sigma=1$ & $\sigma=2$
    & $\sigma=5$ \\
    & \includegraphics[width=.225\linewidth, height=2cm]{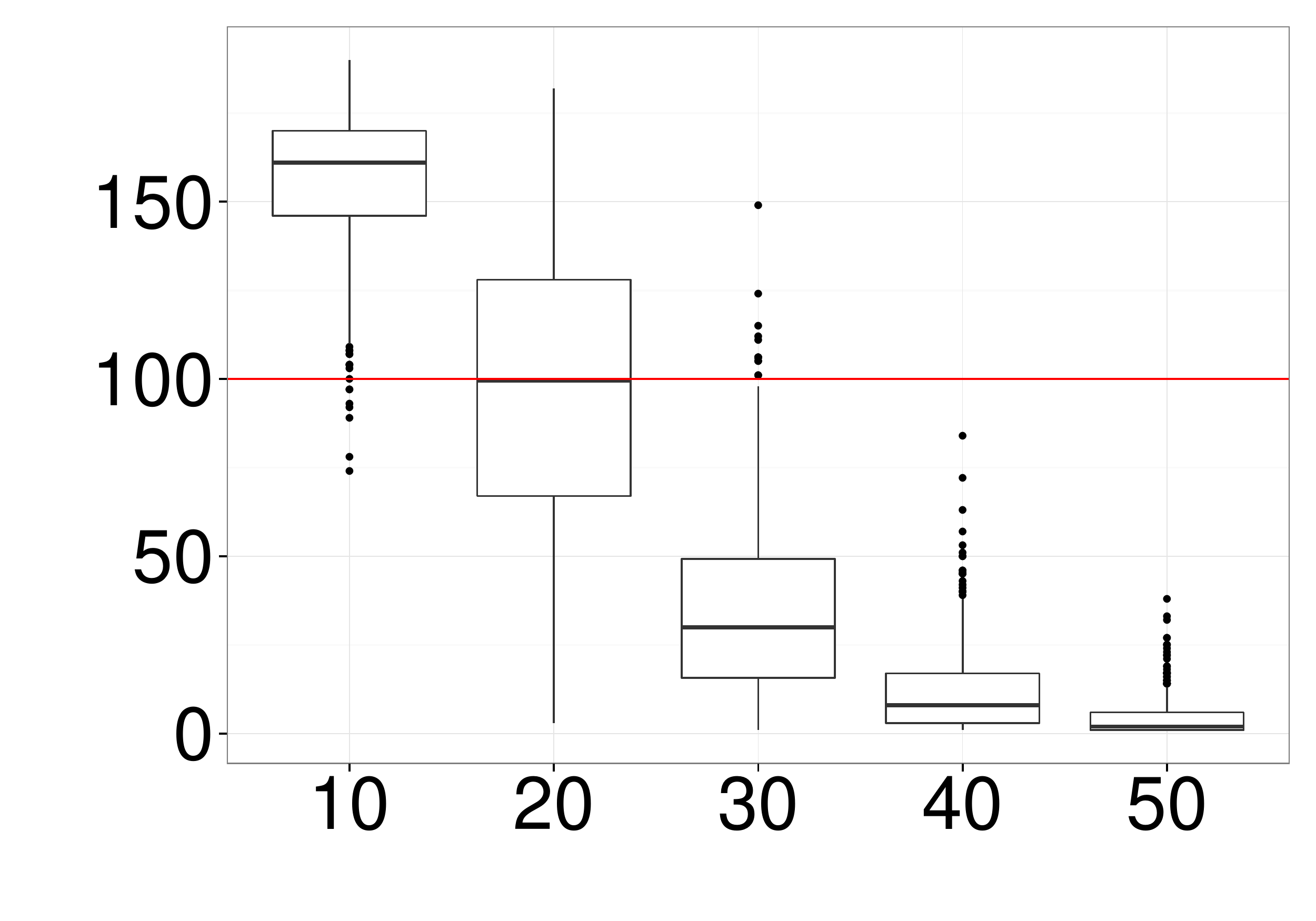}
    & \includegraphics[width=.225\linewidth, height=2cm]{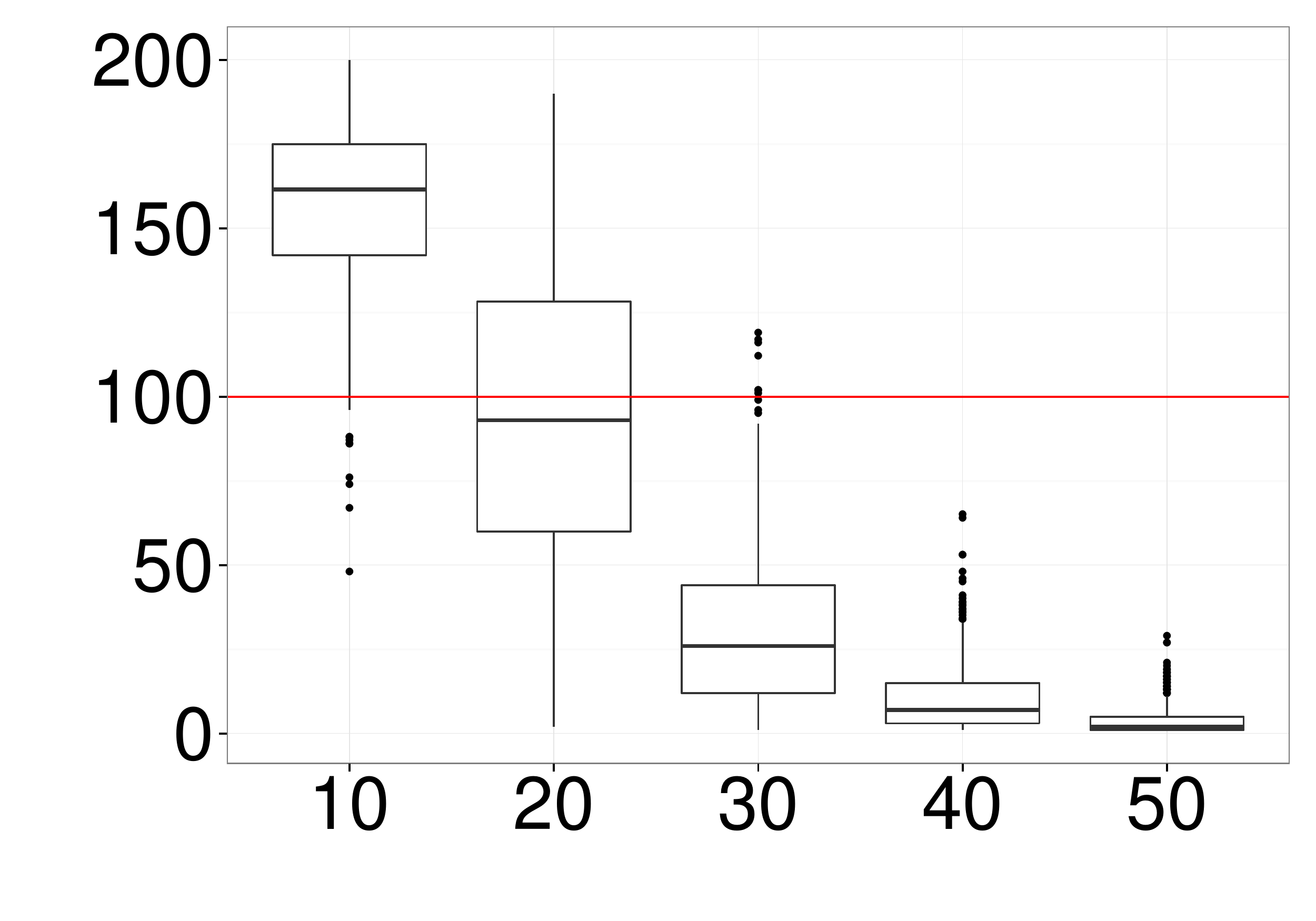}
    & \includegraphics[width=.225\linewidth, height=2cm]{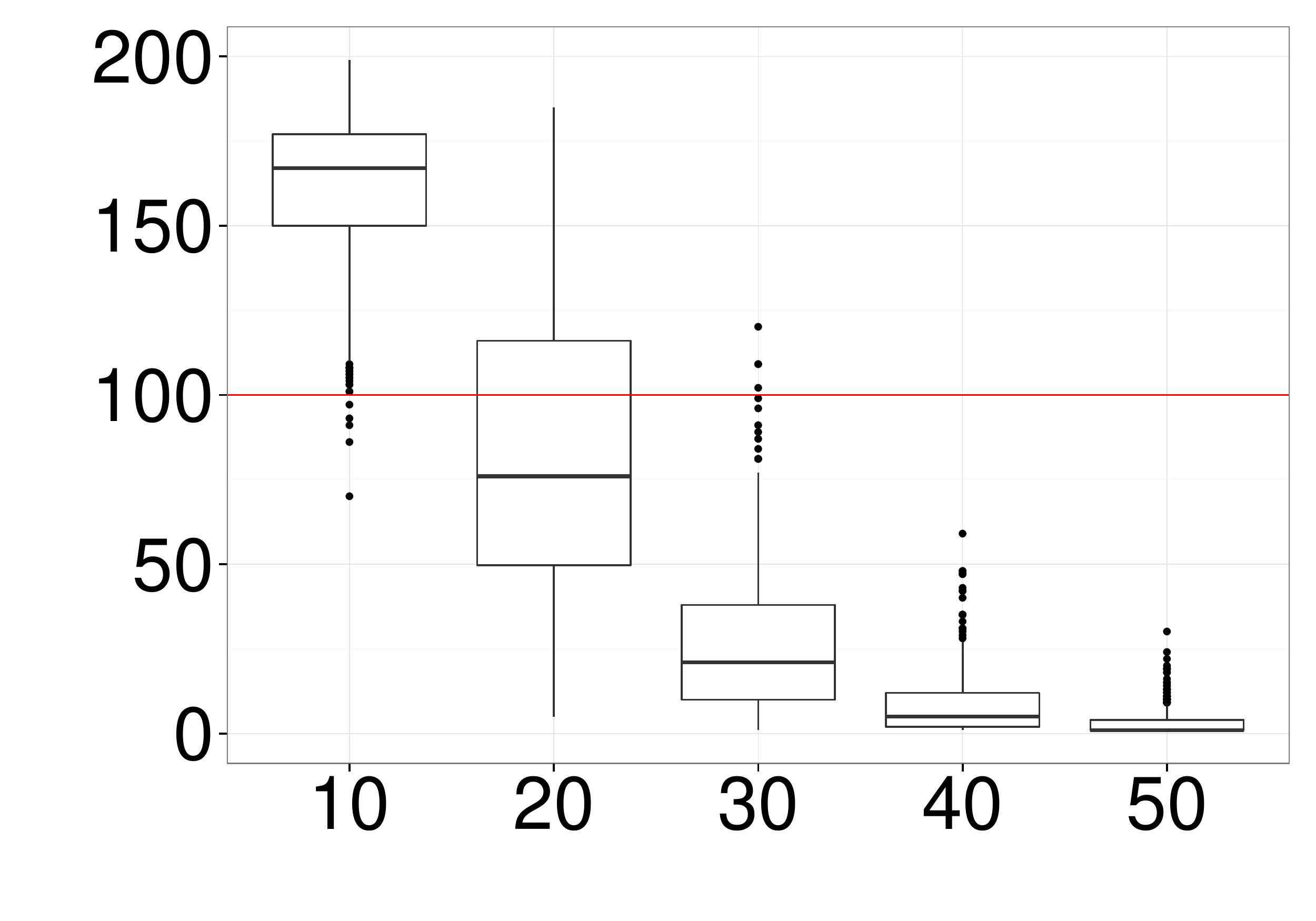} \\
  \end{tabular}
  \caption{\label{fig:SelectionModeln1000K100}
Boxplots of the estimation of $\Ks_1$ for different values of $\sigma$ and thresholds (thresh) after the post-processing step.
}
\end{figure}

\begin{figure}[!h]
  \centering
  \begin{tabular}{@{}l@{}ccc}
    & $\sigma=1$ & $\sigma=2$
    & $\sigma=5$ \\
    \rotatebox{90}{\hspace{2.2em}\small 50$\%$}
    & \includegraphics[width=.225\linewidth, height=2cm]{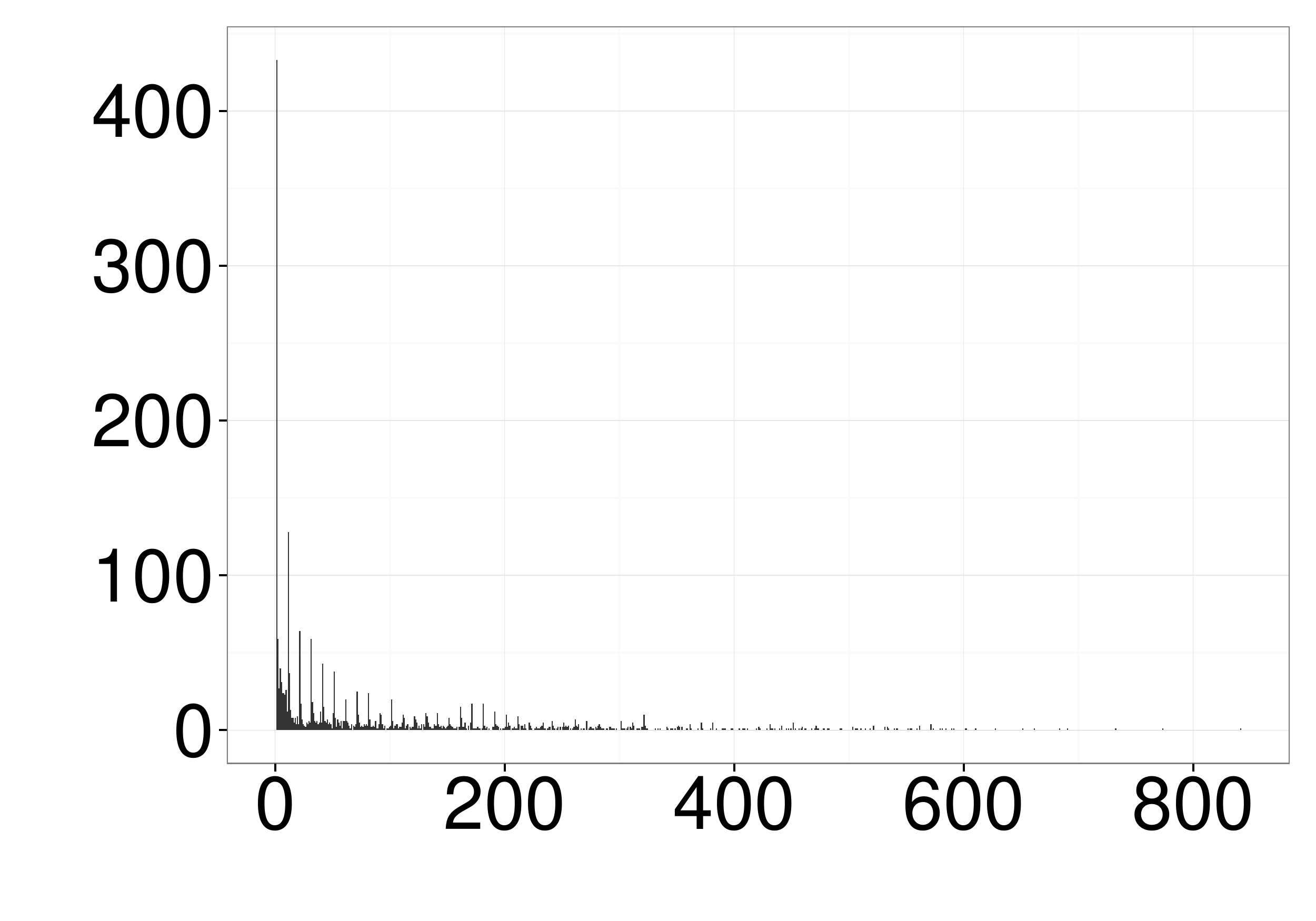}
    & \includegraphics[width=.225\linewidth, height=2cm]{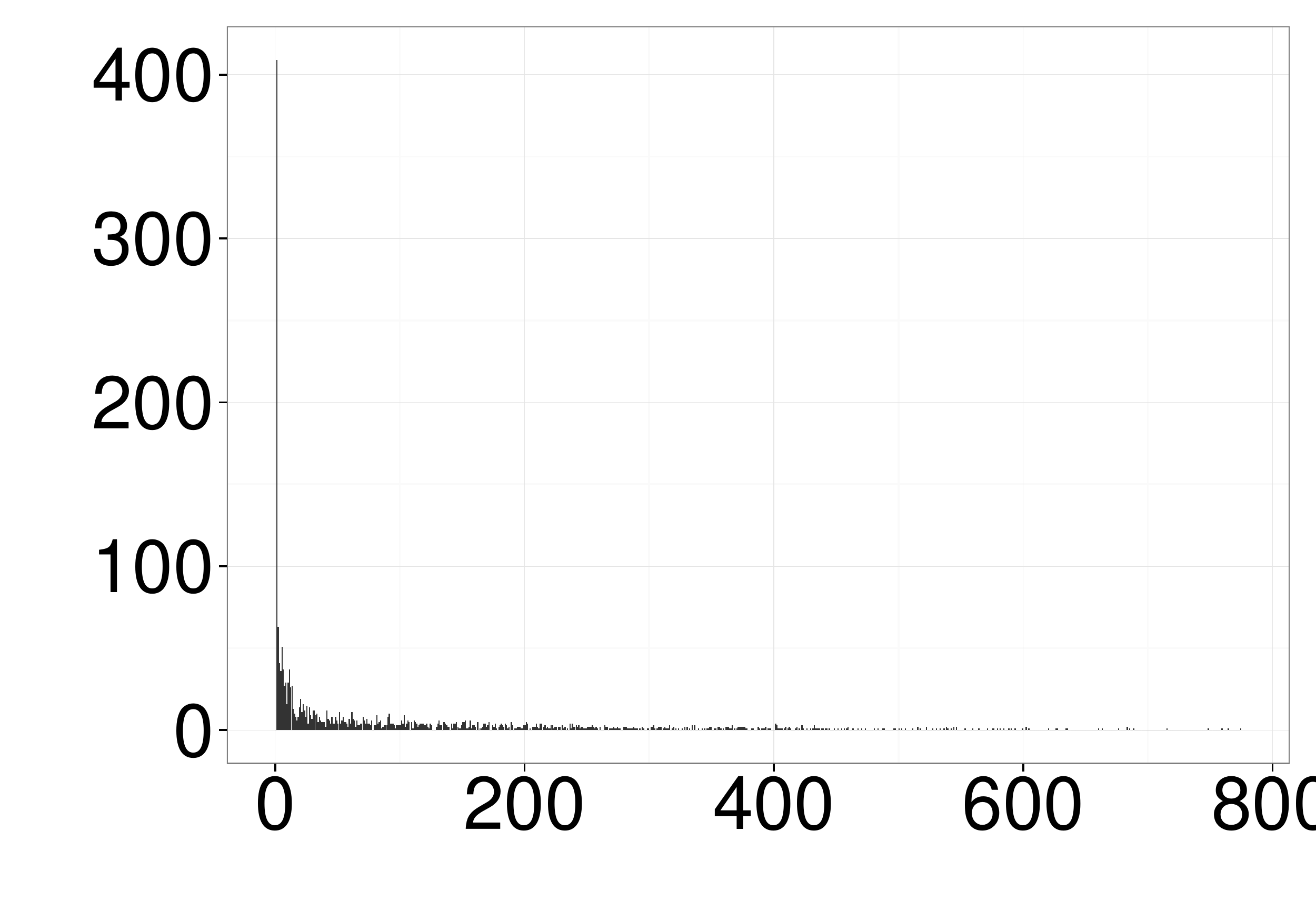}
    & \includegraphics[width=.225\linewidth, height=2cm]{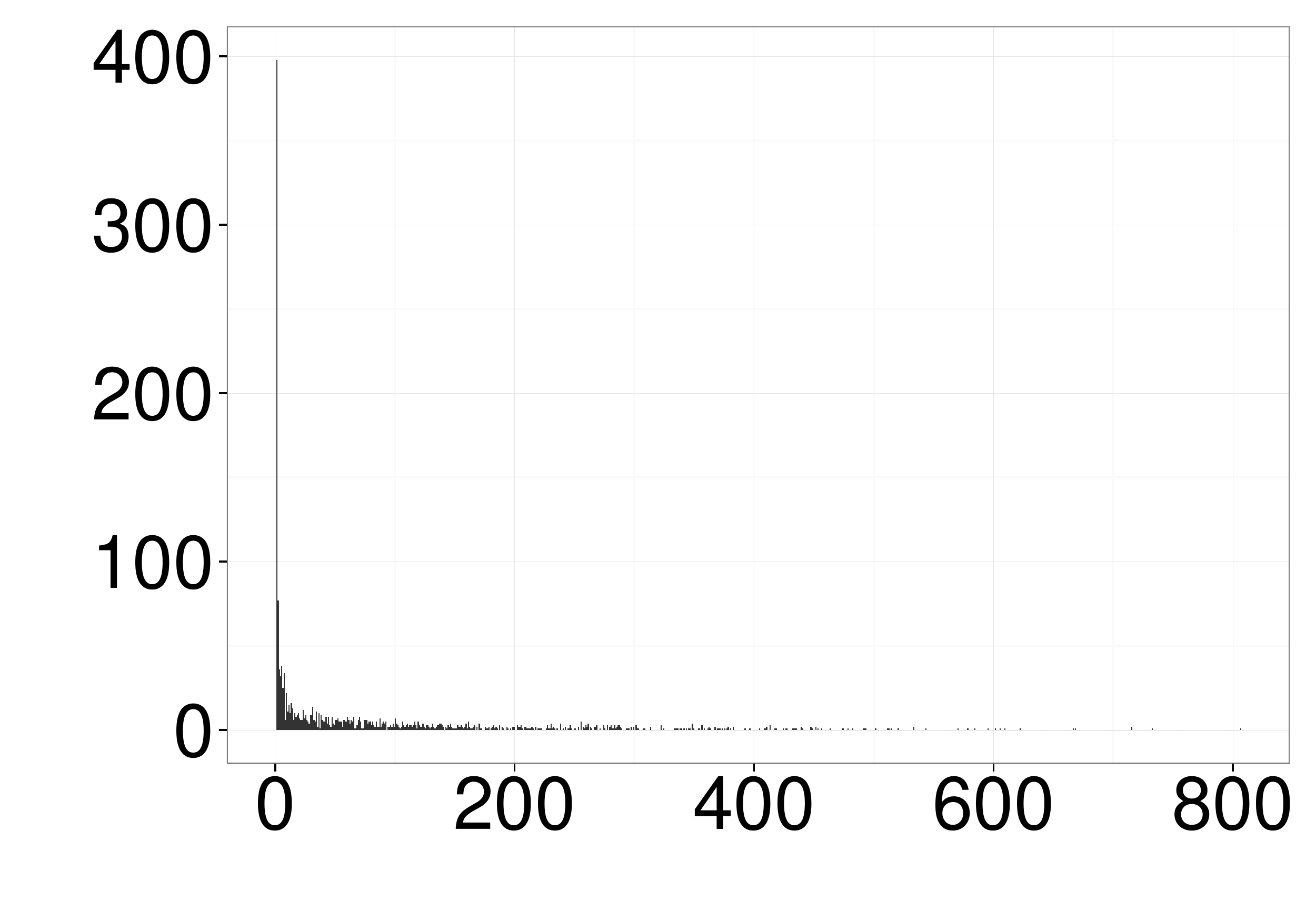} \\
    \rotatebox{90}{\hspace{2.2em}\small 40$\%$}
    & \includegraphics[width=.225\linewidth, height=2cm]{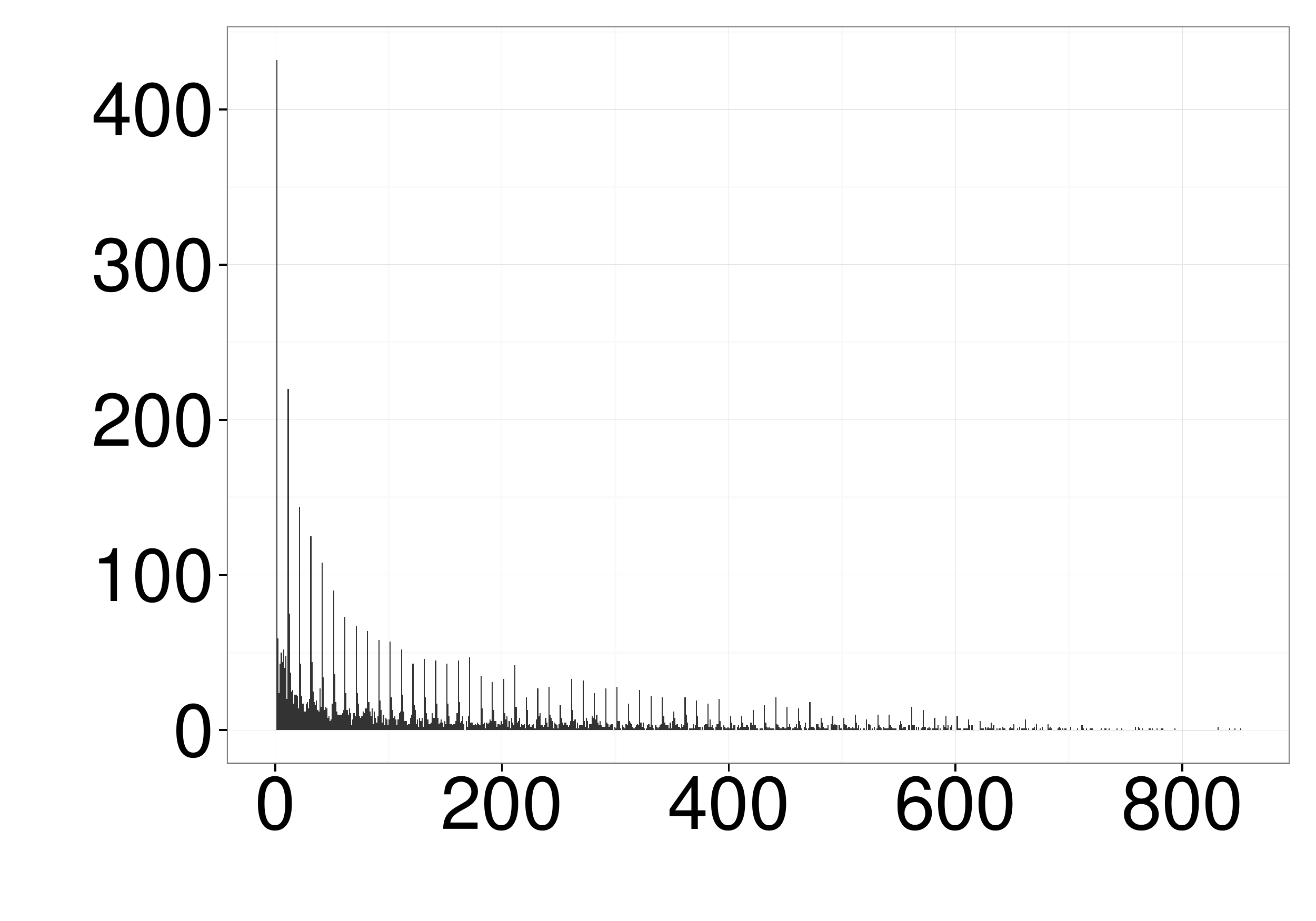}
    & \includegraphics[width=.225\linewidth, height=2cm]{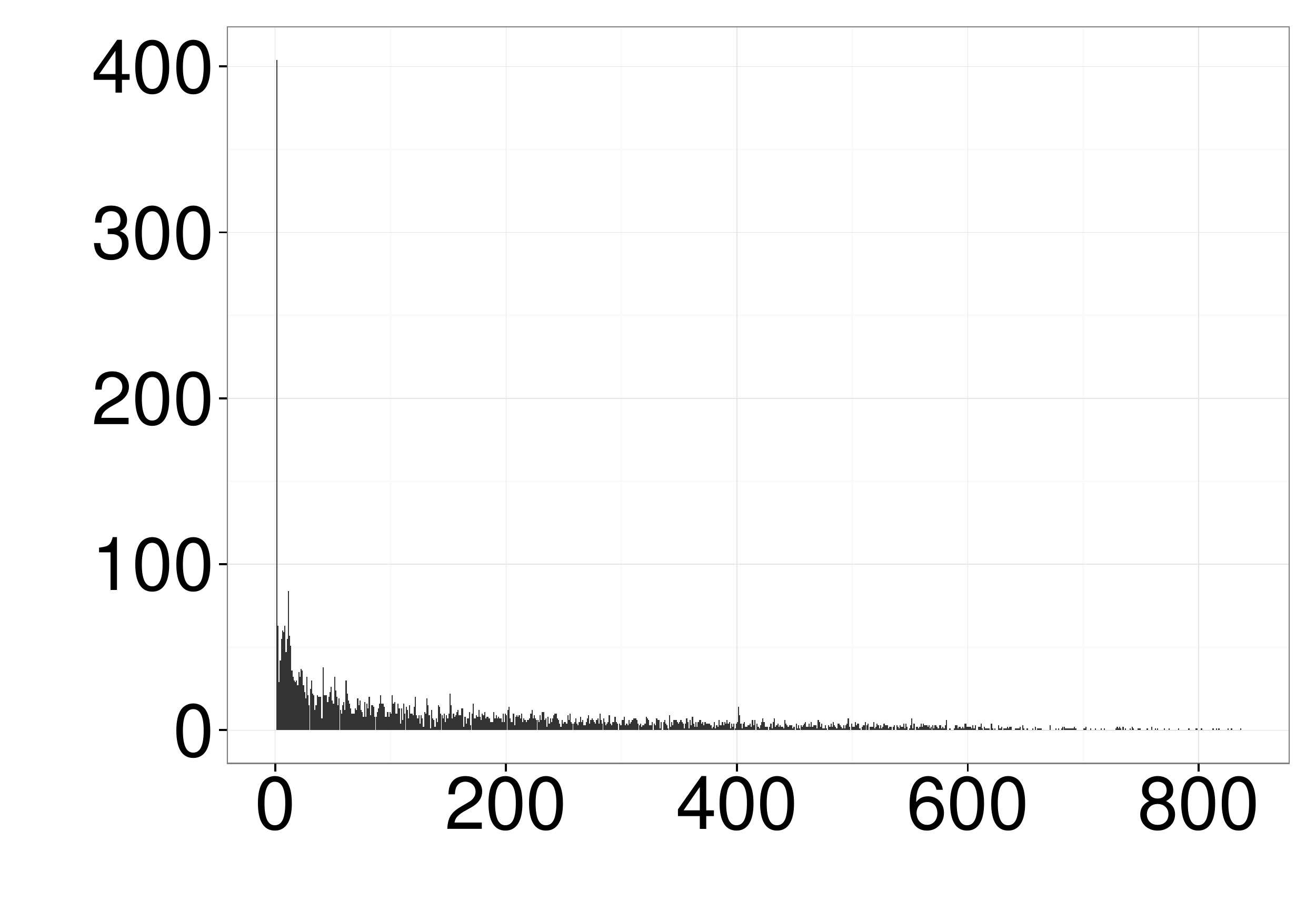}
    & \includegraphics[width=.225\linewidth, height=2cm]{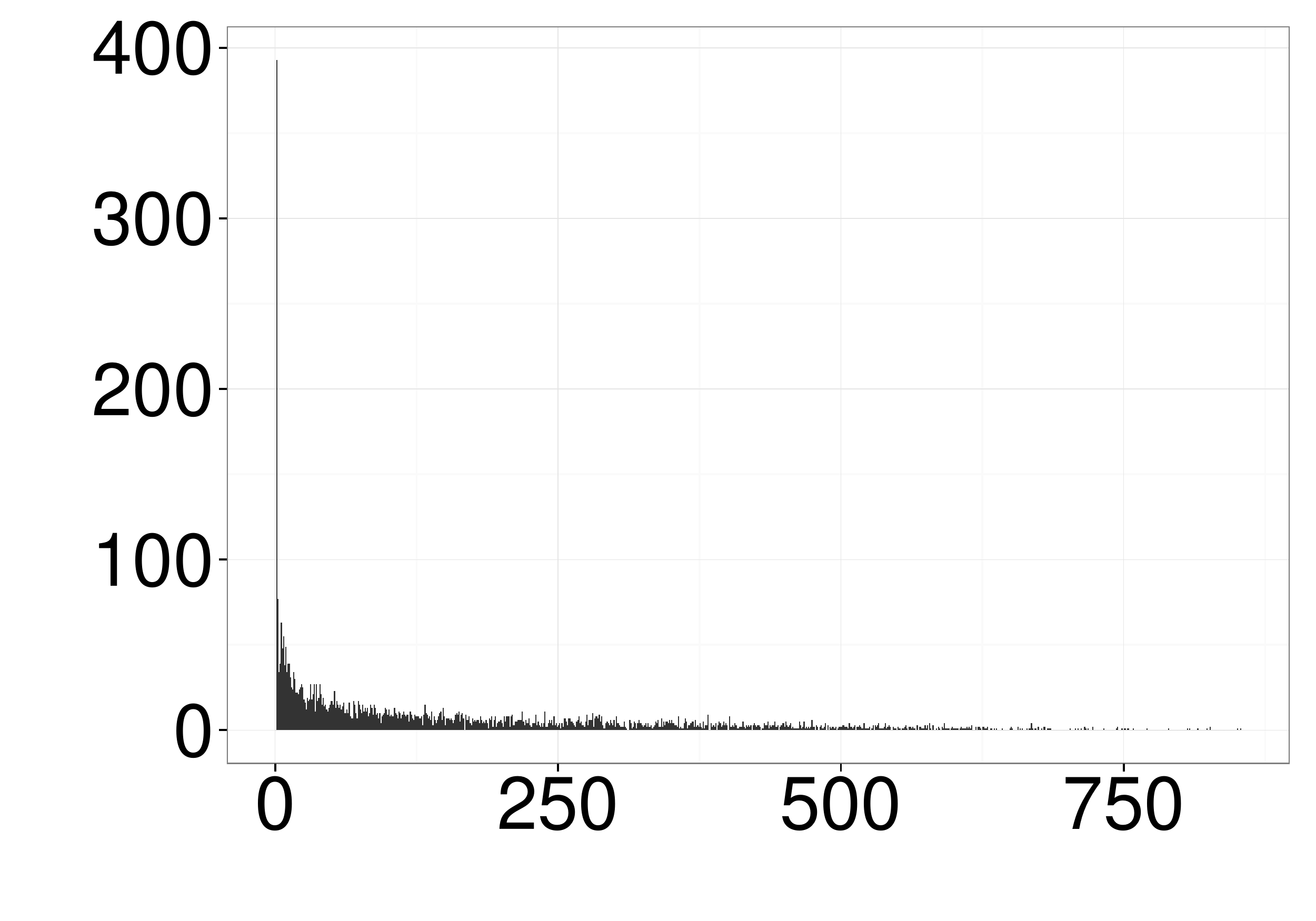} \\
    \rotatebox{90}{\hspace{2.2em}\small 30$\%$}
    & \includegraphics[width=.225\linewidth, height=2cm]{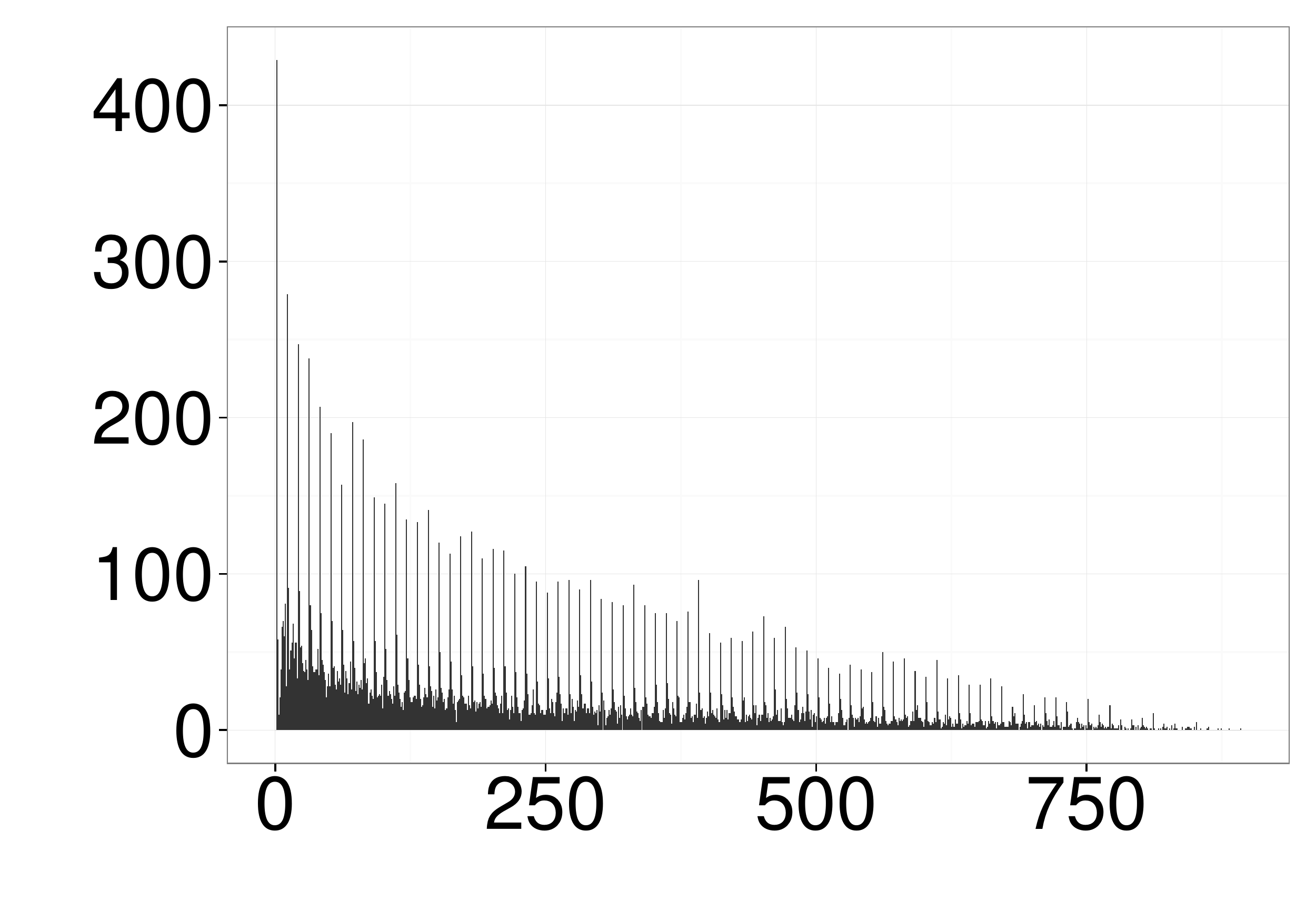}
    & \includegraphics[width=.225\linewidth, height=2cm]{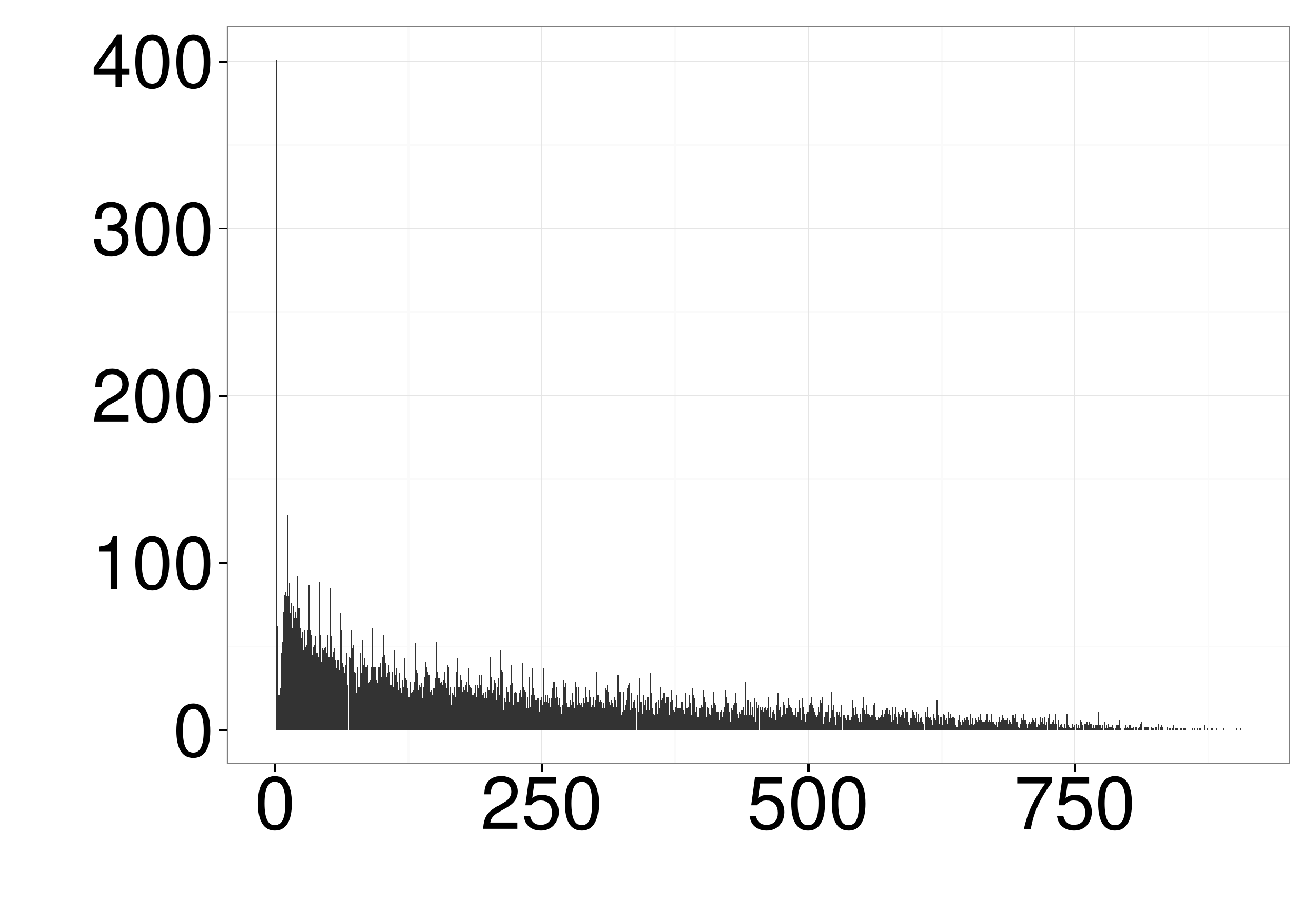}
    & \includegraphics[width=.225\linewidth, height=2cm]{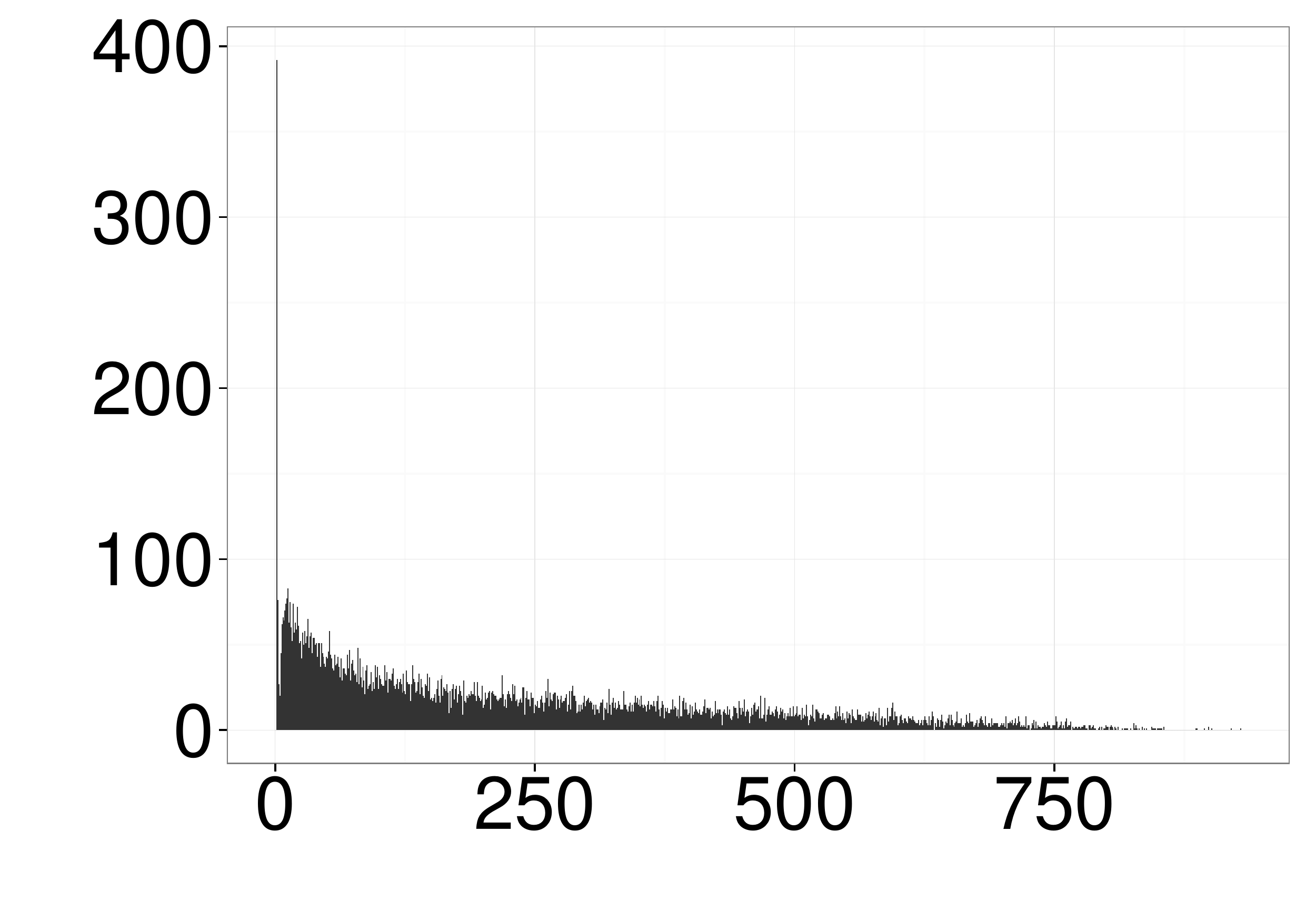} \\
    \rotatebox{90}{\hspace{2.2em}\small 20$\%$}
    & \includegraphics[width=.225\linewidth, height=2cm]{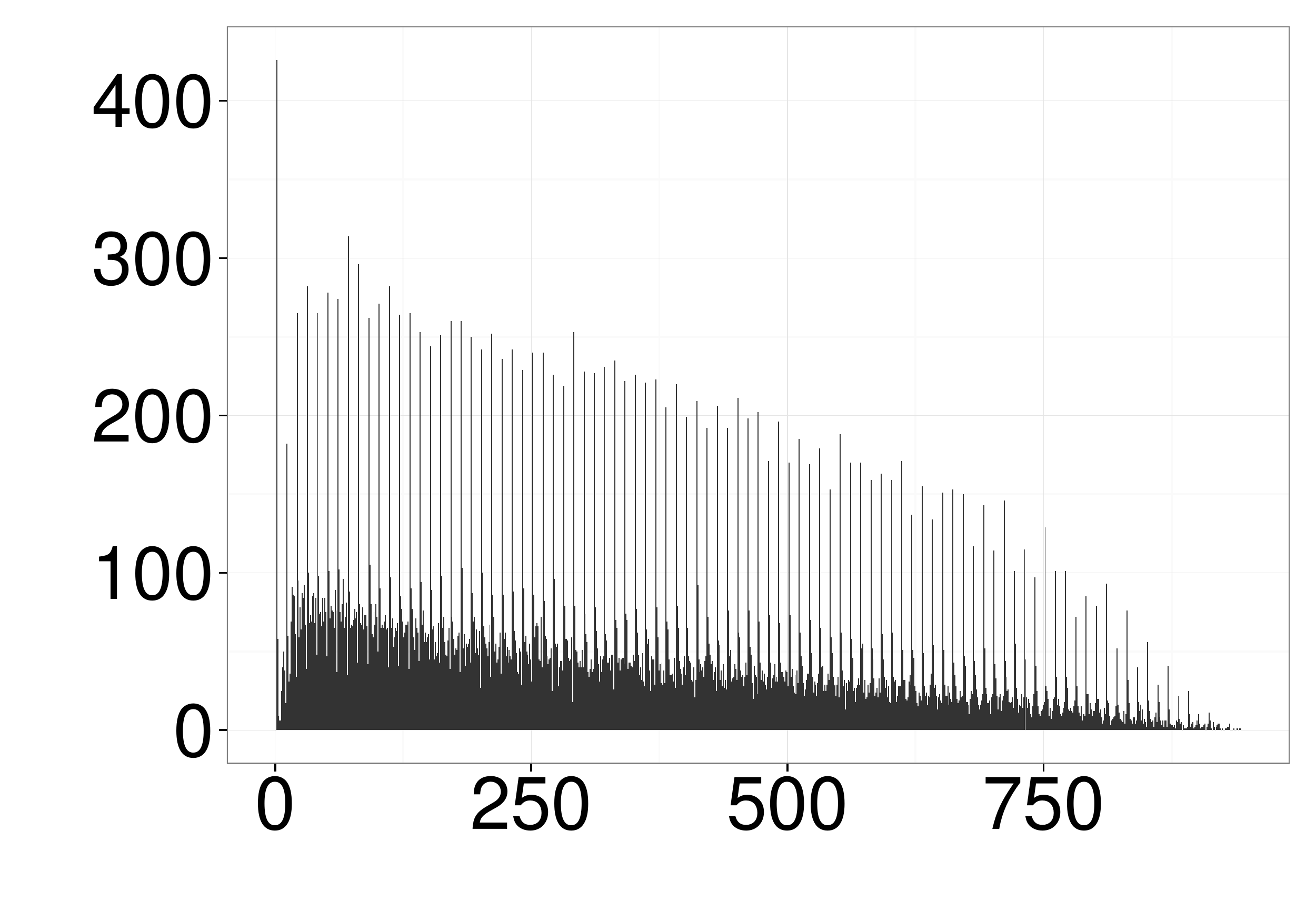}
    & \includegraphics[width=.225\linewidth, height=2cm]{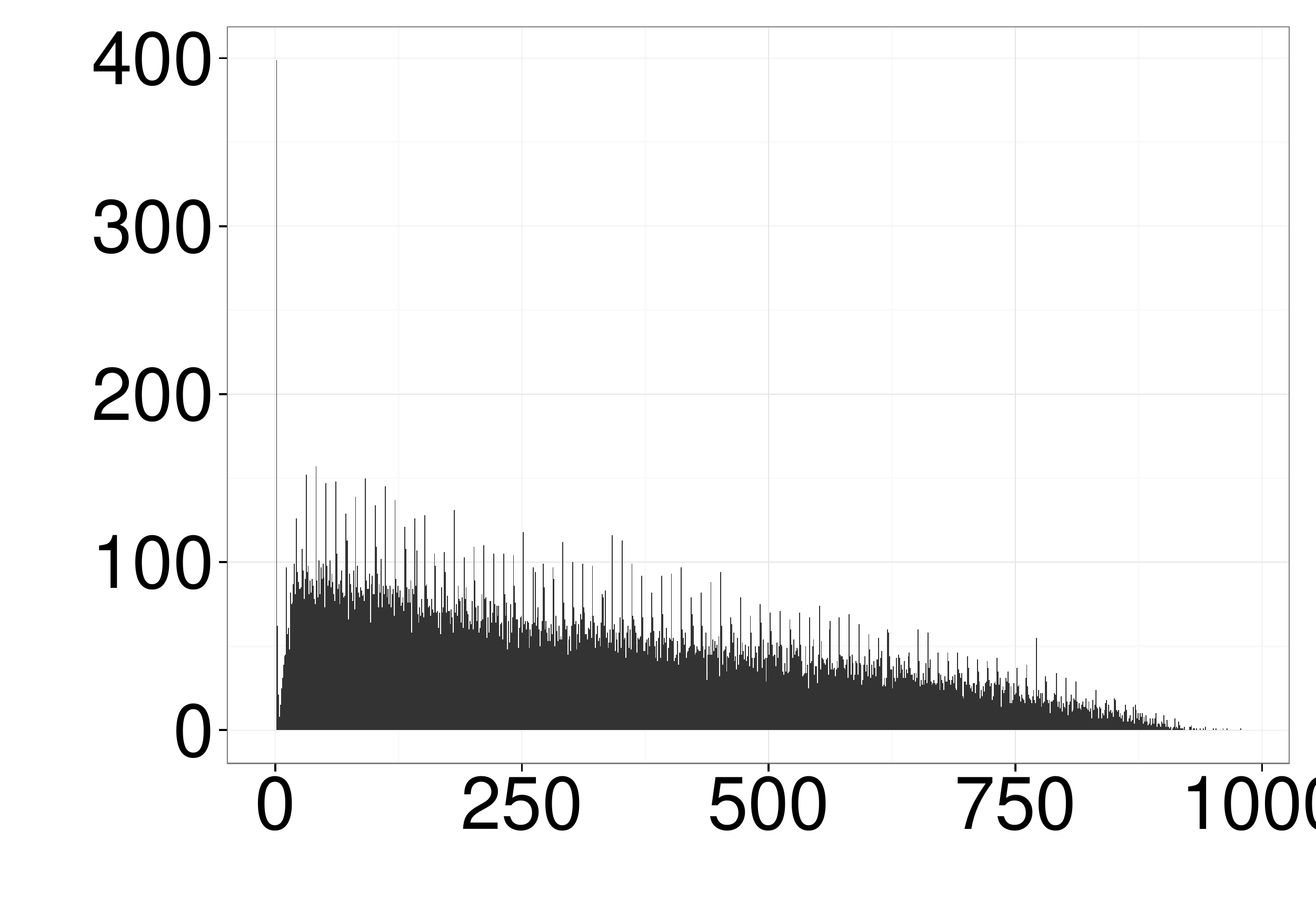}
    & \includegraphics[width=.225\linewidth, height=2cm]{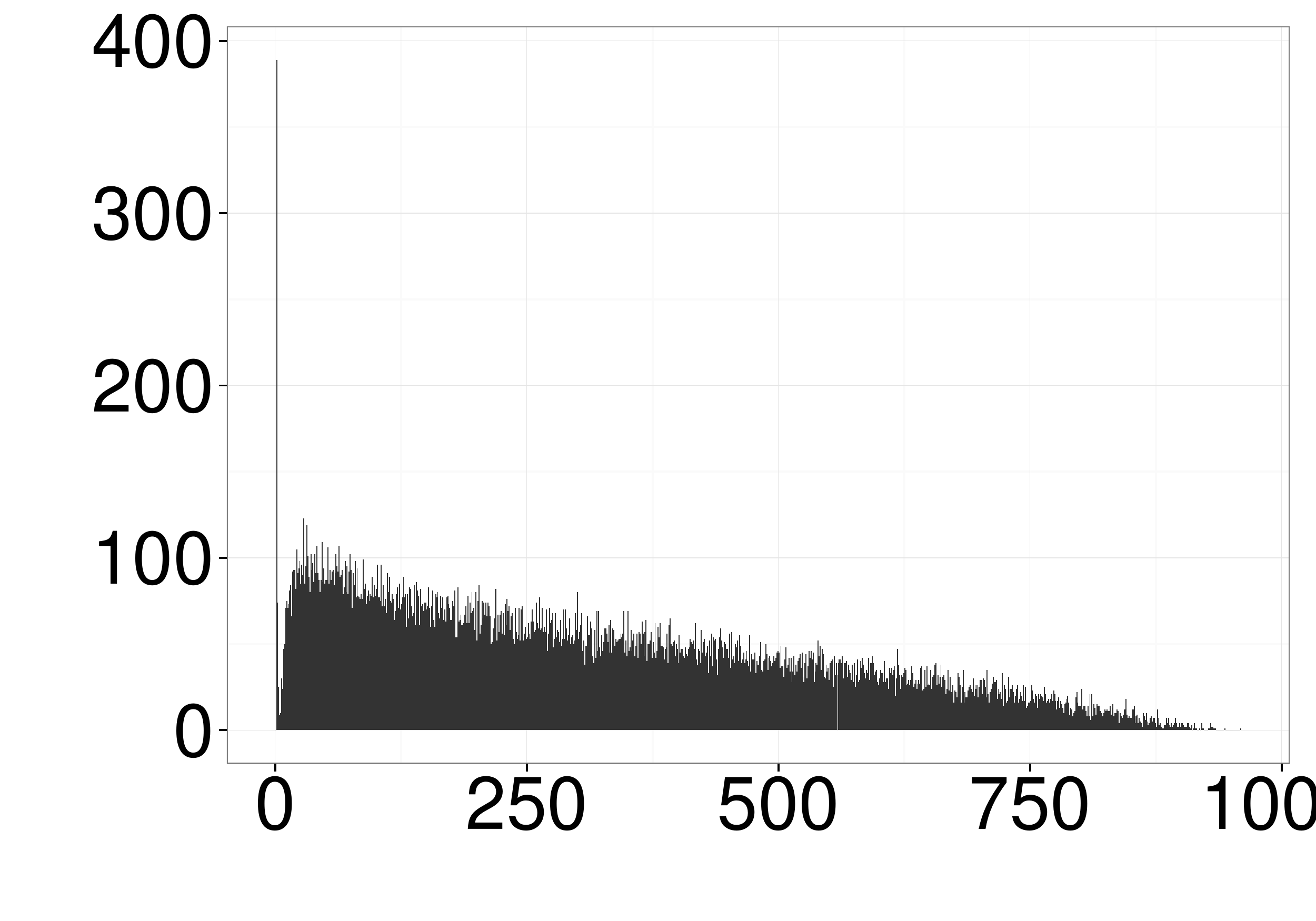} \\
    \rotatebox{90}{\hspace{2.2em}\small 10$\%$}
    & \includegraphics[width=.225\linewidth, height=2cm]{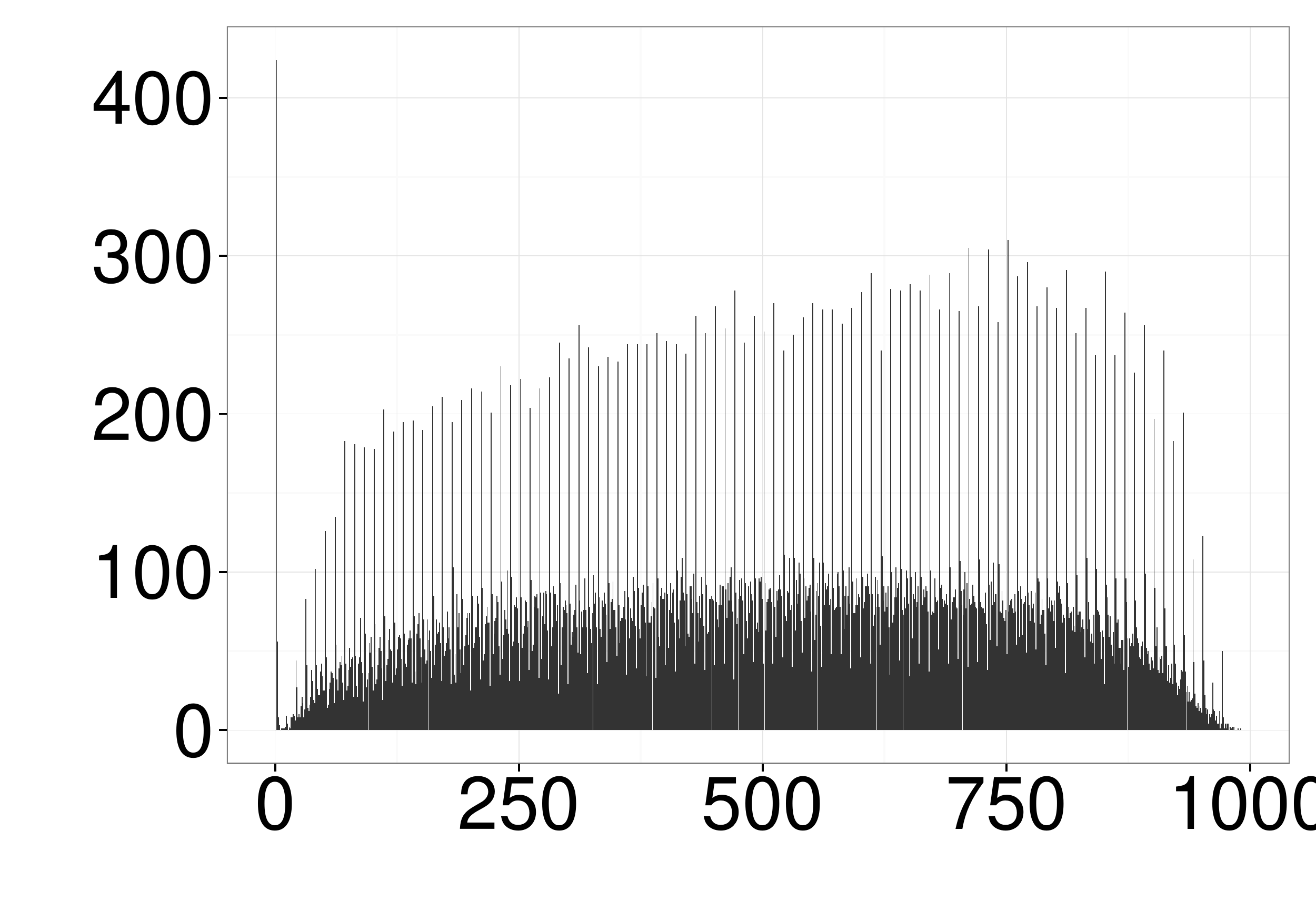}
    & \includegraphics[width=.225\linewidth, height=2cm]{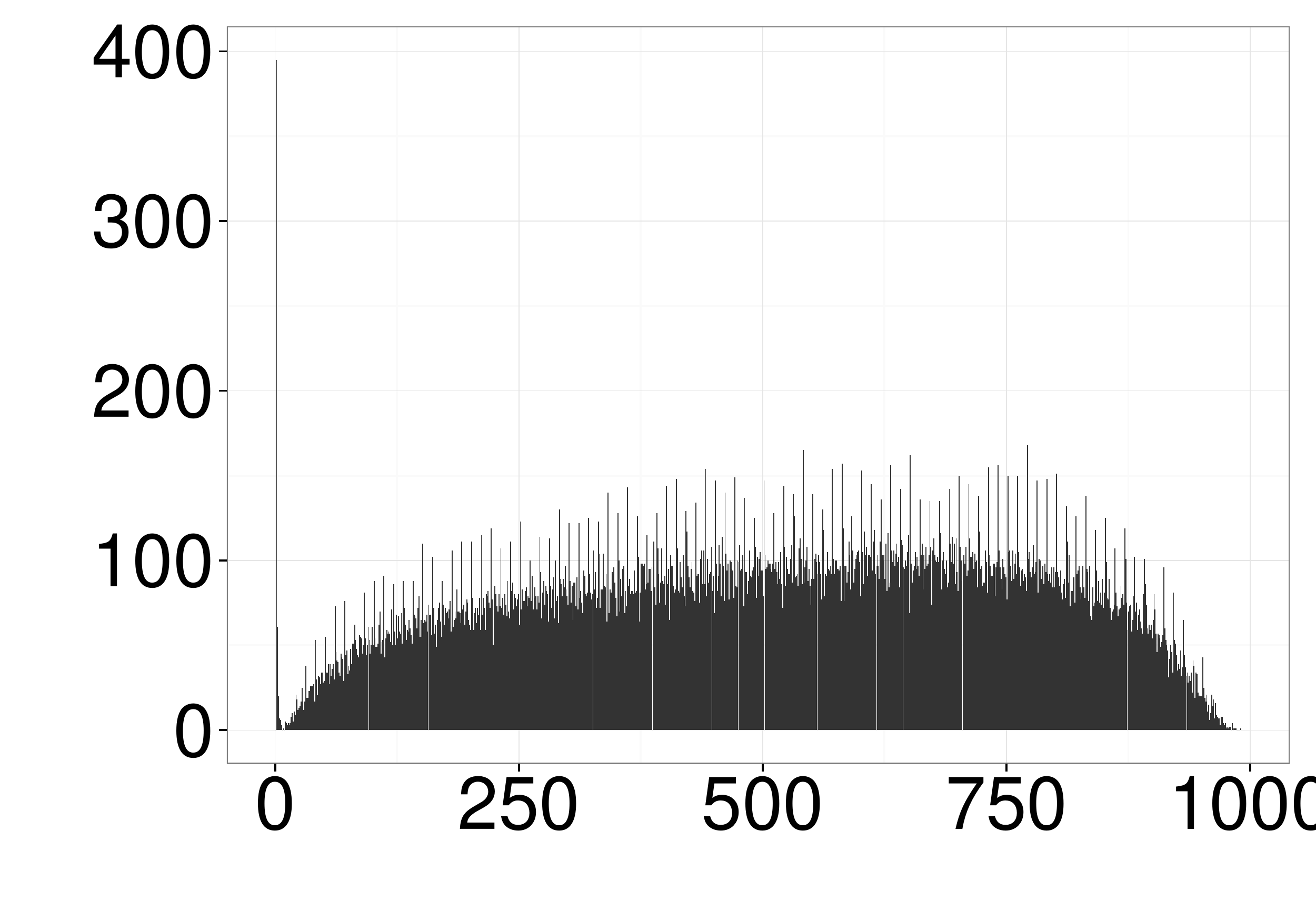}
    & \includegraphics[width=.225\linewidth, height=2cm]{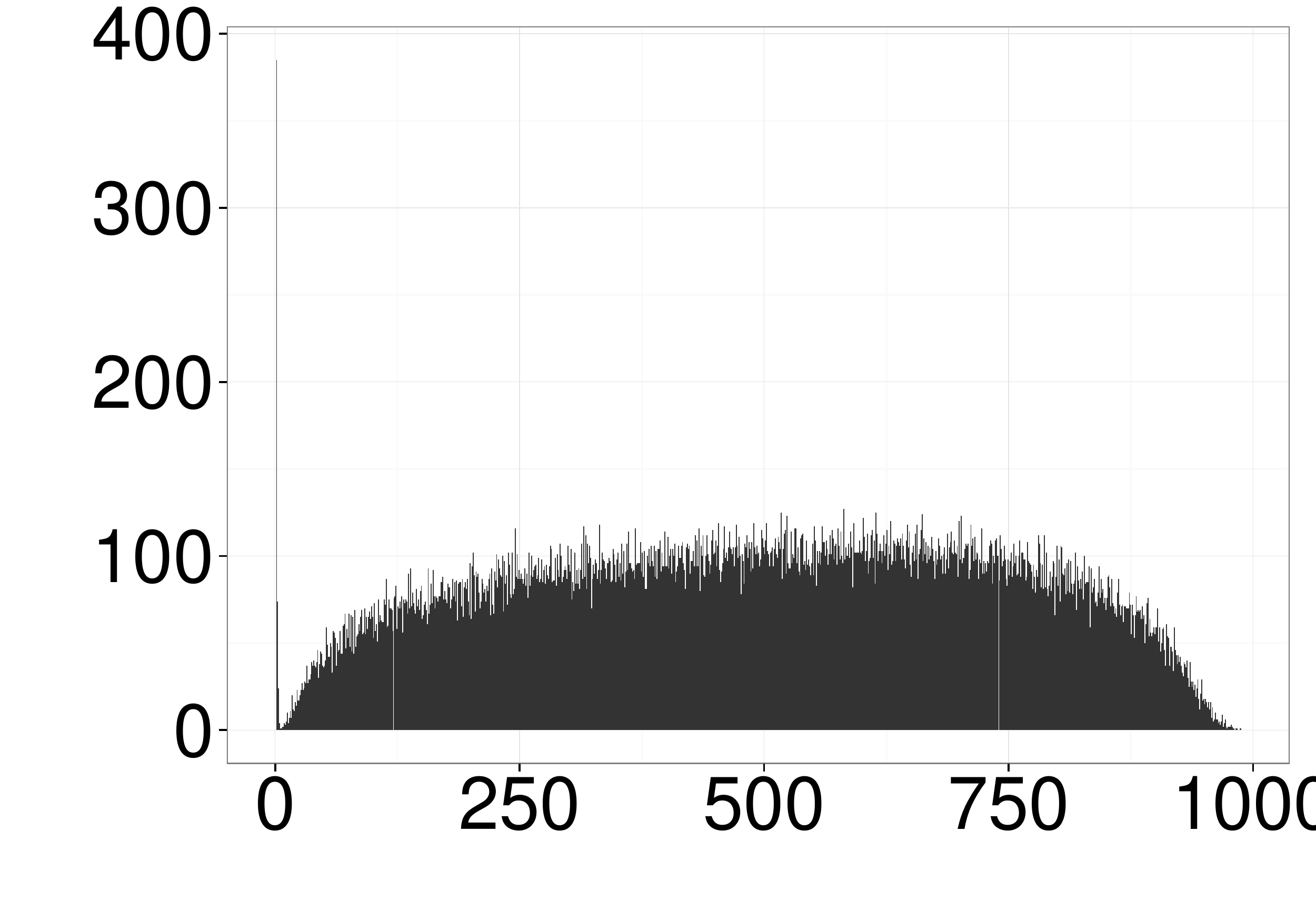} \\
  \end{tabular}
  \caption{\label{fig:SelectionModelHiston1000K100} Barplots of the estimated change-points for different variances (columns)
and different thresholds (rows) in the case where $n=1000$ and $\Ks_1=\Ks_2=100$.}
\end{figure}

\begin{figure}[!h]
  \centering
  \begin{tabular}{@{}l@{}ccc}
    & $\sigma=1$ & $\sigma=2$
    & $\sigma=5$ \\
    \rotatebox{90}{\hspace{2.2em}\small 50$\%$}
    & \includegraphics[width=.225\linewidth, height=2cm]{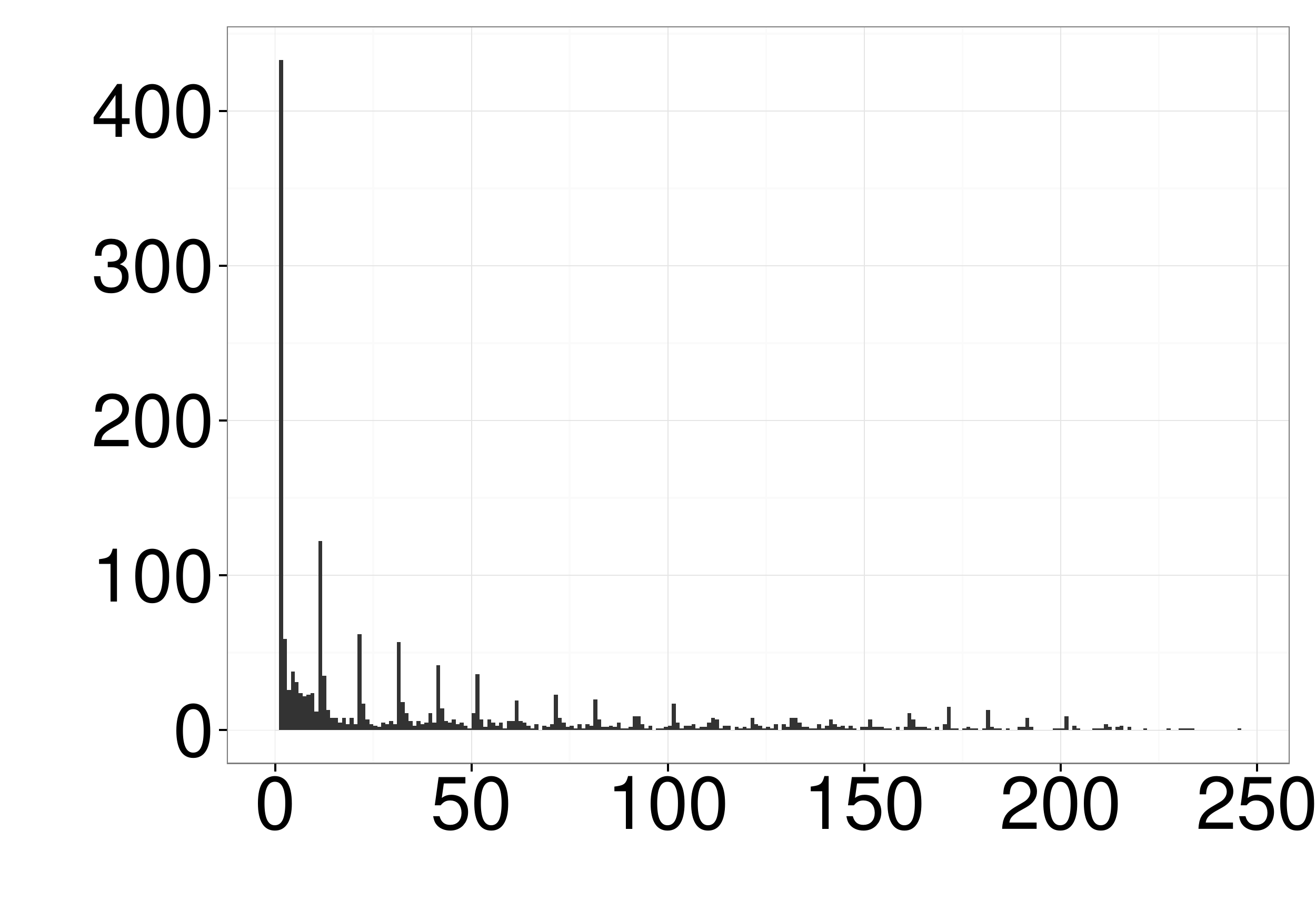}
    & \includegraphics[width=.225\linewidth, height=2cm]{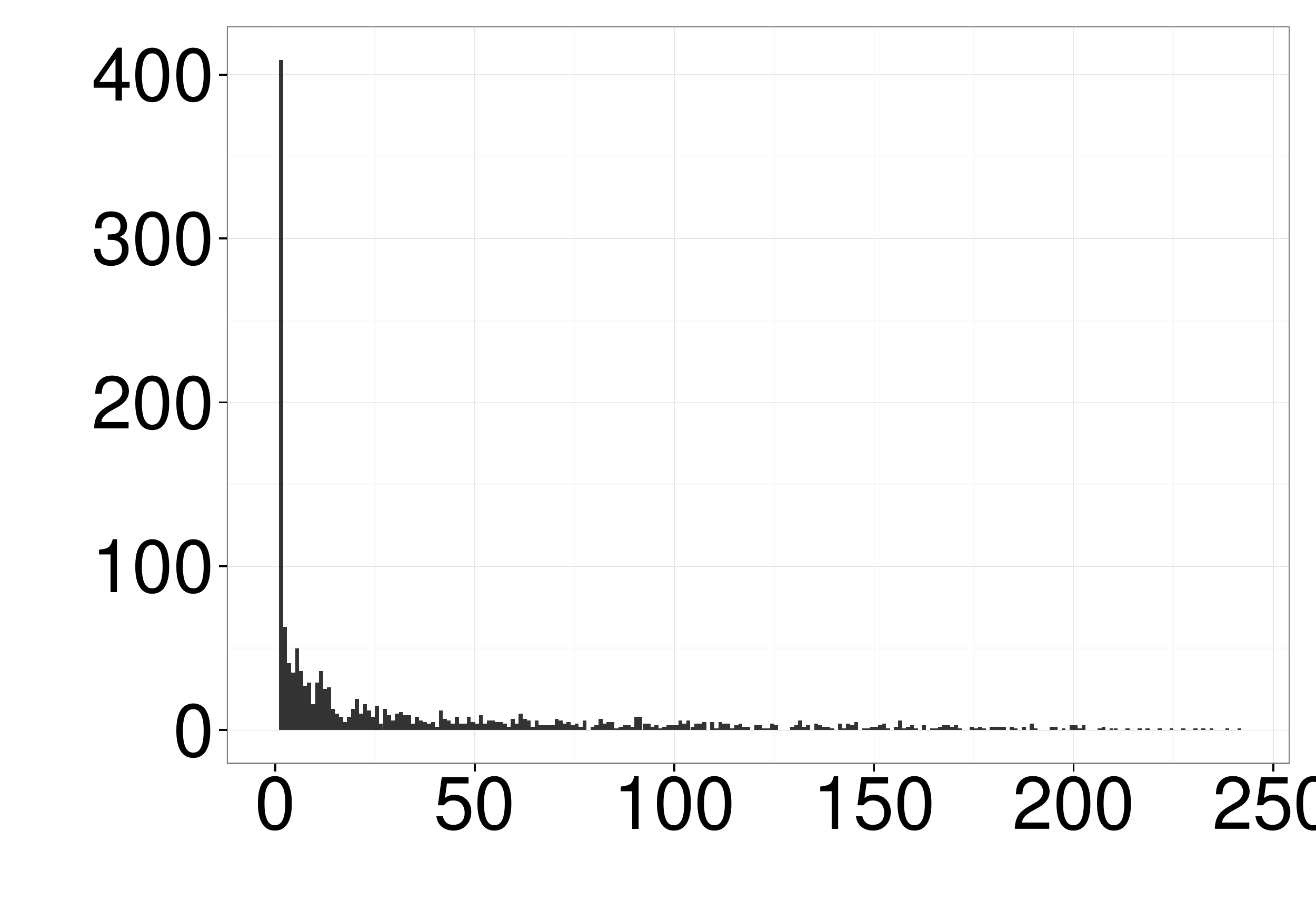}
    & \includegraphics[width=.225\linewidth, height=2cm]{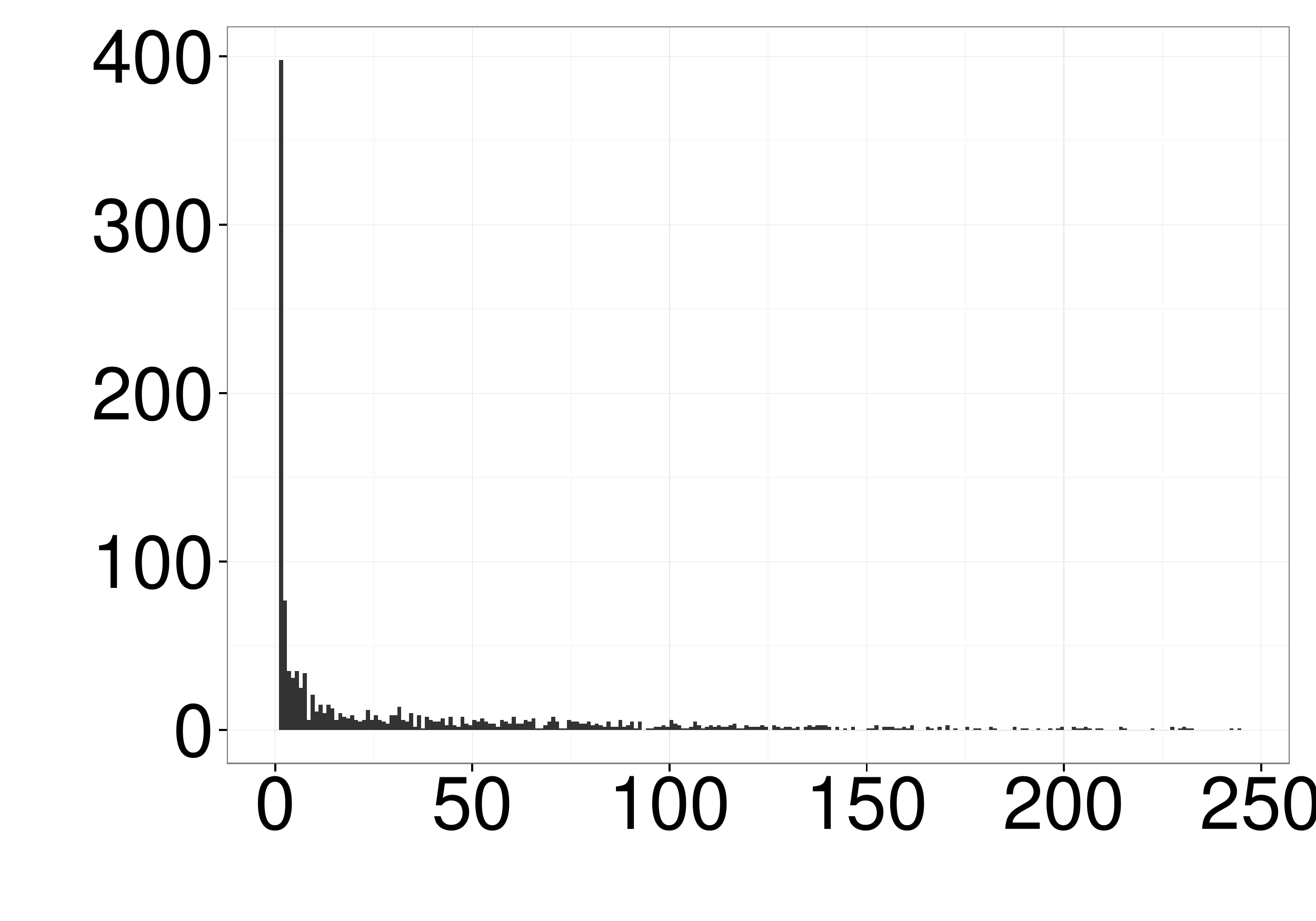} \\
    \rotatebox{90}{\hspace{2.2em}\small 40$\%$}
    & \includegraphics[width=.225\linewidth, height=2cm]{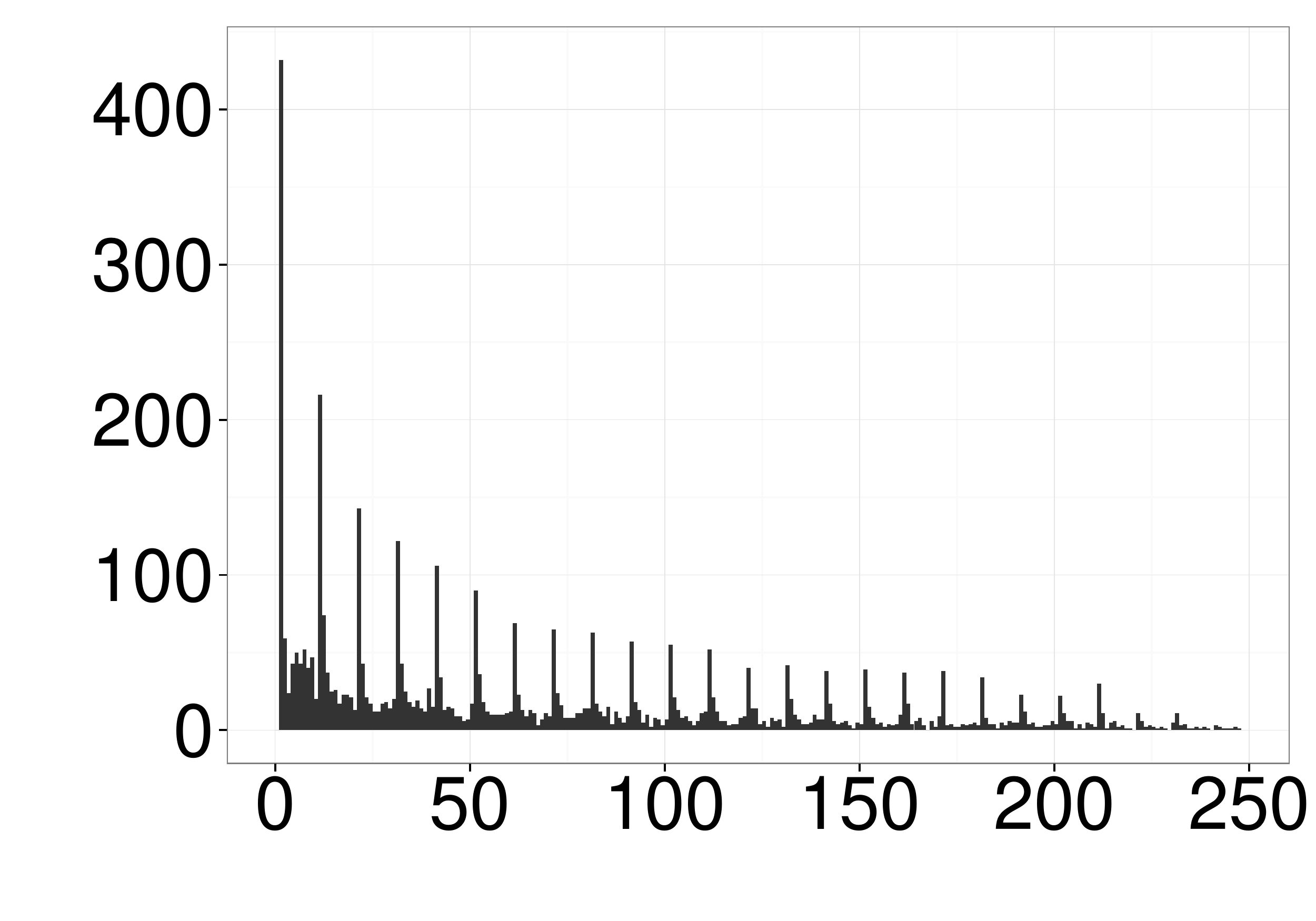}
    & \includegraphics[width=.225\linewidth, height=2cm]{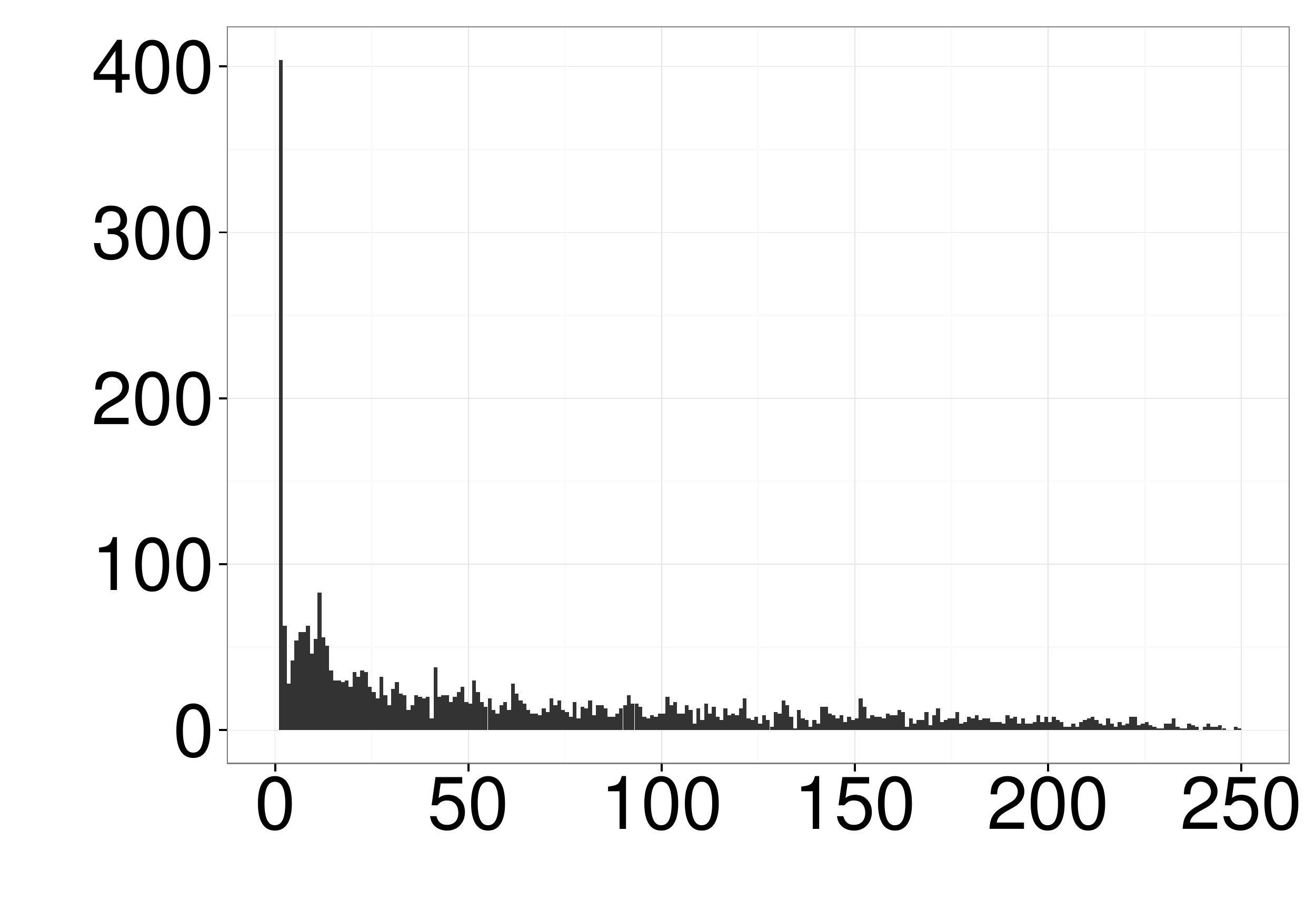}
    & \includegraphics[width=.225\linewidth, height=2cm]{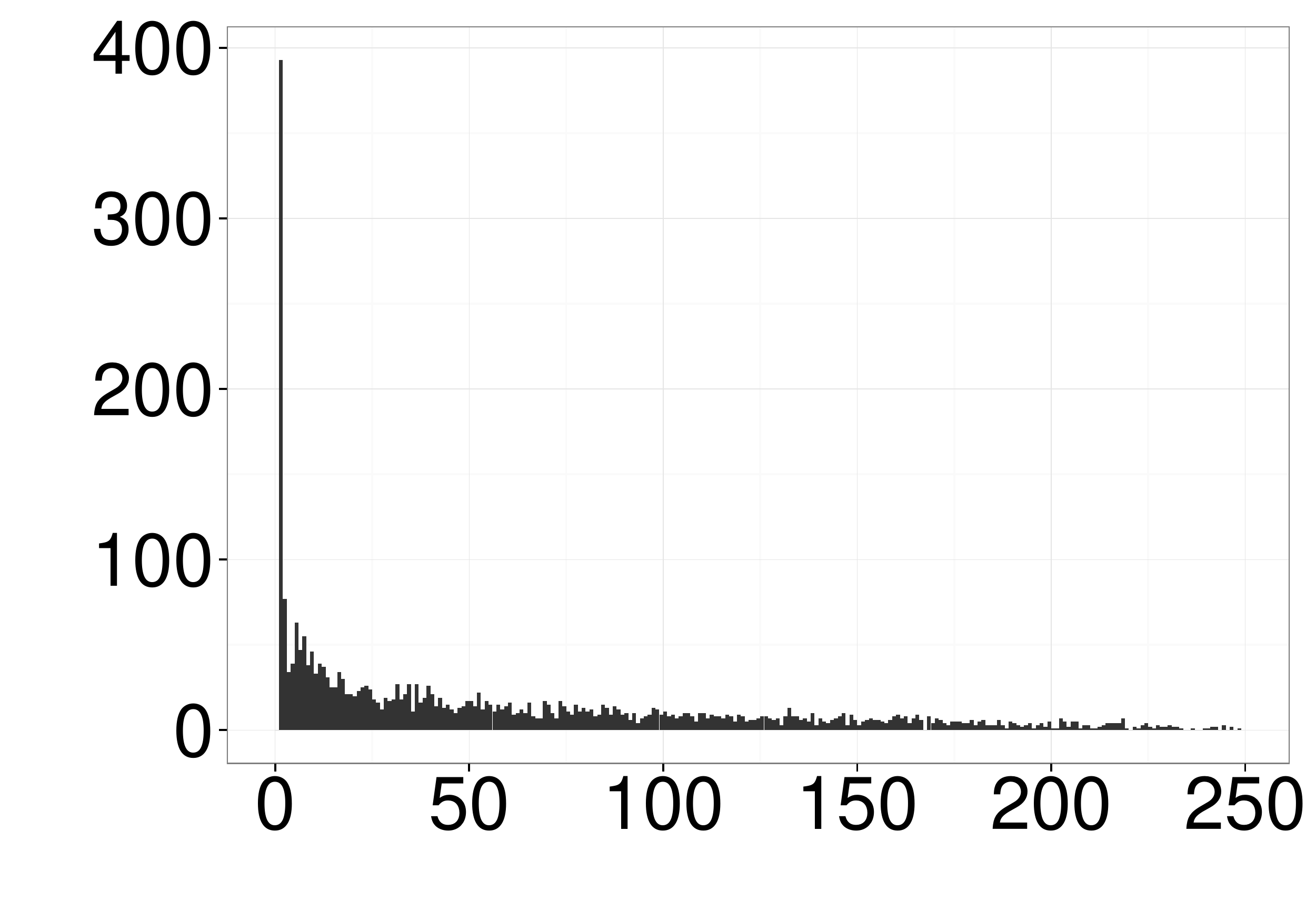} \\
    \rotatebox{90}{\hspace{2.2em}\small 30$\%$}
    & \includegraphics[width=.225\linewidth, height=2cm]{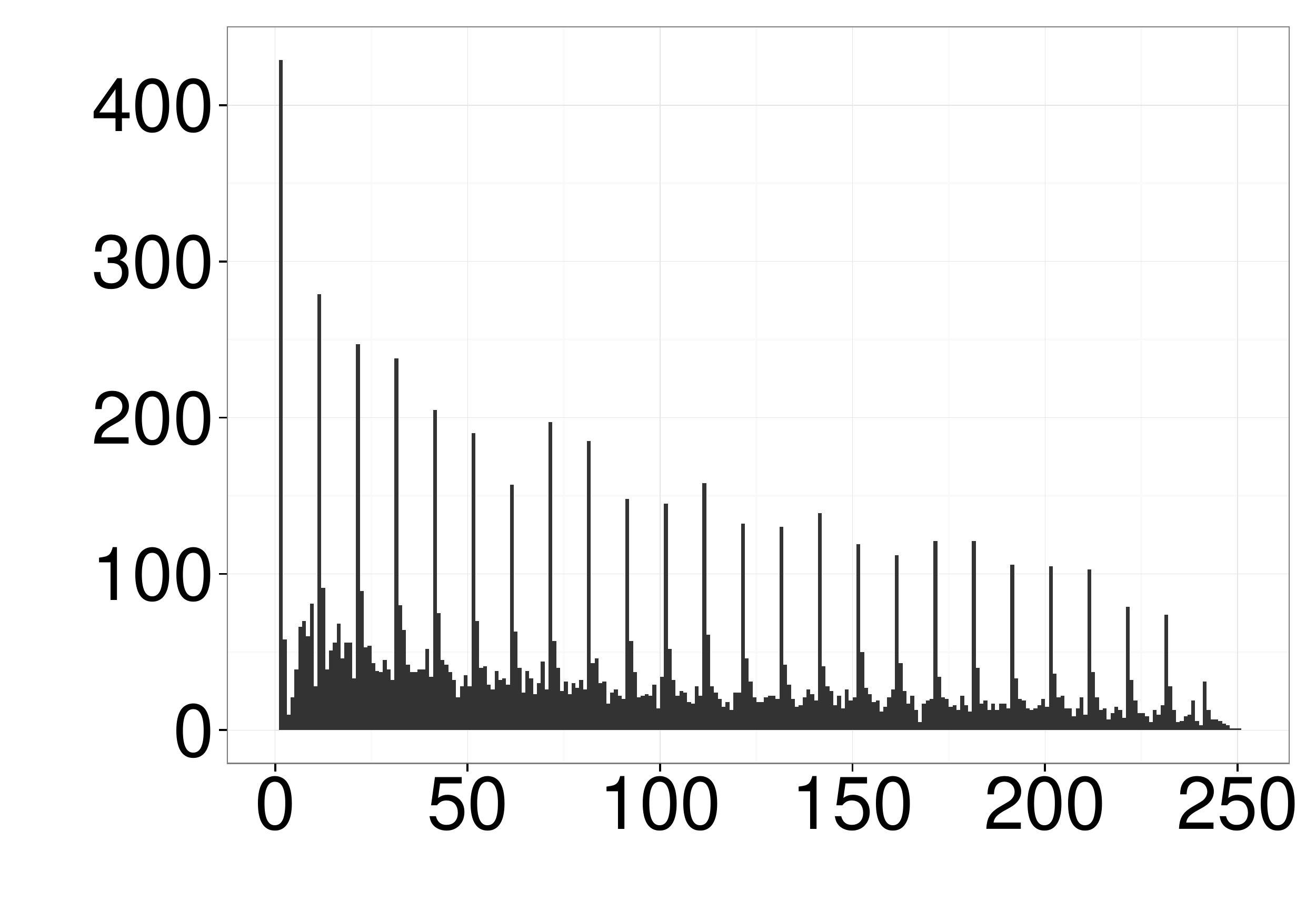}
    & \includegraphics[width=.225\linewidth, height=2cm]{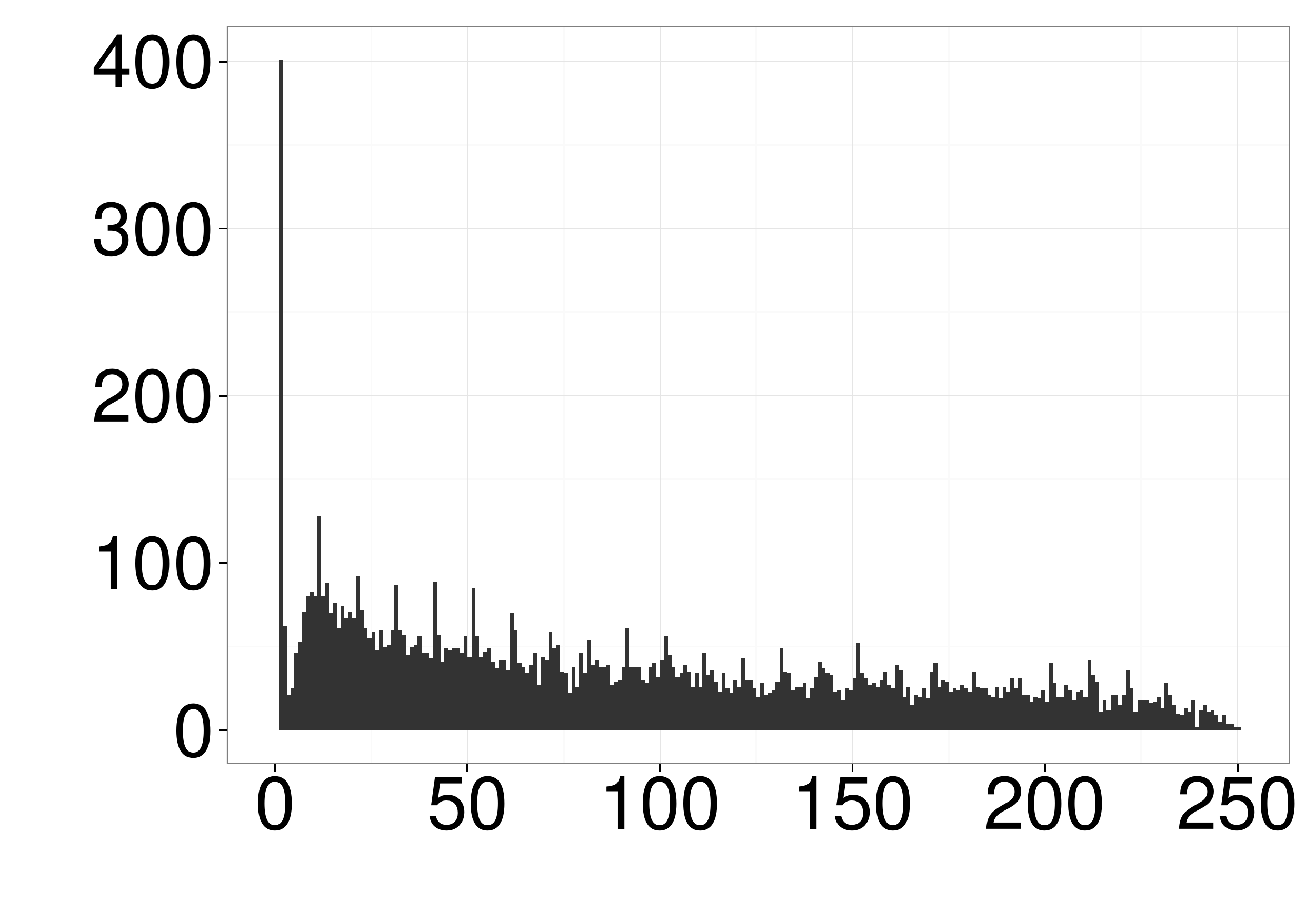}
    & \includegraphics[width=.225\linewidth, height=2cm]{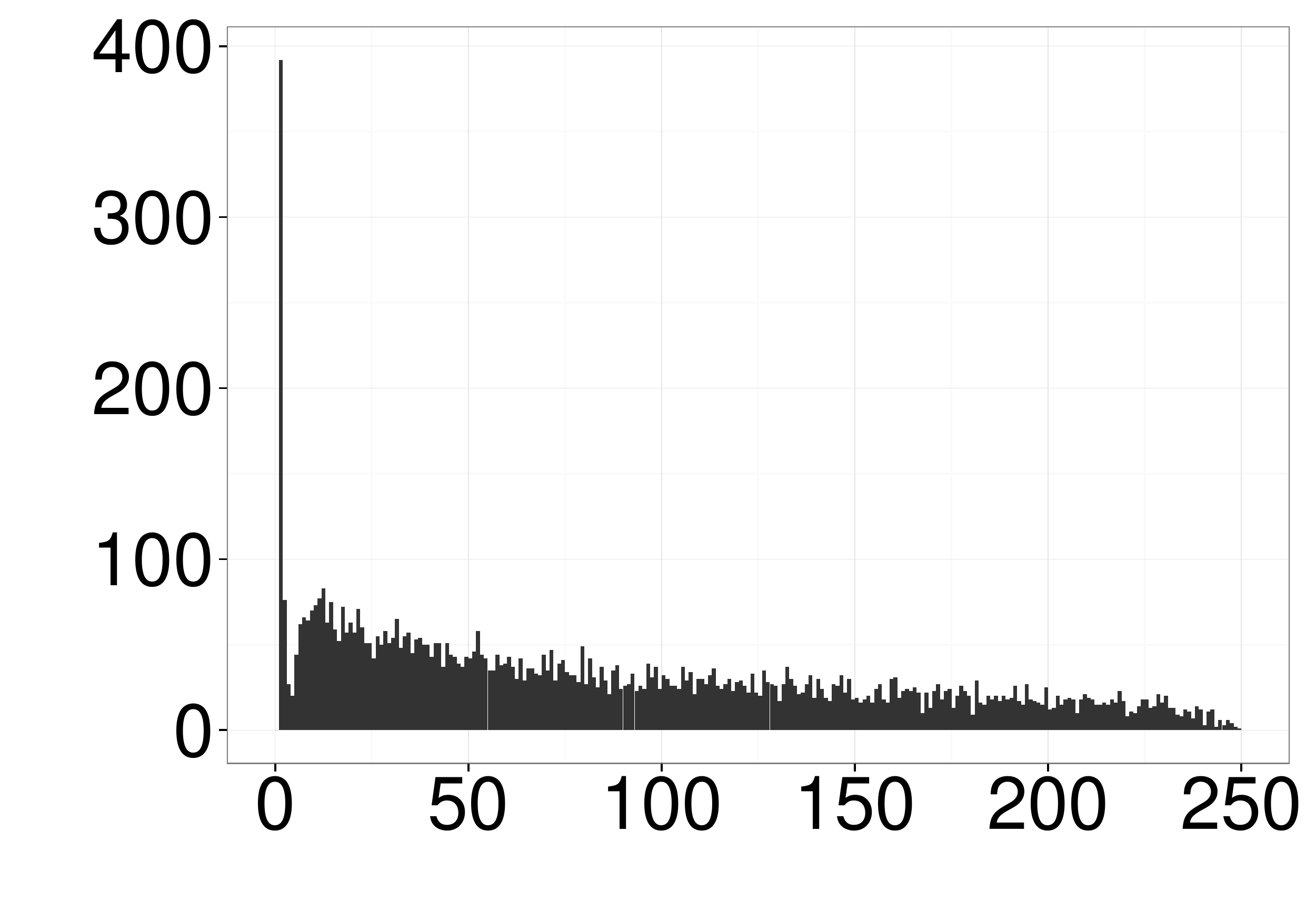} \\
    \rotatebox{90}{\hspace{2.2em}\small 20$\%$}
    & \includegraphics[width=.225\linewidth, height=2cm]{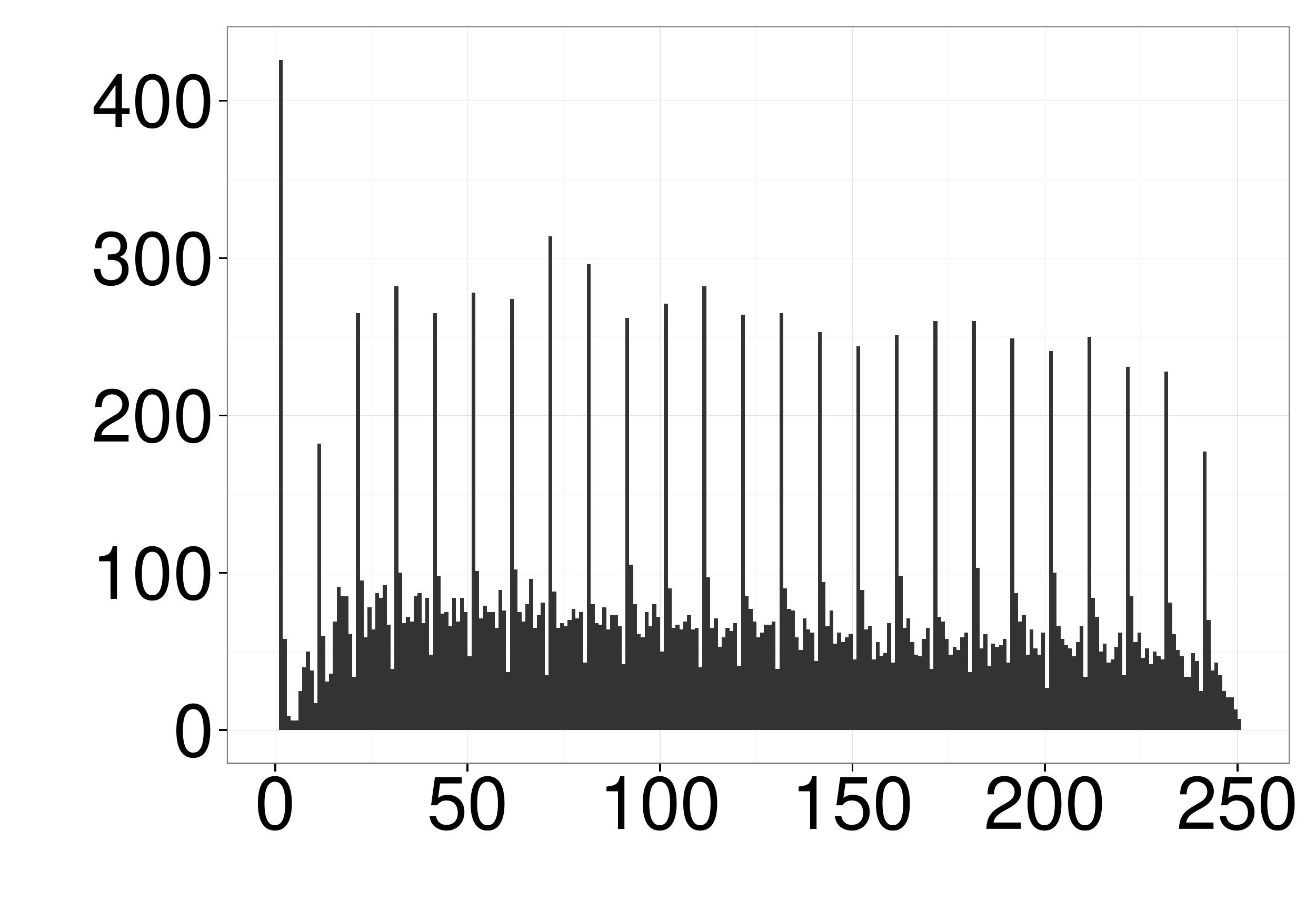}
    & \includegraphics[width=.225\linewidth, height=2cm]{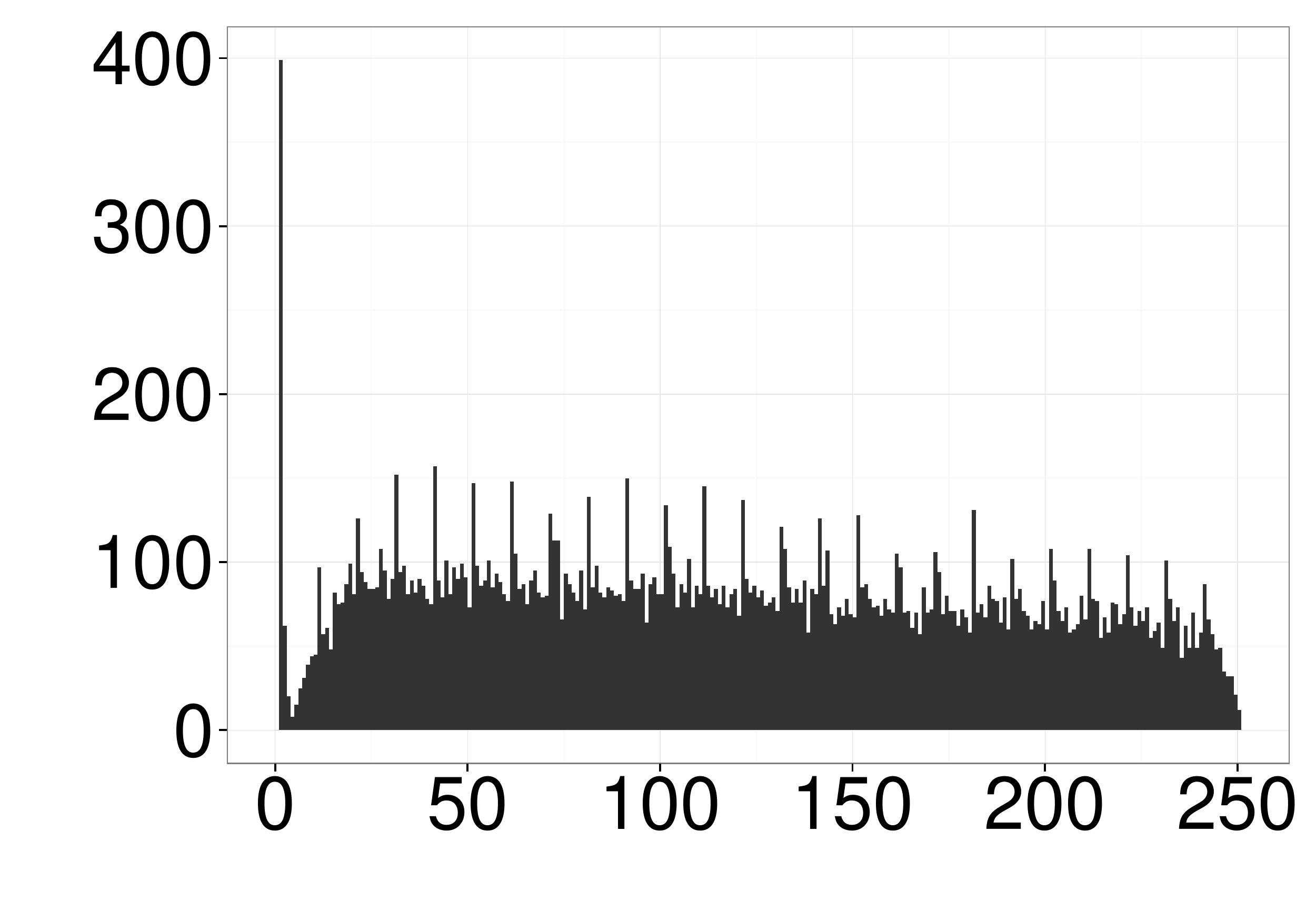}
    & \includegraphics[width=.225\linewidth, height=2cm]{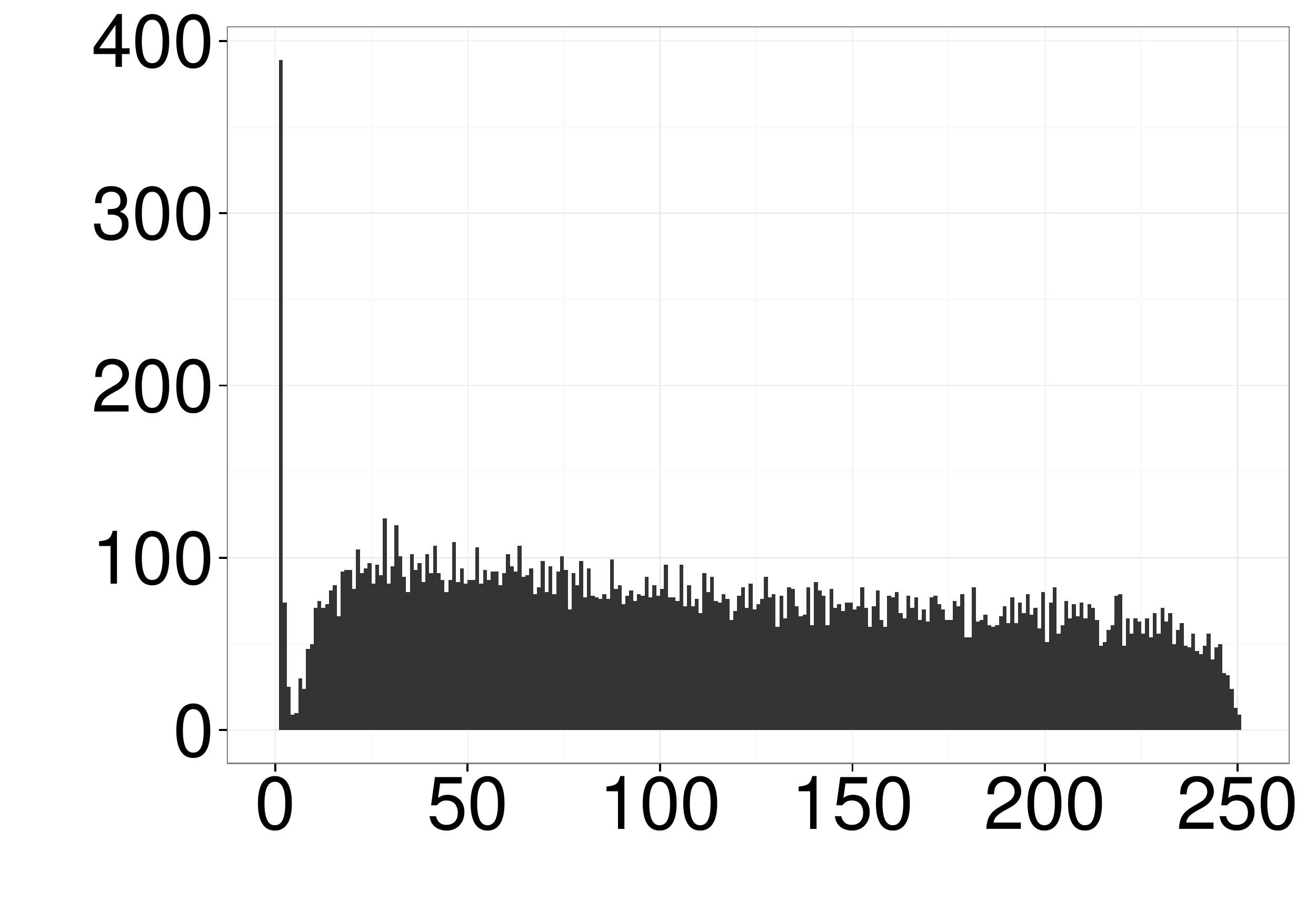} \\
    \rotatebox{90}{\hspace{2.2em}\small 10$\%$}
    & \includegraphics[width=.225\linewidth, height=2cm]{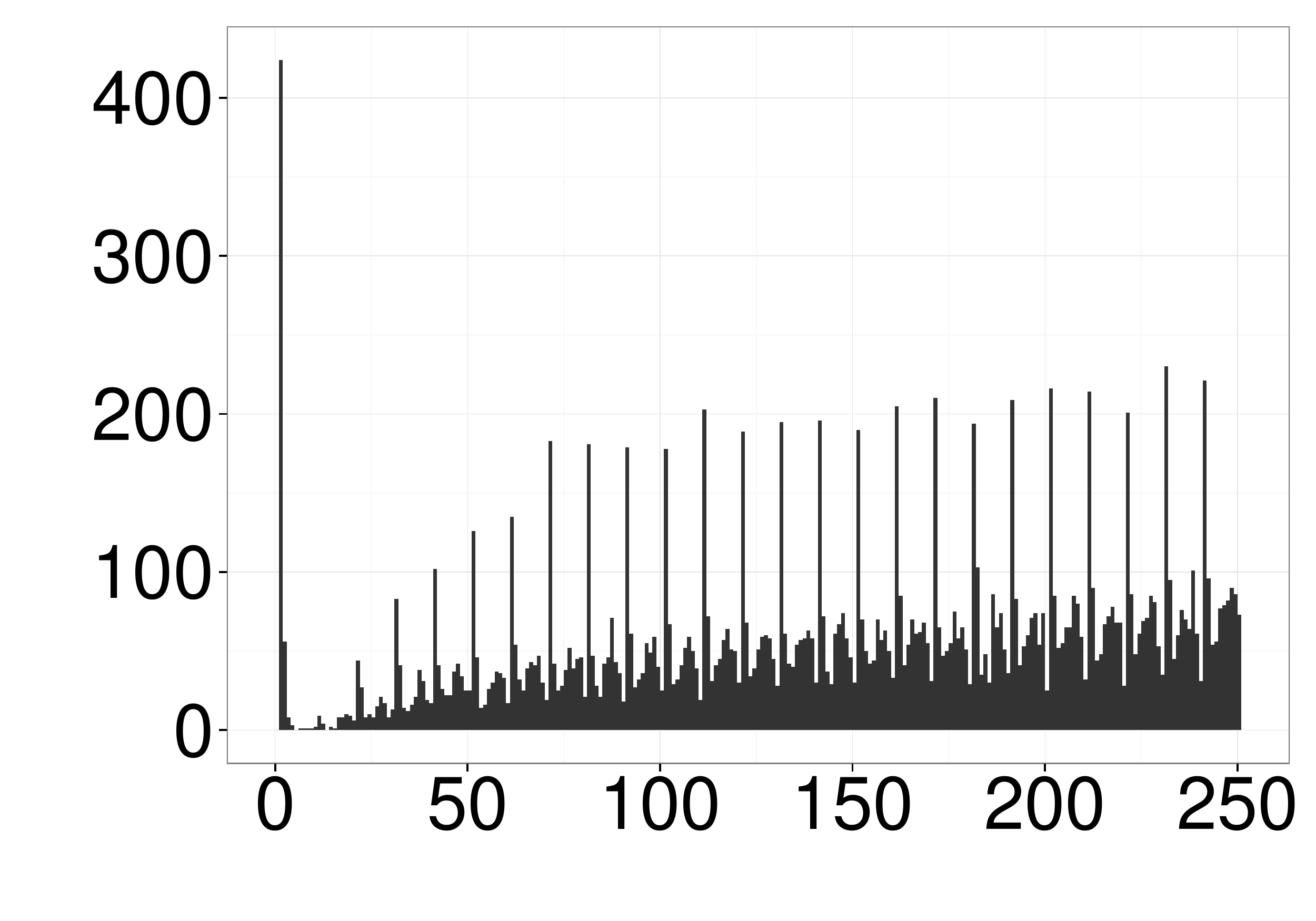}
    & \includegraphics[width=.225\linewidth, height=2cm]{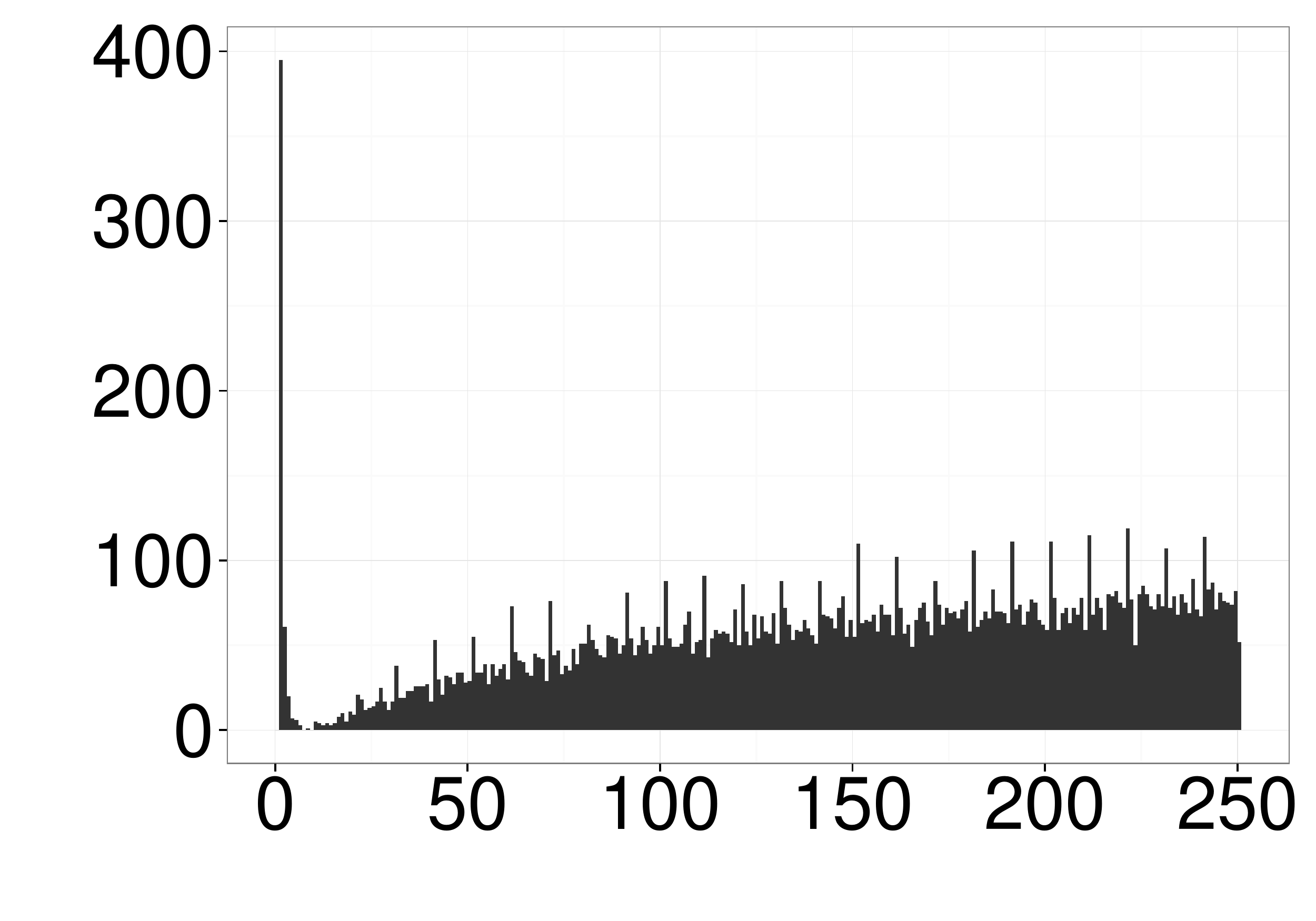}
    & \includegraphics[width=.225\linewidth, height=2cm]{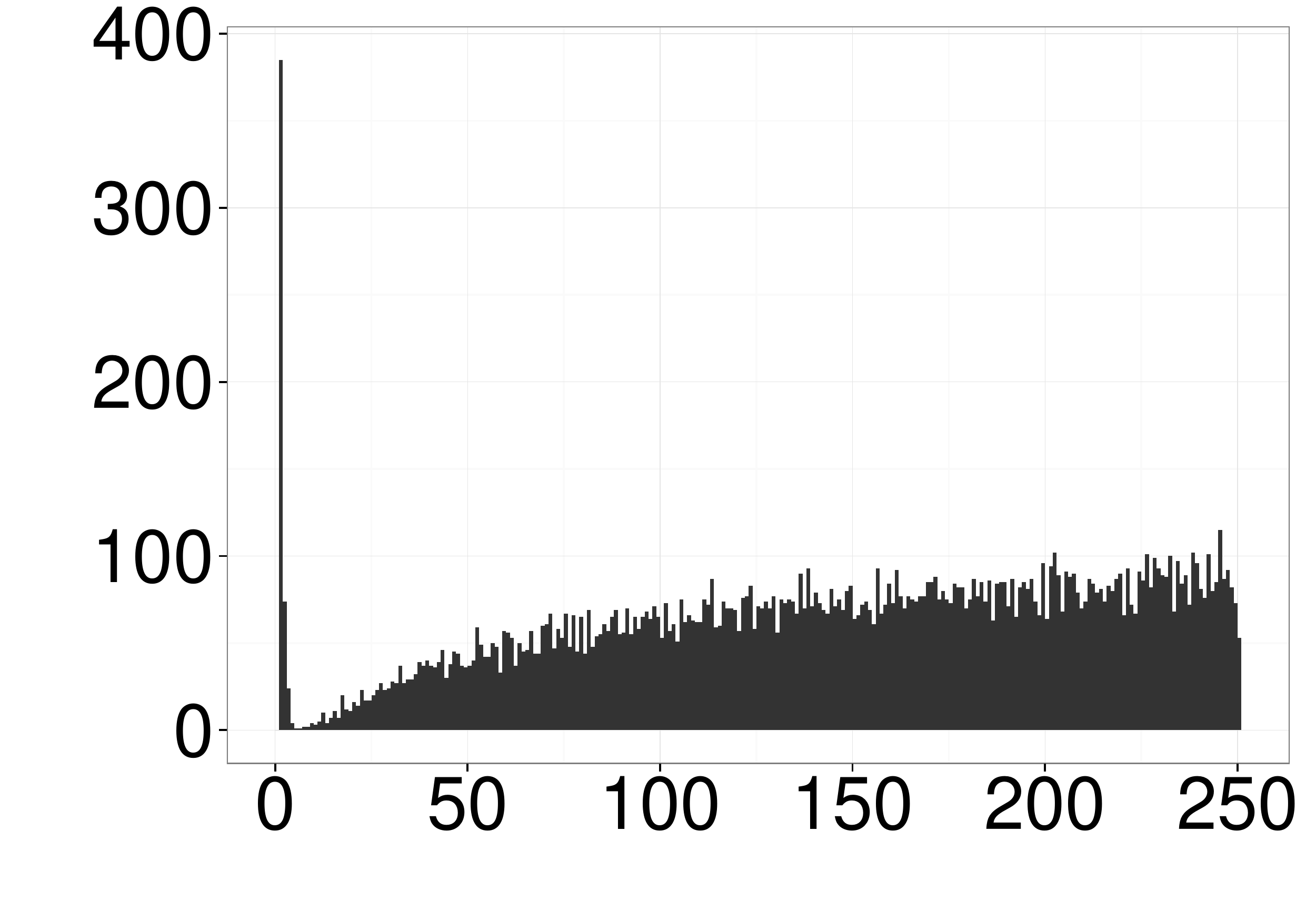} \\
  \end{tabular}
  \caption{\label{fig:zoomSelectionModelHiston1000K100} Zoom of the barplots of Figure \ref{fig:SelectionModelHiston1000K100}.}
\end{figure}

In order to further assess  our methodology including
  the post-processing step and to be in a framework closer to our real
data application, we generated observations following (\ref{eq:model1})
with $n=1000$ and $\Ks_1=\Ks_2=100$
where we used for the matrix $\U$ the same shape as the one of the
matrix $\left({\mu}_{k,\ell}^{\star,(1)}\right)$ except
that $\Ks_1=\Ks_2=100$. In this framework, the proportion of change-points is thus ten times larger than the one of the previous
case. The corresponding results are displayed in Figures \ref{fig:SelectionModeln1000K100},
\ref{fig:SelectionModelHiston1000K100} and \ref{fig:zoomSelectionModelHiston1000K100}. We can see from the last figure that
taking a threshold equal to 20\% provides the best estimations of the number and of the change-point positions.
This threshold corresponds to the lower bound of the thresholds
interval obtained in the previous configuration.
Our package \texttt{blockseg} provides an estimation of the matrix
$\U$ for any threshold given by the user as we shall explain in the
next section.

\FloatBarrier


\section{Application to HiC data}
\label{sec:app_hic}

In this section, we apply our methodology to publicly available HiC data
(\url{http://chromosome.sdsc.edu/mouse/hi-c/download.html})
already studied by \cite{dixon2012topological}.
This technology is based on a deep sequencing approach
and provides read pairs corresponding to pairs of genomic loci that
physically interacts in the nucleus, see \cite{lieberman2009comprehensive} for more details.
The raw measurements provided by HiC data is therefore a list of pairs of
locations along the chromosome, at the nucleotide resolution. These
measurement are often summarized as a square matrix where each entry
at row $i$ and column $j$ stands for the total number of read pairs matching in position
$i$ and position $j$, respectively. Positions refer here to a sequence
of non-overlapping windows of equal sizes covering the genome.
The number of windows may vary from one study to another:
\cite{lieberman2009comprehensive} considered a Mb resolution, whereas
\cite{dixon2012topological} went deeper and used windows of 40kb (called hereafter the resolution).

In our study, we processed the interaction matrices
of Chromosomes 1 and 19 of the mouse cortex at a resolution 40 kb and
we compared the number and the location of the estimated change-points found by our approach with those obtained
by \cite{dixon2012topological} on the same data since no ground truth is available.
More precisely, in the case of Chromosome 1, $n=4930$ and in the case of Chromosome 19, $n=1534$.

\begin{figure}[!h]
  \centering
  \begin{tabular}{@{}l@{}cc}
  \includegraphics[width=.45\linewidth]{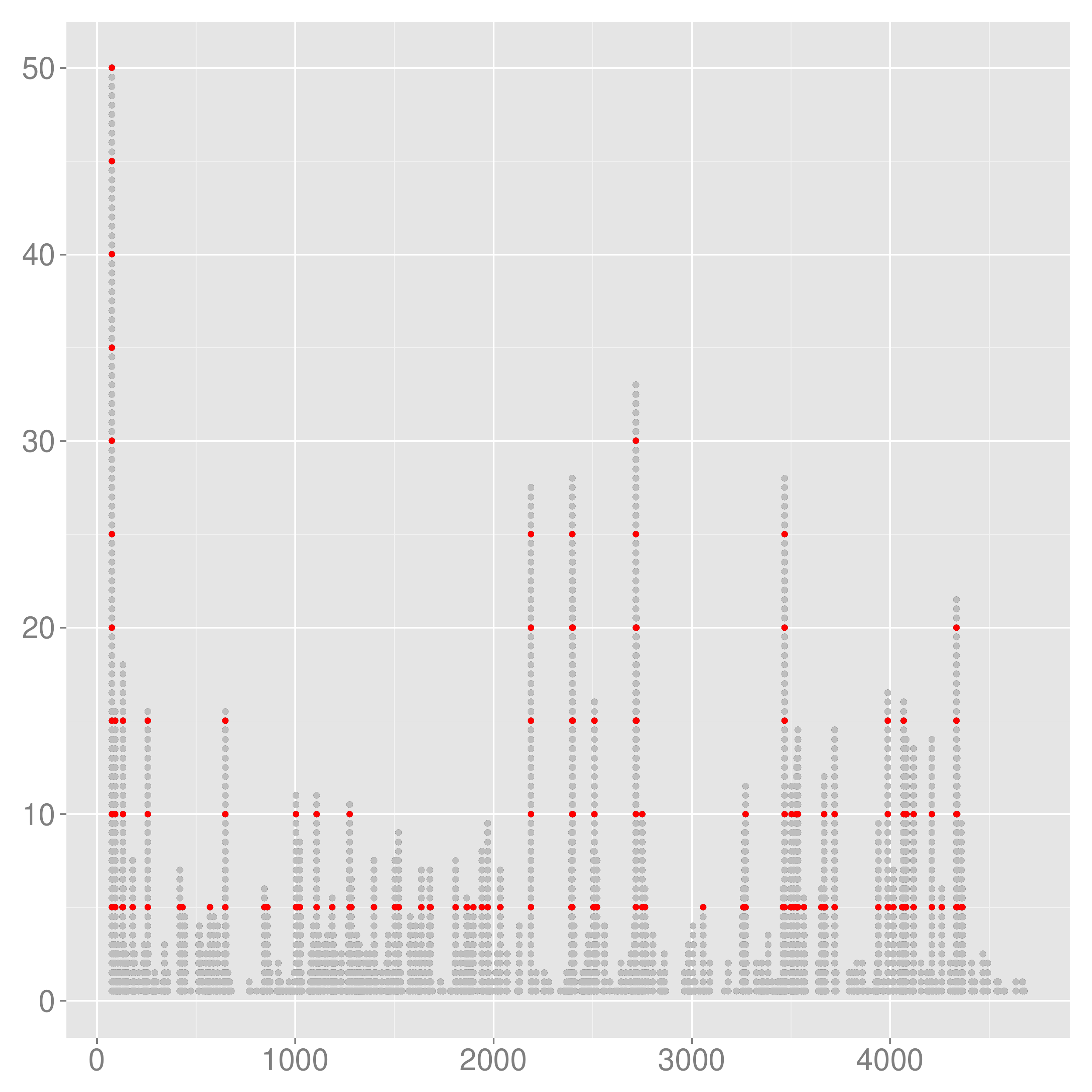}&
  \includegraphics[width=.45\linewidth]{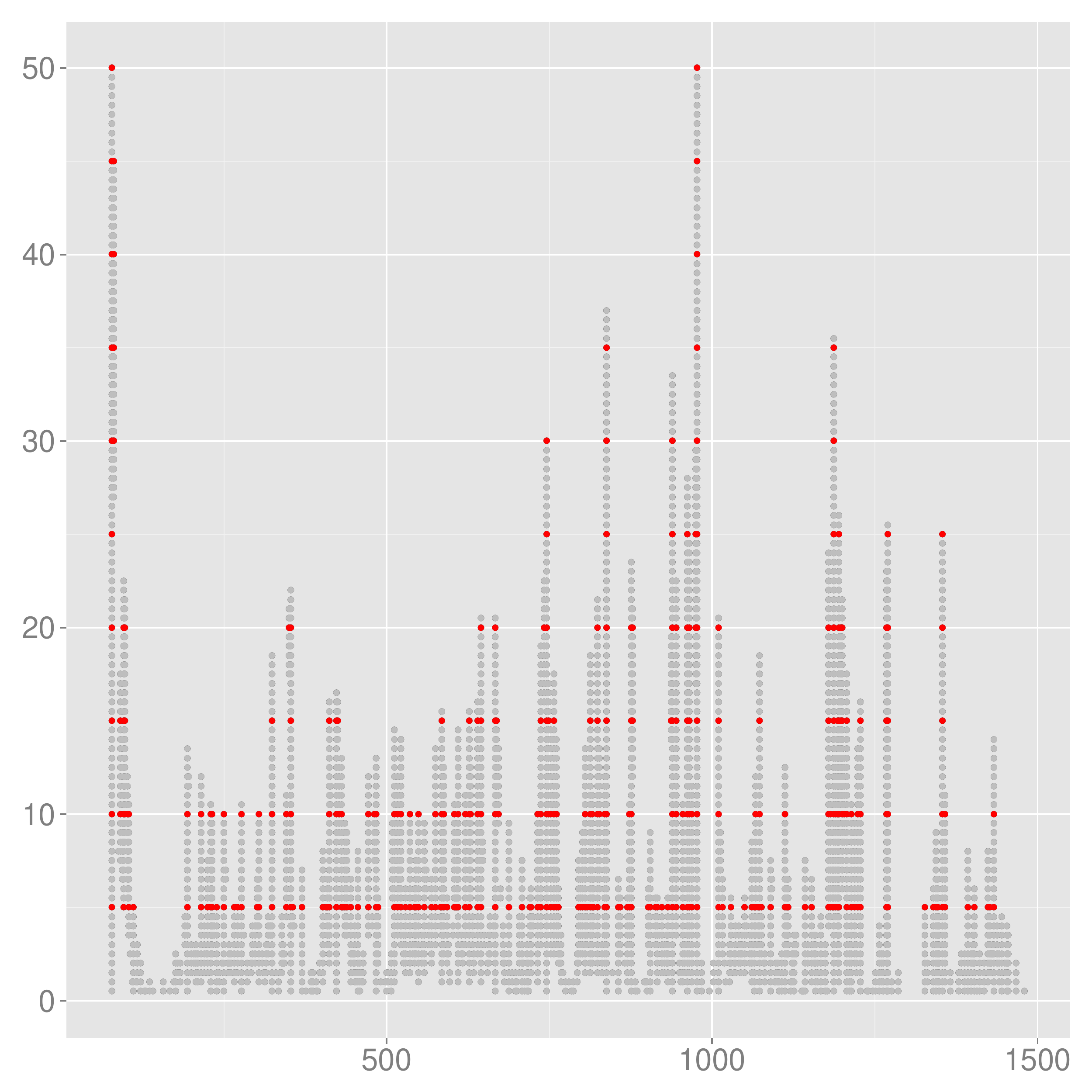}
  \end{tabular}
 \caption{\label{fig:HiCPostRupt} Plots of the estimated change-points locations ($x$-axis) for different thresholds ($y$-axis) from
0.5\% to 50\% by 0.5\% for Chromosome 1 (left) and Chromosome 19
(right). The estimated change-point locations associated to threshold
which are multiples of 5\% are displayed in red.
}
 \end{figure}

\begin{figure}[!h]
\centering
\begin{tabular}{cccc}
&10\%&15\%&20\%\\
\rotatebox{90}{\hspace{2.25em}\small Chromosome 1}&
  \includegraphics[width=.3\linewidth]{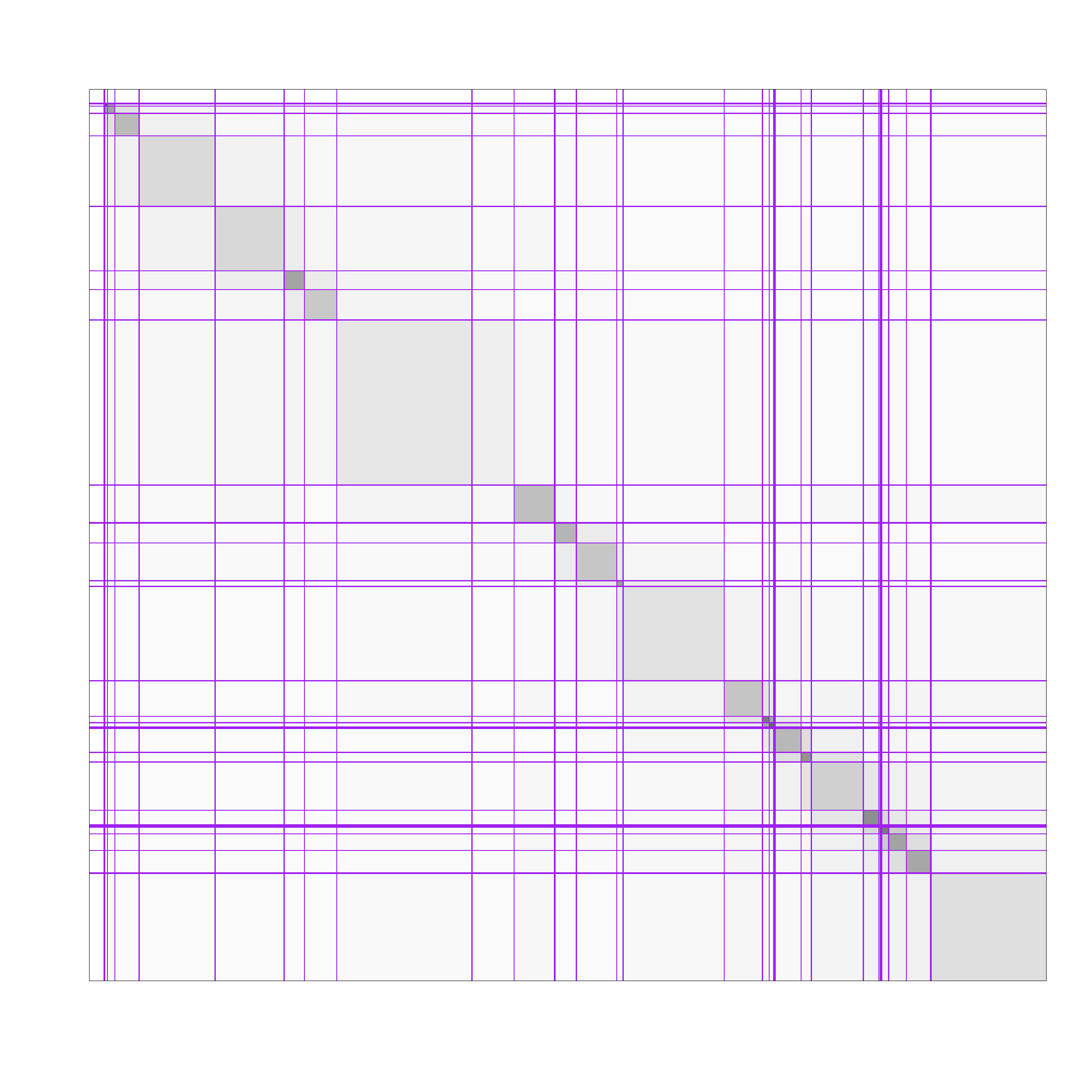}&
  \includegraphics[width=.3\linewidth]{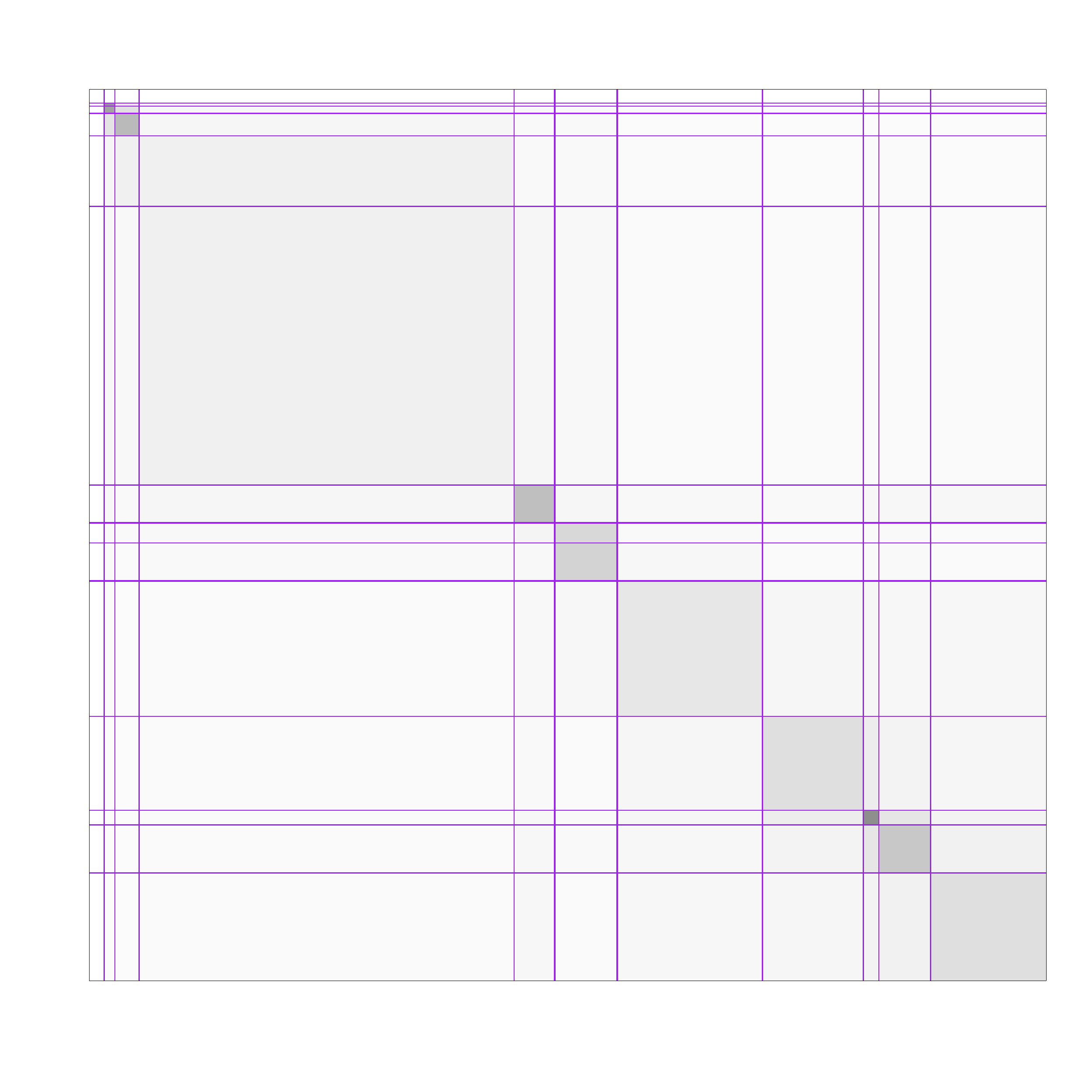}&
  \includegraphics[width=.3\linewidth]{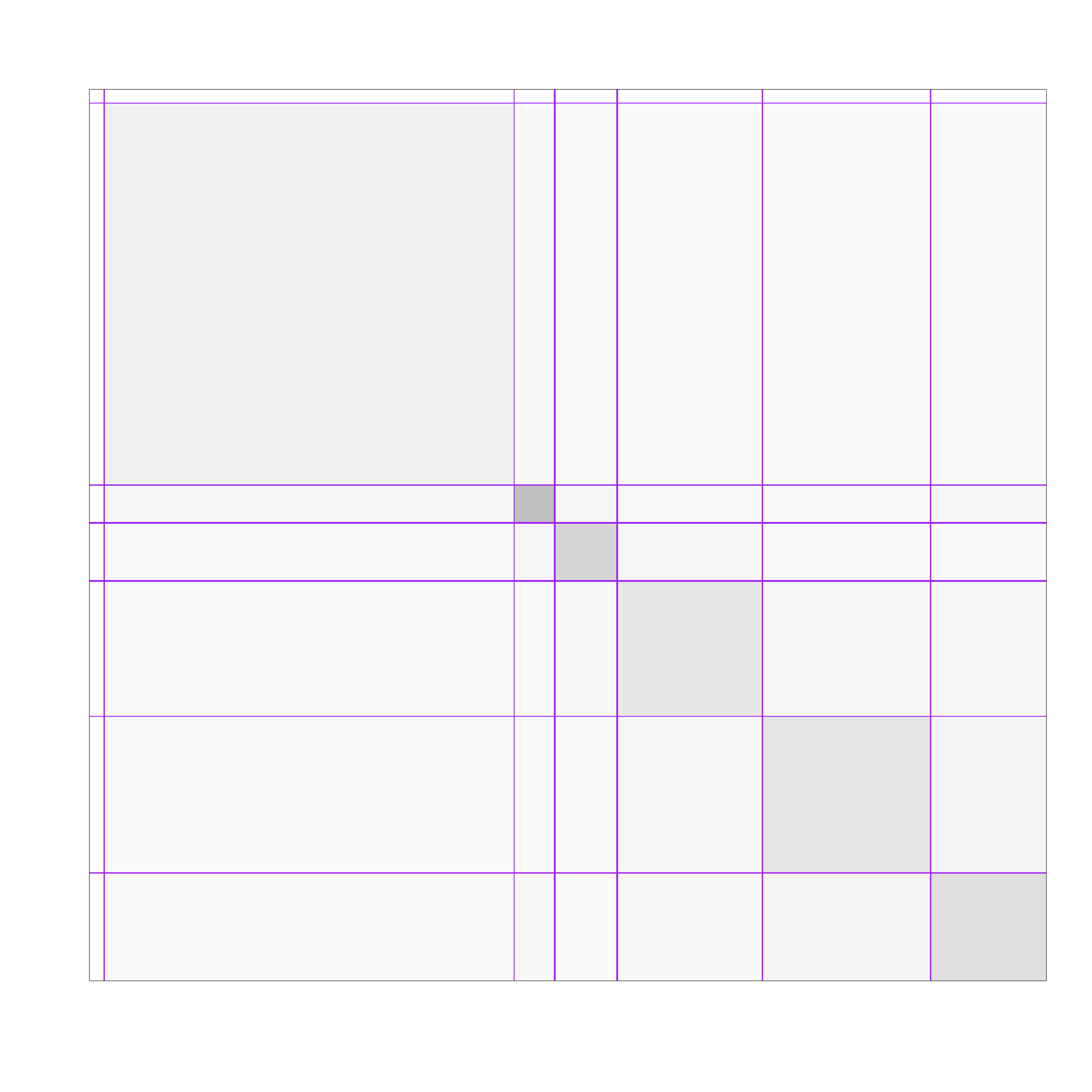}\\
 \rotatebox{90}{\hspace{2.25em}\small Chromosome 19}&
  \includegraphics[width=.3\linewidth]{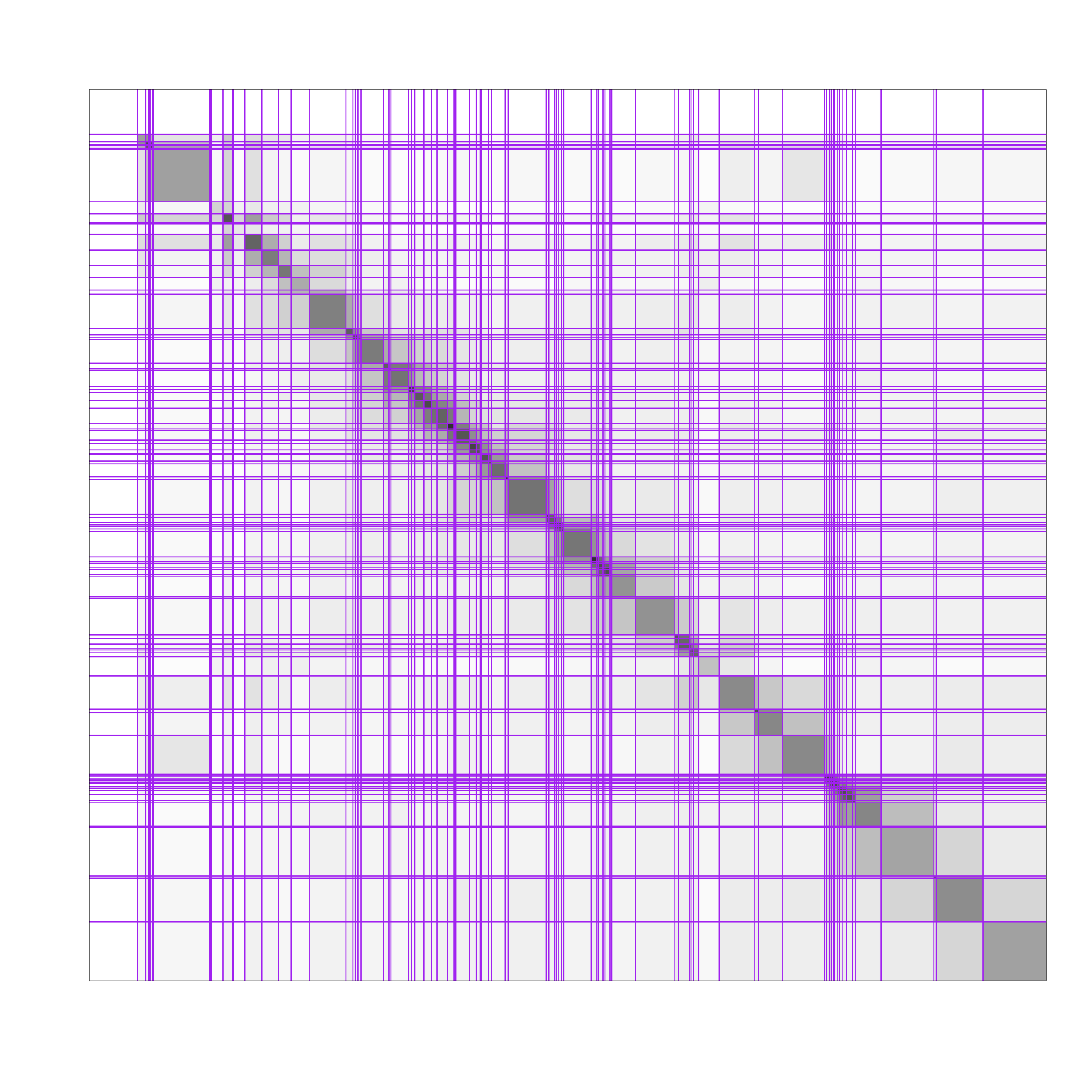}&
  \includegraphics[width=.3\linewidth]{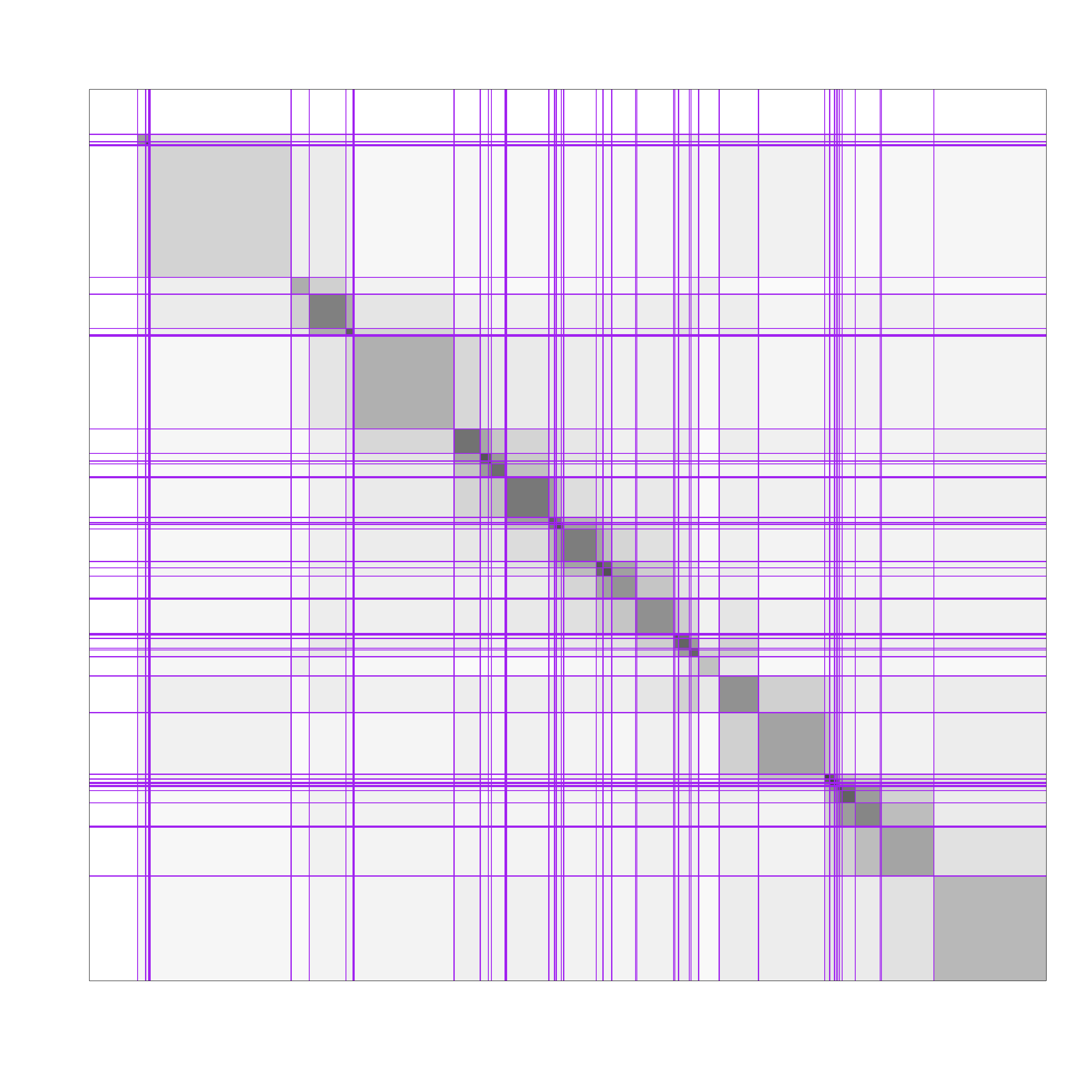}&
  \includegraphics[width=.3\linewidth]{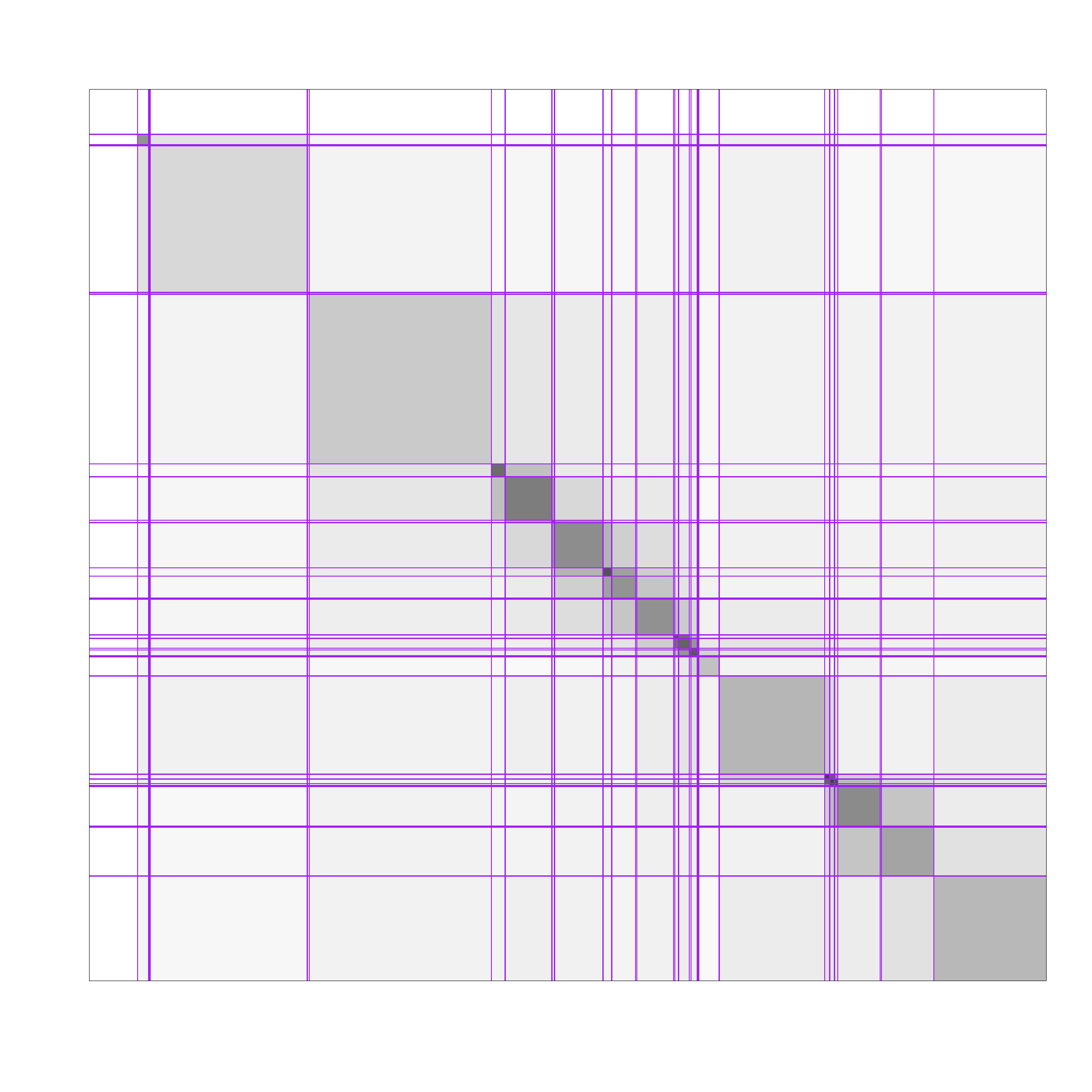}\\
\end{tabular}
\caption{\label{fig:HiCResum} Estimated matrices $\widehat{\Y}=\widehat{\U}$ for Chromosomes 1 and 19 for the thresholds 10, 15 and 20\%.}
\end{figure}

Let us first give the results obtained by using our methodology.
Figure \ref{fig:HiCPostRupt} displays the change-point locations obtained for the different values
of the threshold used in our adaptation of the stability selection approach and defined in Section \ref{sec:model_selection}. 
The corresponding estimated matrices $\widehat{\Y}=\widehat{\U}$ for Chromosome 1 and 19
are displayed in Figure \ref{fig:HiCResum} when the thresholds are equal to 10, 15 and 20\%, which correspond to the red horizontal levels in 
Figure \ref{fig:HiCPostRupt}.

\begin{figure}[!h]
  \centering
  \begin{tabular}{@{}l@{}cc}
  \includegraphics[width=.45\linewidth]{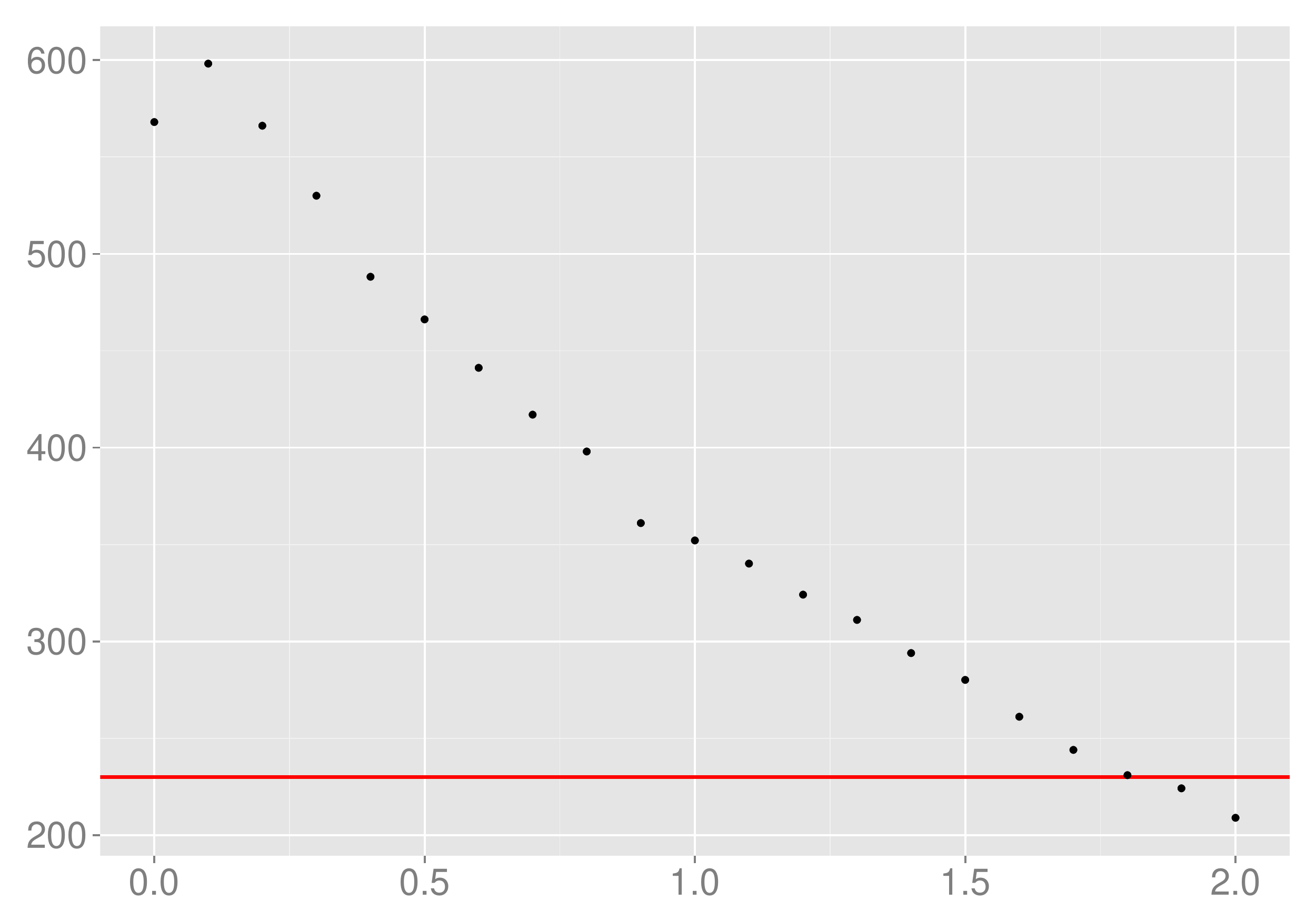}&
  \includegraphics[width=.45\linewidth]{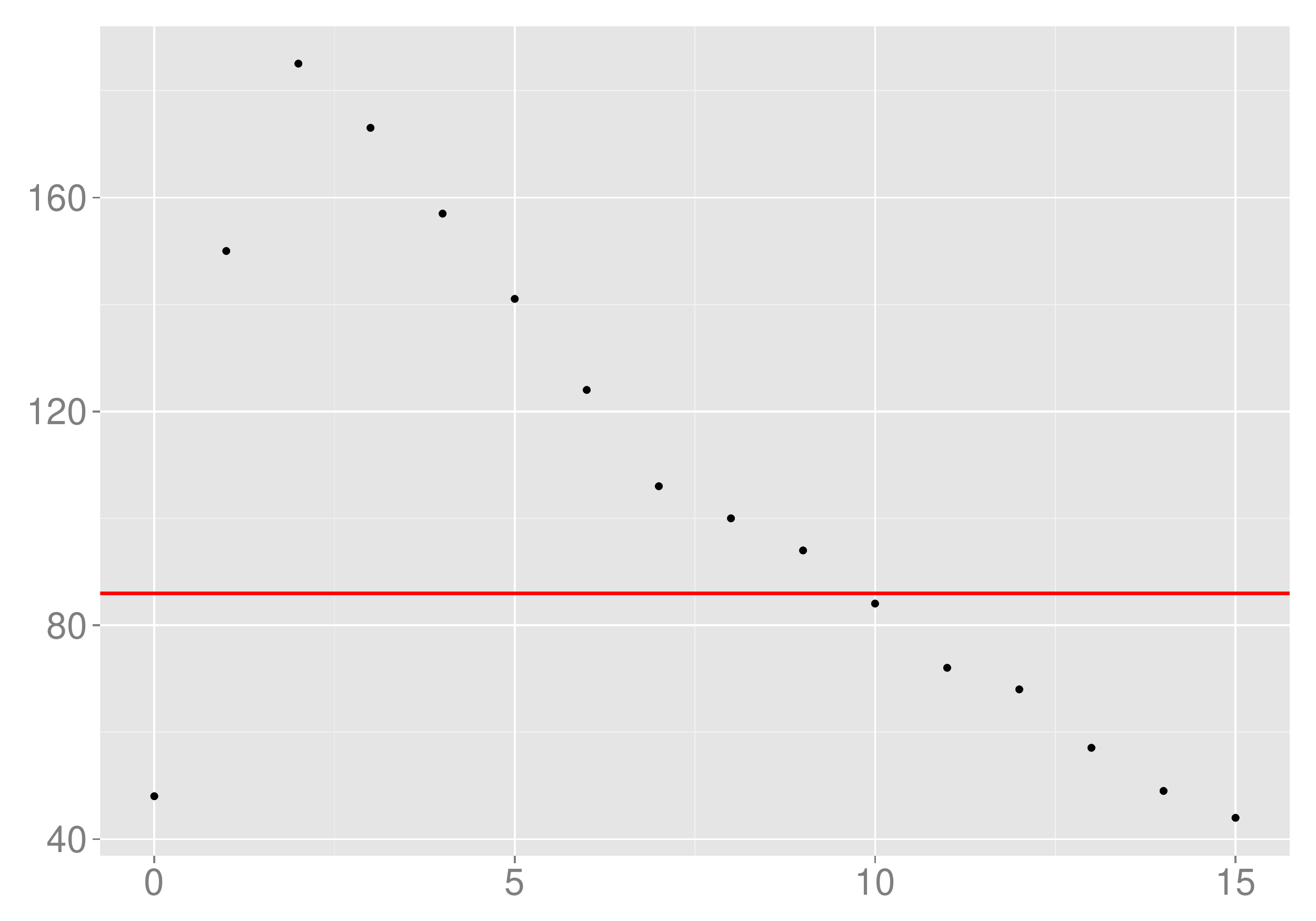}
  \end{tabular}
 \caption{\label{fig:nb_change_points} Number of change-points in rows found by our approach as a function of the threshold (in \%)
for the interaction matrices of Chromosome 1 (left) and Chromosome 19 (right) of the mouse cortex.
The red line corresponds to the number of
change-points found by \cite{dixon2012topological}.
}
\end{figure}

In order to compare our approach with the technique devised by \cite{dixon2012topological},
we display in Figure \ref{fig:nb_change_points} the number of change-points in rows found by our methodology as a function of the
threshold and a red line corresponding to the number of change-points found by
\cite{dixon2012topological}. Note that we did not display the change-points in columns in order to save space since they are similar.

\begin{figure}[!h]
  \centering
  \begin{tabular}{@{}l@{}cc}
\hspace{-5mm}
  \includegraphics[width=.45\linewidth]{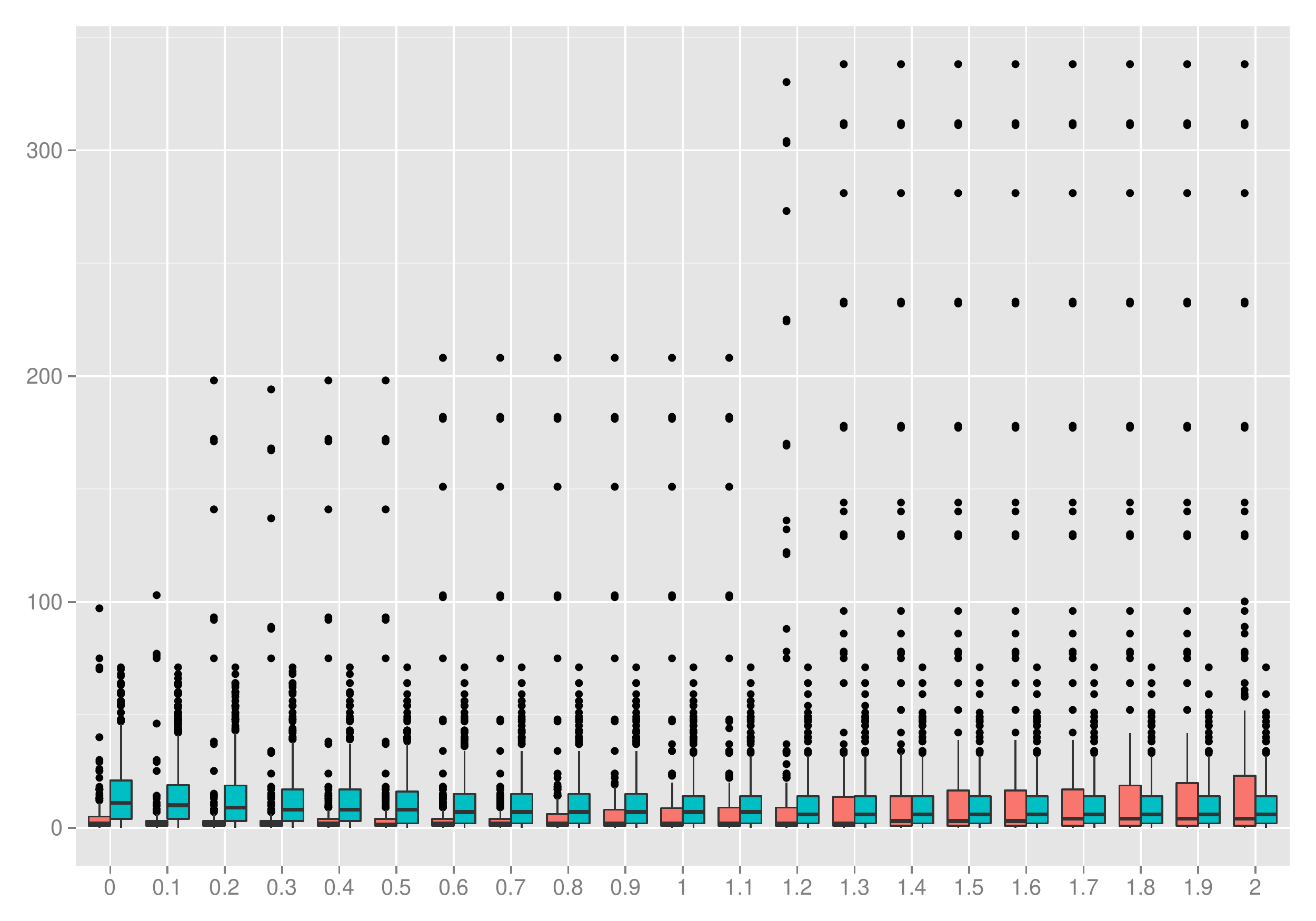}&
  \includegraphics[width=.45\linewidth]{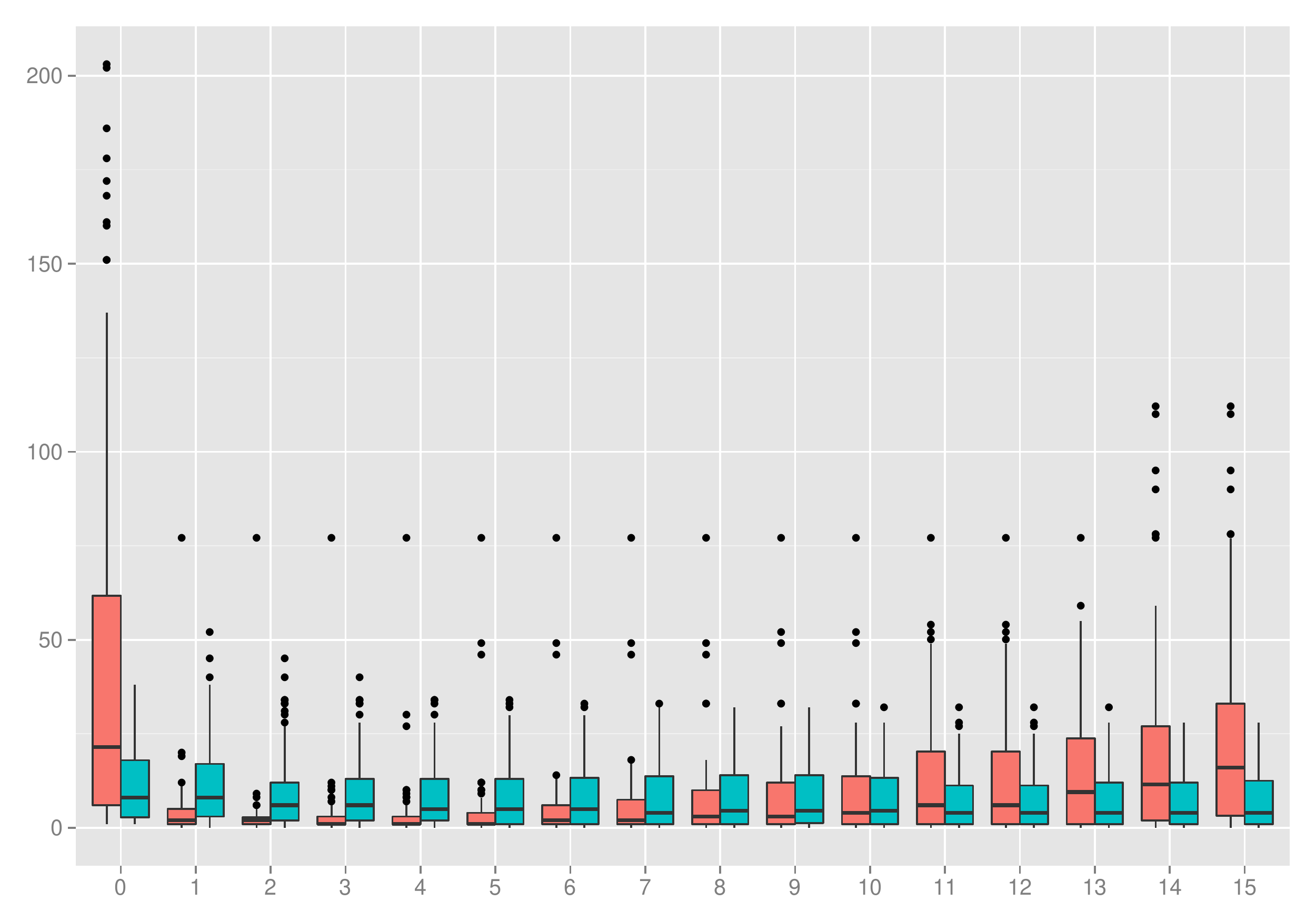}\\
  \end{tabular}
 \caption{\label{fig:hausdorff} Boxplots for the infimum parts of the Hausdorff distances $d_1$ (orange) and $d_2$ (blue)
between the change-points found by \cite{dixon2012topological} and our approach for the Chromosome 1 (left) and the Chromosome 19 (right) of the mouse cortex
for the different thresholds in \%.
}
\end{figure}

We also compute the two parts of the Hausdorff distance
for the change-points in rows which is defined by
\begin{equation}\label{eq:hausdorff}
d\left(\widehat{\boldsymbol{t}}_B , \widehat{\boldsymbol{t}}\right)
= \max\left(
d_1 \left(\widehat{\boldsymbol{t}}_B, \widehat{\boldsymbol{t}}\right),
d_2 \left(\widehat{\boldsymbol{t}}_B, \widehat{\boldsymbol{t}}\right)
\right)\;,
\end{equation}
where $\widehat{\boldsymbol{t}}$ and $\widehat{\boldsymbol{t}}_B$ are the change-points in rows found
by our approach and \cite{dixon2012topological}, respectively. In (\ref{eq:hausdorff}),
\begin{eqnarray}
d_1 \left(\mathbf{a},\mathbf{b}\right) & = & \sup_{b\in\mathbf{b}} \inf_{a\in\mathbf{a}} \left\vert a - b \right\vert \label{eq:hausd_1}, \\
d_2 \left(\mathbf{a},\mathbf{b}\right) & = & d_1 \left(\mathbf{b},\mathbf{a}\right)\label{eq:hausd_2}.
\end{eqnarray}
More precisely, Figure \ref{fig:hausdorff} displays the boxplots of
the $d_1$ and $d_2$ parts of the Hausdorff distance without taking the supremum in orange and blue, respectively.

We can observe from Figure \ref{fig:hausdorff} that
some differences indeed exist between the segmentations produced by the two approaches but that the boundaries of the
blocks are quite close when the number of estimated change-points are the same, which is the case when $\textrm{thresh}=1.8\%$ (left)
and 10\% (right).

\begin{figure}[!h]
  \centering
  \begin{tabular}{@{}l@{}cc}
  \includegraphics[width=.45\linewidth]{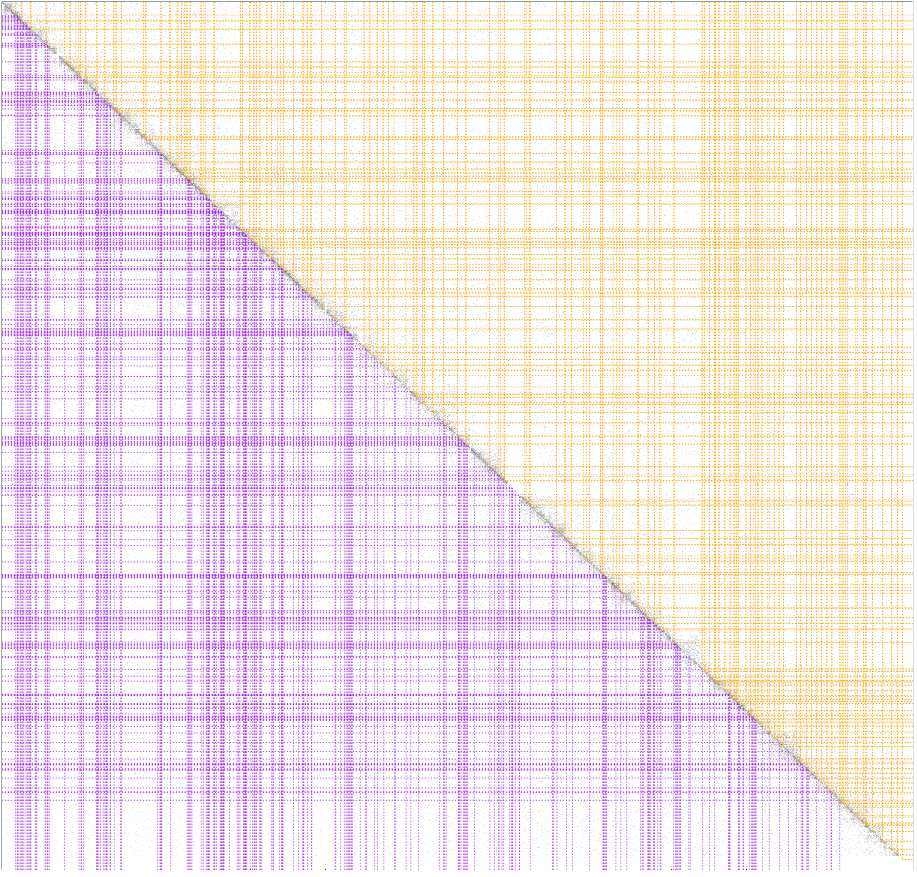}&
  \includegraphics[width=.45\linewidth]{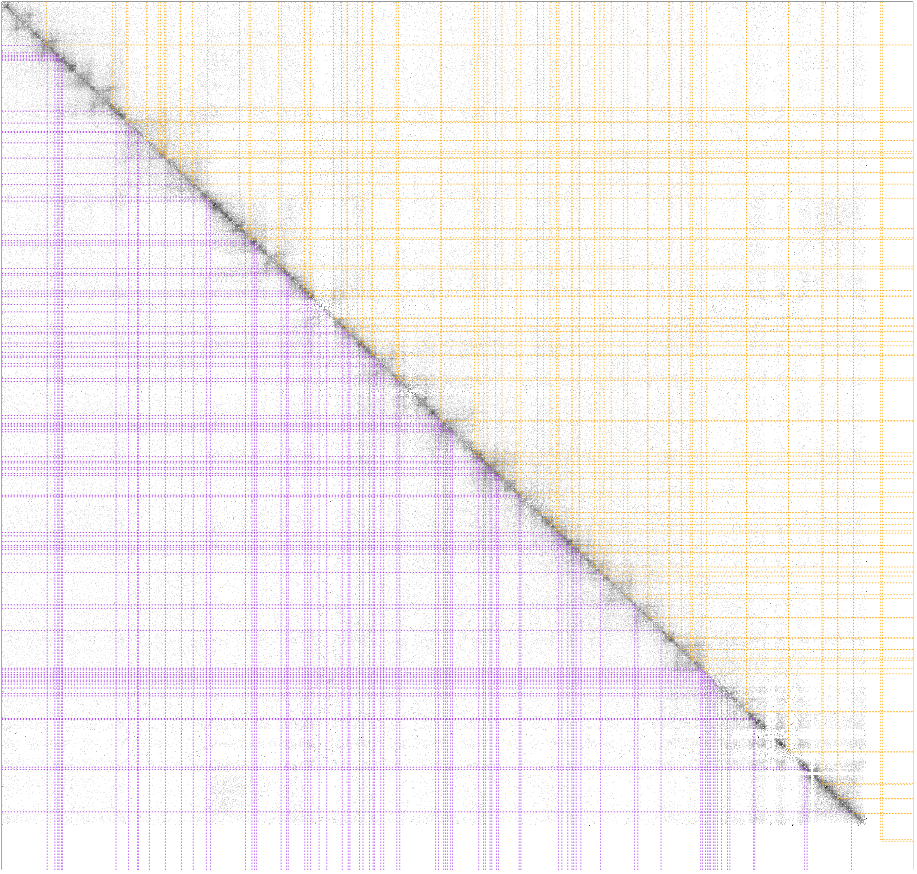}
  \end{tabular}
 \caption{\label{fig:HiC1} Topological domains detected by \cite{dixon2012topological} (upper
triangular part of the matrix) and by our method (lower triangular part of
the matrix) from the interaction matrix of Chromosome 1 (left) and Chromosome 19 (right) of the mouse cortex with a threshold giving
232 (resp 85) estimated change-points in rows and columns.
}
 \end{figure}

In the case where the number of estimated change-points are on a par with those of \cite{dixon2012topological}, we can
see from Figure \ref{fig:HiC1} that the change-points found with our strategy present a lot of similarities
with those found by the HMM based approach of \cite{dixon2012topological}. However, contrary to our method, the approach of 
\cite{dixon2012topological} can only deal with binned data at the resolution of several kilobases
of nucleotides.
The very low computational burden of our strategy paves the way for processing data
collected at a very high resolution, namely at the nucleotide resolution, which is one of the main current challenges of molecular biology.

\section{Conclusion}

In this paper, we proposed a novel approach for retrieving
the boundaries of a block wise constant matrix corrupted with noise by rephrasing this problem
as a variable selection issue. Our approach is implemented in the R package \texttt{blockseg} which is
 available from the Comprehensive R Archive Network (CRAN).
In the course of this study, we have shown that our method has two main features which make it very attractive.
Firstly, it is very efficient
both from the theoretical and practical point of view. Secondly, its very low computational burden
makes its use possible on very large data sets coming from molecular biology.

\appendix

\section{Proofs}\label{sec:proofs}

\subsection{Proofs of statistical results}

\begin{proof}[Proofof Lemma \ref{lem:KKT}]
A necessary and sufficient condition for a vector $\widehat{\calB}$ in $\rset^{n^2}$ to minimize the function
$\Phi$ defined by: $\Phi(\calB)=\sum_{i=1}^{n^2}(\calY_i-(\calX\calB)_i)^2+\lambda_n\sum_{i=1}^{n^2}|\calB_i|$, is that
the zero vector in $\rset^{n^2}$ belongs to the subdifferential of $\Phi$ at $\widehat{\calB}$ that is:
\begin{align*}
\left(\t{\calX}(\calY-\calX\widehat{\calB})\right)_j &=\frac{\lambda_n}{2},\quad \textrm{if }\widehat{\calB}_j\neq 0,\\
\left|\left(\t{\calX}(\calY-\calX\widehat{\calB})\right)_j\right| &\leq \frac{\lambda_n}{2},\quad \textrm{if }\widehat{\calB}_j= 0.
\end{align*}
Using that $\t{\calX}\calY=\t{(\TT\otimes \TT)}\calY=(\t{\TT}\otimes \t{\TT})\calY=\Ve(\t{\TT}\Y \TT)$, where
$(\t{\TT}\Y \TT)_{i,j}=\sum_{k=i}^n\sum_{\ell=j}^n Y_{k,\ell}$, and that $\widehat{\calU}=\calX\widehat{\calB}$, Lemma \ref{lem:KKT} is proved.
\end{proof}

\begin{proof}[Proofof Lemma \ref{lem:noise}]
Note that
\begin{eqnarray*}
\PP\left(\max_{\stackrel{1\leq r_n<s_n\leq n}{|r_n-s_n|\geq v_n}}\left|(s_n-r_n)^{-1}\sum_{j=r_n}^{s_n-1}E_{n,j}\right|\geq x_n\right)&&\\
&&\!\!\!\!\!\!\!\!\!\!\!\!\!\!\!\!\!\!\!\!\!\!\!\!\!\!\!\!\!\!\!\!\!\!\!\!\!\!\!\!\!\!\!\!\!\!\!\!\!\!\!\!\!\!\!\!\!\!\!\!\!\!\!\!\leq\sum_{\stackrel{1\leq r_n<s_n\leq n}{|r_n-s_n|\geq v_n}}\PP\left(\left|(s_n-r_n)^{-1}\sum_{j=r_n}^{s_n-1}E_{n,j}\right|\geq x_n\right).\\
\end{eqnarray*}
By (A\ref{hyp:noise}) and the Markov inequality, we get that for all positive $\eta$,
\begin{align*}
\PP\left((s_n-r_n)^{-1}\sum_{j=r_n}^{s_n-1}E_{n,j}\geq x_n\right)&\leq\exp[-\eta(s_n-r_n)x_n](\PE(\exp(\eta E_{1,1})))^{s_n-r_n}\\
&\leq\exp[-\eta(s_n-r_n)x_n+\beta\eta^2(s_n-r_n)].
\end{align*}
By taking $\eta=x_n/(2\beta)$, we get that
$$
\PP\left((s_n-r_n)^{-1}\sum_{j=r_n}^{s_n-1}E_{n,j}\geq x_n\right)\leq\exp[-x_n^2(s_n-r_n)/(4\beta)].
$$
Since the same result is valid for $-E_{n,j}$, we get that
$$
\PP\left(\max_{\stackrel{1\leq r_n<s_n\leq n}{|r_n-s_n|\geq v_n}}\left|(s_n-r_n)^{-1}\sum_{j=r_n}^{s_n-1}E_{n,j}\right|\geq x_n\right)
\leq 2n^2\exp[-x_n^2 v_n/(4\beta)],
$$
which concludes the proof of Lemma \ref{lem:noise}.
\end{proof}

\begin{proof}[Proofof Proposition \ref{prop:consist}]
Since
\begin{multline}\label{equation1}
\PP\left(\left\{\max_{1\leq k\leq\Ks_1}\left|\that_{1,k}-\ts_{1,k}\right|> n\delta_n\right\}\cup
\left\{\max_{1\leq k\leq\Ks_2}\left|\that_{2,k}-\ts_{2,k}\right|> n\delta_n\right\}\right)\\
\leq \PP\left(\max_{1\leq k\leq\Ks_1}\left|\that_{1,k}-\ts_{1,k}\right|> n\delta_n\right)
+\PP\left(\max_{1\leq k\leq\Ks_2}\left|\that_{2,k}-\ts_{2,k}\right|> n\delta_n\right),
\end{multline}
it is enough to prove that both terms in (\ref{equation1}) tend to zero for proving (\ref{eq:consist}). We shall only prove that
the second term in the rhs of (\ref{equation1}) tends to zero, the proof being the same for the first term.
Since
$\PP(\max_{1\leq k\leq\Ks_2}|\that_{2,k}-\ts_{2,k}|> n\delta_n)\leq\sum_{k=1}^{\Ks_2}\PP(|\that_{2,k}-\ts_{2,k}|> n\delta_n)$, it is enough
to prove that for all $k$ in $\{1,\dots,\Ks_2\}$, $\PP(A_{n,k})\to 0$, where $A_{n,k}=\{|\that_{2,k}-\ts_{2,k}|> n\delta_n\}$.
Let $C_n$ be defined by
\begin{equation}\label{eq:Cn}
C_n=\left\{\max_{1\leq k\leq\Ks_2}|\that_{2,k}-\ts_{2,k}|<\Imindeux /2\right\}.
\end{equation}
It is enough to prove that, for all $k$ in $\{1,\dots,\Ks_2\}$, $\PP(A_{n,k}\cap C_n)$ and $\PP(A_{n,k}\cap \overline{C_n})$ tend to 0, as $n$
tends to infinity.

Let us first prove that for all $k$ in $\{1,\dots,\Ks_2\}$, $\PP(A_{n,k}\cap C_n)\to 0$. Observe that
(\ref{eq:Cn}) implies that $\ts_{2,k-1}<\that_{2,k}<\ts_{2,k+1}$,  for all $k$ in $\{1,\dots,\Ks_2\}$.
For a given $k$, let us assume that $\that_{2,k}\leq\ts_{2,k}$. Applying (\ref{eq:kkt_1}) and (\ref{eq:kkt_2})
with $r_j+1=n$, $q_j+1=\that_{2,k}$ on the one hand and $r_j+1=n$, $q_j+1=\ts_{2,k}$ on the other hand, we get that
$$
\left|\sum_{j=\that_{2,k}}^{\ts_{2,k}-1} Y_{n,j}-\sum_{j=\that_{2,k}}^{\ts_{2,k}-1} \widehat{\calU}_{n,j}\right|\leq \lambda_n.
$$
Hence using (\ref{eq:model1}), the notation: $\E([a,b];[c,d])=\sum_{i=a}^b\sum_{j=c}^d E_{i,j}$ and the definition
of $\widehat{\calU}$ given by Lemma \ref{lem:KKT}, we obtain that
$$
\left|(\ts_{2,k}-\that_{2,k})(\mus_{\Ks_1+1,k}-\widehat{\mu}_{\Ks_1+1,k+1})+\E(n;[\that_{2,k},\ts_{2,k}-1])\right|\leq \lambda_n,
$$
which can be rewritten as follows
\begin{multline*}
\left|(\ts_{2,k}-\that_{2,k})(\mus_{\Ks_1+1,k}-\mus_{\Ks_1+1,k+1})
+(\ts_{2,k}-\that_{2,k})(\mus_{\Ks_1+1,k+1}-\widehat{\mu}_{\Ks_1+1,k+1})\right.\\
\left.+\E(n;[\that_{2,k},\ts_{2,k}-1])\right|\leq \lambda_n.
\end{multline*}
Thus,
\begin{multline}\label{eq:upper_bound_1}
\PP(A_{n,k}\cap C_n)\leq\PP(\lambda_n/(n\delta_n)\geq |\mus_{\Ks_1+1,k}-\mus_{\Ks_1+1,k+1}|/3)\\
+\PP(\{|\mus_{\Ks_1+1,k}-\widehat{\mu}_{\Ks_1+1,k+1}|\geq |\mus_{\Ks_1+1,k}-\mus_{\Ks_1+1,k+1}|/3\}\cap C_n)\\
+\PP(\{|\E(n;[\that_{2,k},\ts_{2,k}-1])|/|\ts_{2,k}-\that_{2,k}|\geq |\mus_{\Ks_1+1,k}-\mus_{\Ks_1+1,k+1}|/3\}\cap A_{n,k}).
\end{multline}
The first term in the rhs of (\ref{eq:upper_bound_1}) tends to 0 by (A\ref{hyp:Jmin}). By Lemma \ref{lem:noise} with
$x_n=\Jmin/3$, $v_n=n\delta_n$ and (A\ref{hyp:delta_n}) the third term in the rhs of (\ref{eq:upper_bound_1}) tends to 0.
Applying Lemma \ref{lem:KKT}
with $r_j+1=n$, $q_j+1=\that_{2,k}$ on the one hand and $r_j+1=n$, $q_j+1=(\ts_{2,k}+\ts_{2,k+1})/2$ on the other hand, we get that
$$
\left|\sum_{j=\ts_{2,k}}^{(\ts_{2,k}+\ts_{2,k+1})/2-1} Y_{n,j}-\sum_{j=\ts_{2,k}}^{(\ts_{2,k}+\ts_{2,k+1})/2-1} \widehat{\calU}_{n,j}\right|\leq \lambda_n.
$$
Since $\that_{2,k}\leq\ts_{2,k}$, $\widehat{\calU}_{n,j}=\widehat{\mu}_{\Ks_1+1,k+1}$ within the interval $[\ts_{2,k},(\ts_{2,k}+\ts_{2,k+1})/2-1]$
and we get that
$$
(\ts_{2,k+1}-\ts_{2,k})|\mus_{\Ks_1+1,k+1}-\widehat{\mu}_{\Ks_1+1,k+1}|/2\leq\lambda_n+|\E(n,|[\ts_{2,k},(\ts_{2,k}+\ts_{2,k+1})/2-1])|.
$$
Therefore the second term in the rhs of (\ref{eq:upper_bound_1}) can be bounded by
\begin{multline*}
\PP\left(\lambda_n\geq (\ts_{2,k+1}-\ts_{2,k}) |\mus_{\Ks_1+1,k}-\mus_{\Ks_1+1,k+1}|/12\right)\\
+\PP\left((\ts_{2,k+1}-\ts_{2,k})^{-1}\left|\E(n,,|[\ts_{2,k},(\ts_{2,k}+\ts_{2,k+1})/2-1])\right|\right.\\\left.\geq|\mus_{\Ks_1+1,k}-\mus_{\Ks_1+1,k+1}|/6\right)
\end{multline*}
By Lemma \ref{lem:noise} and (A\ref{hyp:Jmin}), (A\ref{hyp:delta_n}) and (A\ref{hyp:Imin}), we get that both terms tend to zero
as $n$ tends to infinity. We thus get that $\PP(A_{n,k}\cap C_n)\to 0$, as $n$ tends to infinity.

Let us now prove that $\PP(A_{n,k}\cap \overline{C_n})$ tend to 0, as $n$ tends to infinity.
Observe that
$$
\PP(A_{n,k}\cap \overline{C_n})=\PP(A_{n,k}\cap D_n^{(\ell)})+\PP(A_{n,k}\cap D_n^{(m)})+\PP(A_{n,k}\cap D_n^{(r)}),
$$
where
\begin{align*}
D_n^{(\ell)}&=\left\{\exists p\in\{1,\dots,\Ks\},\; \that_{2,p}\leq\ts_{2,p-1}\right\}\cap\overline{C_n},\\
D_n^{(m)}&=\left\{\forall k\in\{1,\dots,\Ks\},\; \ts_{2,k-1}<\that_{2,k}<\ts_{2,k+1}\right\}\cap\overline{C_n},\\
D_n^{(r)}&=\left\{\exists p\in\{1,\dots,\Ks\},\; \that_{2,p}\geq\ts_{2,p+1}\right\}\cap\overline{C_n}.\\
\end{align*}
Using the same arguments as those used for proving that $\PP(A_{n,k}\cap C_n)\to 0$, we can prove that
$\PP(A_{n,k}\cap D_n^{(m)})\to 0$, as $n$ tends to infinity. Let us now prove that $\PP(A_{n,k}\cap D_n^{(\ell)})\to 0$.
Note that
\begin{eqnarray}
\PP(D_n^{(\ell)})&\leq&\sum_{k=1}^{\Ks_2-1}\PP(\{\ts_{2,k}-\that_{2,k}>\Imin/2\}\cap\{\that_{2,k+1}-\ts_{2,k}>\Imin/2\})\nonumber\\
                  &   &\quad\quad+\PP(\ts_{2,\Ks_2}-\that_{2,\Ks_2}>\Imin/2).\label{eq:Dn_l}
\end{eqnarray}
Applying (\ref{eq:kkt_1}) and (\ref{eq:kkt_2})
with $r_j+1=n$, $q_j+1=\that_{2,k}$ on the one hand and $r_j+1=n$, $q_j+1=\ts_{2,k}$ on the other hand, we get that
$$
\left|\sum_{j=\that_{2,k}}^{\ts_{2,k}-1} Y_{n,j}-\sum_{j=\that_{2,k}}^{\ts_{2,k}-1} \widehat{\calU}_{n,j}\right|\leq \lambda_n.
$$
Thus,
\begin{eqnarray}
& &\!\!\!\!\!\!\!\!\PP(\{\ts_{2,k}-\that_{2,k}>\Imin/2\}\cap\{\that_{2,k+1}-\ts_{2,k}>\Imin/2\})\nonumber\\
&\leq&\PP(\lambda_n/(n\delta_n)\geq |\mus_{\Ks_1+1,k}-\mus_{\Ks_1+1,k+1}|/3)\nonumber\\
&&+\PP(\{|\mus_{\Ks_1+1,k}-\widehat{\mu}_{\Ks_1+1,k+1}|\geq |\mus_{\Ks_1+1,k}-\mus_{\Ks_1+1,k+1}|/3\}\nonumber\\
&&\quad\quad\cap\{\that_{2,k+1}-\ts_{2,k}>\Imin/2\})\nonumber\\
&&+\PP(\{|\E(n;[\that_{2,k},\ts_{2,k}-1])|/(\ts_{2,k}-\that_{2,k})\geq |\mus_{\Ks_1+1,k}-\mus_{\Ks_1+1,k+1}|/3\}\nonumber\\
& &\quad\quad\quad\quad\cap \{\ts_{2,k}-\that_{2,k}>\Imin/2\}).\label{eq1}
\end{eqnarray}
Using the same arguments as previously we get that the first and the third term in the rhs of (\ref{eq1}) tend to zero as $n$ tends to infinity.
Let us now focus on the second term of the rhs of (\ref{eq1}). Applying (\ref{eq:kkt_1}) and (\ref{eq:kkt_2})
with $r_j+1=n$, $q_j+1=\that_{2,k+1}$ on the one hand and $r_j+1=n$, $q_j+1=\ts_{2,k}$ on the other hand, we get that
$$
\left|\sum_{j=\ts_{2,k}}^{\that_{2,k+1}-1} Y_{n,j}-\sum_{j=\ts_{2,k}}^{\that_{2,k+1}-1} \widehat{\calU}_{n,j}\right|\leq \lambda_n.
$$
Hence,
$$
|(\mus_{\Ks_1+1,k}-\widehat{\mu}_{\Ks_1+1,k+1})(\that_{2,k+1}-\ts_{2,k})+\E(n,[\ts_{2,k};\that_{2,k+1}-1])|\leq\lambda_n.
$$
The second term of the rhs of (\ref{eq1}) is thus bounded by
\begin{eqnarray*}
&&\PP(\{\lambda_n(\that_{2,k+1}-\ts_{2,k})^{-1}\geq |\mus_{\Ks_1+1,k}-\mus_{\Ks_1+1,k+1}|/6\}\\
&&\quad\quad\cap\{\that_{2,k+1}-\ts_{2,k}>\Imin/2\})\\
&&+\PP(\{(\that_{2,k+1}-\ts_{2,k})^{-1}|\E(n,[\ts_{2,k};\that_{2,k+1}-1])|\geq |\mus_{\Ks_1+1,k}-\mus_{\Ks_1+1,k+1}|/6\}\\
&&\quad\quad\quad\quad\cap\{\that_{2,k+1}-\ts_{2,k}>\Imin/2\}),
\end{eqnarray*}
which tend to zero by Lemma \ref{lem:noise}, (A\ref{hyp:Jmin}), (A\ref{hyp:delta_n}) and (A\ref{hyp:Imin}).
It is thus proved that the first term in the rhs of (\ref{eq:Dn_l}) tends to zero as $n$ tends to infinity.
The same arguments can be used for addressing the second term in the rhs of (\ref{eq:Dn_l}) since $\that_{2,\Ks_2+1}=n$ and hence $\that_{2,\Ks_2+1}-\ts_{2,\Ks_2}>\Imin/2$.

Using similar arguments, we can prove that $\PP(A_{n,k}\cap D_n^{(r)})\to 0$, which concludes the proof of Proposition \ref{prop:consist}.
\end{proof}

\subsection{Proofs of computational lemmas}

\begin{proof}[Proofof Lemma~\ref{lem:Xtx}.]    Consider   $\calX
  \mathbf{v}$ for instance (the  same reasoning applies for $\t{\calX}
  \mathbf{v}$): we have $\calX \mathbf{v} = (\TT \otimes \TT) \mathbf{v}
  = \Ve(\TT \mathbf{V} \t{\TT})$ where  $\mathbf{V}$ is the $n\times n$
  matrix  such that  $\Ve(\mathbf{V})=  \mathbf{v}$.   Because of  its
  triangular structure, $\TT$ operates as  a cumulative sum operator on
  the columns of  $\mathbf{V}$. Hence, the computations  for the $j$th
  column is done  by induction in $n$ operations.  The  total cost for
  the $n$ columns of $\TT  \mathbf{V}$ is thus $n^2$.  Similarly, right
  multiplying a  matrix by $\t{\TT}$  boils down to  perform cumulative
  sums over the  rows.  The final cost for $\calX  \mathbf{v} = \Ve(\TT
  \mathbf{V}  \t{\TT})$  is thus  $2n^2$  in  case  of a  dense  matrix
  $\mathbf{V}$, and possibly less when $\mathbf{V}$ is sparse.
\end{proof}

\begin{proof}[Proofof Lemma~\ref{lem:cholXtX}.]
Let $\supp=\{a_1,\dots,a_K\}$, then
\begin{equation}\label{eq:tXX_AA}
\left(\t{\calX} \calX\right)_{\supp,\supp}=\t{(\TT\otimes\TT)}_{\bullet,\supp} (\TT\otimes\TT)_{\bullet,\supp},
\end{equation}
where $(\TT\otimes\TT)_{\bullet,\supp}$ (resp. $\t{(\TT\otimes\TT)}_{\bullet,\supp}$) denotes the columns (resp. the rows) of $\TT\otimes\TT$ lying in $\supp$.
For $j$ in $\supp$, let us consider the Euclidean division of $j-1$ by $n$ given by: $(j-1)=nq_j+r_j$, then
$
(\TT\otimes\TT)_{\bullet,j}=\TT_{\bullet,q_j+1}\otimes\TT_{\bullet,r_j+1}.
$
Hence, $(\TT\otimes\TT)_{\bullet,\supp}$ is a $n^2\times K$ matrix defined by:
$$
(\TT\otimes\TT)_{\bullet,\supp}=\left[\TT_{\bullet,q_{a_1+1}}\otimes\TT_{\bullet,r_{a_1+1}};\TT_{\bullet,q_{a_2+1}}\otimes\TT_{\bullet,r_{a_2+1}};\dots;
\TT_{\bullet,q_{a_K+1}}\otimes\TT_{\bullet,r_{a_K+1}}\right].
$$
Thus,
\begin{eqnarray*}
(\TT\otimes\TT)_{\bullet,\supp}=\TT_{\bullet,Q_\supp}*\TT_{\bullet,R_\supp},&\textrm{ where }& Q_\supp=\{q_{a_1}+1,\dots,q_{a_K}+1\},\\
                                                                            &                &R_\supp=\{r_{a_1}+1,\dots,r_{a_K}+1\}\\
\end{eqnarray*}
and $*$ denotes the Khatri-Rao product, which is defined as follows for two  $n\times n$ matrices $A$ and $B$
$$
A*B=\left[a_1\otimes b_1 ; a_2\otimes b_2;\dots a_n\otimes b_n\right],
$$
where the $a_i$ (resp. $b_i$) are the columns of $A$ (resp. B).
Using (25) of Theorem 2 in \cite{liu:trenkler:2008}, we get that
$$
\t{(\TT\otimes\TT)}_{\bullet,\supp} (\TT\otimes\TT)_{\bullet,\supp}
=\left(\t{\TT}_{\bullet,Q_\supp}{\TT}_{\bullet,Q_\supp}\right)\circ\left(\t{\TT}_{\bullet,R_\supp}{\TT}_{\bullet,R_\supp}\right),
$$
where $\circ$ denotes the Hadamard or entry-wise product. Observe that by definition of $\TT$,
$(\t{\TT}_{\bullet,Q_\supp}{\TT}_{\bullet,Q_\supp})_{k,\ell}=n-(q_{a_k}\vee q_{a_\ell})$
and $(\t{\TT}_{\bullet,R_\supp}{\TT}_{\bullet,R_\supp})_{k,\ell}=n-(r_{a_k}\vee r_{a_\ell})$.
By (\ref{eq:tXX_AA}), $\left(\t{\calX} \calX\right)_{\supp,\supp}$ is a Gram matrix which is positive and definite since
the vectors $\TT_{\bullet,q_{a_1+1}}\otimes\TT_{\bullet,r_{a_1+1}}$, $\TT_{\bullet,q_{a_2+1}}\otimes\TT_{\bullet,r_{a_2+1}}$, $\dots$,
$\TT_{\bullet,q_{a_K+1}}\otimes\TT_{\bullet,r_{a_K+1}}$ are linearly independent.

\end{proof}

\begin{proof}[Proofof Lemma~\ref{lem:cholupdate}.]  The operations of
  adding/removing a  column to a Cholesky  factorization are classical
  and  well treated  in  books  of numerical  analysis, see  e.g.
  \cite{golub2012matrix}. An advantage of  our settings is that there
  is  no  additional  computational   cost  for  computing  $\t{\calX}
  \calX_{\centerdot j}$ when entering a new variable $j$ thanks to the
  closed-form expression~\eqref{eq:XtXij}.
\end{proof}

%

\bibliographystyle{abbrvnat}
\bibliography{biblio}

\begin{thebibliography}{20}
\providecommand{\natexlab}[1]{#1}
\providecommand{\url}[1]{\texttt{#1}}
\expandafter\ifx\csname urlstyle\endcsname\relax
  \providecommand{\doi}[1]{doi: #1}\else
  \providecommand{\doi}{doi: \begingroup \urlstyle{rm}\Url}\fi

\bibitem[Bach et~al.(2012)Bach, Jenatton, Mairal, and
  Obozinski]{bach2012optimization}
F.~Bach, R.~Jenatton, J.~Mairal, and G.~Obozinski.
\newblock Optimization with sparsity-inducing penalties.
\newblock \emph{Foundations and Trends{\textregistered} in Machine Learning},
  4\penalty0 (1):\penalty0 1--106, 2012.

\bibitem[Bellman(1961)]{Bellman:1961}
R.~Bellman.
\newblock On the approximation of curves by line segments using dynamic
  programming.
\newblock \emph{Commun. ACM}, 4\penalty0 (6):\penalty0 284--, 1961.
\newblock ISSN 0001-0782.
\newblock \doi{10.1145/366573.366611}.

\bibitem[{Breiman} et~al.(1984){Breiman}, {Friedman}, {Olshen}, and
  {Stone}]{breiman:1984}
L.~{Breiman}, J.~H. {Friedman}, R.~A. {Olshen}, and C.~J. {Stone}.
\newblock \emph{Classification and Regression Trees}.
\newblock Statistics/Probability Series. Wadsworth Publishing Company, Belmont,
  California, U.S.A., 1984.

\bibitem[Dixon et~al.(2012)Dixon, Selvaraj, Yue, Kim, Li, Shen, Hu, Liu, and
  Ren]{dixon2012topological}
J.~R. Dixon, S.~Selvaraj, F.~Yue, A.~Kim, Y.~Li, Y.~Shen, M.~Hu, J.~S. Liu, and
  B.~Ren.
\newblock Topological domains in mammalian genomes identified by analysis of
  chromatin interactions.
\newblock \emph{Nature}, 485\penalty0 (7398):\penalty0 376--380, 2012.

\bibitem[Efron et~al.(2004)Efron, Hastie, Johnstone, Tibshirani,
  et~al.]{efron2004least}
B.~Efron, T.~Hastie, I.~Johnstone, R.~Tibshirani, et~al.
\newblock Least angle regression.
\newblock \emph{The Annals of statistics}, 32\penalty0 (2):\penalty0 407--499,
  2004.

\bibitem[Fisher(1958)]{fisher:1958}
W.~D. Fisher.
\newblock On grouping for maximum homogeneity.
\newblock \emph{Journal of the American Statistical Association}, 53\penalty0
  (284):\penalty0 789--798, 1958.
\newblock ISSN 01621459.

\bibitem[Golub and Van~Loan(2012)]{golub2012matrix}
G.~H. Golub and C.~F. Van~Loan.
\newblock \emph{Matrix computations}.
\newblock JHU Press, 2012.
\newblock 3rd edition.

\bibitem[Harchaoui and L\'evy-Leduc(2010)]{harchaoui:levyleduc:2012}
Z.~Harchaoui and C.~L\'evy-Leduc.
\newblock Multiple change-point estimation with a total variation penalty.
\newblock \emph{Journal of the American Statistical Association}, 105\penalty0
  (492):\penalty0 1480--1493, 2010.

\bibitem[Hoefling(2010)]{hoefling2010path}
H.~Hoefling.
\newblock A path algorithm for the fused lasso signal approximator.
\newblock \emph{J. Comput. Graph. Statist.}, 19\penalty0 (4):\penalty0
  984--1006, 2010.

\bibitem[Kay(1993)]{kay:1993}
S.~Kay.
\newblock \emph{Fundamentals of statistical signal processing: detection
  theory}.
\newblock Prentice-Hall, Inc., 1993.

\bibitem[L{\'e}vy-Leduc et~al.(2014)L{\'e}vy-Leduc, Delattre, Mary-Huard, and
  Robin]{levy2014}
C.~L{\'e}vy-Leduc, M.~Delattre, T.~Mary-Huard, and S.~Robin.
\newblock Two-dimensional segmentation for analyzing hi-c data.
\newblock \emph{Bioinformatics}, 30\penalty0 (17):\penalty0 i386--i392, 2014.

\bibitem[Lieberman-Aiden et~al.(2009)Lieberman-Aiden, Van~Berkum, Williams,
  Imakaev, Ragoczy, Telling, Amit, Lajoie, Sabo, Dorschner,
  et~al.]{lieberman2009comprehensive}
E.~Lieberman-Aiden, N.~L. Van~Berkum, L.~Williams, M.~Imakaev, T.~Ragoczy,
  A.~Telling, I.~Amit, B.~R. Lajoie, P.~J. Sabo, M.~O. Dorschner, et~al.
\newblock Comprehensive mapping of long-range interactions reveals folding
  principles of the human genome.
\newblock \emph{science}, 326\penalty0 (5950):\penalty0 289--293, 2009.

\bibitem[Liu and Trenkler(2008)]{liu:trenkler:2008}
S.~Liu and G.~Trenkler.
\newblock Hadamard, khatri-rao, kronecker and other matrix products.
\newblock \emph{Int. J. Inform. Syst. Sci.}, 4:\penalty0 160--177, 2008.

\bibitem[Maidstone et~al.(2016)Maidstone, Hocking, Rigaill, and
  Fearnhead]{rigaill2016}
R.~Maidstone, T.~Hocking, G.~Rigaill, and P.~Fearnhead.
\newblock On optimal multiple changepoint algorithms for large data.
\newblock \emph{Statistics and Computing}, pages 1--15, 2016.
\newblock ISSN 1573-1375.
\newblock \doi{10.1007/s11222-016-9636-3}.

\bibitem[Meinshausen and B{\"u}hlmann(2010)]{meinshausen2010stability}
N.~Meinshausen and P.~B{\"u}hlmann.
\newblock Stability selection.
\newblock \emph{Journal of the Royal Statistical Society: Series B (Statistical
  Methodology)}, 72\penalty0 (4):\penalty0 417--473, 2010.

\bibitem[Osborne et~al.(2000)Osborne, Presnell, and Turlach]{osborne2000new}
M.~R. Osborne, B.~Presnell, and B.~A. Turlach.
\newblock A new approach to variable selection in least squares problems.
\newblock \emph{IMA journal of numerical analysis}, 20\penalty0 (3):\penalty0
  389--403, 2000.

\bibitem[{R Core Team}(2015)]{rbase}
{R Core Team}.
\newblock \emph{R: A Language and Environment for Statistical Computing}.
\newblock R Foundation for Statistical Computing, Vienna, Austria, 2015.
\newblock URL \url{http://www.R-project.org/}.

\bibitem[Sanderson(2010)]{armadillo}
C.~Sanderson.
\newblock Armadillo: An open source {C}++ linear algebra library for fast
  prototyping and computationally intensive experiments.
\newblock Technical report, NICTA, 2010.

\bibitem[Tibshirani and Taylor(2011)]{tibshirani2011}
R.~J. Tibshirani and J.~Taylor.
\newblock The solution path of the generalized lasso.
\newblock \emph{Ann. Statist.}, 39\penalty0 (3):\penalty0 1335--1371, 2011.

\bibitem[Vert and Bleakley(2010)]{vert2010fast}
J.-P. Vert and K.~Bleakley.
\newblock Fast detection of multiple change-points shared by many signals using
  group lars.
\newblock In \emph{Advances in Neural Information Processing Systems}, pages
  2343--2351, 2010.

\end{thebibliography}

\end{document}